\documentclass[vecphys]{svmult}

\usepackage{epsfig}
\usepackage{psfig}
\usepackage{makeidx}         
\usepackage{graphicx}        
\usepackage{multicol}        
\usepackage[bottom]{footmisc}

\makeindex             

\begin{document}

\title*{Observations of the High Redshift Universe}
\author{Richard S Ellis\inst{1}}
\institute{Astronomy Department, California Institute of Technology
\texttt{rse@astro.caltech.edu}}
%
%
\maketitle

\centerline{\bf ABSTRACT}

In this series of lectures, aimed for non-specialists, I review the considerable 
progress that has been made in the past decade in understanding how galaxies 
form and evolve. Complementing the presentations of my theoretical colleagues, 
I focus primarily on the impressive achievements of observational
astronomers. A credible framework, the $\Lambda$CDM model, now exists 
for interpreting these observations: this is a universe with dominant dark energy
whose structure grows slowly from the gravitational clumping of dark matter 
halos in which baryonic gas cools and forms stars. The standard
model fares well in matching the detailed properties of local galaxies,
and is addressing the growing body of detailed multi-wavelength 
data at high redshift. Both the star formation history and the assembly
of stellar mass can now be empirically traced from redshifts $z\simeq$6 to the
present day, but how the various distant populations relate to one
another and precisely how stellar assembly is regulated by feedback 
and environmental processes remains unclear.  In the latter part of my 
lectures, I discuss how these studies are being extended to locate
and characterize the earliest sources beyond $z\simeq$6. Did
early star-forming galaxies contribute significantly to the reionization
process and over what period did this occur? Neither theory
nor observations are well-developed in this frontier topic but
the first results are exciting and provide important guidance on how we 
might use more powerful future facilities to fill in the details.

\newpage

\section{Role of Observations in Cosmology \& Galaxy Formation}
\label{sec:1}

\subsection{The Observational Renaissance}

These are exciting times in the field of cosmology and galaxy formation! To justify this
claim it is useful to review the dramatic progress made in the subject over the past 
$\simeq$25 years. I remember vividly the first distant galaxy conference I
attended: the IAU Symposium 92 {\it Objects of High Redshift}, held in Los Angeles 
in 1979. Although the motivation was strong and many observers were
pushing their 4 meter telescopes to new limits, most imaging detectors were still
photographic plates with efficiencies of a few percent and there was no
significant population of sources beyond a redshift of $z$=0.5, other than
some radio galaxies to $z\simeq$1 and more distant quasars. 

In fact, the present landscape in the subject would have been barely 
recognizable even in 1990. In the cosmological arena, convincing angular 
fluctuations had not yet been detected in the cosmic microwave background nor 
was there any consensus on the total energy density $\Omega_{TOT}$. Although 
the role of dark matter in galaxy formation was fairly well appreciated, neither its 
amount nor its power spectrum were 
particularly well-constrained. The presence of dark energy had not been 
uncovered and controversy still reigned over one of the most basic parameters 
of the Universe: the current expansion rate as measured by Hubble's constant.
In galaxy formation, although evolution was frequently claimed in the counts and 
colors of galaxies, the physical interpretation was confused. In particular, there was 
little synergy between observations of faint galaxies and models of structure formation. 

In the present series of lectures, aimed for non-specialists, I hope to show that 
we stand at a truly remarkable time in the history of our subject, largely (but 
clearly not exclusively) by virtue of a growth in observational capabilities. By the
standards of all but the most accurate laboratory physicist, we have `precise' 
measures of the form and energy content of our Universe and a detailed physical understanding of how structures grow and evolve.  We have successfully charted 
and studied the distribution and properties of hundreds of thousands of nearby 
galaxies in controlled surveys and probed their luminous precursors out to 
redshift $z\simeq$6 - corresponding to a period only 1 Gyr after the Big Bang. 
Most importantly, a standard model has emerged which, through detailed numerical simulations, is capable of detailed predictions and interpretation of observables. 
Many puzzles remain, as we will see, but the progress is truly impressive.

This gives us confidence to begin addressing the final frontier in galaxy evolution: 
the earliest stellar systems and their influence on the intergalactic medium.
When did the first substantial stellar systems begin to shine? Were they responsible
for reionizing hydrogen in intergalactic space and what physical processes
occurring during these early times influenced the subsequent evolution of
normal galaxies?

Let's begin by considering a crude measure of our recent progress. Figure 1 shows 
the rapid pace of discovery in terms of the {\em relative fraction} of 
the refereed astronomical literature in two North American journals pertaining to 
studies of galaxy evolution and cosmology. These are cast alongside some 
milestones in the history of 
optical facilities and the provision of widely-used datasets. The figure raises the
interesting question of whether more publications in a given field means most of 
the key questions are being answered. Certainly, we can conclude that more 
researchers are being drawn to work in the area. But some might argue that 
new students should move into other, less well-developed, fields. Indeed, the 
progress in cosmology, in particular, is so rapid that some have raised the 
specter that the subject may soon reaching some form of natural conclusion 
(c.f. Horgan 1998). 

\begin{figure}
\centering
\includegraphics[height=9cm]{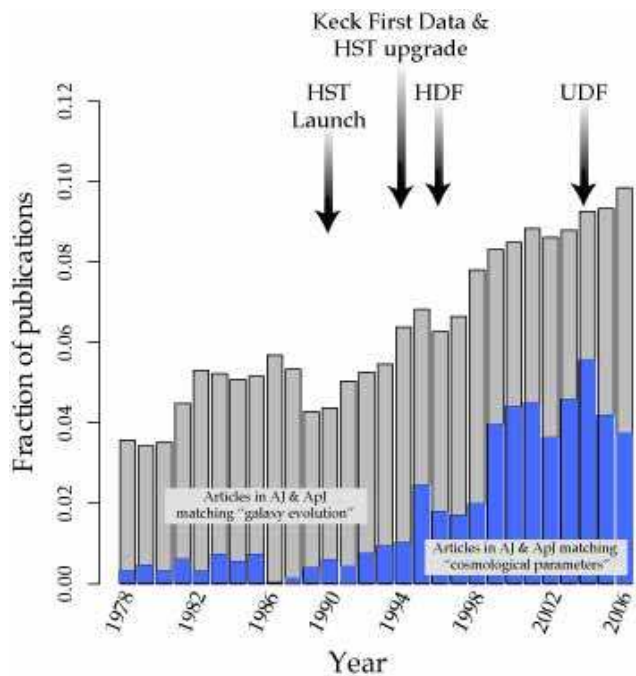}
\caption{Fraction of the refereed astronomical literature in two North American
journals related to galaxy evolution and the cosmological parameters. The 
survey implies more than a doubling in fractional share over the past 15 years. 
Some possibly-associated milestones in the provision of unique facilities and 
datasets are marked (courtesy: J. Brinchmann).}
\label{fig:1}      
\end{figure}

I believe, however, that the rapid growth in the share of publications is largely a 
reflection of new-found observational capabilities. We are witnessing an expansion
of {\em exploration} which will most likely be followed with a more detailed 
{\em physical phase} where we will be concerned with {\it understanding} how 
galaxies form and evolve.  


\subsection{Observations Lead to Surprises}

It's worth emphasizing that many of the key features which define our current 
view of the Universe were either not anticipated by theory or initially rejected
as unreasonable. Here is my personal short list of surprising observations which 
have shaped our view of the cosmos:

\begin{enumerate}

\item{} The cosmic expansion discovered by Slipher and Hubble during the period
1917-1925 was not anticipated and took many years to be accepted. Despite the 
observational evidence and the prediction from General Relativity for evolution 
in world models with gravity, Einstein maintained his preference for a static 
Universe until the early 1930's.

\item{} The hot Big Bang picture received widespread support only in 1965
upon the discovery of the cosmic microwave background (Penzias \& Wilson 1965).
Although many supported the hypothesis of a primeval atom, Hoyle and others
considered an unchanging `Steady State' universe to be a more natural
solutuion.

\item{} Dark matter was inferred from the motions of galaxies in clusters over seventy
years ago (Zwicky 1933) but no satisfactory explanation of this puzzling problem
was ever presented. The ubiquity of dark matter on galactic scales was 
realized much later (Rubin et al 1976). The dominant role that dark matter 
plays in structure formation only followed the recent observational evidence 
(Blumenthal et al 1984).\footnote{For an amusing musical history of the role of 
dark matter in cosmology suitable for students of any age check out 
{\tt http://www-astronomy.mps.ohio-state.edu/~dhw/Silliness/silliness.html}}

\item{} The cosmic acceleration discovered independently by two distant 
supernovae teams (Riess et al 1998, Perlmutter et al 1999) was a complete
surprise (including to the observers, who set out to measure the deceleration).
Although the cosmological constant, $\Lambda$, had been invoked many times 
in the past, the presence of dark energy was completely unforeseen.

\end{enumerate}

Given the observational opportunities continue to advance. it seems 
reasonable to suppose further surprises may follow!

\subsection{Recent Observational Milestones}

Next, it's helpful to examine a few of the most significant observational 
achievements in cosmology and structure formation over the past $\simeq$15 
years. Each provides the basis of knowledge from which we can move
forward, eliminating a range of uncertainty across a wide field of research.

\subsubsection{The Rate of Local Expansion: Hubble's Constant} 

The Hubble Space Telescope (HST) was partly launched to resolve the
puzzling dispute between various observers as regards to the 
value of Hubble's constant $H_0$, normally quoted in kms sec$^{-1}$ Mpc$^{-1}$,
or as $h$, the value in units of 100 kms sec$^{-1}$ Mpc$^{-1}$. During
the planning phases, a number of scientific {\em key projects} were defined
and proposals invited for their execution.

A very thorough account of the impasse reached by earlier ground-based
observers in the 1970's and early 1980's can be found in 
Rowan-Robinson (1985) who reviewed the field and concluded 
a compromise of 67 $\pm$ 15 kms sec$^{-1}$ Mpc$^{-1}$, surprisingly
close to the presently-accepted value. Figure 2 nicely illustrates
the confused situation.

\begin{figure}
\centering
\includegraphics[height=12cm,angle=270]{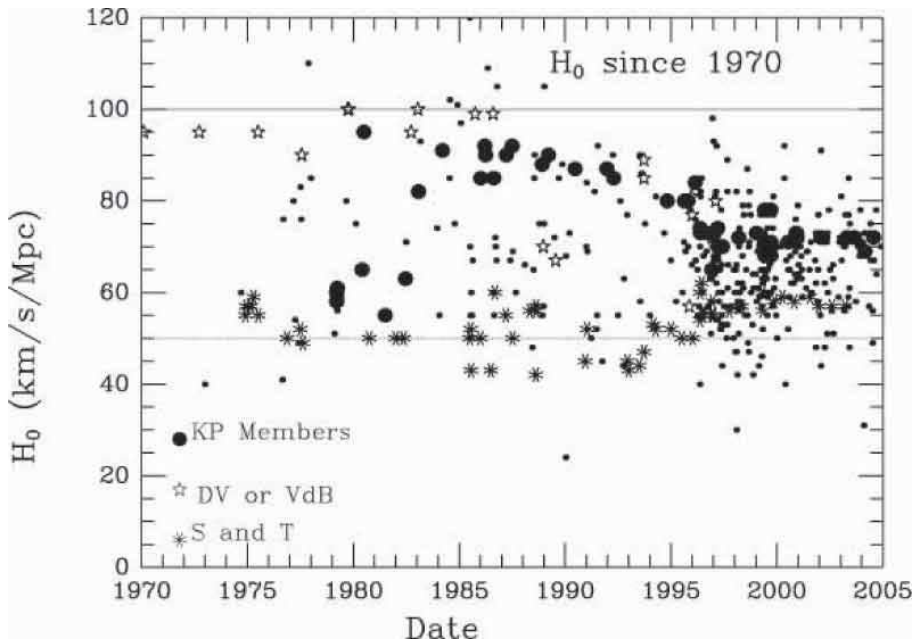}
\caption{Various values of Hubble's constant in units of kms sec$^{-1}$ Mpc$^{-1}$
plotted as a function of the date of publication. Labels refer to estimates by Sandage
\& Tammann, de Vaucouleurs, van den Bergh and their respective collaborators.
Estimates from the HST Key Project group (Freedman et al 2001) are labeled KP.
From an initial range spanning 50$<H_0<$100, a gradual convergence to the
presently-accepted value is apparent. (Plot compiled and kindly made available 
by J. Huchra) }
\label{fig:2}      
\end{figure}

Figure 3 shows the two stage `step-ladder' technique used by Freedman et al 
(2001) who claim a final value of 67 $\pm$ 15 kms sec$^{-1}$ Mpc$^{-1}$.
`Primary' distances were estimated to a set of nearby galaxies via the 
measured brightness and periods of luminous Cepheid variable stars
located using HST's WFPC-2 imager. Over the distance range
across which such individual stars can be seen ($<$25 Mpc), the
leverage on $H_0$ is limited and seriously affected by the peculiar
motions of the individual galaxies. At $\simeq$20 Mpc, the smooth
cosmic expansion would give $V_{exp}\simeq$1400 kms sec$^{-1}$ and
a 10\% error in $H_0$ would provide a comparable contribution, at this 
distance, to the typical peculiar motions of galaxies of $V_{pec}\simeq$50-100 
kms sec$^{-1}$. Accordingly, a secondary distance scale was
established for spirals to 400 Mpc distance using the empirical relationship 
first demonstrated by Tully \& Fisher (1977) between the $I$-band 
luminosity and rotational velocity. At 400 Mpc, the effect of $V_{pec}$
is negligible and the leverage on $H_0$ is excellent. Independent
distance estimators utilizing supernovae and elliptical galaxies were
used to verify possible systematic errors.

\begin{figure}
\centerline{\hbox{
\psfig{file=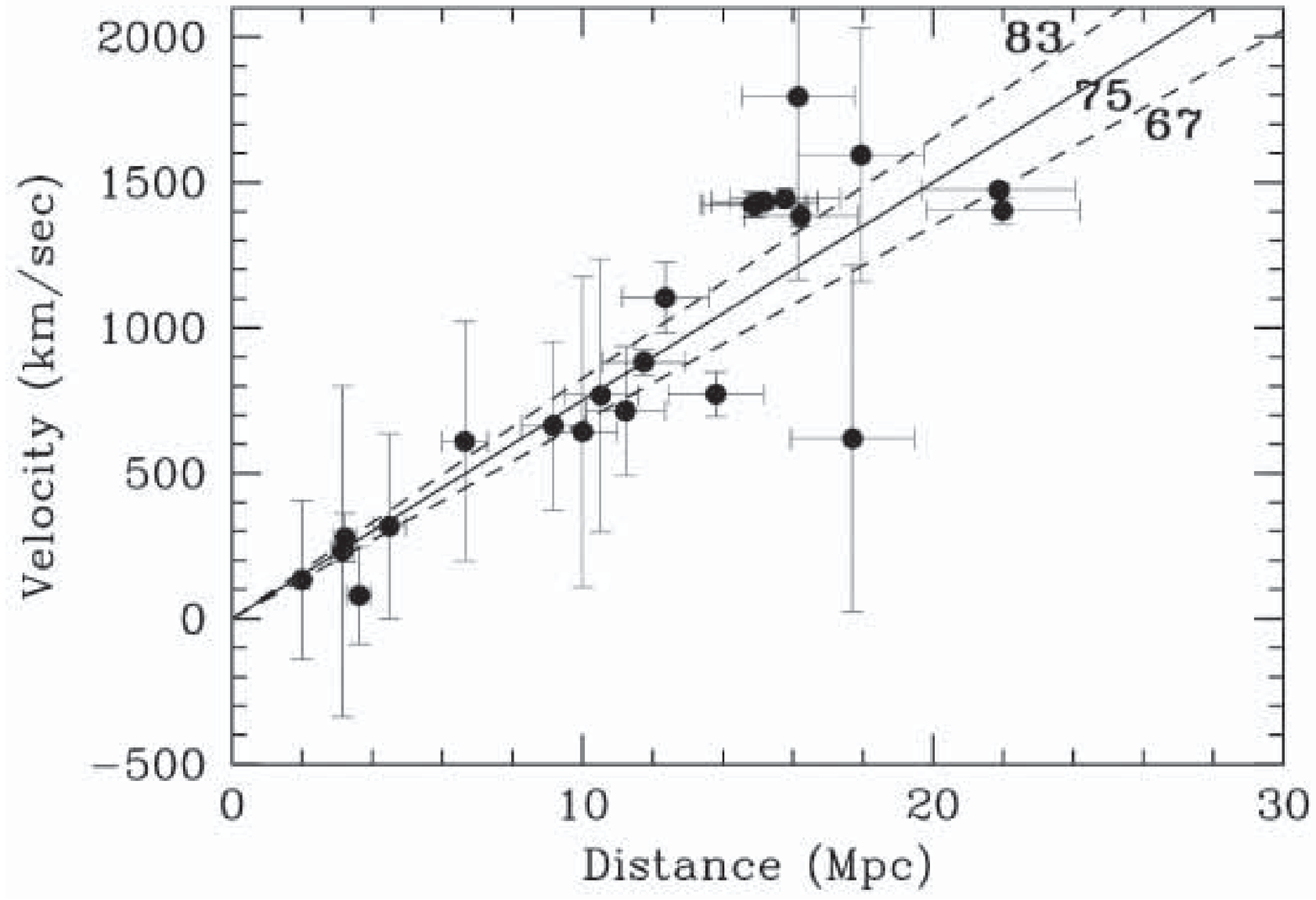,height=1.8in,angle=0}
\psfig{file=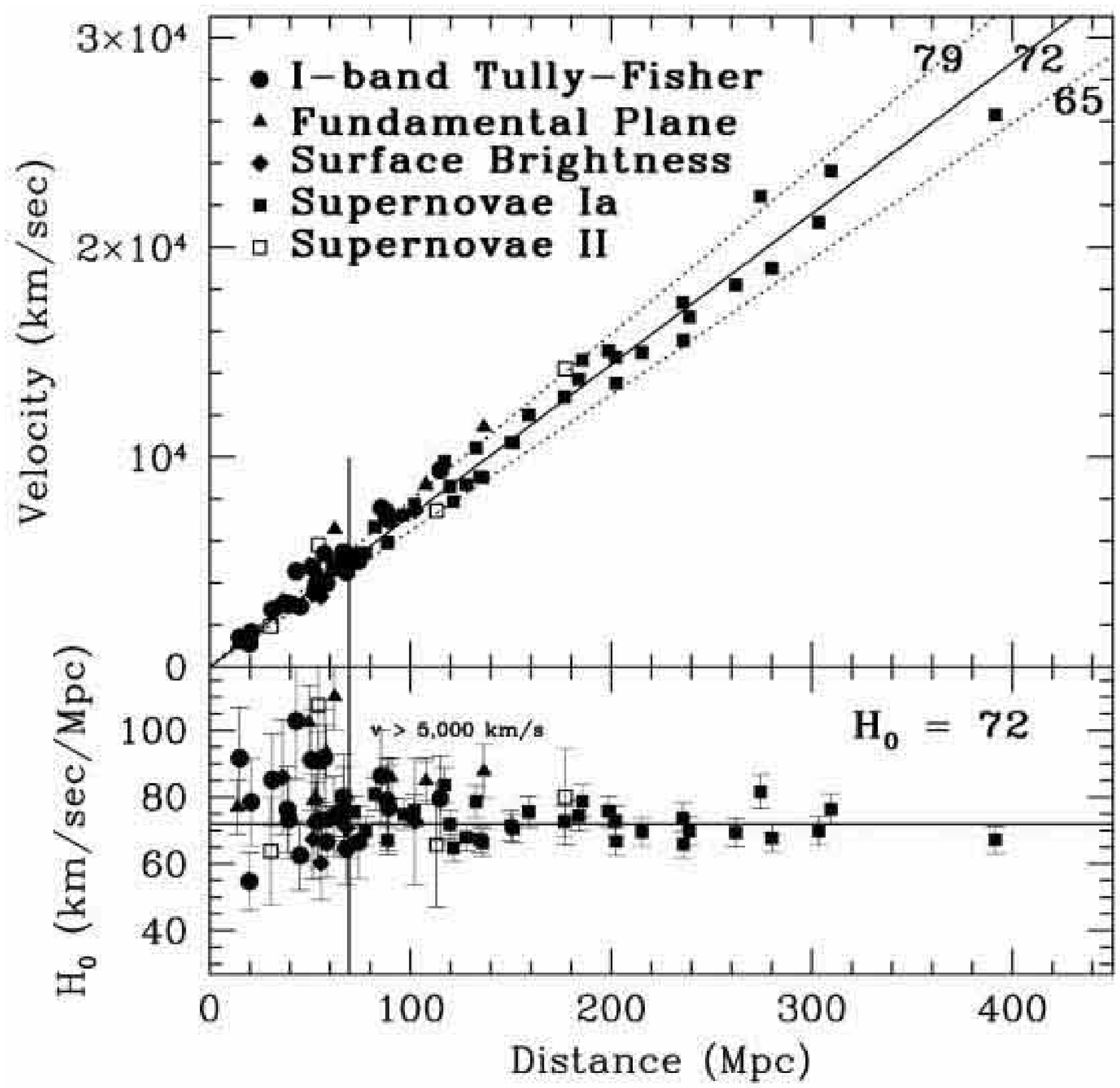,height=1.9in,angle=0}}}
\caption{Two step approach to measuring Hubble's constant $H_0$ - the local
expansion rate (Freedman et al 2001). (Left) Distances to nearby galaxies 
within 25 Mpc were 
obtained by locating and monitoring Cepheid variables using HST's WFPC-2
camera; the leverage on $H_0$ is modest over such small distances and
affected seriously by peculiar motions. (Right) Extension of the distance-velocity
relation to 400 Mpc using the I-band Tully-Fisher relation and other techniques. The
absolute scale has been calibrated using the local Cepheid scale.}
\label{fig:3}      
\end{figure}

\subsubsection{Cosmic Microwave Background: Thermal Origin and Spatial Flatness}

The second significant milestone of the last 15 years is the improved understanding
of the cosmic microwave background (CMB) radiation, commencing with the 
precise black body nature of its spectrum (Mather et al 1990) indicative of
its thermal origin as a remnant of the cosmic fireball, and the subsequent detection 
of fluctuations (Smoot et al 1992), both realized with the COBE satellite data. The improved 
angular resolution of later ground-based and balloon-borne experiments led
to the isolation of the acoustic horizon scale at the epoch of recombination (de Bernadis
et al 2000, Hanany et al 2000).  Subsequent improved measures of the
angular power spectrum by the Wilkinson Microwave Anisotropy Probe (WMAP,
Spergel et al 2003, 2006) have refined these early observations. The location
of the primary peak in the angular power spectrum at a multiple moment
$l\simeq$200 (corresponding to a physical angular scale of $\simeq$1 degree)
provides an important constraint on the total energy density $\Omega_{TOT}$
and hence spatial curvature.

The derivation of spatial curvature from the angular location of the first acoustic (or `Doppler') peak, $\theta_H$, is not completely independent of other cosmological parameters.
There are dependences on the scale factor via $H_0$ and the contribution
of gravitating matter $\Omega_M$, viz:

$$\theta_H \propto (\Omega_M\, h^{3.4})^{0.14}\, \Omega_{TOT}^{1.4}$$

where $h$ is $H_0$ in units of 100 kms sec$^{-1}$ Mpc$^{-1}$.

However, in the latest WMAP analysis, combining with distant supernovae
data, space is flat to within 1\%.

\subsubsection{Clustering of Galaxies: Gravitational Instability}

Galaxies represent the most direct tracer of the rich tapestry of structure
in the local Universe. The 1970's  saw a concerted effort to introduce a formalism 
for describing and interpreting their statistical distribution through angular and
spatial two point correlation functions (Peebles 1980). This, in turn, led to an
observational revolution in cataloging their distribution, first in 2-D from 
panoramic photographic surveys aided by precise measuring machines, 
and later in 3-D from multi--object spectroscopic redshift surveys.

The angular correlation function $w(\theta)$ represents the excess
probability $\delta\,P$ of finding a pair of galaxies separated by an angular separation
$\theta$ (degrees).

In a catalog averaging $N$ galaxies per square degree, the probability
of finding a pair separated by $\theta$ can be written:

$$\delta\,P = N [1 + w(\theta)]\delta\,\Omega$$

where $\delta\,\Omega$ is the solid angle of the counting bin, (i.e.
$\theta$ to $\theta + \delta\,\theta$).

The corresponding spatial equivalent, $\xi(r)$ in a catalog of mean
density $\rho$ per Mpc$^3$ is thus:

$$\delta\,P = \rho [ 1 + \xi(r) ] \delta\,V$$

One can be statistically linked to the other if the overall redshift distribution
of the sources is available. 

Figure 4 shows a pioneering detection of the angular correlation function $w(\theta)$
for the Cambridge APM Galaxy Catalog (Maddox et al 1990). This was one of
the first well-constructed panoramic 2-D catalogs from which the large scale 
nature of the galaxy distribution could be discerned. A power law form is evident:

$$w(\theta) =  A \theta^{-0.8}$$ 

where, for example, $\theta$ is measured in degrees. The amplitude $A$ 
decreases with increasing depth due to both increased projection from
physically-uncorrelated pairs and the smaller projected physical scale
for a given angle.

\begin{figure}
\centering
\includegraphics[height=7.1cm,angle=0]{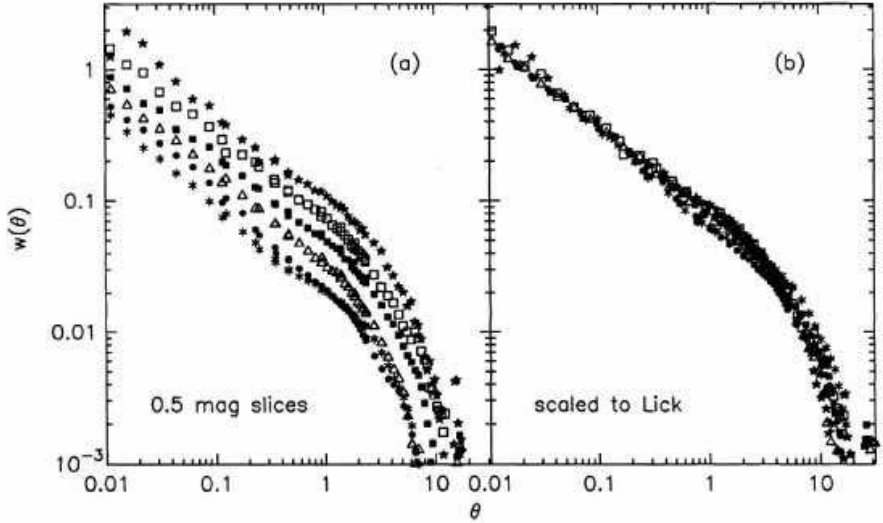}
\caption{Angular correlation function for the APM galaxy catalog - a photographic
survey of the southern sky (Maddox et al 1990) - partitioned according to 
limiting magnitude (left). The amplitude of the clustering decreases with increasing
depth due to an increase in the number of uncorrelated pairs and a smaller
projected physical scale for a given angle. These effects can be corrected in order
to produce a high signal/noise function scaled to a fixed depth clearly illustrating
a universal power law form over nearly 3 dex (right). }
\label{fig:4}      
\end{figure}

Highly-multiplexed spectrographs such as the 2 degree field instrument
on the Anglo-Australian Telescope (Colless et al 2001) and the Sloan
Digital Sky Survey (York et al 2001) have led to the equivalent progress
in 3-D surveys (Figure 5). In the early precursors to these grand surveys,
the 3-D equivalent of the angular correlation function, was also found
to be a power law:

$$\xi(r) = (\frac{r}{r_0})^{-1.8}$$

where $r_o$ (Mpc) is a valuable {\it clustering scale length} for the
population. 

As the surveys became more substantial, the power spectrum $P(k)$ has 
become the preferred analysis tool because its form can be readily predicted for
various dark matter models. For a given density field $\rho({\bf x})$,
the fluctuation over the mean is $\delta = \rho\,/\,\overline{\rho}$ and
for a given wavenumber $k$, the power spectrum becomes:

$$P(k) = <|\delta_k^2|> = \int \xi(r) exp(i{\bf k.r})d^3r$$

The final power spectrum for the completed 2dF survey is
shown in Figure 6 (Cole et al 2005) and is in remarkably good
agreement with that predicted for a cold dark matter spectrum
consistent with that which reproduces the CMB angular fluctuations.

\begin{figure}
\centering
\includegraphics[height=6.5cm,angle=0]{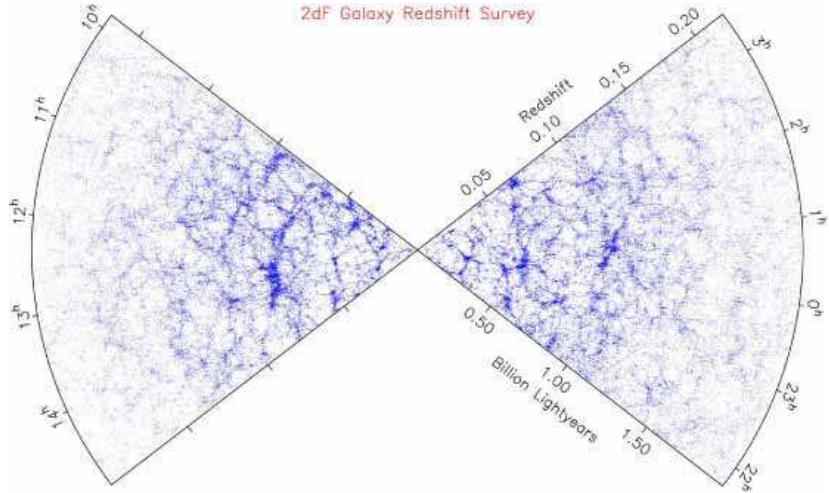}
\caption{Galaxy distribution from the completed 2dF redshift survey
(Colless et al 2001). }
\label{fig:5}      
\end{figure}

\begin{figure}
\centering
\includegraphics[height=7cm,angle=0]{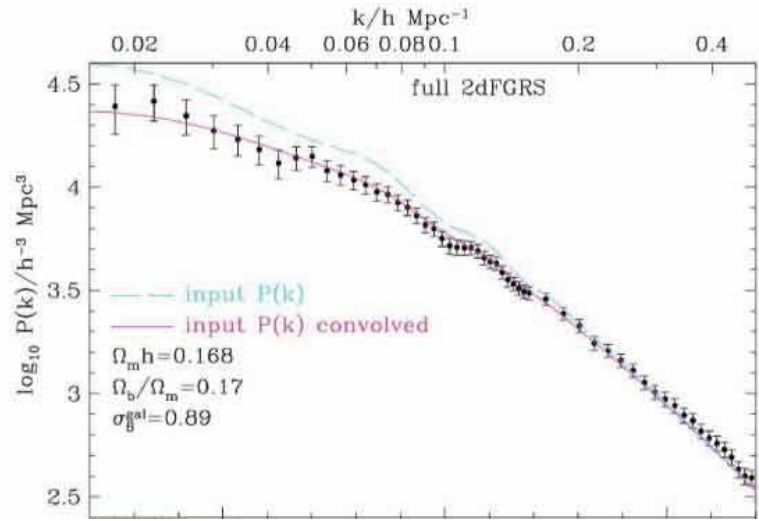}
\caption{Power spectrum from the completed 2dF redshift survey
(Cole et al 2005). Solid lines refer to the input power spectrum for
a  dark matter model with the tabulated parameters and that
convolved with the geometric `window function'  which affects 
the observed shape on large scales.}
\label{fig:6}      
\end{figure}

\subsubsection{Dark Matter and Gravitational Instability}

We have already mentioned the ubiquity of dark matter
on both cluster and galactic scales. The former was recognized
as early as the 1930's from the high line of sight velocity
dispersion $\sigma_{los}$ of galaxies in the Coma cluster (Zwicky
1933). Assuming simple virial equilibrium and isotropically-arranged
galaxy orbits, the cluster mass contained with some physical scale $R_{cl}$ 
is:

$$M = 3 <\sigma_{los}^2> R_{cl}/G$$

which far exceeds that estimated from the stellar populations in
the cluster galaxies. High cluster masses can also be confirmed
completely independently from {\em gravitational lensing} where
a background source is distorted to produce a `giant arc' - in effect
a partial or incomplete `Einstein ring' whose diameter $\theta_E$
for a concentrated mass $M$ approximates:

$$\theta_E = \frac{4GM^\frac{1}{2}}{c^2}\,D^\frac{1}{2}$$

and $D=D_s\,D_l\/,/\,D_{ds}$ where the subscripts $s$ and $l$
refer to angular diameters distances of the background source
and lens respectively.

On galactic scales, extended rotation curves of gaseous emission lines
in spirals (see review by Rubin 2000) can trace the mass distribution
on the assumption of circular orbits, viz:

$$\frac{GM(<R)}{R^2} = \frac{V^2}{R}$$

Flat rotation curves ($V\sim$constant) thus imply $M(<R)\propto\,R$.
Together with arguments based on the question on the stability of
flattened disks (Ostriker \& Peebles 1973), such observations were
critical to the notion that all spiral galaxies are embedded in dark extensive
`halos'. 

The evidence for halos around local elliptical galaxies is less convincing
largely because there are no suitable tracers of the gravitational potential
on the necessary scales (see Gerhard et al 2001). However, by combining 
gravitational lensing with stellar dynamics for intermediate redshift
ellipticals, Koopmans \& Treu (2003) and Treu et al (2006) have mapped 
the projected dark matter distribution and show it to be closely fit by an 
isothermal profile $\rho(r) \propto r^{-2}$.

The presence of dark matter can also be deduced statistically from
the distortion of the galaxy distribution viewed in {\em redshift space},
for example in the 2dF survey (Peacock et al 2001). The original idea was
discussed by Kaiser et al (1987). The spatial correlation
function $\xi(r)$ is split into its two orthogonal components,
$\xi(\sigma, \pi)$ where $\sigma$ represents the projected separation
perpendicular to the line of sight (unaffected by peculiar
motions) and $\pi$ is the separation along the line of sight (inferred
from the velocities and hence used to measure the effect). The
distortion of $\xi(\sigma,\pi)$ in the $\pi$ direction can be measured
on various scales and used to estimate the line of sight velocity
dispersion of pairs of galaxies and hence their mutual gravitational
field. Depending on the extent to which galaxies are biased tracers of the
density field, such tests indicate $\Omega_M$=0.25.

On the largest scales, {\em weak gravitational lensing} can trace
the overall distribution and dark matter content of the Universe
(Blandford \& Narayan 1992, Refregier 2003). Recent surveys
are consistent with these estimates (Hoekstra et al 2005).

\subsubsection{Dark Energy and Cosmic Acceleration}

Prior to the 1980's observational cosmologists were obsessed
with two empirical quantities though to govern the cosmic
expansion history -- $R(t)$: Hubble's constant $H_0 =  dR/dt $
and a second derivative, the deceleration parameter $q_0$,
which would indicate the fate of the expansion:

$$q_0 = - \frac{d^2R/dt^2}{(dR/dt)^2}$$

In the presence only of gravitating matter, Friedmann cosmologies
indicate $\Omega_M = 2\,q_0$. The distant supernovae searches
were begun in the expectation of measuring $q_0$ independently
of $\Omega_M$ and verifying a low density Universe.

As we have discussed, Type Ia supernovae (SNe) were found to be
{\em fainter} at a given recessional velocity than expected
in a Universe with a low mass density; Figure 7 illustrates
the effect for the latest results from the Canada-France SN
Legacy Survey (Astier et al 2006). In fact the results cannot
be explained even in a Universe with {\em no} gravitating
matter! A formal fit for $q_0$ indicates a negative value
corresponding to a cosmic acceleration.

Acceleration is permitted in Friedmann models with a non-zero
cosmological constant $\Lambda$. In general (Caroll et al
1992):

$$q_0 =\frac{\Omega_M}{2} - 3\,\frac{\Omega_\Lambda}{2}$$

where $\Omega_\Lambda=\Lambda/8\pi\,G$ is the energy density
associated with the cosmological constant.

\begin{figure}
\centering
\includegraphics[height=15cm,angle=0]{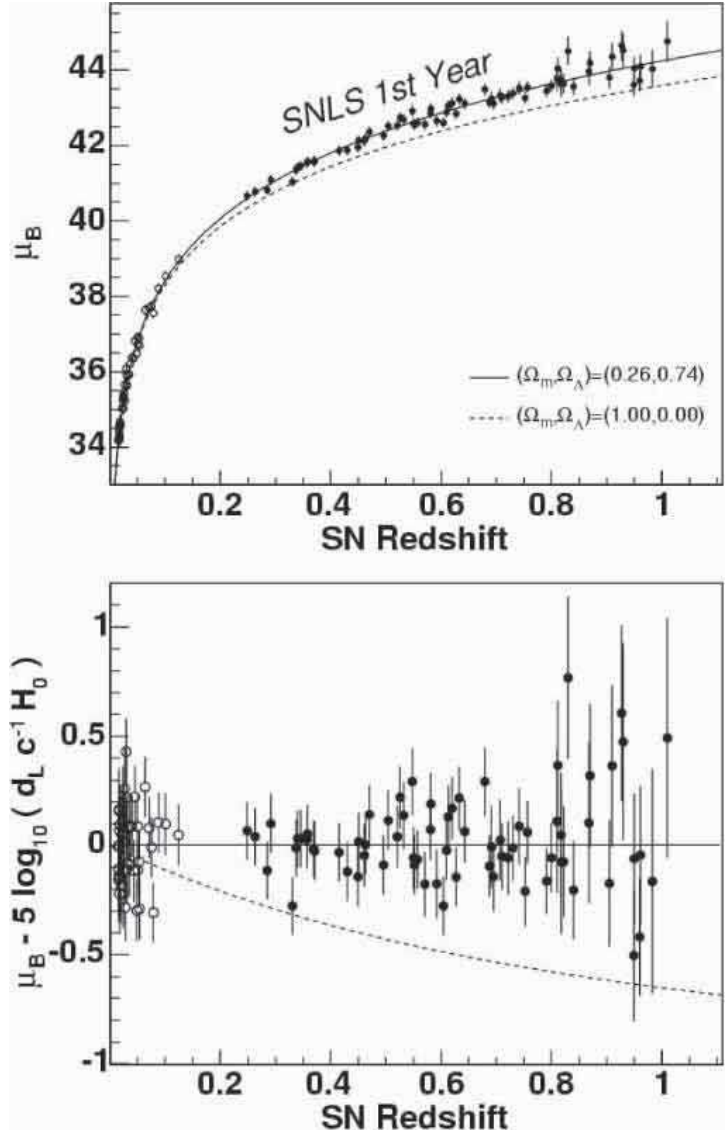}
\caption{Hubble diagram (distance-redshift relation) for calibrated
Type Ia supernovae from the first year data taken by the Canada
France Supernova Legacy Survey (Astier et al 2006). Curves
indicate the relation expected for a high density Universe without
a cosmological constant and that for the concordance cosmology (see
text)}
\label{fig:7}      
\end{figure}

The appeal of resurrecting the cosmological constant is not only its ability
to explain the supernova data but also the spatial flatness in the
acoustic peak in the CMB through the combined energy densities
$\Omega_M + \Omega_\Lambda$ - the so-called {\em Concordance
Model} (Ostriker \& Steinhardt 1995, Bahcall et al 1999).

However, the observed acceleration raises many puzzles. The
absolute value of the cosmological constant cannot be understood
in terms of physical descriptions of the vacuum energy density,
and the fact that $\Omega_M\simeq\Omega_\Lambda$ implies
the accelerating phase began relatively recently (at a redshift of 
z$\simeq$0.7). Alternative physical descriptions of the phenomenon
(termed `dark energy') are thus being sought which can be generalized 
by imagining the vacuum obeys an equation of state where the
negative pressure $p$ relates to the energy density $\rho$ via an
index $w$,

$$p=w\,\rho$$

in which case the dependence on the scale factor $R$ goes as

$$\rho \propto R^{-3(1+w)}$$

The case $w$=-1 would thus correspond to a constant term
equivalent to the cosmological constant, but in principle any
$w<$-1/3 would produce an acceleration and conceivably
$w$ is itself a function of time. The current SNLS data indicate
$w$=-1.023 $\pm$ 0.09 and combining with the WMAP data
does not significantly improve this constraint.

\subsection{Concordance Cosmology: Why is such a curious model acceptable?}

According to the latest WMAP results (Spergel et al 2006) and the
analysis which draws upon the progress reviewed above (the HST
Hubble constant Key Project, the large 2dF and SDSS redshift surveys, 
the CFHT supernova survey and the first weak gravitational lensing
constraints), we live in a universe with the constituents listed in Table 1.

\begin{table}[htdp]
\caption{Cosmic Constituents}
\begin{center}
\begin{tabular}{|l|c|c|} \hline
Total Matter & $\Omega_M$ & 0.24 $\pm$ 0.03 \\
Baryonic Matter & $\Omega_B$ & 0.042 $\pm$ 0.004 \\
Dark Energy & $\Omega_\Lambda$ & 0.73 $\pm$ 0.04 \\
\hline
\end{tabular}
\end{center}
\label{tab:1}
\end{table}

Given only one of the 3 ingredient is physically understood
it may be reasonably questioned why cosmologists are triumphant
about having reached the era of `precision cosmology'! Surely we
should not confuse measurement with understanding?

The underlying reasons are two-fold. Firstly, many independent
probes (redshift surveys, CMB fluctuations and lensing) indicate the low
matter density. Two independent probes not discussed (primordial
nucleosynthesis and CMB fluctuations) support the baryon
fraction. Finally, given spatial flatness, even if the supernovae
data were discarded, we would deduce the non-zero dark energy
from the above results alone. 

Secondly, the above parameters reconcile the growth of structure
from the CMB to the local redshift surveys in exquisite detail.
Numerical simulations based on 10$^{10}$ particles (e.g. Springel et al 2005) 
have reached the stage where they can predict the non-linear growth
of the dark matter distribution at various epochs over a dynamic range of 3-4 dex
in physical scales. Although some input physics is needed to predict
the local galaxy distribution, the agreement for the concordance model 
(often termed $\Lambda$CDM) is impressive. In short, a low
mass density and non-zero $\Lambda$ both seem necessary to
explain the present abundance and mass distribution of galaxies.
Any deviation would either lead to too much or too little structure. 

This does not mean that the scorecard for $\Lambda$CDM should
be considered perfect at this stage. As discussed, we have little idea 
what the dark matter or dark energy might be. Moreover,  there are 
numerous difficulties in reconciling the distribution of dark matter with
observations on galactic and cluster scales and frequent challenges
that the mass assembly history of galaxies is inconsistent with
the slow hierarchical growth expected in a $\Lambda$-dominated 
Universe. However, as we will see in later lectures, most of these 
problems relate to applications in environments where dark matter 
co-exists with baryons. Understanding how to incorporate baryons
into the very detailed simulations now possible is an active area
where interplay with observations is essential. It is helpful to view
this interplay as a partnership between theory and observation
rather than the oft-quoted `battle' whereby observers challenge
or call into question the basic principles.

\subsection{Lecture Summary}

I have spent my first lecture discussing largely cosmological
progress and the impressive role that observations have played
in delivering rapid progress.

All the useful cosmological functions - e.g. time, distance and
comoving volume versus redshift, are now known to high accuracy
which is tremendously beneficial for our task in understanding
the first galaxies and stars. I emphasize this because even a
decade ago, none of the physical constants were known well
enough for us to be sure, for example, the cosmic age
corresponding to a particular redshift.

I have justified $\Lambda\,CDM$ as an acceptable standard
model, despite the unknown nature of its two dominant constituents,
partly because there is a concordance in the parameters
when viewed from various observational probes, and partly
because of the impressive agreement with the distribution
of galaxies on various scales in the present Universe.

Connecting the dark matter distribution to the observed
properties of galaxies requires additional physics relating
to how baryons cool and form stars in dark matter halos.
Detailed observations are necessary to `tune' the models
so these additional components can be understood.

All of this will be crucial if we are correctly predict and
interpret signals from the first objects.

\newpage


\section{Galaxies \& the Hubble Sequence}
\label{sec:2}

\subsection{Introduction: Changing Paradigms of Galaxy Formation}

We now turn to the interesting history of how our
views of galaxy formation have changed over the past
20-30 years. It is convenient to break this into 3 eras

\begin{enumerate}

\item{} The classical era (pre-1985) as articulated for example
in the influential articles by Beatrice Tinsley and others. Galaxies 
were thought to evolve in isolation
with their present-day properties governed largely by one
function - the time-dependent star formation rate $\psi(t)$. Ellipticals suffered
a prompt conversion of gas into stars, whereas spirals
were permitted a more gradual consumption rate leading
to a near-constant star formation rate with time.

\item{} The dark matter-based era (1985-): in hierarchical
models of structure formation involving gravitational instability,
the ubiquity of dark matter halos means that merger
driven assembly is a key feature. If mergers redistribute 
angular momentum, galaxy morphologies are transformed.

\item{} Understanding feedback and the environment (1995-): In 
the most recent work, the evolution of the morphology-density
relation (Dressler et al 1997) and the dependence of
the assembly history on galactic mass (`downsizing', Cowie et al 1996)
have emphasized that star formation is regulated by
processes other than gas cooling and infall associated
with DM-driven mergers.

\end{enumerate}

\subsection{Galaxy Morphology - Valuable Tool or Not?}

In the early years, astronomers placed great stock on understanding 
the origin of the {\it morphological} distribution of galaxies, sometimes
referred to as the Hubble sequence (Hubble 1936). Despite this
simple categorization 70 years ago, the scheme is evidently still in
common use. In its support, Sandage (e.g. 2005) has commented on this 
classification scheme  as describing `a true order among the galaxies, 
not one imposed by the classifier'. However, many contemporary modelers 
and observers have paid scant attention to morphology and placed 
more emphasis on understanding stellar population differences. What 
value should we place on accurately measuring and reproducing the 
morphological distribution?

The utility of Hubble's scheme, at least for local galaxies, lies in its 
ability to distinguish dynamically distinct structures - spirals and
S0s are rotating stellar disks, whereas luminous spheroids are
pressure-supported ellipsoidal or triaxial systems with anisotropic 
velocity fields. This contains key information on the degree of
dissipation in their formation (Fall \& Efstathiou 1980).

There are also physical variables that seem to underpin the sequence,
including (i) gas content and color which relate to the ratio of the
current to past average star formation ratio $\psi(t_0)/\,\overline{\psi}$ 
(Figure 8) and (ii) inner structures including the bulge-to-disk ratio. Various
modelers (Baugh et al 1996) have argued that the bulge-to-disk
ratio is closely linked to the merger history and attempted to
reproduce the present distribution as a key test of hierarchical
assembly.

\begin{figure}
\centerline{\hbox{
\psfig{file=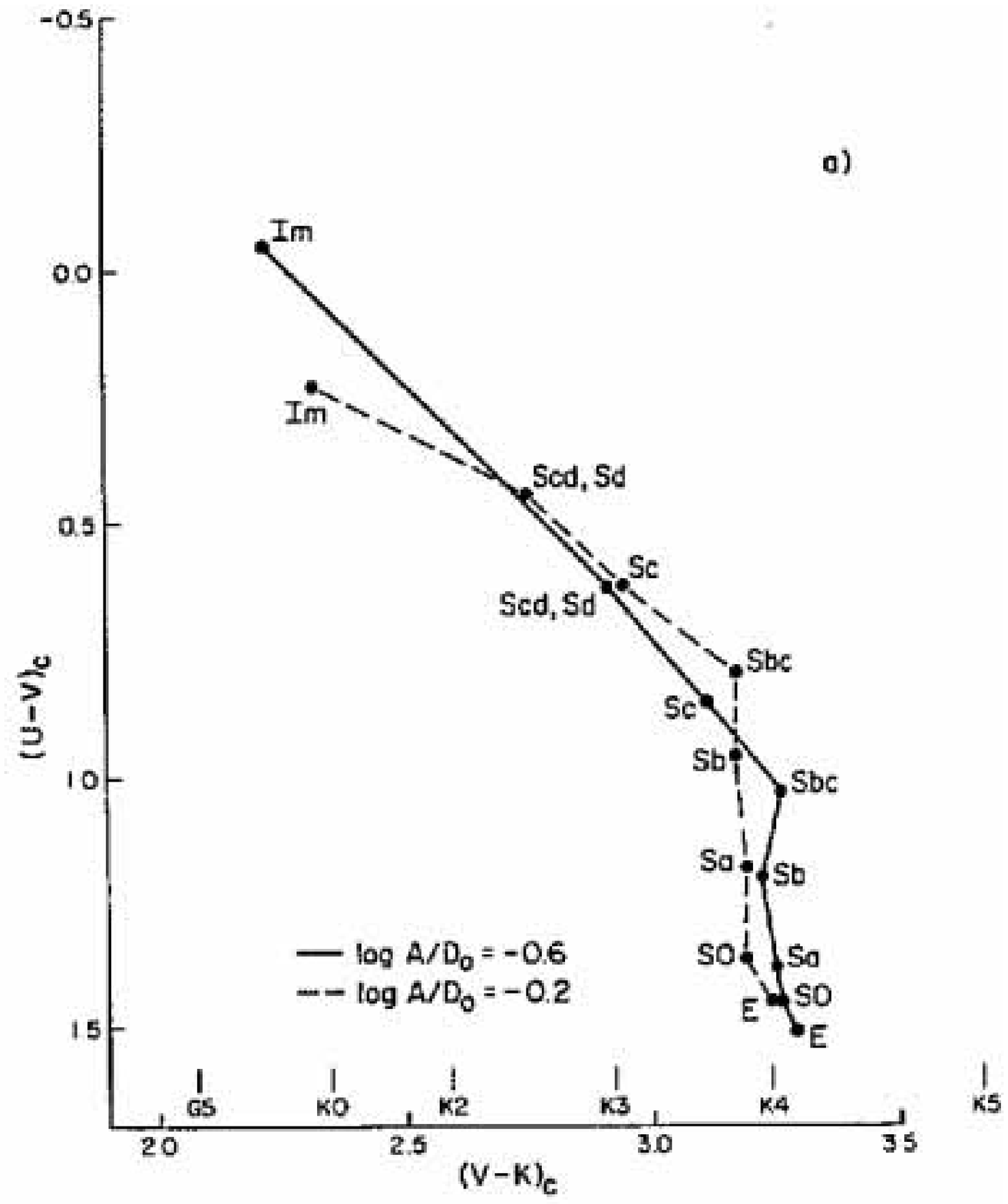,height=2.5in,angle=0}
\psfig{file=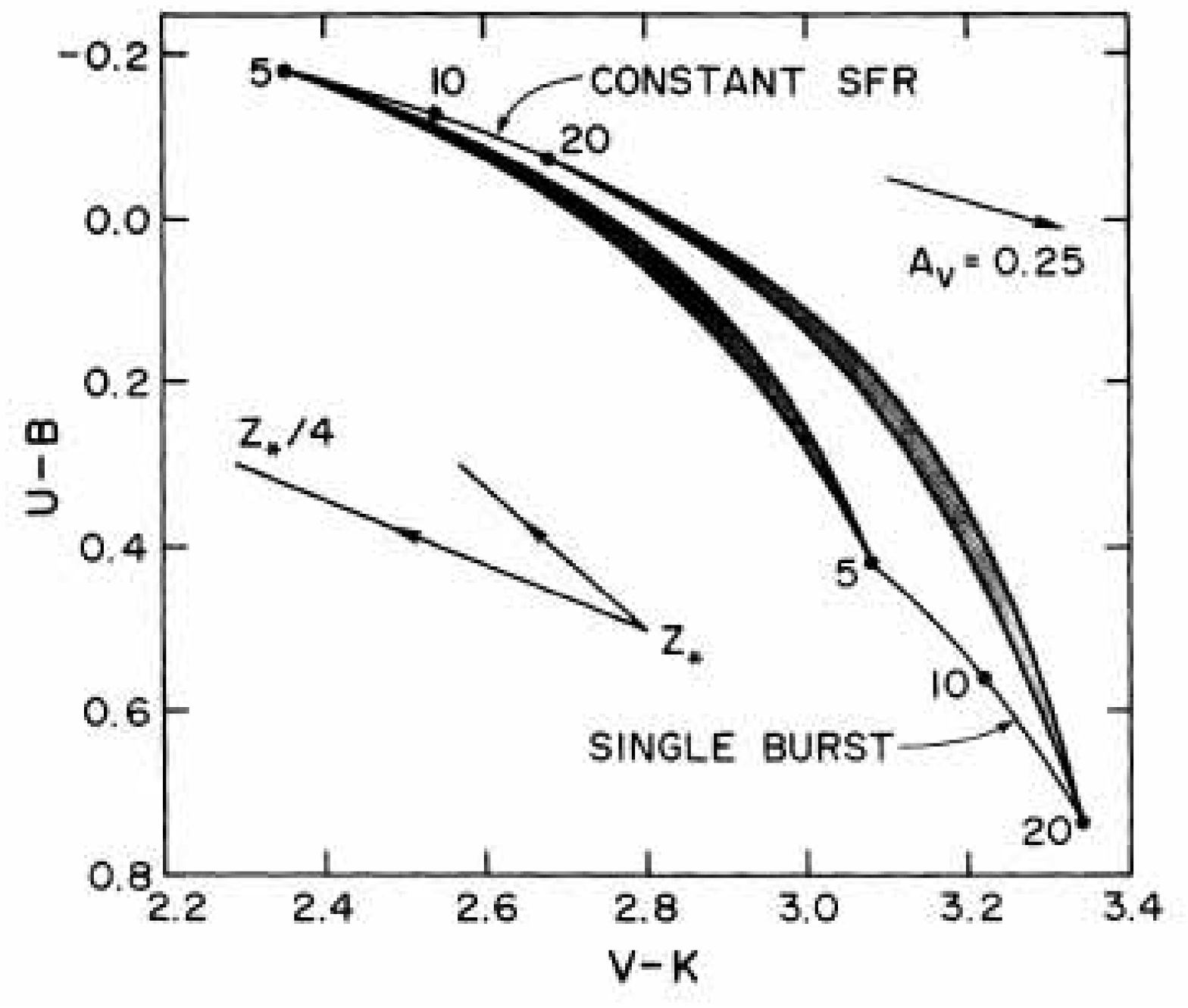,height=2.0in,angle=0}}}
\caption{A succinct summary of the classical view of galaxy formation
(pre-1985): (Left) The monotonic distribution of Hubble sequence 
galaxies in the $U-V$ vs $V-K$ color plane (Aaaronson 1978). (Right)
A simple model which reproduces this trend by changing only
the ratio of the current to past average star formation rate (Struck-Marcell
\& Tinsley 1978). Galaxies with constant star formation permanently
occupy the top left (blue) corner; galaxies with an initial burst rapidly evolve
to the bottom right (red) corner.}
\label{fig:8}      
\end{figure}

Much effort has been invested in attempting to classify galaxies
at high redshift, both visually and with automated algorithms. This 
is a challenging task because the precise
appearance of diagnostic features such as spiral arms and 
the bulge/disk ratio depends on the rest-wavelength of the
observations. An effect termed the `morphological k-correction'
can thus shift galaxies to apparently {\em later} types as the redshift
increases for observations conducted in a fixed band. A further
limitation, which works in the opposite sense, is surface brightness dimming, 
which proceeds as $\propto\,(1+z)^4$, rendering disks less prominent 
at high redshift and shifting some galaxies to apparent {\em earlier} types.

The most significant achievements from this effort has
been the realization that, despite the above quantitative uncertainties,
faint star-forming galaxies are generally more irregular in
their appearance than in local samples (Glazebrook et al 1995, 
Driver et al 1995). Moreover, HST images suggest on-going mergers
with an increasing frequency at high redshift (LeFevre et al
2000) although quantitative estimates of the merging fraction
as a function of redshift remain uncertain (see Bundy et al 2004).

The idea that morphology is driven by mergers took some time
for the observational community to accept. Numerical simulations
by Toomre \& Toomre (1972) provided the initial theoretical inspiration, 
but the observational evidence supporting the notion that spheroidal galaxies 
were simple collapsed systems containing old stars was strong
(Bower et al 1992). Tell-tale signs of mergers in local ellipticals
include the discovery of orbital shells (Malin \& Carter 1980) and
multiple cores revealed only with 2-D dynamical studies
(Davies et al  2001).

\subsection{Semi-Analytical Modeling}

As discussed by the other course lecturers and briefly in $\S$1, our
ability to follow the distribution of dark matter and its growth
in numerical simulations is well-advanced (e.g. Springel et al
2005). The same cannot be said of understanding how the
baryons destined, in part, to become stars are allocated to
each DM halo. This remains the key issue in interfacing
theory to observations.

Progress has occurred in two stages - according to the eras
discussed in \S2.1. {\em Semi analytic} codes were 
first developed in the 1990's to introduce baryons into 
DM n-body simulations using prescriptive methods for 
star formation, feedback and morphological assembly
(Figure 9). These codes were initially motivated to demonstrate
that the emerging DM paradigm was consistent with the
abundance of observational data (Kauffmann et al 1993, 
Somerville \& Primack 1999, Cole et al 2000). Prior to 
development of these codes, evolutionary 
predictions were based almost entirely on the `classical'
viewpoint with stellar population modeling based on variations
in the star formation history $\psi(t)$ for galaxies
evolving in isolation e.g. Bruzual (1980).

\begin{figure}
\centering
\includegraphics[height=6cm,angle=0]{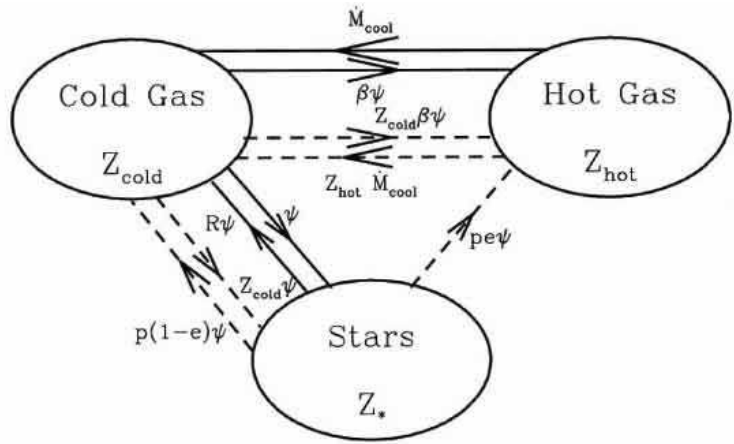}
\caption{Schematic of the ingredients inserted into
a semi-analytical model (from Cole et al 2000). Solid
lines refer to mass transfer, dashed lines to the
transfer of metals according to different compositions
$Z$. Gas cooling ($\dot{M}$) and star formation ($\psi$) 
is inhibited by the effect of supernovae ($\beta$). Stars 
return some fraction of their mass to the interstellar 
medium ($R$) and to the hot gas phase ($e$) according
to a metal yield $p$.}
\label{fig:9}      
\end{figure}

Initially these feedback prescriptions were adjusted
to match observables such as the luminosity function
(whose specific details we will address below), as well 
as specific attributes of
various surveys (counts, redshift distributions, colors 
and morphologies). In the recent versions, more
elaborate physically-based models for feedback processes
are being considered (e.g. Croton et al 2006)

The observational community was fairly skeptical
of the predictions from the first semi-analytical
models since it was argued that the parameter
space implied by Figure 9 enabled considerable freedom 
even for a fixed primordial fluctuation spectrum and
cosmological model. Moreover, where different
codes could be compared, considerably different 
predictions emerged (Benson et al 2002). Only
as the observational data has moved from colors and 
star formation rates to physical variables more closely 
related to galaxy assembly (such as stellar masses) have
the limitations of the early semi-analytical models 
been exposed.

\subsection{A Test Case: The Galaxy Luminosity Function}

One of the most straightforward and fundamental predictions 
a theory of galaxy formation can make is the present 
distribution of galaxy luminosities - the luminosity function (LF)
$\Phi(L)$ whose units are normally per comoving Mpc$^3$
\footnote{If Hubble's constant is not assumed, it is quoted
in units of $h^{-3}$\, Mpc$^{-3}$}.

As the contribution of a given luminosity bin $dL$ to the integrated
luminosity density per unit volume is $\propto\Phi(L)\,L\,dL$,
an elementary calculation shows that all luminosity functions
(be they for stars, galaxies or QSOs) must have a bend at some
characteristic luminosity, otherwise they would yield an
infinite total luminosity (see Felton 1977 for a cogent early discussion 
of the significance and intricacies of the LF). Recognizing this, 
Schechter (1976) proposed the product of a power law and 
an exponential as an appropriate analytic representation of the LF,
viz:

$$\Phi(L) \frac{dL}{L^{\ast}} = \Phi^{\ast} (\frac{L}{L^{\ast}})^{-\alpha}\,
exp(-\frac{L}{L^{\ast}})\,\frac{dL}{L^{\ast}}$$

where $\Phi^{\ast}$ is the overall normalization corresponding to the
volume density at the turn-over (or characteristic) luminosity $L^{\ast}$,
and $\alpha$ is the faint end slope which governs the relative abundance 
of faint and luminous galaxies.

The total abundance of galaxies per unit volume is then:

$$N_{TOT}= \int \Phi(L)\,dL = \Phi^{\ast}\,\Gamma(\alpha + 1)$$

and the total luminosity density is:

$$\rho_L =  \int \Phi(L)\,L\,dL = \Phi^{\ast}\,\Gamma(\alpha + 2)$$

where $\Gamma$ is the incomplete gamma function which can be
found tabulated in most books with integral tables (e.g. Gradshteyn
\& Ryzhik 2000). Note that $N_{TOT}$ diverges if $\alpha<$-1, 
whereas $\rho_L$ diverges only if $\alpha<$-2.

Recent comprehensive surveys by the 2dF team (Norberg et al 2002)
and by the Sloan Digital Sky Survey (Blanton et al 2001) have provided
definitive values for the LF in various bands. Encouragingly, when
allowance is made for the various photometric techniques, the two
surveys are in excellent agreement. Figure 10 shows the Schechter
function is a reasonably good (but not perfect) fit to the 2dF data
limited at apparent magnitude $b_J<$19.7. Moreover there is no 
significant difference between the LF derived independently for 
the two Galactic hemispheres. The slight excess of intrinsically
faint galaxies in the northern cap is attributable to a local
inhomogeneity in the nearby Virgo supercluster.

\begin{figure}
\centering
\includegraphics[height=12cm,angle=0]{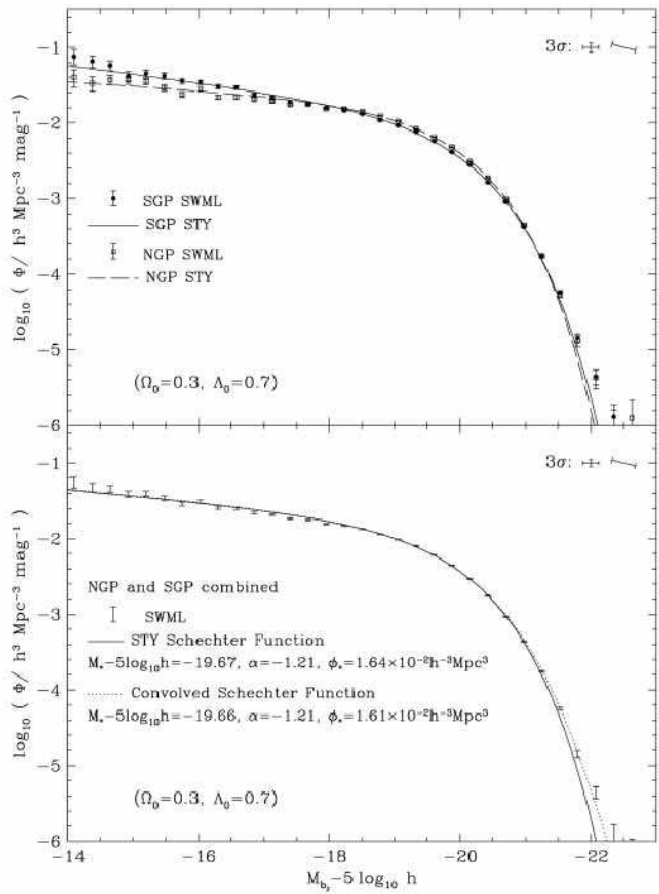}
\caption{Rest-frame $b_J$ luminosity function from the 2dF
galaxy redshift survey (Norberg et al 2002). (Top) a comparison 
of results across the northern and southern Galactic caps; there 
is only a marginal difference in the abundance of intrinsically faint 
galaxies. (Bottom) combined from both hemispheres indicating
the Schechter function is a remarkably good fit except at the
extreme ends of the luminosity distribution.}
\label{fig:10}      
\end{figure}

Fundamental though this function is, despite ten years of
semi-analytical modeling, reproducing its form has proved
a formidable challenge (as discussed by Benson et al 2003,
Croton et al 2006 and de Lucia et al 2006). Early predictions
also failed to reproduce the color distribution along the LF.
The halo mass distribution does not share the sharp bend
at $\L^{\ast}$ and too much star formation activity is retained 
in massive galaxies. These early predictions produced too many
luminous blue galaxies and too many faint red galaxies (Bower
et al 2006).

As a result, more specific feedback recipes have been created
to resolve this discrepancy. Several physical processes have
been invoked to regulate star formation as a function of mass,
viz:

\begin{itemize}

\item{} Reionization feedback: radiative heating from the first
stellar systems at high redshift which increases the Jeans mass,
inhibiting the early formation of low mass systems,

\item{} Supernova feedback: this was considered in the early
semi-analytical models but is now more precisely implemented so
as to re-heat the interstellar medium, heat the halo gas or even
eject the gas altogether from low mass systems,

\item{} Feedback from active galactic nuclei: the least well-understood
process with various modes postulated to transfer energy from
an active nucleus to the halo gas.

\end{itemize}

Benson et al (2003) and Croton et al (2006) illustrate
the effects of these more detailed prescriptions for these feedback
modes on the predicted LF and find that supernova and
reionization feedback largely reduce the excess of intrinsically
faint galaxies but, on grounds of energetics, only AGN can
inhibit star formation and continued growth in massive galaxies.
There remains an excess at the very faint end (Fig.~11).

\begin{figure}
\centerline{\hbox{
\psfig{file=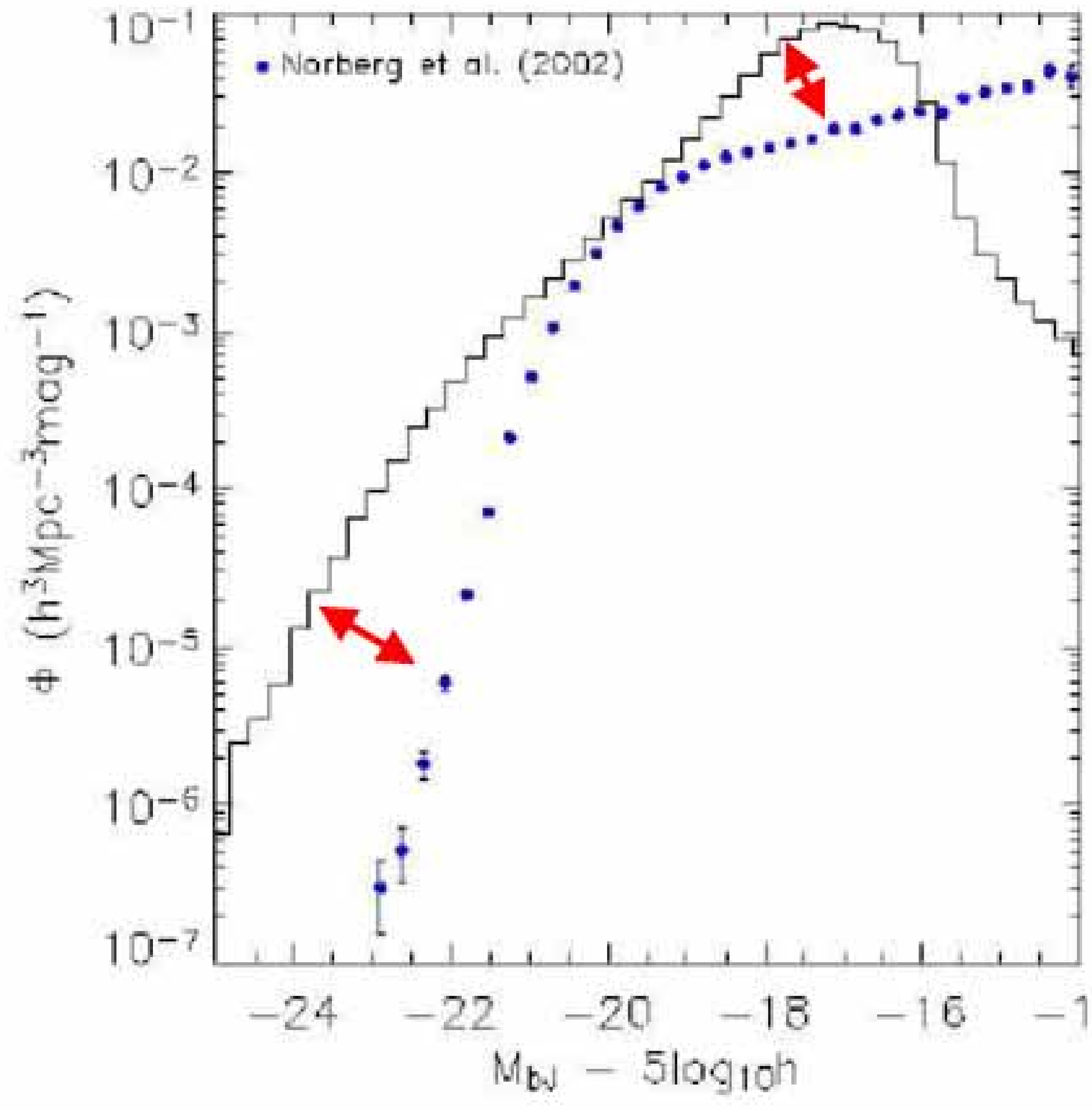, height=1.6in}
\psfig{file=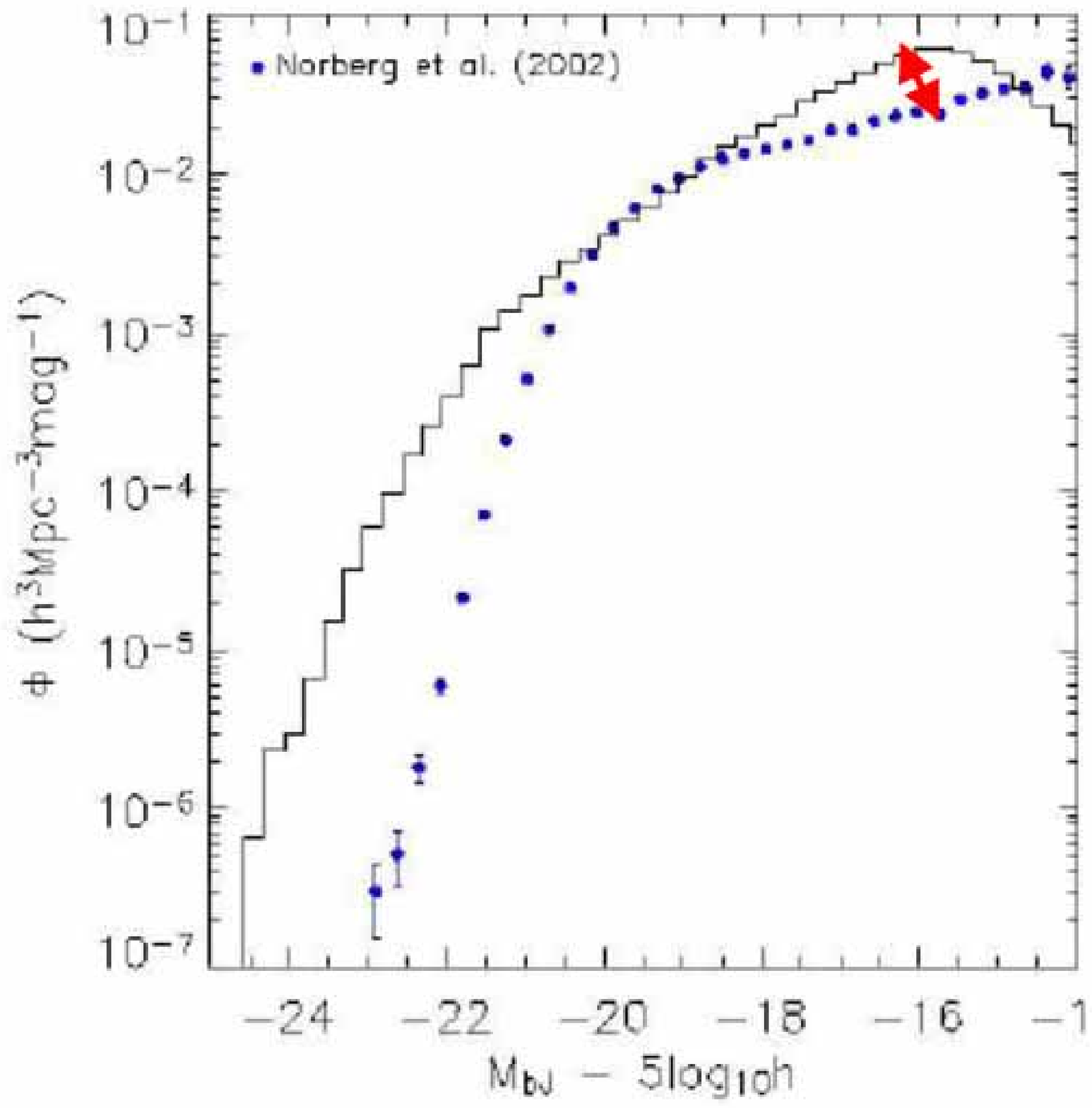, height=1.6in}
\psfig{file=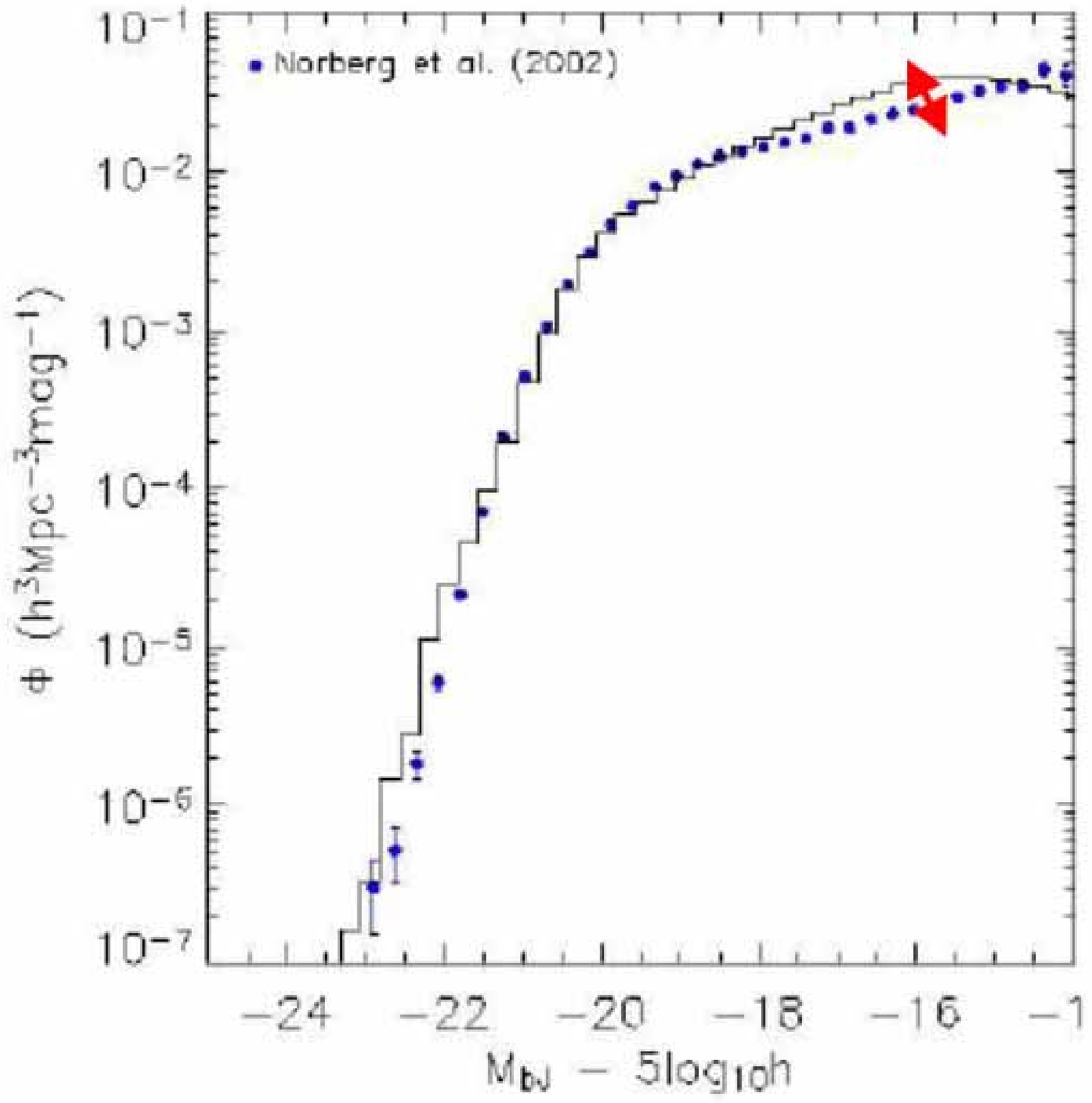, height=1.6in}}}
\caption{The effect of various forms of feedback (dots) on the
resultant shape of the blue 2dF galaxy luminosity function 
(Norberg et al 2002). From left to right: no feedback, supernova
feedback only, supernova, reionization \& AGN feedback (Courtesy:
Darren Croton).}
\label{fig:11}      
\end{figure}

\subsection{The Role of the Environment}

In addition to recognizing that more elaborate modes of 
feedback need to be incorporated in theoretical models,
the key role of the environment has also emerged as an
additional feature which can truncate star formation and
alter galaxy morphologies. 

The preponderance of elliptical and S0 galaxies in rich clusters 
was noticed in the 1930's but the first quantitative study of this
effect was that of Dressler (1980) who correlated the fraction of galaxies 
of a given morphology $T$ above some fixed luminosity with the 
projected galaxy density, $\Sigma$, measured in galaxies Mpc$^{-2}$. 

The local $T-\Sigma$ relation was used to justify two 
rather different possibilities. In the first, the {\em nature} hypothesis, 
those galaxies which formed in high density peaks at 
early times were presumed to have consumed their gas
efficiently, perhaps in a single burst of star formation. Galaxies in lower 
density environments continued to accrete gas and thus
show later star formation and disk-like morphologies.
In short, segregation was established at birth and the present relation
simply represents different ways in which galaxies formed 
according to the density of the environment at the time of 
formation. In the second, the {\em nurture} hypothesis, 
galaxies are transformed at later times from spirals into spheroidals by 
environmentally-induced processes. 

Work in the late 1990's, using morphologies determined using Hubble Space 
Telescope, confirmed a surprisingly rapid evolution in the 
$T-\Sigma$ relation over 0$<z<$0.5 (Couch et al 1998, Dressler et al 1997)
strongly supporting environmentally-driven evolution along the
lines of the {\em nurture} hypothesis. Impressive Hubble images 
of dense clusters at quite modest redshifts ($z\simeq$0.3-0.4) 
showed an abundance of spirals in their cores whereas few or 
none exist in similar environs today.

What physical processes drive this relation and how has
the $T-\Sigma$ relation evolved in quantitative detail?
Recent work (Smith et al 2005, Postman et al 2005, Figure 12) has
revealed that the basic relation was in place at $z\simeq$1, but
that the fraction $f_{E+S0}$ of Es and S0s has doubled
in dense environments since that time. Smith et al suggest
that a continuous, density-dependent, transformation of spirals
into S0s would explain the overall trend. Treu et al (2003)
likewise see a strong dependence of the fraction as a function
of $\Sigma$ (and to a lesser extent with cluster-centric radius) within 
a well-studied cluster at $z\simeq$0.4; they review the various
physical mechanisms that may produce such a transformation. 

\begin{figure}
\centering
\includegraphics[height=7cm,angle=0]{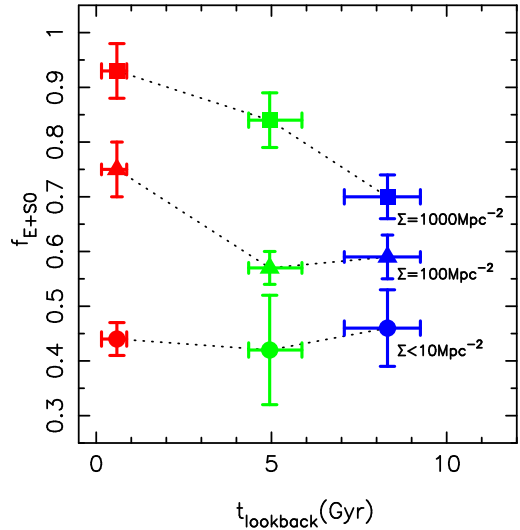}
\caption{The fraction of observed E and S0 galaxies down to a
fixed rest-frame luminosity as a function of lookback time and projected
density $\Sigma$ from the study of Smith et al (2005). Although
the morphology density relation was already in place at $z\simeq$1,
there has been a continuous growth in the fraction subsequently,
possibly as a result of the density-dependent transformation of spirals 
into S0s.}
\label{fig:12}      
\end{figure}

Figure 12 encapsulates much of what we now know about
the role of the environment on galaxy formation. The early
development of the $T-\Sigma$ relation implies dense peaks in
the dark matter distribution led to accelerated evolution in gas
consumption and stellar evolution and this is not dissimilar to 
the {\em nature} hypothesis. However, the subsequent
development of this relation since $z\simeq$1 reveals the  
importance of environmentally-driven morphological transformations. 

\subsection{The Importance of High Redshift Data}

This glimpse of evolving galaxy populations to $z\simeq$1 has 
emphasized the important role of high redshift data. In the
case of the local morphology-density relation,
Hubble morphologies of galaxies in distant clusters have given
us a clear view of an evolving relationship, partly driven by 
environmental processes. Indeed, the data seems to confirm
the {\em nurture} hypothesis for the origin of the morphology-density
relation.

Although we can place important constraints on the past star 
formation history from detailed studies of nearby galaxies,
as the standard model now needs several additional ingredients
(e.g. feedback) to reproduce even the most basic local properties 
such as the luminosity function (Figure 11), data at significant
look-back times becomes an essential way to test the validity
of these more elaborate models.

Starting in the mid-1990's, largely by virtue of the arrival of
the Keck telescopes - the first of the new generation of 8-10m
class optical/infrared telescopes - and the refurbishment of
the Hubble Space Telescope, there has been an explosion
of new data on high redshift galaxies.

It is helpful at this stage to introduce three broad classes
of distant objects which will feature significantly in the next few
lectures. Each gives a complementary view of the galaxy
population at high redshift and illustrates the challenge
of developing a unified vision of galaxy evolution. 

\begin{figure}
\centerline{\hbox{
\psfig{file=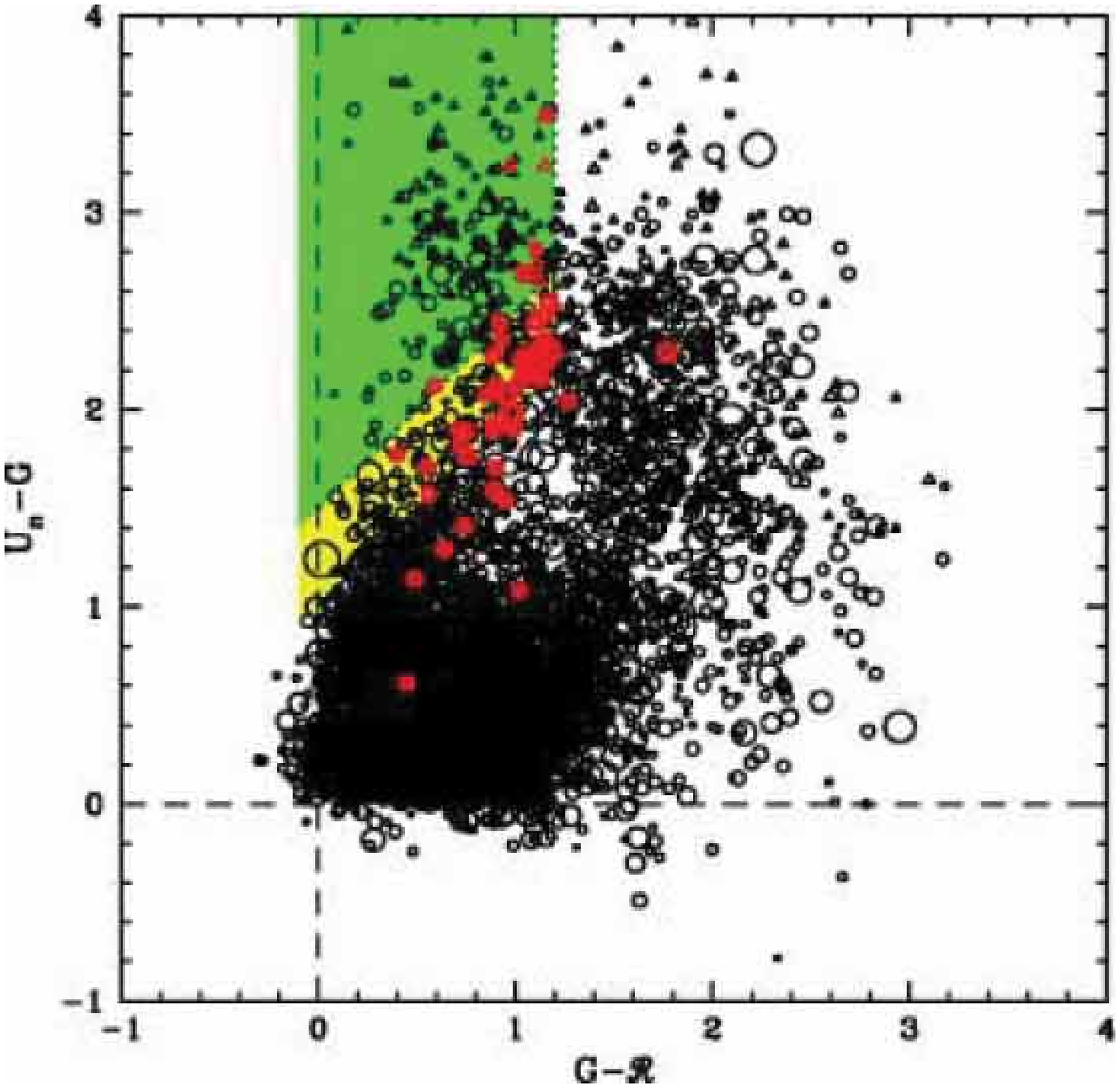,height=2.5in,angle=0}
\psfig{file=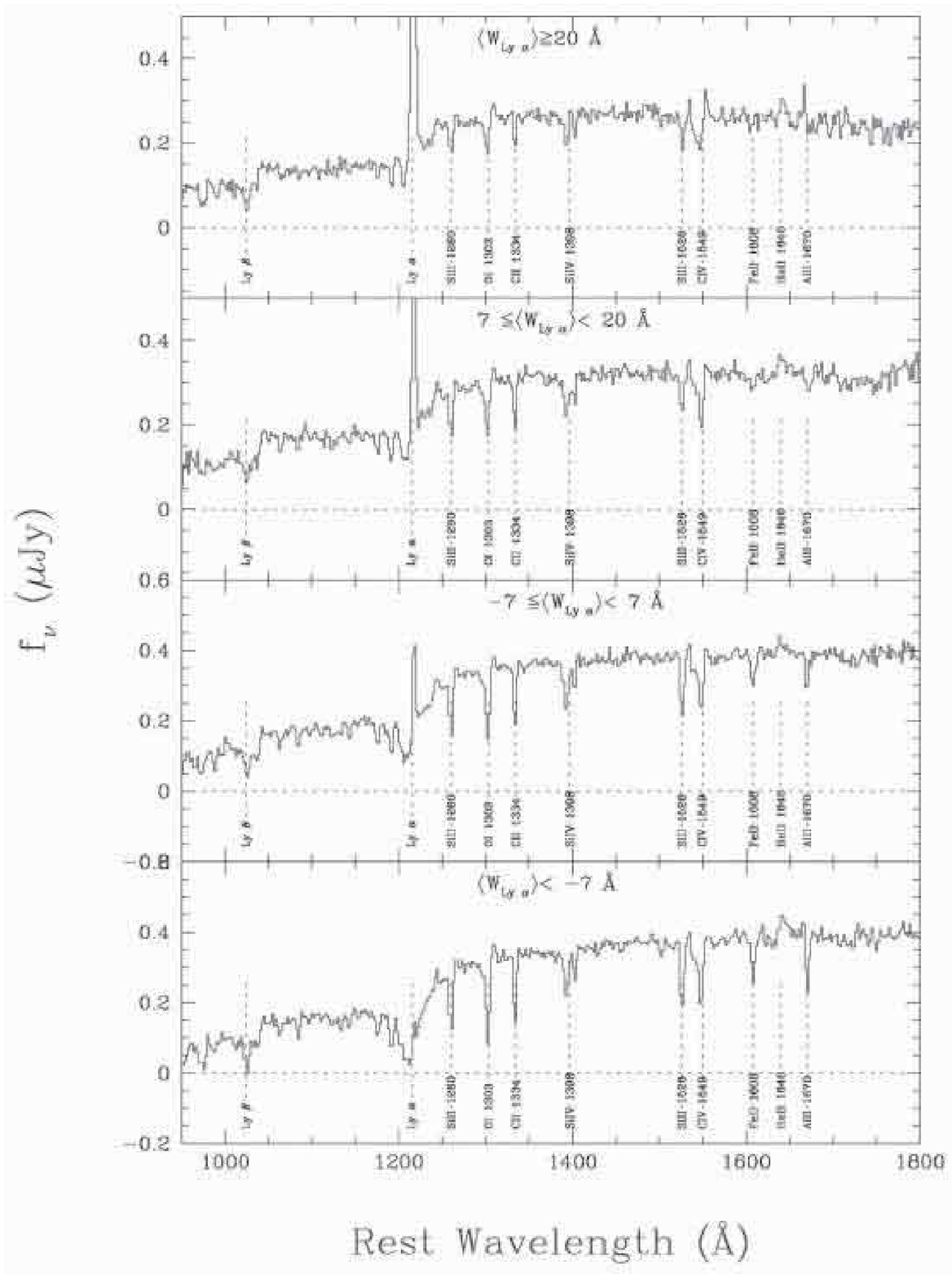,height=3.0in,angle=0}}}
\caption{Location and verification of the Lyman break population
in the redshift range 2.7$<z<$3.4 (Steidel et al 2003). (Left) 
$UGR$ color-color plane for a single field;
green and yellow shading refers to variants on the color
selection of high redshift candidates. Contaminating Galactic
stars cut across the lower right corner of this selection; those
confirmed spectroscopically are marked in red. (Right) Coadded
rest-frame Keck spectra for samples of typically 200 Lyman break
galaxies binned according to the strength of Lyman $\alpha$ emission.}
\label{fig:13}      
\end{figure}

\begin{itemize}

\item{}{\em Lyman-break galaxies: color-selected luminous star
forming galaxies at $z>$2}. First located spectroscopically
by Steidel et al (1996, 1999a, 1999b, 2003), these sources are selected
by virtue of the increased opacity shortward of the Lyman
limit ($\lambda=$912\AA\ ) arising from the combined effect
of neutral hydrogen in hot stellar atmospheres, the interstellar
gas and the intergalactic medium. When redshifted beyond
z$\simeq$2, the characteristic `drop out' in the Lyman
continuum moves into the optical (Figure 13). We will review
the detailed properties of this, the most well-studied, distant
galaxy population over 2$<z<$5 in subsequent lectures.

\begin{figure}
\centering
\includegraphics[height=8cm,angle=0]{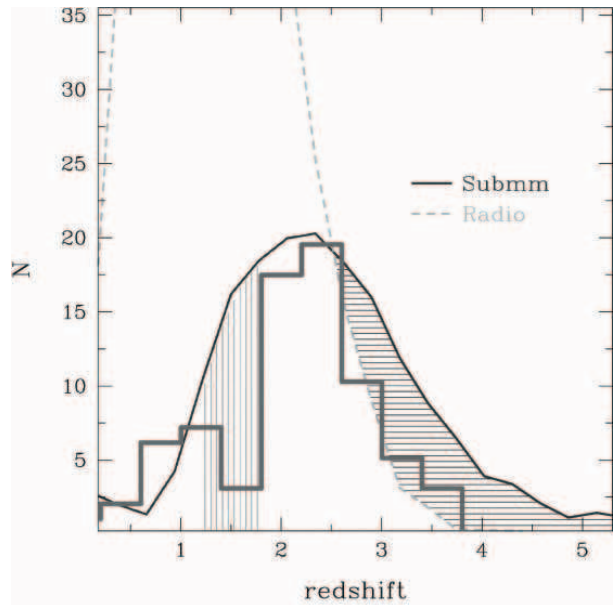}
\caption{Redshift distribution of 73 radio-identified SCUBA
sources from Chapman et al (2005). To illustrate the
possible bias arising from the necessary condition of a radio
position for a Keck redshift, the solid curve represents
a model prediction for the entire $>$5mJy sub-mm population. 
The sub-mm population does not seem to extend significantly 
beyond z$\simeq$4 and has a median redshift of $z$=2.2. }
\label{fig:14}      
\end{figure}

\item{}{\em Sub-millimeter star forming sources:} The
SCUBA 850$\mu$m array on the 15 meter James Clerk
Maxwell Telescope and other sub-mm imaging devices 
have also been used to locate distant
star forming galaxies (Smail et al 1997, Hughes et al 1998). In 
this case, emission is detected from dust, heated either by 
vigorous star formation or an active nucleus. Remarkably,
their visibility does not fall off significantly with redshift
because they are detected in the Rayleigh-Jeans tail 
of the dust blackbody spectrum (Blain et al 2002).

Progress in understanding the role and nature of this 
population has been slower because sub-mm sources are 
often not visible at optical and near-infrared wavelengths 
(due to obscuration) and the positional accuracy of the sub-mm
arrays is too coarse for follow-up spectroscopy.
The importance of sub-mm sources lies in the
fact that they contribute significantly to the star 
formation rate at high redshift. Regardless of their
redshift, the source density at faint limits is 1000 times
higher than a no-evolution prediction based on
the local abundance of dusty IRAS sources. For several
years the key issue was to nail the redshift distribution.

Progress has been made by securing accurate
positions using radio interferometers such as the
VLA (Frayer et al 2000). About 70\% of those
brighter than 5 mJy have VLA detections and
spectroscopic redshift have now been determined for 
a significant fraction of this population (Chapman et al
2003, 2005, Figure 14).

\begin{figure}
\centering
\includegraphics[height=8cm,angle=0]{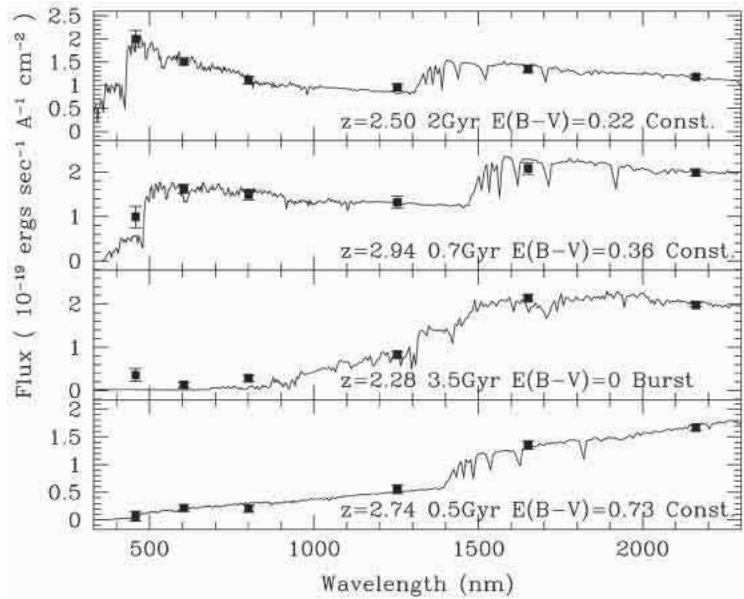}
\caption{Observed spectral energy distribution of distant
red sources selected with a red $J-K$ color superimposed
on model spectra (Franx et al 2003). Although some sources reveal modest
star formation and can be spectroscopically confirmed to
lie above $z\simeq$2, others (such as the lower two examples) 
appear to be passively-evolving with no active star formation. }
\label{fig:15}      
\end{figure}

\item{}{\em Passively-Evolving Sources:} The Lyman-break
and sub-mm sources are largely star-forming galaxies.
The arrival of panoramic near-infrared cameras has 
opened the possibility of locating quiescent sources
that are no longer forming stars. Such sources would not 
normally be detected via the other techniques and
so understanding their contribution to the integrated
stellar mass at, say, $z\simeq$2, is very important.

The nomenclature here is confusing with intrinsically red sources 
being termed `extremely red objects (EROs)' or `distant
red galaxies (DRGs)' with no agreed selection
criteria (see McCarthy 2004 for a review).
When star formation is complete, stellar evolution continues 
in a passive sense with main sequence dimming;
the galaxy fades and becomes redder.

Of particular interest are the most distant examples
which co-exist alongside the sub-mm and Lyman-break
galaxies, i.e. at $z>$2 selected according to their infrared
$J-K$ color (van Dokkum et al 2003, Figure 15). 

\end{itemize}

\subsection{Lecture Summary}

We have seen in this brief tour that galaxy formation is a
process involving gravitational instability driven by the hierarchical
assembly of dark matter halos; this component we understand
well. However, additional complexities arise from star formation,
dynamical interactions and mergers, environmental processes
and various forms of feedback which serve to regulate how
star formation continues as galaxies grow in mass.

Theorists have attempted to deal with this complexity by
augmenting the highly-successful numerical (DM-only) simulations
with semi-analytic tools for incorporating these complexities.
As the datasets have improved so it is now possible to consider
`fine-tuning' these semi-analytical ingredients. {\em Ab initio}
modeling is never likely to be practical.

I think it fair to say that many observers have philosophical 
reservations about this `fine-tuning' process in the sense that
although it may be possible to reach closure on models and
data, we seek a deeper understanding of the physical reality
of many of the ingredients. This is particularly the case for
feedback processes. Fortunately, high redshift data forces this reality check
as it gives us a direct measure of the galaxy assembly history
which will be the next topics we discuss. 

As a way of illustrating the importance of high redshift data,
I have introduced three very different populations of
galaxies each largely lying in the redshift range 2$<z<$4. When these
were independently discovered, it was (quite reasonably) claimed 
by their discoverers that their category represented a major, if
not the most significant, component of the distant galaxy
population. We now realize that UV-selected, sub-mm selected and 
non-star forming galaxies each provide a complementary view of the complex
history of galaxy assembly and the challenge is to complete the
`jig-saw' from these populations.

\newpage


\section{Cosmic Star Formation Histories}

\subsection{When Did Galaxies Form? Searches for Primeval Galaxies}

The question of the appearance of an early forming galaxy goes back
to the 1960's. Partride \& Peebles (1967) imagined the free-fall
collapse of a 700 $\L^{\ast}$ system at $z\simeq$10 and predicted
a diffuse large object with possible Lyman $\alpha$ emission. Meier
(1976) considered primeval galaxies might be compact and intense
emitters such as quasars.

In the late 1970's and 1980's when the (then) new generation of 4 meter 
telescopes arrived, astronomers sought to discover the distinct era 
when galaxies formed. Stellar synthesis models 
(Tinsley 1980, Bruzual 1980) suggested present-day passive
systems (E/S0s) could have formed via a high redshift luminous
initial burst. Placed at $z\simeq$2-3, sources of the same stellar
mass would be readily detectable at quite modest magnitudes,
$B\simeq$22-23, and provide an excess population of blue galaxies.

In reality, the (now well-studied) excess of faint blue galaxies over locally-based
predictions is understood to be primarily a phenomenon associated with a 
gradual increase in star formation over $0<z<$1 rather than one 
due to a distinct new population of intensely luminous sources at high redshift 
(Koo \& Kron 1992, Ellis 1997). Moreover, dedicated searches for suitably
intense Lyman $\alpha$ emitters were largely unsuccessful. Pritchet (1994)
comprehensively reviews a decade of searching.

Our thinking about primeval galaxies changed in two respects in the late
1980's. Foremost, synthesis models such as those developed by Tinsley 
and Bruzual assumed isolated systems; dark matter-based models
emphasized the gradual assembly of massive galaxies. This change meant
that, at $z\simeq$2-3, the abundance of massive galaxies should be much 
reduced. Secondly, the flux limits searched for primeval galaxies were 
optimistically bright; we slowly realized the more formidable challenge
of finding these enigmatic sources.

\subsection{Local Inventory of Stars}

An important constraint on the past star formation history is the present-day
stellar density. The former must, when integrated, yield the latter.
Fukugita et al (1998) and Fukugita \& Peebles (2004) have considered this
important problem based on local survey data provided by the SDSS (Kauffmann
et al 2003) and 2dF (Cole et al 2001) redshift samples. 
 
The derivation of the integrated density of stars involves many assumptions
and steps but is based primarily on the local infrared ($K$-band) luminosity
function of galaxies. The rest-frame $K$ luminosity of a galaxy is a much 
more reliable proxy for its stellar mass than that at a shorter (e.g. optical) 
wavelength because its value is largely irrespective of the past star formation
history - a point illustrated by Kauffmann \& Charlot (1998, Figure 16).
Another way to phrase this is to say that the infrared mass/light ratio ($M/L_K$)
is fairly independent of the star formation history, so that the stellar mass
can be derived from the observed $K$-band luminosity by a multiplicative factor.

\begin{figure}
\centering
\includegraphics[height=9cm,angle=0]{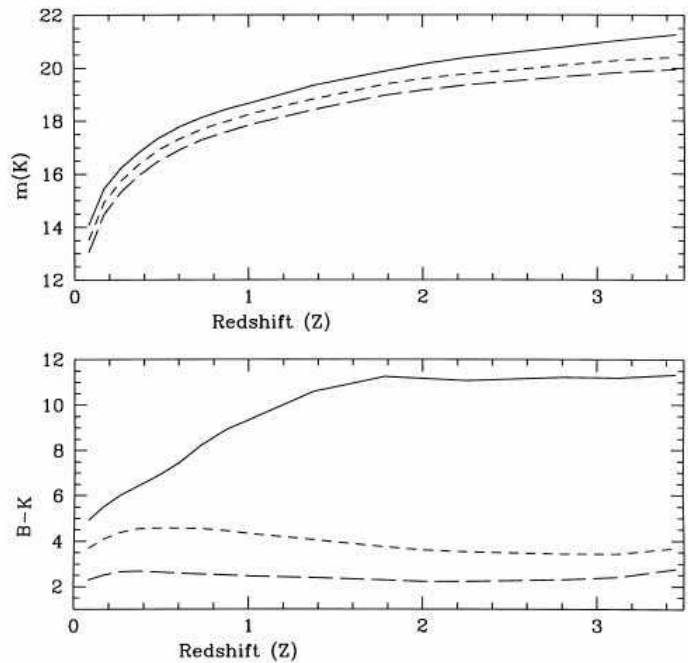}
\caption{The robustness of the $K$-band luminosity of a galaxy
as a proxy for stellar mass (Kauffmann \& Charlot 1998). The upper 
panel shows the $K$-band apparent 
magnitude of a galaxy defined to have a fixed stellar mass of $10^{11} 
M_{\odot}$ when placed at various redshifts. The different curves represent
extreme variations in the way the stellar population was created. Whereas the 
$B-K$ color is strongly dependent upon the star formation history, the $K$-band 
luminosity is largely independent of it. }
\label{fig:16}      
\end{figure}

In practice the mass/light ratio depends on the assumed distribution of
stellar masses in a stellar population. The zero age or initial mass function
is usually assumed to be some form of power law which can only be
determined reliable for Galactic stellar populations, although constraints
are possible for extragalactic populations from colors and nebular
line emission (see reviews by Scalo 1986, Kennicutt 1998, Chabrier 2003)  

In its most frequently-used form the IMF is quoted in mass fraction
per logarithmic mass bin: viz:

$$\xi(log\, m) = \frac{dn}{d\,log\, m} \propto m^{-x}$$

or, occasionally,

$$\xi(m) = \frac{dn}{dm} = \frac{1}{m(ln\,10)}\xi(log m) \propto m^{-\alpha}$$

where $x=\alpha-1$.

In his classic derivation of the IMF, Salpeter (1955) determined a pure power
law with $x$=1.35. More recently adopted IMFs are compared in 
Figure 17. They differ primarily in how to restrict the low mass contribution, 
but there is also some dispute on the high mass slope (although the Salpeter
value is supported by various observations of galaxy colors and H$\alpha$
distributions, Kennicutt 1998).  

\begin{figure}
\centering
\includegraphics[height=8cm,angle=0]{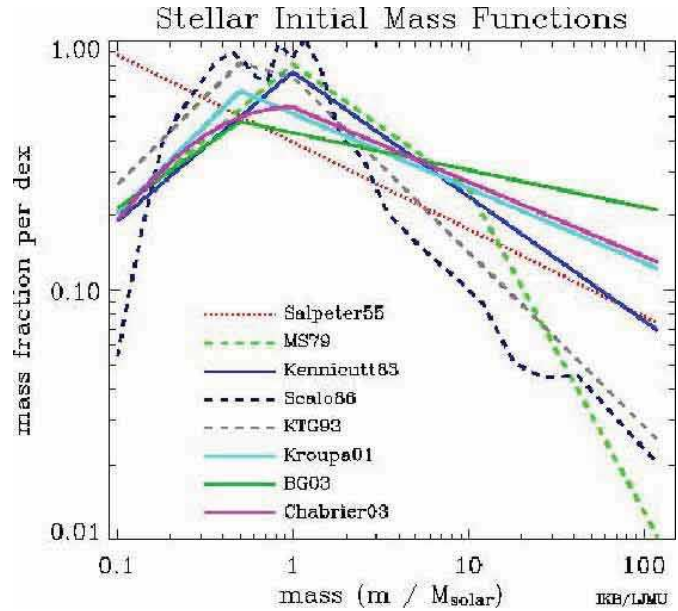}
\caption{A comparison of popular stellar initial mass functions (courtesy of
Ivan Baldry). }
\label{fig:17}      
\end{figure}

\begin{table}[htdp]
\caption{K-band Stellar Mass/Light Ratios}
\begin{center}
\begin{tabular}{|l|c|c|c|} \hline
{\bf Source} & {\bf Stars} & {\bf Stars + WDs/BHs} & {\bf  Total (Past SFR)} \\ \hline
Salpeter (1955) & 1.15 & 1.30 & 1.86 \\
Miller Scalo (1979) & 0.46 & 0.60 & 0.99 \\
Kennicutt (1983) & 0.46 & 0.60 & 1.06  \\
Scalo (1986) & 0.52 & 0.61 & 0.84 \\
Kroupa et al (1993) & 0.65 & 0.76 & 1.09 \\
Kroupa (2001) & 0.67 & 0.83 & 1.48 \\
Baldry \& Glazebrook (2003) & 0.67 & 0.86 & 1.76 \\
Chabrier (2003) & 0.59 & 0.75 & 1.42 \\  \hline
\end{tabular}
\end{center}
\label{tab:2}
\end{table}

The IMF has a direct influence on the assumed $M/L_K$ (as discussed
by Baldry \& Glazebrook 2003, Chabrier, 2003 and Fukugita \& Peebles, 
2004) in a manner which depends on the age, composition and 
past star formation history. The adopted mass/light ratio is then a
crucial ingredient for computing both stellar masses (LectureLecture 4)
and galaxy colors.

Baldry\footnote{\rm http://www.astro.livjm.ac.uk/~ikb/research/imf-use-in-cosmology.html} has undertaken a very useful comparative study
of the impact of various IMF assumptions using the PEGASE 2.0 stellar synthesis
code for a population 10 Gyr old with solar metallicity, integrating 
between stellar masses of 0.1 and 120 $M_{\odot}$ (Table 2). Stellar masses 
have been defined in various ways as represented by
the 3 columns in Table 2. Typically we are 
interested in the {\it observable stellar mass} at a given time (i.e. main sequence 
and giant branch stars), but it is interesting to also compute the total mass
which is not in the interstellar medium, which includes that locked in 
evolved degenerate objects (white dwarfs and black holes). The most
inclusive definition of stellar mass (total) is the integral of the past star formation 
history. Depending on the definition, and chosen IMF, the uncertainties
range almost over a factor of 4 for the most popularly-used functions,
quite apart from the unsettling question of whether the {\it form} of the IMF
might vary with epoch or type of object. 

Although the stellar mass function for a galaxy survey can be derived 
assuming a fixed mass/light ratio, the useful stellar density is that corrected 
for the fractional loss, $R$, of stellar material due to winds and supernovae. 
Only with this correction ($R$=0.28 for a Salpeter IMF), does the present-day 
value represent the integral of the past star formation.

Figure 18 shows the $K$-band luminosity and derived stellar mass function 
for galaxies in the 2dF redshift survey from the analysis of Cole et al (2001).
$K$-band measures were obtained by correlation with the $K<$13.0 catalog
obtained by the 2MASS survey. 

\begin{figure}
\centerline{\hbox{
\psfig{file=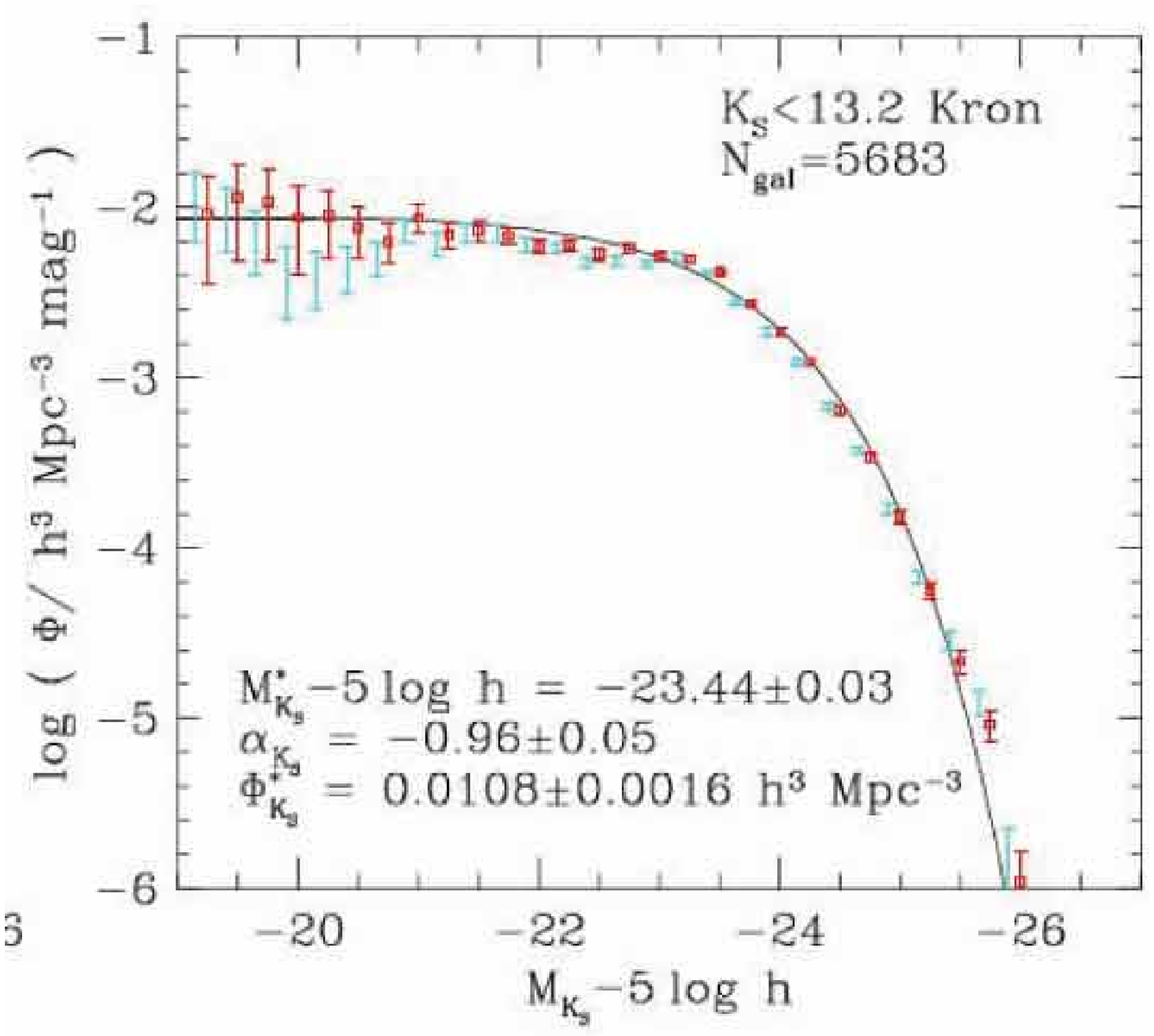,height=2.2in,angle=0}
\psfig{file=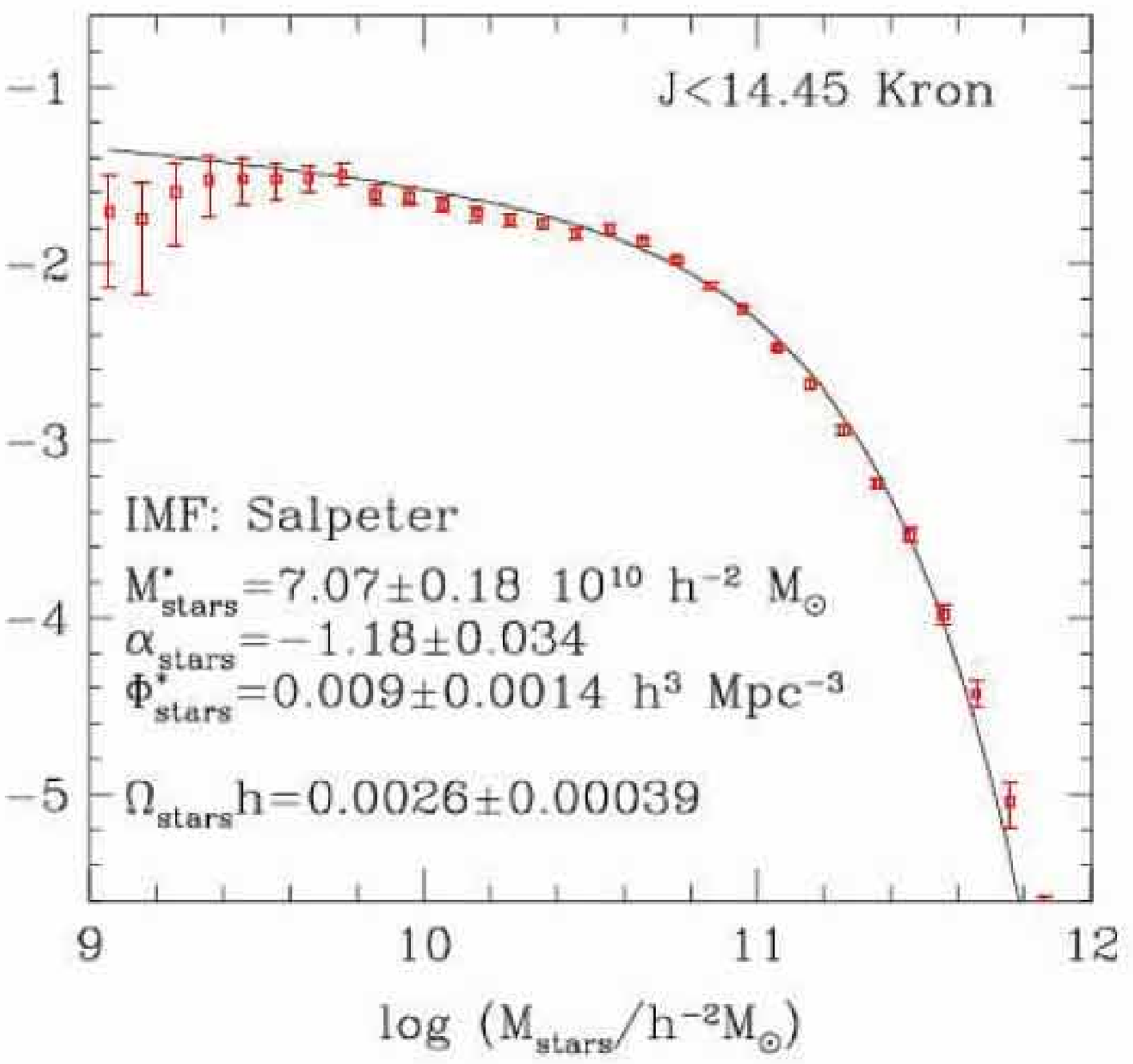,height=2.3in,angle=0}}}
\caption{(Left) Rest-frame $K$-band luminosity function derived from
the combination of redshifts from the 2dF survey with photometry from
2MASS (Cole et al 2001); Schechter parameter fits are shown. (Right) Derived
stellar mass function assuming a Salpeter IMF corrected for lost
material assuming $R$=0.28 (see text).}
\label{fig:18}      
\end{figure}

The integrated stellar density, corrected for stellar mass loss, is (Cole et al 2001):

$$\Omega_{stars}\,h = 0.0027 \pm 0.00027 $$

for a Salpeter IMF, a value very similar to that derived independently
by Fukugita \& Peebles (2004). By comparison the local mass fraction in neutral
HI + He I gas is:

$$\Omega_{gas}\,h = 0.00078 \pm 0.00016$$

Thus only 5\% of all baryons are in stars with the bulk in ionized gas.
 
\subsection{Diagnostics of Star Formation in Galaxies}

When significant redshift surveys became possible at intermediate and
high redshift through the advent of multi-object spectrographs, so it became
possible to consider various probes of the star formation rate (SFR) at different
epochs. As in the formalism for calculating the integrated luminosity
density, $\rho_L$, per comoving Mpc$^3$, so for a given population
various diagnostics of on-going star formation can yield an equivalent
global star formation rate $\rho_{SFR}$ in units of $M_{\odot}$ yr$^{-1}$ 
Mpc$^{-3}$.

Such integrated measures average over a whole host
of important details, such as differences in evolutionary behavior 
between luminous and sub-luminous galaxies and, of course, morphology.
Moreover, in any survey at high redshift, only a portion of the population
is rendered visible so uncertain corrections must be made to compare
results at different epochs. The importance of the cosmic star formation
history, i.e. $\rho_{SFR}(z)$, is it displays, in a simple manner, the
epoch and duration of galaxy growth. By integrating the function, one
should recover the present stellar density (\S3.5).

There are various probes of star formation in galaxies, each with its
advantages and drawbacks. Not only is there no single `best' method
to gauge the current star formation rate of a chosen galaxy,
but as each probe samples the effect of young stars in different initial
mass ranges, so each averages the star formation rate over a different
time interval. If, as is often the case in the most energetic sources,
the star formation is erratic or burst-like, one would not expect
different diagnostics to give the same measure of the instantaneous
SFR even for the same galaxies.

Four diagnostics are in common use (see review by Kennicutt 1998).

\begin{itemize}

\item{} The rest-frame ultraviolet continuum ($\lambda\lambda\simeq$1250-1500\AA\
) has the advantage of being directly connected to well-understood high mass 
($>5M_{\odot}$) main sequence stars. Large datasets are available for high redshift
star-forming galaxies, including some to $z\simeq$6. Via the GALEX satellite and 
earlier balloon-borne experiments, local data is also available. The disadvantage 
of this diagnostic lies in the uncertain (and significant) corrections necessary for 
dust extinction and a modest sensitivity to the assumed initial mass function. 
Obscured populations are completely missed in UV samples.
Kennicutt suggests the following calibration for the UV luminosity:

$$SFR (M_{\odot} yr^{-1}) = 1.4\, 10^{-28} L_\nu (ergs \, s^{-1} Hz^{-1})$$

\item{} Nebular emission lines such as $H\alpha$ and [O II] are 
also available for a range of redshifts ($z<2.5$), for example as a 
natural by-product of faint redshift surveys. Gas clouds are
photo-ionized by very massive ($>10 M_{\odot}$) stars.
 Dust extinction can often be evaluated
from higher order Balmer lines under various radiative assumptions
depending on the escape fraction of ionizing photons. The sensitivity
to the initial mass function is strong. 

$$SFR (M_{\odot} yr^{-1}) = 7.9\, 10^{-42} L(H\alpha) (ergs \,  s^{-1}) $$

$$SFR (M_{\odot} yr^{-1}) = 1.4 \pm 0.4 10^{-41} L(O II) (ergs \, s^{-1}) $$

\item{} Far infrared emission (10-300 $\mu$m) arises from dust heated 
by young stars. It is clearly only a tracer in the most dusty systems
and thus acts as a valuable complementary probe to the UV continuum.
As we have seen in \S2, luminous far infrared galaxies are also seen 
to high redshift. However, not all dust heating is due to young stars
and the bolometric far infrared flux, $L_{FIR}$, is needed for an accurate 
measurement.

$$SFR (M_{\odot} yr^{-1}) = 4.5 \, 10^{-44} L_{FIR} (ergs \, s^{-1}) $$

\item{} Radio emission, e.g. at 1.4GHz, is thought to arise from 
synchrotron emission generated by relativistic electrons accelerated
by supernova remnants following the rapid evolution of the most
massive stars. Its great advantage is that it offers a dust-free measure 
of the recent SFR. Current radio surveys do not have the sensitivity to 
see emission beyond $z\simeq$1, so its promise has yet to be fully 
explored. This process is also the least well-understood and
calibrated. Sullivan et al (2001) discuss this point in some
detail and conclude:

$$SFR (M_{\odot} yr^{-1}) = 1.1 \, 10^{-28} L_{1.4} (ergs \, s^{-1} Hz^{-1}) $$

for bursts of duration $>$100 Myr.

\end{itemize}

The question of the time-dependent nature of the SFR is an important
point (Sullivan et al 2000, 2001). For an instantaneous burst of star 
formation, Figure 19a shows the
`response' of the various diagnostics. Clearly if the SF is erratic
on 0.01-0.1 Gyr timescales, each will provide a different sensitivity.
Sullivan et al (2000) compared UV and H$\alpha$ diagnostics
for a large sample of nearby galaxies and found a scatter beyond
that expected from the effects of dust extinction or observational
error, presumably from this effect (Figure 19b).

\begin{figure}
\centerline{\hbox{
\psfig{file=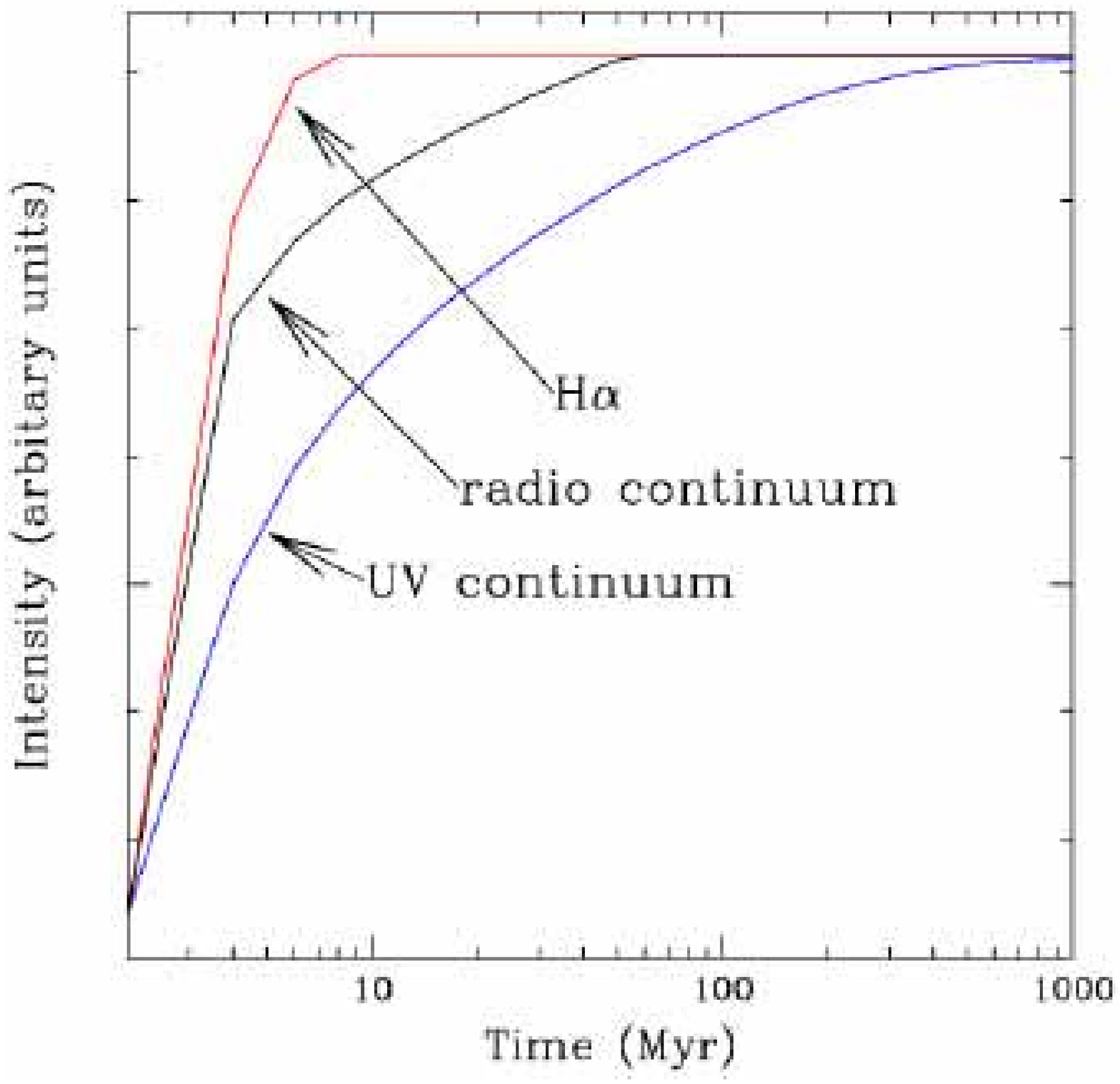,height=2.2in,angle=0}
\psfig{file=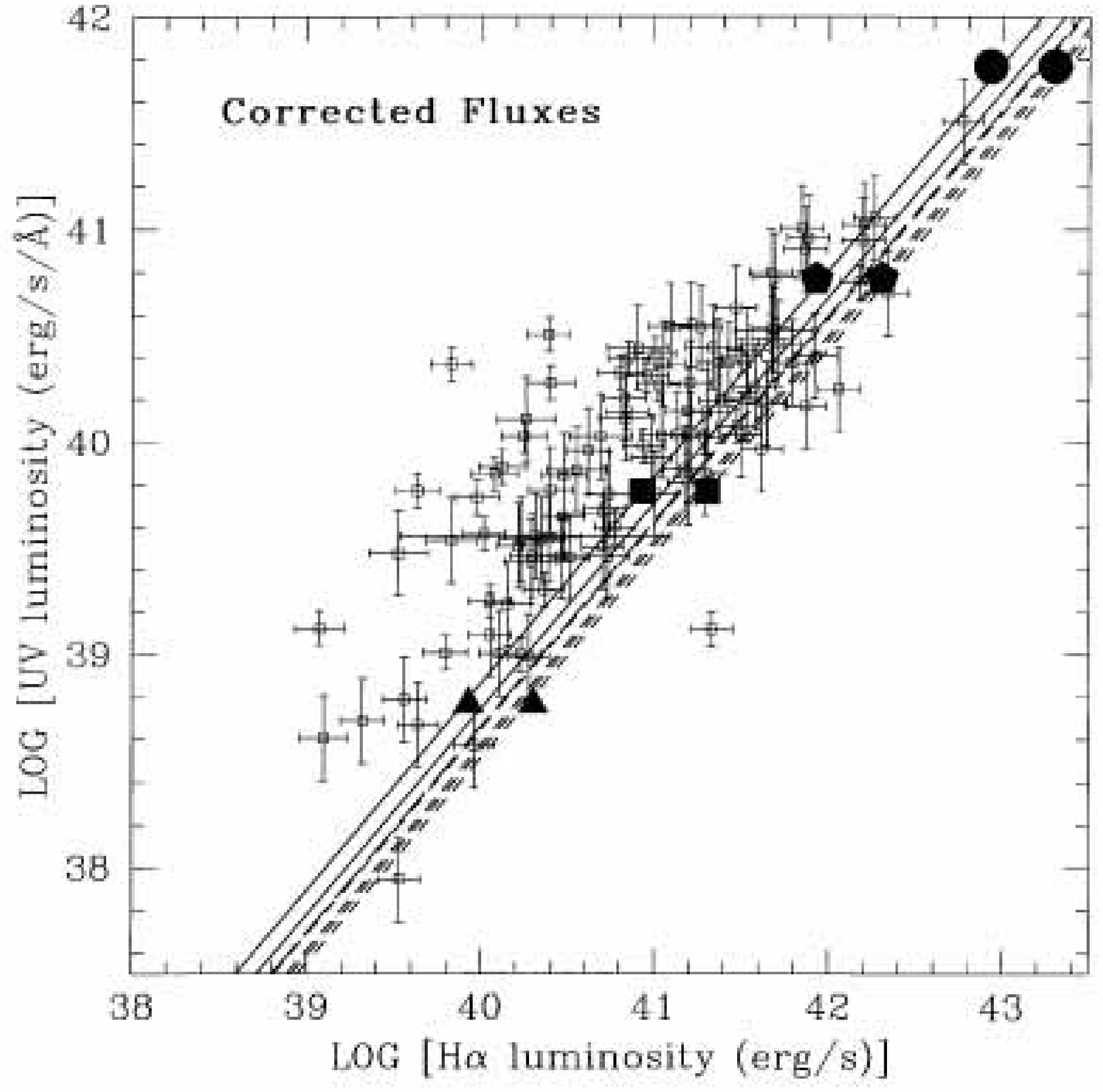,height=2.2in,angle=0}}}
\caption{Time dependence of various diagnostics of star formation
in galaxies. (Left) sensitivities for a single burst of star formation. 
(Right) scatter in the UV and H$\alpha$ galaxy luminosities
in the local survey of Sullivan et al (2000); lines represent model
predictions for various (constant) star formation rates and
metallicities. It is claimed that some fraction of local galaxies must 
undergo erratic periods of star formation in order to account for 
the offset and scatter.}
\label{fig:19}      
\end{figure}

In addition to the initial mass function (already discussed), a
key uncertainty affecting the UV diagnostic is the selective dust 
extinction law. Over the wavelength range 0.3$<\lambda<$1 
$\mu$m, differences between laws deduced for the Milky Way, 
the Magellan clouds and local starburst galaxies (Calzetti et al 2000) 
are quite modest. Significant differences occur around the
2200 \AA\ feature (dominant in the Milky Way but absence in
Calzetti's formula) and shortward of 2000 \AA\ where the
various formulae differ by $\pm$2 mags in $A(\lambda)/E(B-V)$.

\subsection{Cosmic Star Formation - Observations}

Early compilations of the cosmic star formation history
followed the field redshift surveys of Lilly et al (1996), Ellis
et al (1996) and the abundance of U-band drop outs in the
early deep HST data (Madau et al 1996). The pioneering
papers in this regard include Lilly et al (1996), Fall et al (1996)
and Madau et al (1996, 1998).

Hopkins (2004) and Hopkins \& Beacom (2006) have undertaken
a valuable recent compilation, standardizing all measures to
the same initial mass function, cosmology and extinction law.
They have also integrated the various luminosity functions for
each diagnostic in a self-consistent manner (except at very
high redshift). Accordingly, their articles give us a valuable
summary of the state of the art. 

Figure 20 summarizes their findings. Although at first
sight somewhat confusing, some clear trends are evident
including a systematic increase in star formation rate per
unit volume out to $z\simeq$1 which is close to (Hopkins
2004):

$$\rho_{SFR}(z) \propto (1+z)^{3.1}$$

A more elaborate formulate is fitted in Hopkins \& Beacom (2006).

There is a broad peak somewhere in the region 2$<z<$4 
where the UV data is consistently an underestimate and
the growing samples of sub-mm galaxies are valuable.
The dispersion here is only a factor of $\pm$2 or so,
which is a considerable improvement on earlier work.
We will return to the question of a possible decline in the
cosmic SFR beyond $z\simeq$3-4 in later sections. 

\begin{figure}
\centering
\includegraphics[height=8.5cm,angle=270]{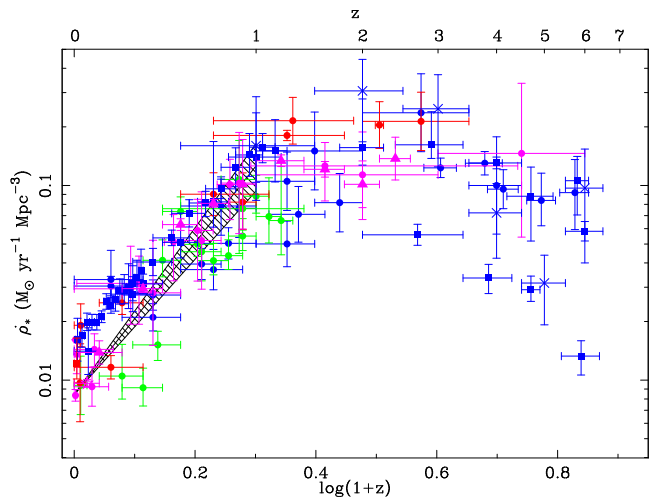}
\caption{Recent compilations of the cosmic star formation history. 
Circles are data from Hopkins (2004) color-coded by method: 
blue: UV, green: [O II], red: H$\alpha/\beta$, magenta: non-optical 
including sub-mm and radio. New data from Hopkins
\& Beacom (2006), represented by various triangles, stars and 
squares, include Spitzer FIR measures (magenta triangles).
The solid lines represent a range of the best fitting parametric form  
for $z<$1. }
\label{fig:20}      
\end{figure}

In their recent update, Hopkins \& Beacom (2006) also parametrically 
fit the resulting $\rho_{SFR}(z)$ in two further redshift sections,
beyond $z\simeq$1, and they use this to predict the growth of the 
absolute stellar mass density, $\rho_{\ast}$, via integration (Figure 21).
Concentrating, for now, on the reproduction of the {\em present
day} mass density (Cole et al 2001), the agreement is
remarkably good. 

Although in detail the result depends on an assumed initial
mass function and the vexing question of whether extinction
might be luminosity-dependent, this is an important result in 
two respects: firstly, as an absolute comparison it confirms that
most of the star formation necessary to explain the presently-observed 
stellar mass has already been detected through various complementary
surveys. Secondly, the study allows us to predict fairly precisely the 
epoch by which time half the present stellar mass was in place; this 
is $z_{\frac{1}{2}}=2.0\pm0.2$. In \S4 we will discuss this conclusion
further attempting to verify it by measuring stellar masses of distant
galaxies directly. 

\begin{figure}
\centering
\includegraphics[height=8cm,angle=0]{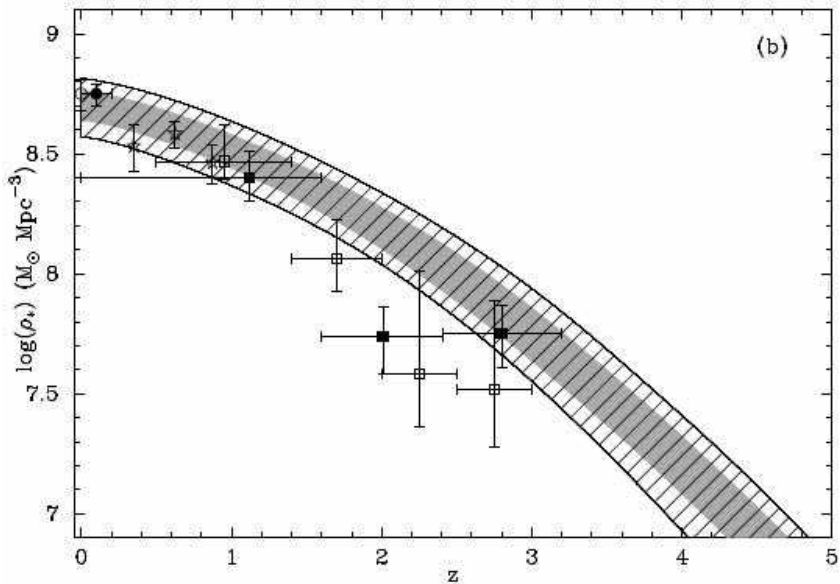}
\caption{Growth of stellar mass density, $\rho_{\ast}$, with redshift
obtained by direct integration of a parametric fit to the cosmic
star formation history deduced by Hopkins \& Beacom (2006, see
Figure 20). The integration accurately reproduces the local stellar
mass density observed by Cole et al (2001) and suggests half
the present density was in place at $z$=2.0 $\pm$ 0.2. }
\label{fig:21}      
\end{figure}

\subsection{Cosmic Star Formation - Theory}

As we have discussed, semi-analytical models have had a hard
time reproducing and predicting the cosmic star formation history.
Amusingly, as the data has improved, the models have largely
done a `catch-up' job (Baugh et al 1998, 2005a). To their credit,
while many observers were still convinced galaxies formed the bulk
of their stars in a narrow time interval (the `primeval galaxy' hypothesis),
CDM theorists were the first to suggest the extended star formation
histories now seen in Figure 20. 

A particular challenge seems to be that of reproducing the abundance
of energetic sub-mm sources whose star formation rates exceed 100-200 
$M_{\odot}$ yr$^{-1}$. Baugh et al (2005b) have suggested it may
require a combination of quiescent and burst modes of star formation,
the former involving an initial mass function steepened towards high mass
stars. Although there is much freedom in the semi-analytical models,
recent models suggest $z_{\frac{1}{2}}\simeq1.3$. By contrast, for the 
same cosmological models, hydrodynamical simulations (Nagamine et al 2004) 
predict much earlier star formation, consistent with $z_{\frac{1}{2}}\simeq2.0-2.5$.

The flexibility of these models is considerable so my personal view
is that not much can be learned from these comparisons either way.
It is more instructive to compare galaxy masses at various
epochs with theoretical predictions. Although we are still some ways from doing this in
a manner that includes both baryonic and dark components, progress
is already promising and will be reviewed in \S4.

\subsection{Unifying the Various High Redshift Populations}

Integrating the various star-forming populations at high redshift to
produce Figure 20 avoids the important question of the physical relevance and roles
of the seemingly-diverse categories of high redshift galaxies. In
the previous lecture (\S2), I introduced three broad categories:
the Lyman break (LBG), sub-mm and passively-evolving
sources (DRGs) which co-exist over 1$<z<$3. What is the relationship
between these objects?

As the datasets on each has improved, we have secured important
physical variables including masses, star formation rates and ages.
We can thus begin to understand not only their relative contributions
to the SFR at a given epoch, but the degree of overlap among the
various populations. Several recent articles have begun to evaluate 
the connection between these various categories (Papovich et al 2006, 
Reddy et al 2005).

A particularly valuable measure is the clustering scale, $r_0$, for each 
population, as defined in \S1.3. This is closely linked to the halo mass 
according to CDM and thus
sets a marker for connecting populations observed at different epochs.
Adelberger et al (1998) demonstrated the strong clustering, $r_0\simeq$3.8
Mpc, of luminous LBGs at $z\simeq$3. Baugh et al (1998) claimed
this was consistent with the progenitor halos of present-day massive
ellipticals. The key to the physical nature of LBGs depends the
origin of their intense star formation. At $z\simeq$3, the bright end
of the UV luminosity function is $\simeq$1.5 mags brighter than its
local equivalent; the mean SFR is 45 $M_{\odot}$ yr$^{-1}$. Is this 
due to prolonged activity, consistent with the build up of the bulk of 
stars which reside in present-day massive ellipticals, or is it a temporary
phase due to merger-induced starbursts (Somerville et al 2001).

Shapley et al (2001, 2003) investigated the stellar population and
stacked spectra of a large sample of $z\simeq$3 LBGs and find 
younger systems with intense SFRs are dustier with weaker Ly$\alpha$ 
emission while outflows (or `superwinds') are present in virtually all 
(Figure 22a).  For the young LBGs, a brief period of elevated star formation 
seems to coincide with a large dust opacity hinting at a possible overlap
with the sub-mm sources. During this rapid phase, gas and dust is depleted
by outflows leading to eventually to a longer, more quiescent phase 
during which time the bulk of the stellar mass is assembled.

If young dusty LBGs with SFRs $\simeq 300 M_{\odot}$ yr$^{-1}$ 
represent a transient phase, we might expect sub-mm sources to 
simply be a yet rarer, more extreme version of the same phenomenon. 
The key to testing this connection lies in the relative clustering scales
of the two populations (Figure 22b). Blain et al (2004) find sub-mm 
galaxies are indeed more strongly clustered than the average LBGs, 
albeit with some uncertainty given the much smaller sample size. 

\begin{figure}
\centerline{\hbox{
\psfig{file=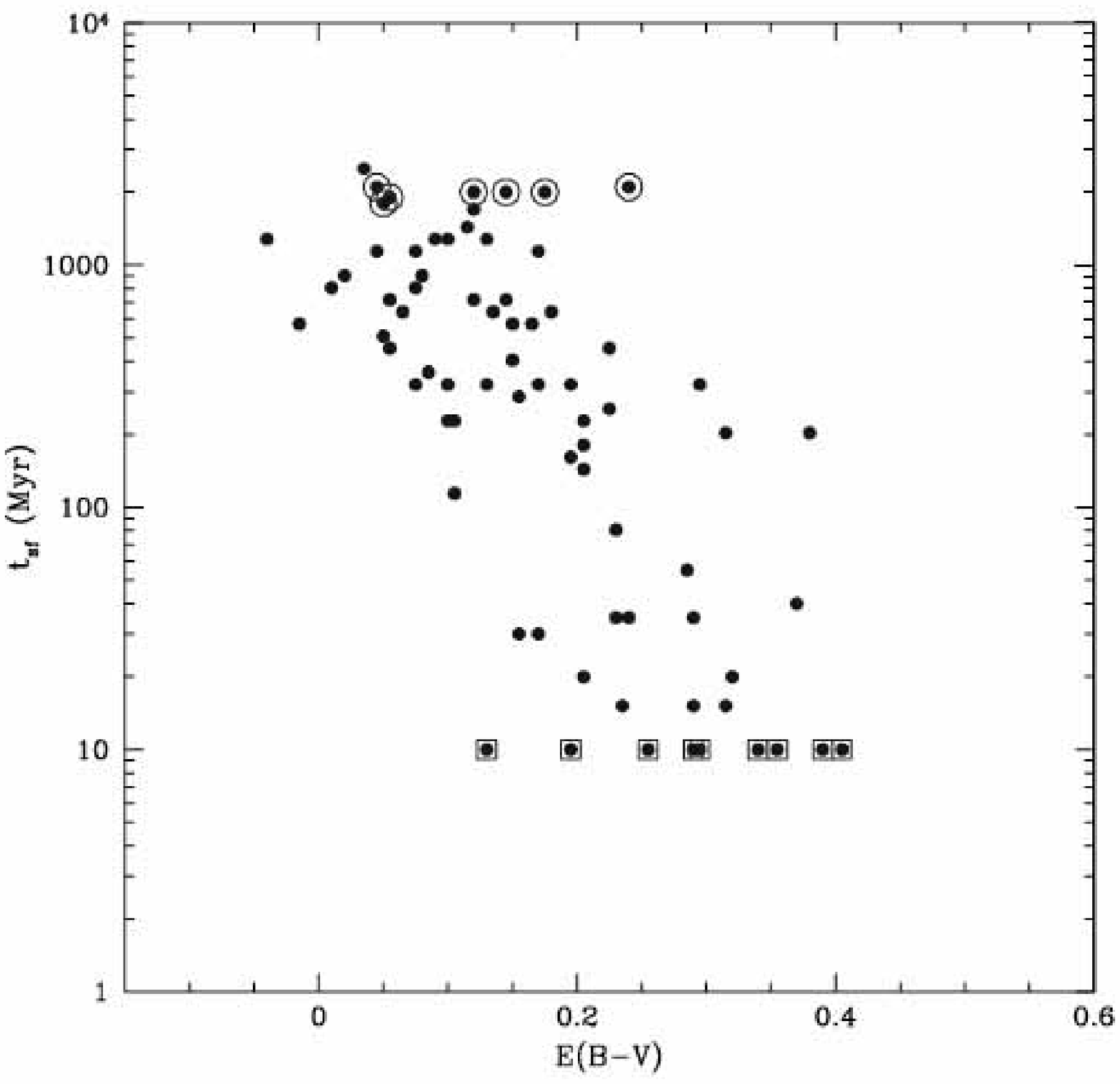,height=2.4in,angle=0}
\psfig{file=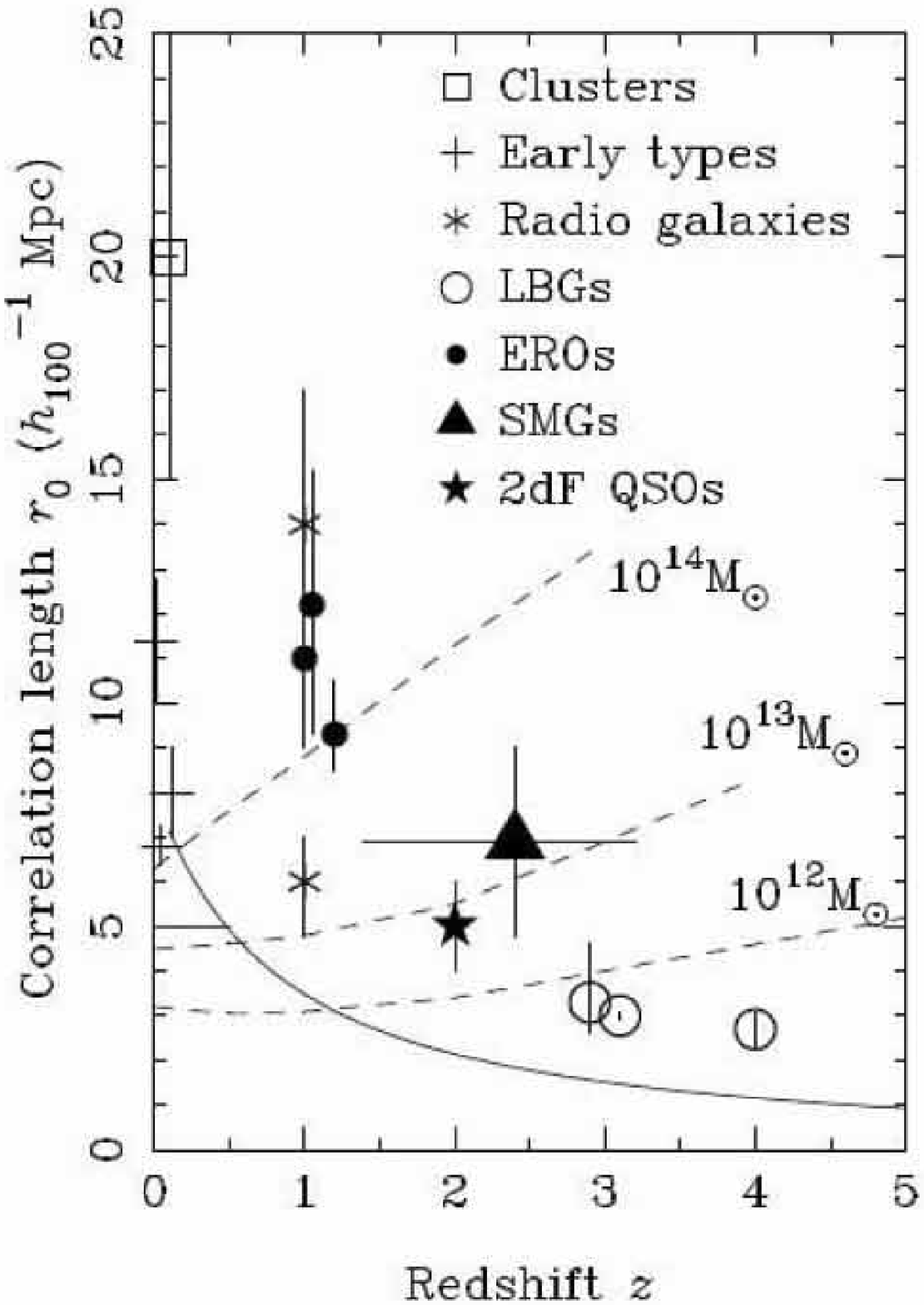,height=2.4in,angle=0}}}
\caption{Connecting Lyman break and sub-mm sources. (Left)
Correlation between the mean age (for a constant SFR) and 
reddening for a sample of $z\simeq$3 LBGs from the analysis
of Shapley et al (2001). The youngest LGBs seem to occupy a 
brief dusty phase limited eventually by the effect of powerful galaxy-scale 
outflows. (Right) The LBG-submm connection can be tested through 
reliable measures of their relative clustering scales, $r_0$ (Blain et al 2004). }
\label{fig:22}      
\end{figure}

Turning to the passively-evolving sources, although McCarthy (2004)
provides a valuable review of the territory, the observational situation is 
rapidly changing. For many years, CDM theorists predicted a fast decline with
redshift in the abundance of red, quiescent sources. Using a large
sample of photometrically-selected sources in the COMBO-17 survey,
Bell et al (2004) claimed to see this decline in abundance by witnessing 
a near-constant luminosity density in red sources to $z\simeq$1 (Figure 23a). 
The key point to understand here is that a passively-evolving galaxy 
fades in luminosity so that the red luminosity density should {\em increase} 
with redshift unless the population is growing. Bell et al surmises
the abundance of red galaxies was 3 times less at $z\simeq$1 as
predicted in early semi-analytical models (Kauffmann et al 1996).

By contrast, the Gemini Deep Deep Survey (Glazebrook et al 2004) 
finds numerous examples of massive red
galaxies with $z>1$ in seeming contradiction with the decline
predicted by CDM supported by Bell et al (2004). Of particular significance
is the detailed spectroscopic analysis of 20 red galaxies with $z\simeq$1.5
(McCarthy et al 2004) whose inferred ages are 1.2-2.3 Gyr implying
most massive red galaxies formed at least as early as $z\simeq$2.5-3
with SFRs of order 300-500 $M_{\odot}$ yr$^{-1}$. Could the most
massive red galaxies at $z\simeq$1.5 then be the descendants of
the sub-mm population? One caveat is that not all the stars whose
ages have been determined by McCarthy et al need necessarily
have resided in single galaxies at earlier times. The key question 
relates to the reliability of the abundance
of early massive red systems. Using a new color-selection technique,
Kong et al (2006) suggest the space density of quiescent
systems with stellar mass $>10^{11} M_{\odot}$ at $z\simeq$1.5-2
is only 20\% of its present value. 

As we will see in \S4, the key to resolving the apparent discrepancy 
between the declining red luminosity density of Bell et al and the
presence of massive red galaxies at $z\simeq$1-1.5, lies in
the mass-dependence of stellar assembly (Treu et al 2005).

Finally, a new color-selection has been proposed to uniformly
select all galaxies lying in the strategically-interesting redshift
range 1.4$<z<$2.5. Daddi et al (2004) have proposed the
`BzK' technique, combining $(z-K)$ and $(B-z)$ to locate
both star-forming and passive galaxies with $z>$1.4; such
systems are termed `sBzK' and `pBzK' galaxies respectively.
Reddy et al (2005) claim there is little distinction between the 
star-forming sBzK and Lyman break galaxies - both contribute similarly
to the star formation density over 1.4$<z<$2.6 and the overlap
fractions are at least 60-80\%.

More interestingly, both Reddy et al (2005) and Kong et al (2006) suggest
significant overlap between the passive and actively star-forming
populations. Kong et al find the angular clustering is similar and
Reddy et al find the stellar mass distributions overlap. 

\begin{figure}
\centerline{\hbox{
\psfig{file=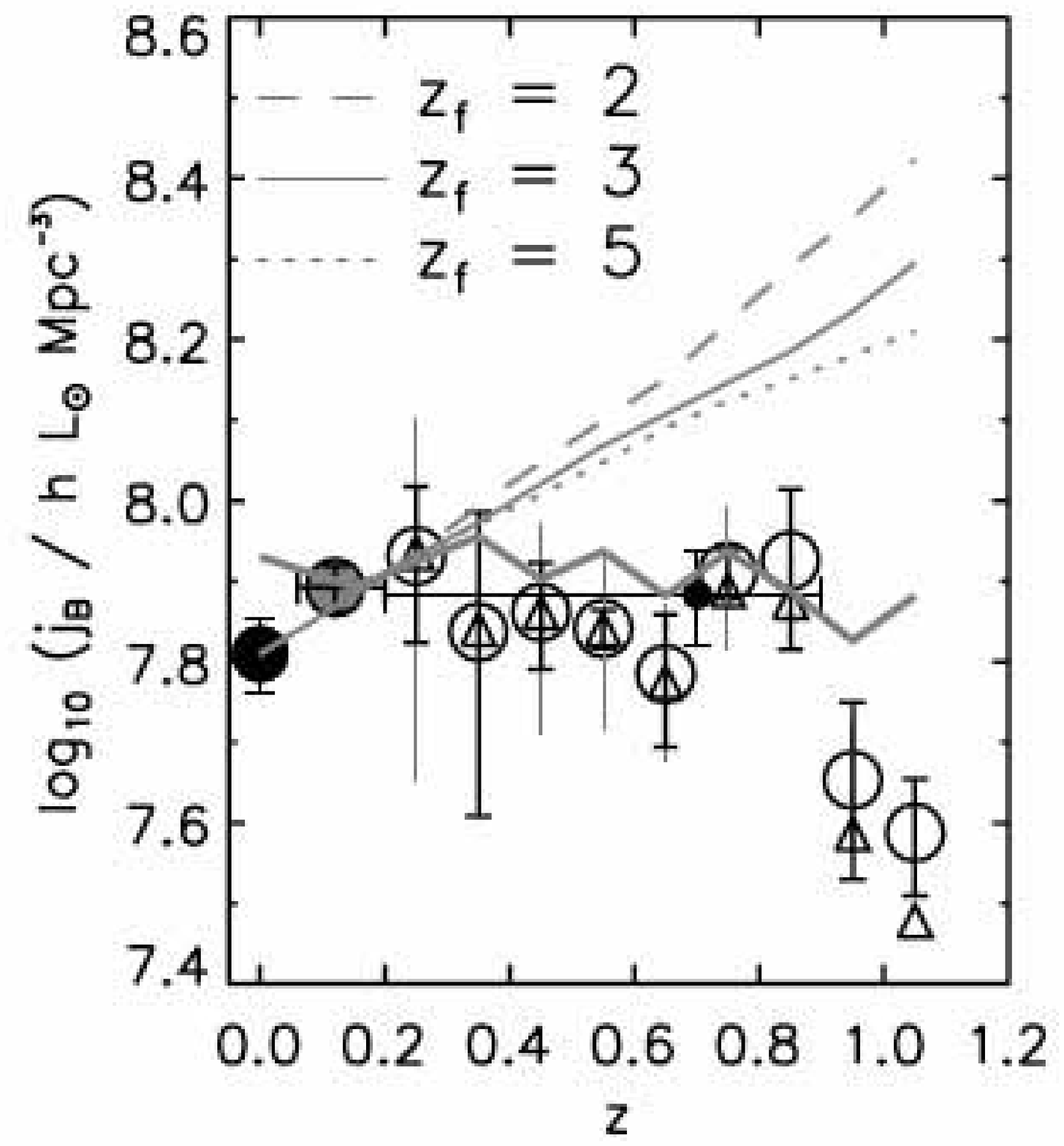,height=2.3in,angle=0}
\psfig{file=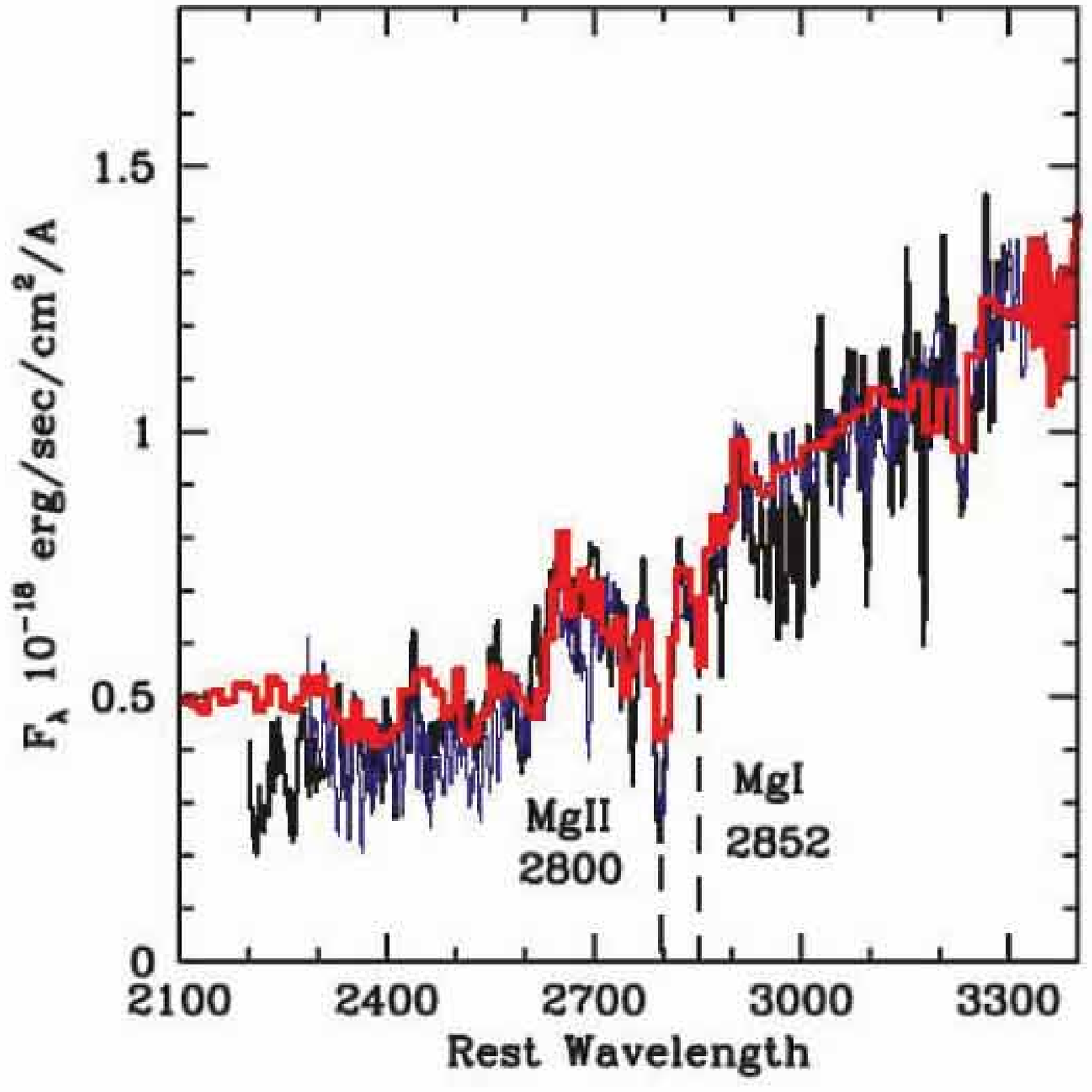,height=2.3in,angle=0}}}
\caption{Left: The rest-frame blue luminosity density of `red
sequence' galaxies as a function of redshift from the COMBO-17
analysis of Bell et al (2004). Since such systems should brighten
in the past the near-constancy of this density implies 3 times fewer red
systems exist at $z\simeq$1 than locally - as expected by standard
CDM models. Right: Composite spectra of red galaxies with 1.3$<z<$1.4 (blue),
1.6$<z<$1.9 (black) from the Gemini Deep Deep Survey (McCarthy et al
2004). A stellar synthesis model with age 2 Gyr is overlaid in red. The analysis
suggests the most massive red galaxies with $z\simeq$5 formed
with spectacular SFRs at redshifts $z\simeq$2.4-3.3. }
\label{fig:23}      
\end{figure}

\subsection{Lecture Summary}

Clearly multi-wavelength data is leading to a revolution in tracking
the history of star formation in the Universe. Because of the vagaries
of the stellar initial mass function, dust extinction and selection biases,
we need multiple probes of star formation in galaxies.

The result of the labors of many groups is a good understanding
of the comoving density of star formation since a redshift $z\simeq$3.
Surprisingly, the trends observed can account with reasonable precision
for the stellar mass density observed today. The implication of this
result is that half the stars we see today were in place by a redshift $z\simeq$2.

What, then, are we to make of the diversity of galaxies we observe
during the redshift range ($z\simeq$2-3) of maximum growth?
Through detailed studies some connections are now being made between
both UV-emitting Lyman break galaxies and dust-ridden sub-mm sources.

More confusion reigns in understanding the role and decline with redshift
in the contribution of passively-evolving red galaxies. Some observers
claim a dramatic decline in their abundance whereas others demonstrate
clear evidence for the presence of a significant population of old, massive
galaxies at $z\simeq$1.5. We will return to this enigma in \S4.

\newpage
 

\section{Stellar Mass Assembly}

\subsection{Motivation}

Although we may be able to account for the present stellar mass density by 
integrating the comoving star formation history (\S3), this represents only
a small step towards understanding the history of galaxy assembly. 
A major limitation is the fact that the star formation density
averages over a range of physical situations and luminosities; we are
missing a whole load of important physics. As we have seen, a single
value for the stellar mass density, e.g. $\rho_{\ast}(z=2)$, is useful when 
considered as a global quantity (e.g. compared to an equivalent estimate 
at $z$=0), but it  does not describe whether the observed star formation is steady or 
burst-like in nature, or even whether the bulk of activity within a given 
volume arises from a large number of feeble sources or a small 
number of intense objects. Such details matter if we are trying to 
construct a clear picture of how galaxies assemble.

Of course we could extend our study of time-dependent star formation 
to determine the {\it distribution functions} of star formation at various 
epochs (e.g. the UV continuum, $H\alpha$ or sub-mm luminosity functions),
but making the integration check only at $z$=0 is second-best to measuring 
the {\it assembled mass} and its distribution function at various redshifts. 
This would allow us to directly witness the growth rate of galaxies of 
various masses at various times and, in some sense, is a more profound 
measurement, closer to theoretical predictions.

Ideally we would like to measure both baryonic and non-baryonic masses
for large numbers of galaxies but, at present at least, we can only make
dynamical or lensing-based total mass estimates for specific types of distant
galaxy and crude estimates for the gaseous component. The bulk of the
progress made in the last few years has followed attempts
to measure {\it stellar masses} for large populations of galaxies. We will
review their achievements in this lecture.

\subsection{Methods for Estimating Galaxy Masses}

What are the options available for estimating galaxy masses of any
kind at intermediate to high redshift? Basically, we can think of three
useful methods. 

\begin{itemize}

\item{} Dynamical methods based on resolved rotation curves for
recognisable disk systems (Vogt et al  1996,1997) or stellar velocity 
dispersions for pressure-supported spheroidals (van Dokkum \& Ellis
2003, Treu et al 2005). These methods only apply for systems known
or assumed {\it a priori} to have a particular form of velocity field. Interesting 
constraints are now available for a few hundred galaxies in the field and in 
clusters out to redshifts of $z\simeq$1.3. Key issues relate to biases
associated with preferential selection of systems with `regular' appearance
and how to interpret mass dynamically-derived over a limited physical
scale (c.f. Conselice et al 2005). In the absence of resolved rotation
curves, sometimes emission linewidths are considered a satisfactory
proxy (Newman \& Davis 2000).

\item{} Gravitational lensing offers the cleanest probe of the total
mass distribution but, as a geometric method, is restricted in its
application to compact, dense lenses (basically, spheroidals)
occupying cosmic volumes typically half way to those probed
by faint star-forming background field galaxies. In practice this
means $z_{lens}<$1. Even so, by sifting through spectra
of luminous red galaxies in the SDSS survey and locating cases
where an emission line from a background lensed galaxy 
enters the spectroscopic fiber, Bolton et al (2006) have identified a 
new and large sample of Einstein rings enabling us to gain
valuable insight into the relative distribution of dark and visible
mass over 0$<z<$1.

\item{} Stellar masses derived from near-infrared photometry 
represents the most popular technique in use at the current time.
The idea has its origins in the recognition (Broadhurst, Ellis \&
Glazebrook 1992, Kauffmann \& Charlot 1998) that the rest-frame 
$K$-band luminosity of a galaxy is less affected by recent star formation 
than its optical equivalent (Figure 16), and thus can act as a closer proxy to the 
well-established stellar population. A procedure for fitting the rest-frame 
optical-infrared spectral energy distribution of a distant galaxy, deriving a 
stellar mass/light ratio ($M/L_K$) and hence the stellar mass if the 
redshift is known, was introduced by Brinchmann \& Ellis (2000). 
The popularity of the technique follows from the fact it can easily be 
applied to large catalogs of galaxies in panoramic imaging surveys and 
extended to very high redshift if IRAC photometry is available.
The main difficulties relate to the poor precision of the method, particularly
if the same photometric data is being used to estimate the redshift (Bundy et al
2006), plus degeneracies arising from poor knowledge of the past star formation
history (Shapley et al 2005, van der Wel et al 2006). 

\end{itemize}

For many galaxies, an important and usually neglected component is the mass locked
up in both ionized and cool gas. In nearby systems amenable to study of hot
ionized gas (from nebular emission lines) and its usually dominant cooler 
neutral component (probed by 21cm studies), as much as 20\% of the 
baryonic mass of a luminous star-forming galaxy can be found in this form 
(Zwaan et al 2003). At present, it is not possible to routinely use radio techniques
to reliably estimate gaseous masses of distant galaxies although approximate 
gas masses have been derived assuming the projected surface density of nebular 
emission correlates with the gas mass within some measured physical scale 
(Erb et al 2006).

\subsection{Results: Regular Galaxies 0$<z<$1.5}

Because of the simplicity of their stellar populations, velocity fields and the 
lack of confusing gaseous components, rather more is known about the
mass assembly history of ellipticals than for spirals. Concerning ellipticals,
one of the key challenges is separating the {\it age of the stars} from the 
{\it age of the assembled mass}. 

The Fundamental Plane (FP, Dressler et al 1987, Djorgovski \& Davis 1987,
Bender, Burstein \& Faber 1992, Jorgensen et al 1996) represents an
empirical correlation between the dynamical mass (via the central
stellar velocity dispersion $\sigma_0$), the effective radius ($R_E$) and 
light distribution (via the enclosed surface brightness $\mu_E$) for
ellipticals, viz:

$$ log R_E = a \, log \, \sigma_0 \, + \, b \, \mu_E \, + \, c$$

For example, with $\sigma$ in km sec$^{-1}$, $R_E$ in kpc and $\mu_E$ in
mags arcsec$^{-2}$ in the $B$ band, $a$=1.25, $b$=0.32, $c$=-8.970 for
$h$=0.65.
 
These observables define an effective dynamical mass $M_E \propto \sigma^2\,R_E/G$
which correlates closely with that deduced from lensing (Treu et al 2006). 
Deviations from the local FP at a given redshift $z$ can be used to 
deduce the change in mass/light ratio $\Delta\,log(M/L)$.

The most comprehensive studies of the evolving FP come from two
independent and consistent studies of field spheroidals to $z\simeq$1 
(Treu et al 2005, van der Wel et al 2005). The evolution in mass/light
ratio $\Delta\,log(M/L)$ deduced from the GOODS-N survey of Treu
et al (2005) is shown in Figure 24. These authors find as little as 
1-3\% by stellar mass of the present-day population in massive ($>10^{11.5}
M_{\odot}$) galaxies formed since $z$=1.2, whereas for low mass systems
($<10^{11}M_{\odot}$) the growth fraction is 20-40\%. This result, confirmed
independently by van der Wel et al (2005), is an important illustration
of the mass-dependent growth in galaxies with the most massive systems
shutting off earliest. 

\begin{figure}
\centering
\includegraphics[height=7cm,angle=0]{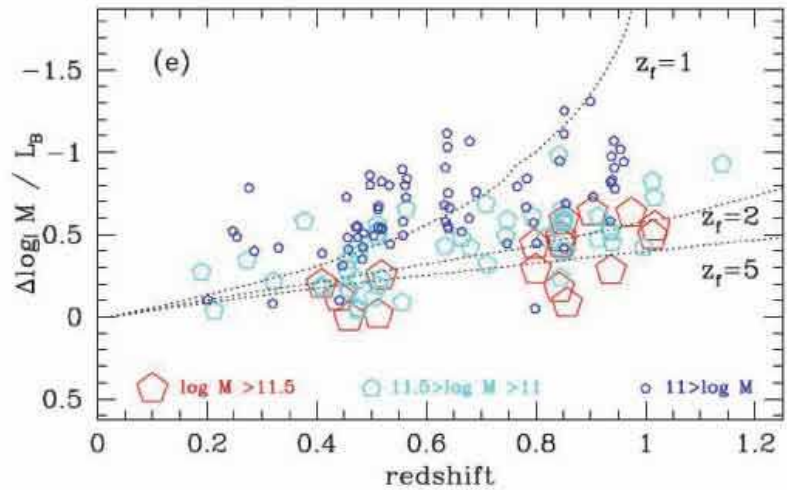}
\caption{Change in $B$-band mass/light ratio with redshift deduced from the Keck
dynamical survey of over 150 field spheroidals in GOODS-N (Treu et al 2005). 
The plot shows the change in mass/light ratio for galaxies of different effective
dynamical masses (red: massive, cyan: intermediate, blue: low mass). Curves
illustrate the change in mass/light ratio expected as a function of redshift
for galaxies assembled monolithically with simply evolving stellar populations
since a redshift of formation $z_f$. Clearly the stars in the most massive
galaxies largely formed at high redshift whereas assembly continued apace
in lower mass systems.}
\label{fig:24}      
\end{figure}

Of course, one should not confuse the {\it age of stars}, as probed by the FP,
with the {\it age of the assembled mass}. van Dokkum (2005) has argued
that if spheroidals preferentially merge with similar gas-poor systems (a process
called `dry mergers') the FP analyses could well indicate early eras of major 
star formation even though the bulk of the assembled mass in individual
systems occurred at $z<$1 \footnote{The reason observers have gone somewhat 
out of their way to consider such complicated scenarios is because late assembly 
of massive spheroidals was, until recently, a fundamental tenet of the CDM 
hierarchy to be salvaged at all costs (Kauffmann et al 1996).}. Bell et al 
(2006) and Tran et al (2005) have cataloged individual cases of dry mergers in both
field and cluster samples, respectively. Their occurrence is not in dispute;
however, only via morphological or other measures of the global mass
assembly can the role of dry mergers as a major feature of galaxy assembly
be addressed.

Related insight into this problem arises from the relative distributions of 
baryonic and dark matter deduced from the combination of lensing and 
stellar dynamics for the recently-located SDSS-selected Einstein rings 
(Treu et al 2006, Figure 25). Although it might be thought that lensing 
preferentially selects the most massive and compact sources, Treu et al
compare the FP of such lenses with those in the larger field sample (Fig. ~25)
and deduce otherwise. An important result from the study of the first set of 
such remarkable lenses is how well the total mass profile can be represented 
by an isothermal form with mass tracing light, viz:

$$\rho_{tot} = \rho\,( \frac{r}{r_0})^{-\gamma}$$

Even over 0$<z<$1, the mass slope $\gamma$ is constant at 2.0 to within 
2\% precision indicating rather precise collisional coupling of dark matter and
gas. 

\begin{figure}
\psfig{file=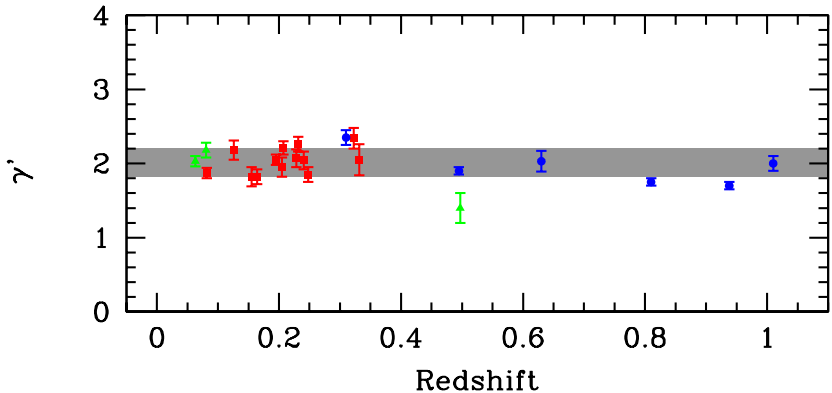,height=2.3in,angle=0}
\caption{Distribution of the gradient $gamma$ in the total mass profile with
redshift as deduced from the combination of stellar dynamics and lensing
for ellipticals in the SLAC survey (Treu et al, in prepn). The remarkable
constancy of the profile slope indicates massive relaxed systems were in place
at $z\simeq$1 and that `dry merging' cannot be a prominent feature of
the assembly history of large galaxies. }
\label{fig:25}      
\end{figure}

Sadly, far less is known about the mass assembly history of regular spirals.
The disk scaling law equivalent to the FP for ellipticals, the Tully-Fisher relation (Tully \&
Fisher 1977) which links rotational velocity to luminosity gives ambiguous
information without additional input. Modest evolution in the TF
relationship was deduced from the pioneering Keck study of $z\simeq$1 spirals
by Vogt et al (1997) but this could amount to $\simeq$0.6 mag of $B$-band
luminosity brightening in sources of a fixed rotational velocity to $z\simeq$1,
or more enhancement if masses were reduced. 

Additional variables capable of breaking the degeneracy between dynamical
mass and luminosity include physical size and stellar mass.
Lilly et al (1998) examined the {\it size-luminosity} relation for several hundred
disks to $z\simeq$1 in a HST-classified redshift survey sample (see Sargent
et al 2006 for an update) and found no significant growth for the largest systems. 
Conselice et al (2005) correlated stellar and dynamical masses for $\simeq$100 
spirals with resolved dynamics in the context of a simple halo formulation. 
Although their deduced halo masses must be highly uncertain, they likewise
deduced that growth must be modest since $z\simeq$1, occurring in a self-similar
fashion for the baryonic and dark components.

\subsection{Stellar Masses from Multi-Color Photometry} 

Bundy et al (2005, 2006) give a good summary and critical analysis of  
the now well-established practice of estimating stellar masses from multi-color 
optical-infrared photometry. Figure 26 gives a practical illustration of the 
technique where it can be seen that even for low $z$ galaxies with good
photometry, the precision in mass is only $\simeq\pm$0.1-0.2 dex. In
most cases, even random uncertainties are at the $\pm$0.2-0.3 dex level
and systematic errors are likely to be much higher.

\begin{figure}
\psfig{file=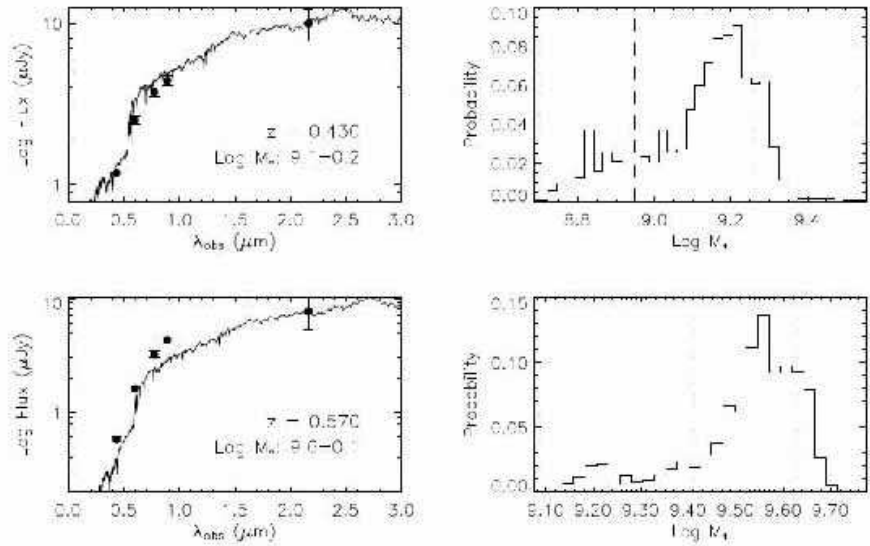,height=3.0in,angle=0}
\caption{Deriving stellar masses for galaxies of known redshift via multi-color
optical photometry (after Bundy et al 2006). (Left) Rest-frame spectral
energy distribution of two DEEP2 galaxies of known redshift from broad-band 
$BRIK$ photometry with a  fit deduced by fitting from a stellar synthesis
library. The fit yields the most likely mass/light ratio of the observed population. 
(Right) Likelihood distribution of the stellar mass for the same two
galaxies derived by Bayesian analysis. Dashed lines indicate the eventual
range of permitted solutions. }
\label{fig:26}      
\end{figure}

Since analysing stellar mass functions is now a major industry in the community,
it is worth spending some time considering the possible pitfalls.
A significant fraction of the large datasets being used are purely photometric, 
with both redshifts and stellar masses being simultaneously deduced from 
multi-color photometry (Drory et al 2002, Bell et al 2004, Fontana et al 2004). 
Few large surveys have extensive spectroscopy from which to check that this 
procedure works.

Bundy et al (2006) address this important question in the context of their
extensive DEEP2 spectroscopic sample by artificially `switching off' the knowledge
of the spectroscopic $z$: how does the stellar mass deduced when simultaneously
deriving the photometric redshift from the same data compare with that derived when 
the spectroscopic value is externally input? Figure 27 illustrates
that a significant component of error is one beyond that expected solely
from the error in the derived redshift (as measured from the $z_{spec}$
vs. $z_{photo}$ comparison).

\begin{figure}
\centerline{\psfig{file=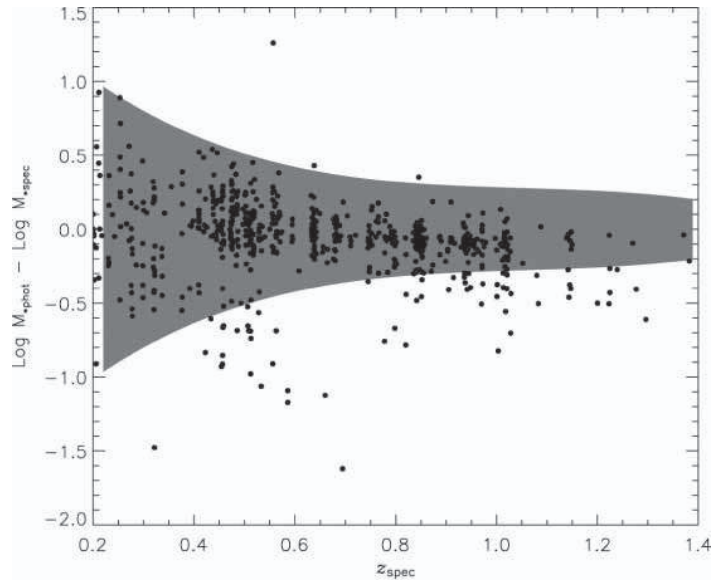,height=3.0in,angle=0}}
\caption{Error in log stellar mass when the known spectroscopic redshift
is ignored (Bundy et al 2005). The shaded region defines the expected
scatter in mass arising solely from that in the photometric-spectrosopic redshift
comparison. Clearly in addition to this component, some catastrophic
failures are evident.}
\label{fig:27}      
\end{figure}

A second restriction in many stellar mass determinations is the absence
of any near-infrared photometry. This may seem surprising given the
classic papers (Kauffmann \& Charlot 1998, Brinchmann \& Ellis 2000)
stressed its key role. However, panoramic near-infrared imaging is
much more expensive in telescope time as few observatories, until
recently, had large format infrared cameras. Inevitably, some groups
have attempted to get by without the near-infrared data.

Figure 28 illustrates how the precision degrades when the $K$ band
data is dropped in the stellar mass fit (Bundy et al 2006)\footnote{A similar
analysis was conducted by Shapley et al (2005) at $z\simeq$2, excluding
the relevant IRAC data.}. Although the systematic error is contained, the
noise increases significantly for redshifts $z>$0.7 where the optical
photometry fails to adequately sample the older, lower mass stars.

\begin{figure}
\psfig{file=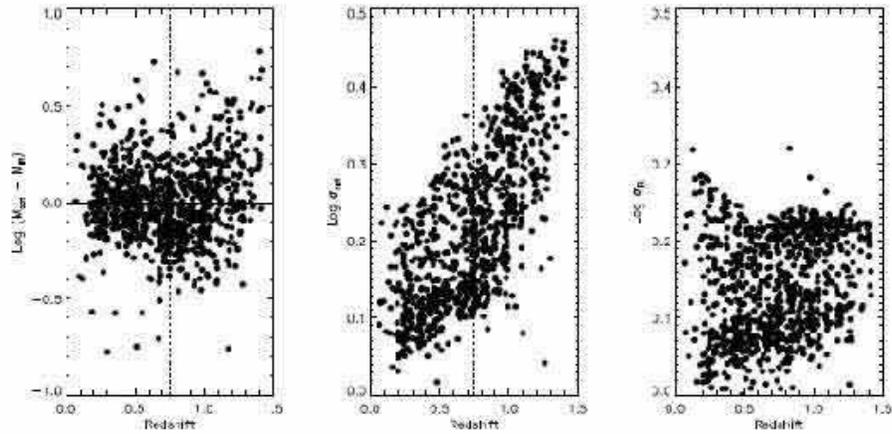,height=2.5in,angle=0}
\caption{Effect of deriving stellar masses from optical photometry
alone (Bundy et al 2006). (Left) Difference in masses derived using
$BRIK$ and $BRI$ photometry. (Center) Distribution of uncertainties
deduced from Bayesian fitting for the $BRI$ fits. (Right) as center for
the $BRIK$ fitting.  Although the systematic offset is small, the uncertainty
in deduced stellar mass increases dramatically for galaxies with $z>$0.7.}
\label{fig:28}      
\end{figure}

Shapley et al (2005) explore a further uncertainty, which is particularly
germane to the analysis of their $z\simeq$2 sources. Using IRAC to represent 
the rest-frame infrared at these redshifts, they consider the role that
recent bursts of star formation might have on the deduced stellar
masses. Although bursts predominantly affect short-wavelength 
luminosities, one might imagine little effect at the longer wavelengths
(c.f. Kauffmann \& Charlot 1998). Shapley et al consider a wider range
of star formation histories and show that the derived stellar mass depends 
{\it less strongly} on the long wavelength luminosity $L$ (4.5$\mu$m)
than expected. The scatter is consistent with variations in mass/light ratios of
$\times$15 (Figure 29). The strong correlation between mass and 
optical-infrared color emphasizes the importance of secondary activity.

\begin{figure}
\centerline{\hbox{
\psfig{file=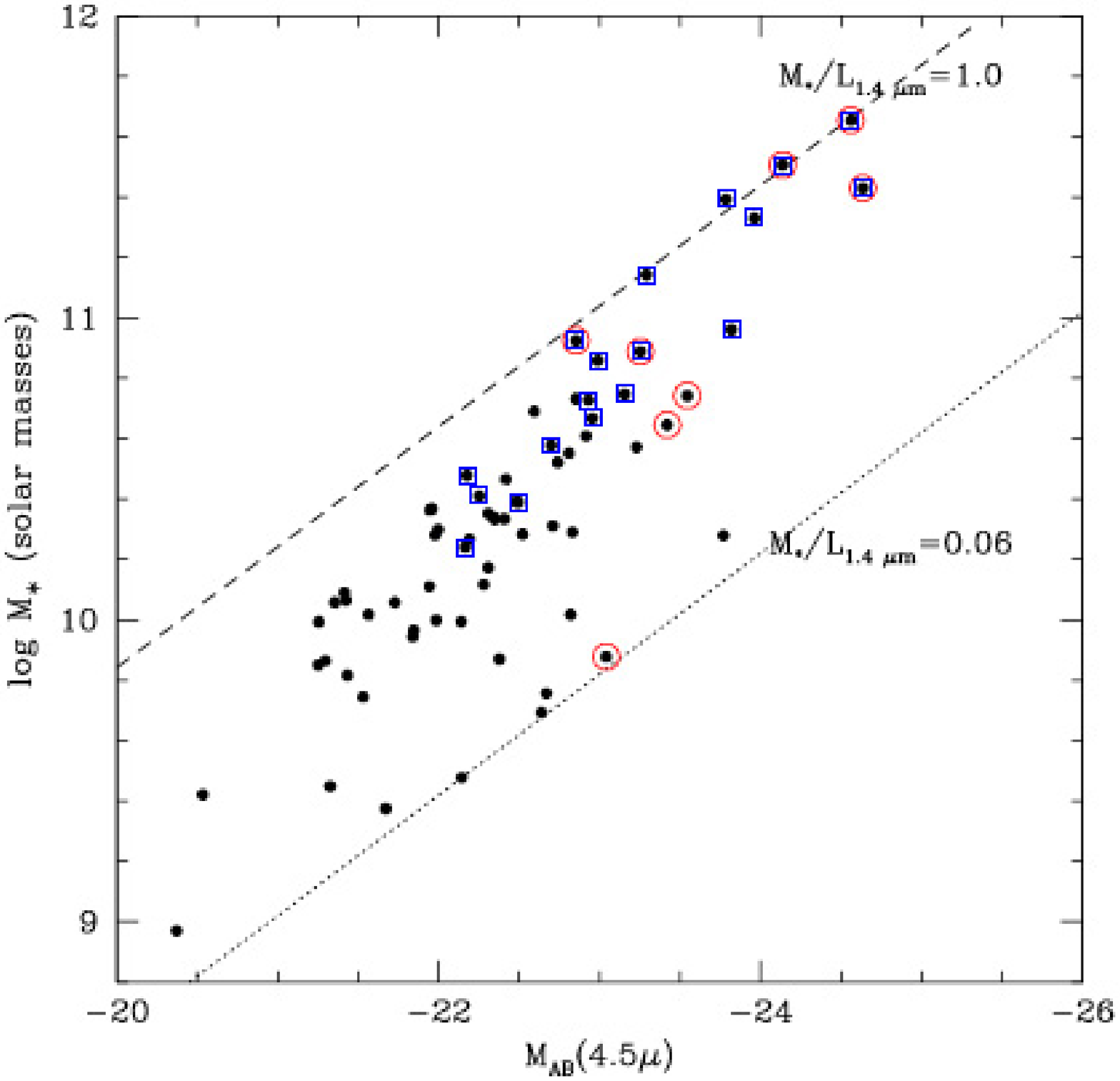,height=2.2in,angle=0}
\psfig{file=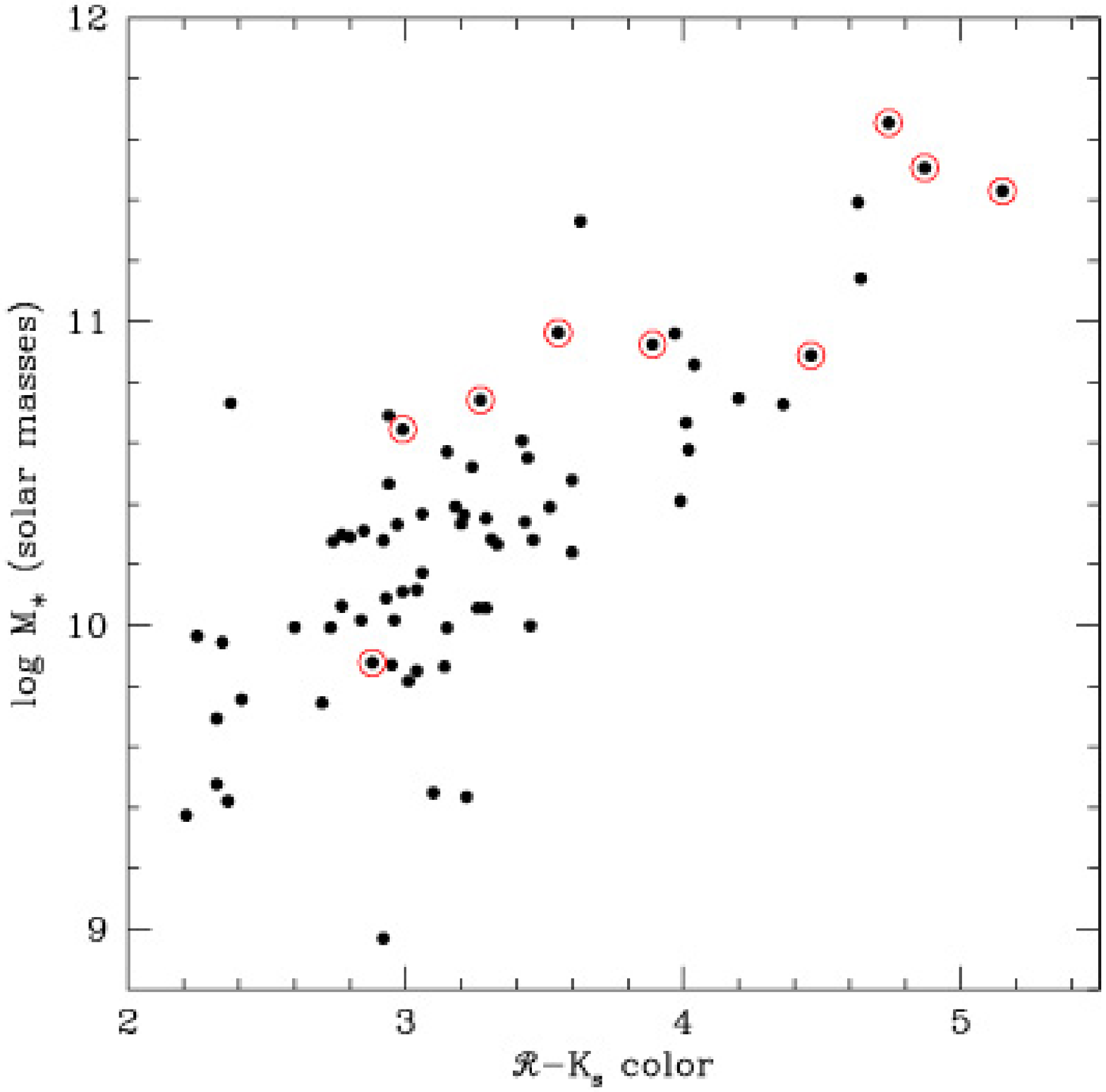,height=2.2in,angle=0}
}}
\caption{Left: Stellar masses versus observed IRAC 4.5$\mu$m luminosities
for the sample of $z\simeq$2 galaxies studied by Shapley et al 
(2005). Allowing a wide range of star formation histories illustrates
a weaker dependence of mass on long wavelength luminosity than
seen in earlier tests at lower redshift based on simpler star
formation histories. The range in rest-frame 1.6$\mu$m mass/light 
ratio is $\simeq\times$15.  (Right) The existence of a fairly tight
correlation between mass and $R-K$ color demonstrates the
importance of considering secondary bursts of star formation in
deriving stellar masses.}
\label{fig:29}      
\end{figure}

\subsection{Results: Stellar Mass Functions 0$<z<$1.5}

Noting the above uncertainties, we now turn to results compiled from 
the stellar mass distributions of galaxies undertaken in both
spectroscopic and photometric surveys. A recent discussion of
the various results can be found in Bundy et al (2006).

Brinchmann \& Ellis (2000) were the first to address the
global evolution of stellar mass in this redshift range using a 
morphological sample of 350 galaxies with spectroscopic
redshifts and infrared photometry. The small sample did
not permit consideration of detailed stellar mass functions, but 
the integrated mass density was partitioned in 3 redshift bins
for spheroidals, spirals and irregulars. Surprisingly little growth
was seen in the overall mass density from $z\simeq$1 to today; 
the strongest evolutionary signal seen is a {\it redistribution}
of mass amongst the morphologies dominated by a declining
mass density in irregulars with time.  

The ratio of the galactic stellar mass $M_{star}$ to the current 
star formation rate $SFR$, e.g. as deduced spectroscopically or 
from the rest-frame optical colors, is sometimes termed the 
{\it specific star formation rate}, $R$. This quantity allows us to 
address the question of whether galaxies have 
been forming stars for a significant fraction of the Hubble time, at a rate 
commensurate with explaining their assembled mass. A low value for $R$ 
implies a quiescent object whose growth has largely ended; the mean stellar 
age is quite large. A high value of $R$ implies an active object which has
assembled recently. A frequently-used alternative is the `doubling time'
- that period over which, at the current SFR, the observed stellar mass would
double. This time would be quite short for active objects.

Cowie et al (1996), Brinchmann \& Ellis (2000), and most recently Juneau et al
(2005), found a surprising trend whereby most massive galaxies 
over $z\simeq$0.5-1.5 are quiescent, having presumably formed 
their stars well before $z\simeq$2, whereas low mass galaxies remain 
surprisingly active. The term `downsizing' - a signature 
of continued growth in lower mass galaxies after that in the high mass galaxies has
been completed - was first coined by Cowie et al (1996) and has been used
rather loosely in the recent literature to imply any signature of anti-hierarchical
activity. In particular, it is important to distinguish between {\it downsizing
in star formation activity}, which presumably represents some physical
process that permits continued star formation in lower mass systems
when that in massive galaxies has concluded, from {\it downsizing
in mass assembly}, a truly `anti-hierarchical' process whereby new mass
is being added to lower mass galaxies at later times (see discussion
in Bundy et al 2006).

Before trying to understanding in more detail what causes this mass-dependent
star formation, it is worth returning to the issue of {\it dry mergers} raised in $\S4.3$.
A `downsizing' signature was also seen in the growth of spheroidal galaxies
as analysed by their location on the evolving Fundamental Plane (Treu et al
2005, van der Wel et al 2005). However, van Dokkum (2005) argued that 
while this may reflect older stars in the more massive galaxies, the
{\it assembled mass} may still be young if merging preferentially
occurs between quiescent objects. The only way to test this
hypothesis is to directly measure the growth rate in spheroidal
systems.

Using the deeper and more extensive sample of galaxies available 
in the GOODS fields, Bundy et al (2005) produced 
type-dependent {\it stellar mass functions} (Figure 30). Here, for
the first time, one can see the morphological evolution detected by 
Brinchmann \& Ellis largely arises via the transformation of intermediate 
mass  ($\simeq$5. 10$^{10}$ - 2.10$^{11} M_{\odot}$)  irregulars and spirals 
into spheroidals. If merging is responsible for this transformation, it is
predominantly occurring between gas-rich and active systems even
at the very highest masses. The bulk of the evolution below $z\simeq$1 is 
simply a redistribution of star formation activity, perhaps as a result
of mergers or feedback processes.

\begin{figure}
\psfig{file=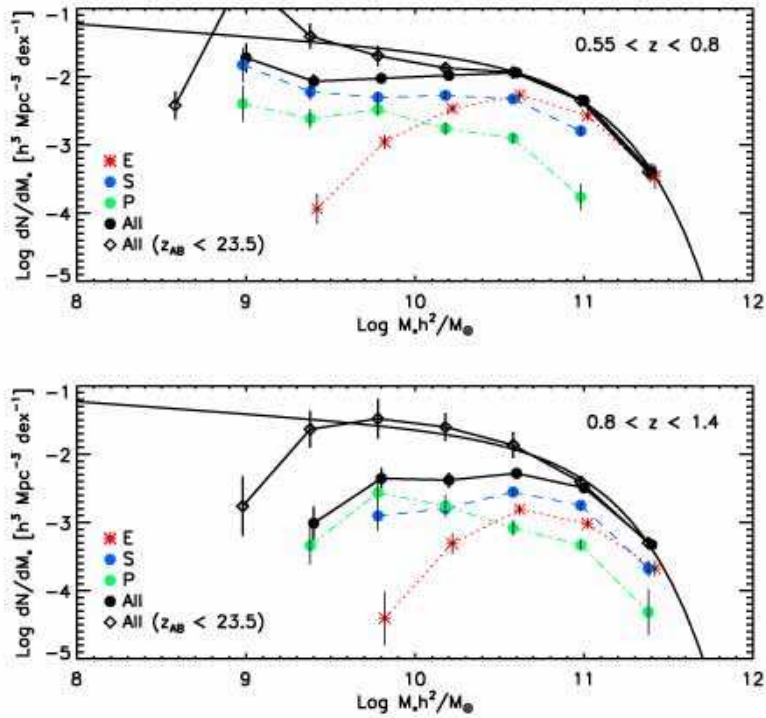,height=4.1in,angle=0}
\caption{Stellar mass function arranged by morphology in two redshift
bins from the analysis of Bundy et al (2005) for $h$=1. The solid curve in
both plots represents the present day mass function from the 2dF
survey (Cole et al 2003). Type-dependent mass functions are
color-coded with black representing the total. The solid line connecting
the filled black circles represents the sample with spectroscopic redshifts, the
dotted line connecting the open black diamonds includes masses 
derived from sources with photometric redshifts. The vertical dashed
lines represent completeness limits for all types.}
\label{fig:30}      
\end{figure}

Interestingly, in Figure 30 there is almost no change in the
total mass function with time above  5. 10$^{10}M_{\odot}$,
suggesting little growth in the mass spheroidals that dominate
the high mass end. However, it worth remembering that the two 
GOODS fields are limited in size (0.1 deg$^2$) and suffer from cosmic 
variance at the 20\% level at high redshift increasing to 60\% at low 
redshift (Bundy et al 2005). 

Stellar mass functions for a much larger sample of field galaxies
of known redshift have been analyzed by Bundy et al (2006)
utilizing the combination of extensive spectroscopy and Palomar 
$K$-band imaging in four DEEP2 fields (totalling 1.5 deg$^2$).
This sample has the benefit of being much less affected by
cosmic variance although, as there is no complete coverage
with HST, morphological classifications are not possible.   
Color bimodality has been analyzed in the DEEP2 sample (Weiner
et al 2005) and Bundy et al use the rest-frame $U-B$ color
and spectroscopic [O II] equivalent width to separate quiescent
and active galaxies. 

Figure 31 shows a direct comparison of the integrated stellar
mass functions from this large survey alongside other, less
extensive surveys, most of which are based only on photometric
redshifts. Although each is variance limited in different ways,
one is struck again by the quite modest changes in the abundance
of massive galaxies since $z\simeq$1. 
 
\begin{figure}
\centerline{\psfig{file=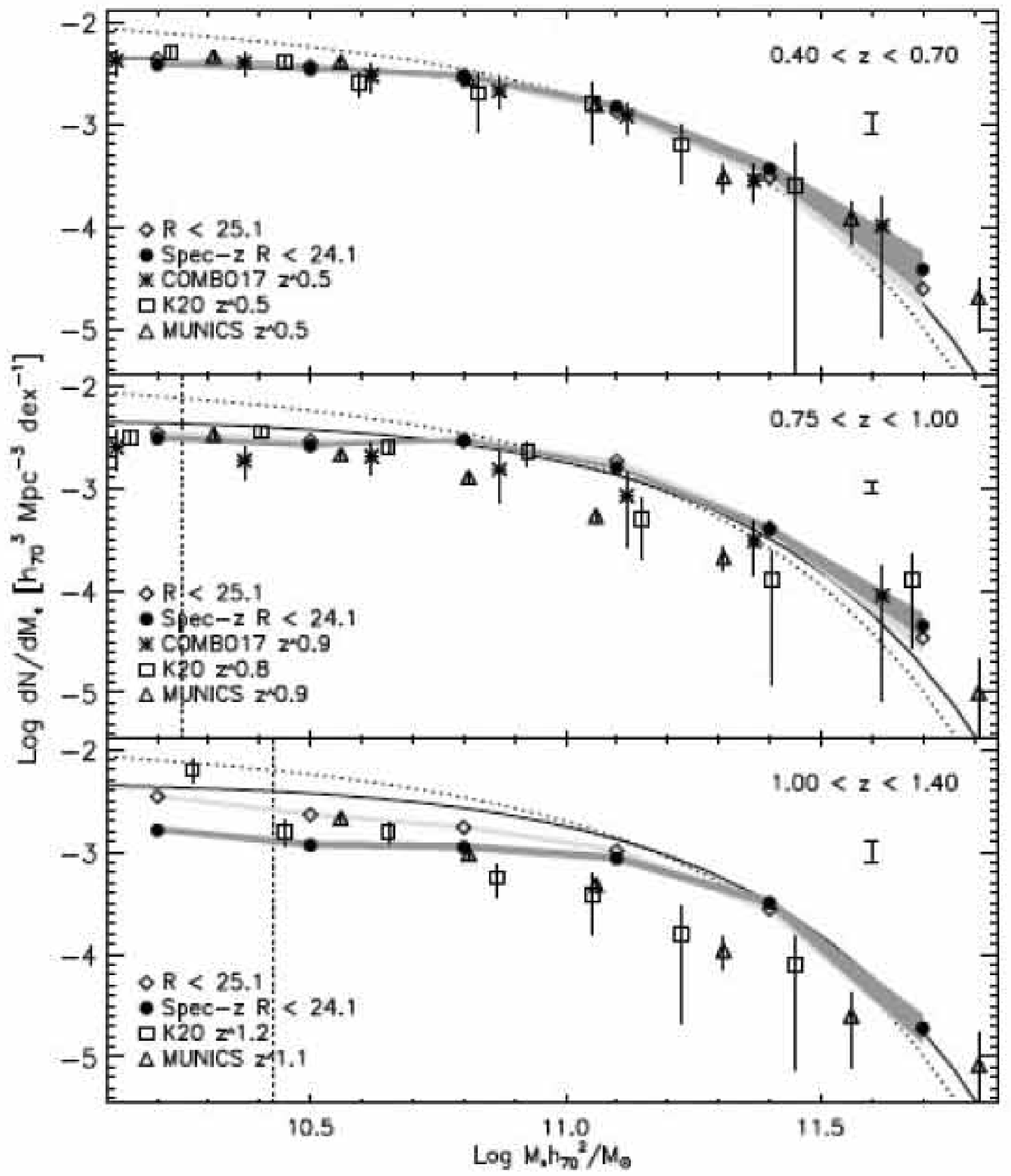,height=4.4in,angle=0}}
\caption{The evolving stellar mass function from the comparison
of Bundy et al (2006). }
\label{fig:31}      
\end{figure}

Next we consider (Figure 32) the stellar mass functions for the quiescent
and active star-forming galaxies independently, partitioned according
to the rest-frame $U-B$ color. The surprising result here is the
existence of a threshold or {\it quenching mass}, $M_{Q}$ above 
which there are no active systems. This is implied independently 
in Figure 30 where, within the redshift bin $0.55<z<0.8$, there 
are no spirals or irregulars with mass $>10^{11}M_{\odot}$. Interestingly,
there is also a modest downward transition in $M_Q$ with time.
 
\begin{figure}
\psfig{file=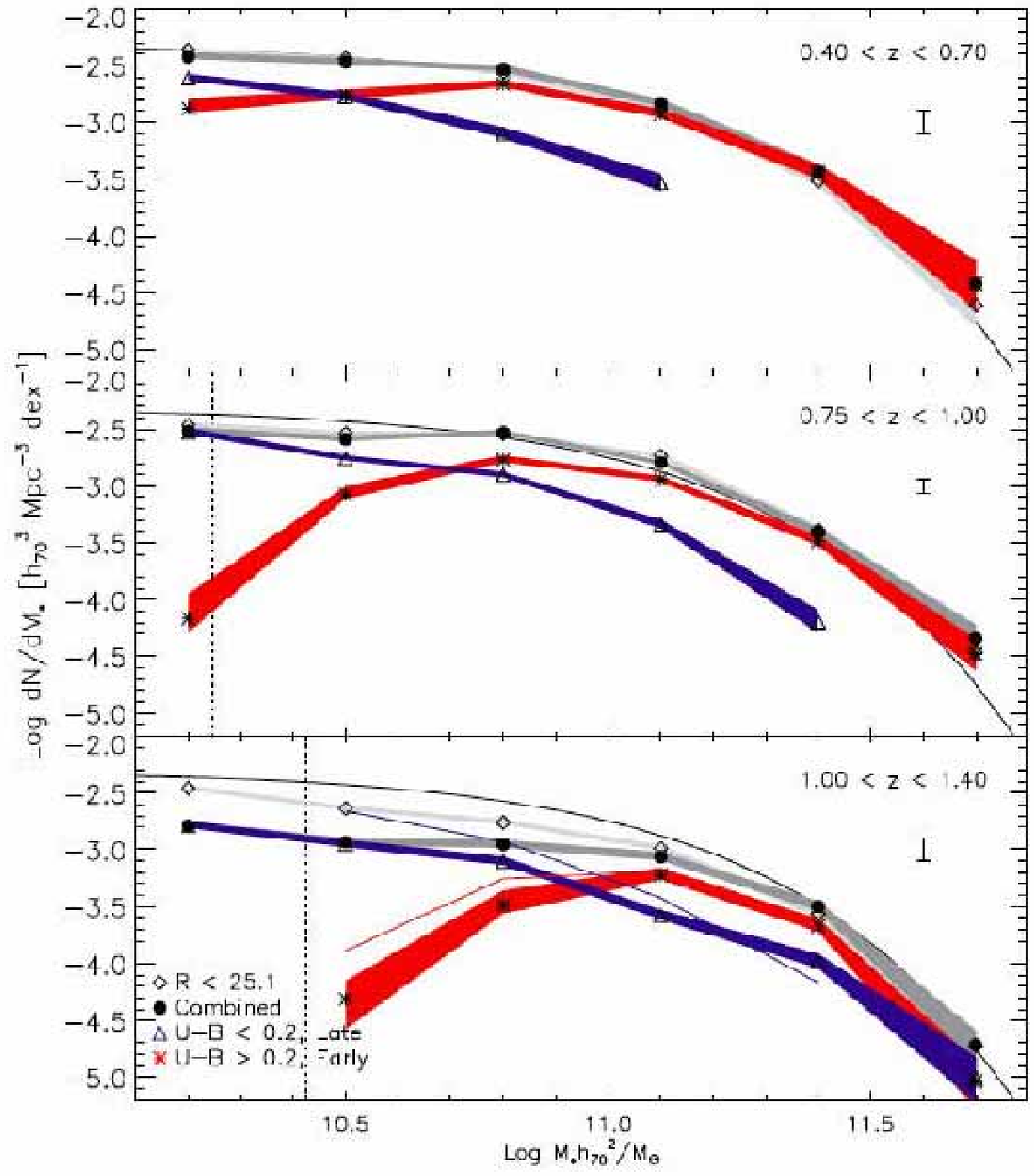,height=4.4in,angle=0}
\caption{The evolving stellar mass function split into quiescent (red)
and active (blue) sources from the survey of Bundy et al (2006). 
Uncertainties arising from counting statistics and errors in the
stellar masses are indicated by the associated shading.
The thin solid line represents the local 2dF function (Cole et al 2003)
and the dark and light grey regions represent total mass functions
using only spectroscopic and including photometric redshifts 
respectively.}
\label{fig:32}      
\end{figure}

A remarkably consistent picture emerges from these studies.
Over the redshift range 0$<z<$1.5, the stellar mass function has 
changed very little at the high mass end. Substantial growth,
in terms of a mass doubling on $\simeq$5-8 Gyr timescale,
is only possible for galaxies with stellar masses below 10$^{10}\,M_{\odot}$.
Above this mass, the basic evolutionary signal is a {\it quenching}
of star formation in well-established systems. This quenching
progresses toward lower mass systems at later times,
consistent with the mass-dependent trends seen in the ages of 
stellar populations in the quiescent spheroidals.

The physical origin of the quenching of star formation, 
the fundamental origin of the various downsizing signals,
is unclear. Although the merger-induced production of active 
galactic nuclei may lead to the temporary expulsion and heating 
of the gaseous halos that surround galaxies (Springel et al 2005, Croton
et al 2006), key tests of this hypothesis include sustaining the
quenching, the weak environmental dependence of the observational 
trends and the surprisingly clear redshift dependence of the effect
(Figure 32). 

\subsection{Results: Stellar Mass Functions z$>$1.5}

With current facilities, the stellar mass data beyond $z\simeq$1.5 
generally probes only the higher mass end of the distribution
and relies on photometric rather than spectroscopic data.
Nonetheless, the results emerging have received as much 
attention as those at lower redshift. Unlike the complexities of 
understanding downsizing and the redistribution of mass
and morphology in the $z<$1 data, the basic question at stake here
is simply whether the {\it abundance} of massive $z\simeq$2
galaxies is larger than expected in the standard model.

Testing the decline with redshift in the comoving abundance of, say, 
systems with stellar mass greater than 10$^{11}\,M_{\odot}$,
expected in $\Lambda\,CDM$ has been a frustrating story
for the observers for two reasons. Firstly, the most massive
systems are rare and clustered, and so determining reliable
density estimates beyond $z\simeq$1.5 has required panoramic
deep infrared data which has only recently arrived. For a
review of the early observational efforts see Benson,
Ellis \& Menanteau (2002)

Secondly, there has been considerable confusion in the theoretical
literature on the expected rate of decline in $>10^{11}\,M_{\odot}$
systems. Early predictions (Kauffmann, Charlot \& White 1996) claimed
a 3-fold decline in the comoving abundance to $z\simeq$1 in
apparent agreement with the large photometric sample analyzed
by Bell et al (2004). However, careful comparisons of independent 
semi-analytical predictions (Kauffmann et al 1999, Cole et al 2000) 
reveal substantial differences (by an order of magnitude for the same 
world model) in the rate of decline even to $z\simeq$1 (Benson,
Ellis \& Menanteau 2002). In reality, the predictions depend
on many parameters where differences can have a large
effect for the region of the mass function where the slope $dN\,/\,dM$
is steep.

Dickinson et al (2003) undertook a pioneering study to evaluate
the growth of stellar mass with photometric redshift using deep 
$H<$26.5 NICMOS data in the Hubble Deep Field North. Although 
a tiny field (5 arcmin$^2$), the work was the first to demonstrate the 
existence of $10^{11}\,M_{\odot}$ galaxies at $z\simeq$1.5-2 as well 
as the challenges of estimating reliable abundances. A similar HDF-S 
analysis was undertaken by Rudnick et al (2003).

A more representative area of 4$\times$30 arcmin was probed 
spectroscopically in the Gemini Deep Deep Survey (Glazebrook et al 2004), 
albeit to a much shallower depth. These authors claimed a `surprising'
abundance of $10^{11}\,M_{\odot}$ systems to $z\simeq$2. Although
an {\it absolute} comparison of the abundances with semi-analytical 
predictions is not likely to be illuminating for the reasons mentioned 
above, the redshift-dependent {\it mass growth rate} over 1$<z<$2, 
derived empirically from both the stellar masses and the integrated 
star formation rate, seems much slower than expected in the 
semi-analytical models (Granato et al 2000). Of particular 
interest is the fact that the mass growth rate seems not depend strongly 
on the mass range in question in apparent disagreement with
the model predictions (Figure 33).

\begin{figure}
\psfig{file=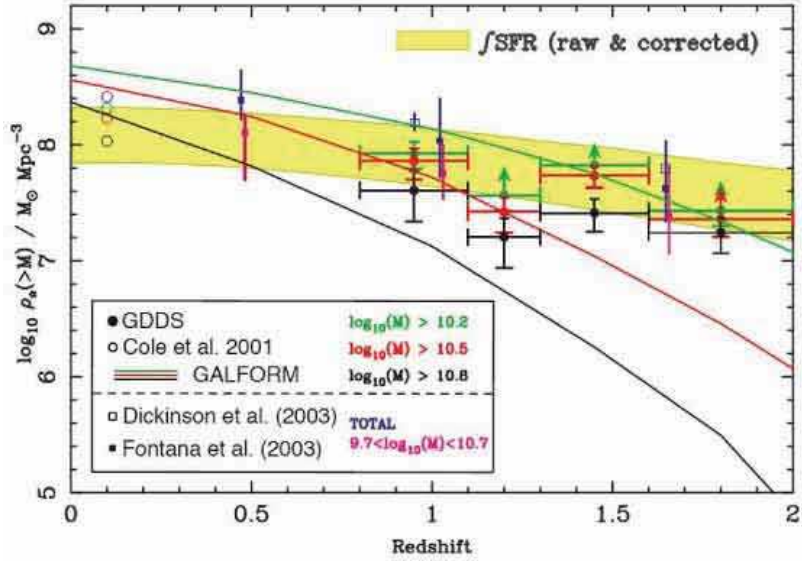,height=3.0in,angle=0}
\caption{The mass assembly history derived from stellar masses
in the Gemini Deep Deep spectroscopic survey (Glazebrook et al
2004). Colored data points refer to observed cumulative densities
$\rho\,(\,>M)$ for various mass ranges and the green shaded region
to estimates derived from the integrated star formation history. 
The colored lines refer to predicted growth rates according to the
GALFORM semi-analytic models.}
\label{fig:33}      
\end{figure}

Although many of the observers associated with the Gemini
Deep Deep Survey have claimed the abundance and slow
growth rate of massive galaxies over 1.5$<z<$2 poses a crisis 
for the standard model, alongside the puzzling `anti-hierarchical' 
behavior observed for $z<$1.5, the plain fact is that there is
considerable uncertainty in the semi-analytical predictions.

\subsection{Quiescent Galaxies with 2$<z<$3}

Finally, it is illustrative to consider the dramatic effect that
infrared data from panoramic ground-based cameras and
the Spitzer Space Telescope is having, not only on our knowledge 
of the distribution of stellar masses at high redshift, but also how 
stellar mass is distributed among quiescent and star-forming
populations. Until recently, there was widespread belief
that the bulk of the star formation in this era, and probably a significant
fraction of the stellar mass, lay in the Lyman break population.
By contrast, van Dokkum et al (2006) have examined the
rest-frame $U-V$ colors of a sample of 300 $>10^{11}\,M_{\odot}$
galaxies with 2$<z_{photo}<3$ (Figure 34) and claim almost
80\% are quiescent. The definition of `quiescent' here is somewhat
important to get correct, given it now emerges that many of the
originally-selected distant red galaxies with $J-K>$2.3
(\S2.6) turn out to have quite respectable star formation rates.
Infrared spectroscopy sensitive to $H\alpha$ emission and
the 4000\AA\ break is making great strides in clarifying this
question (Kriek et al 2006).

\begin{figure}
\psfig{file=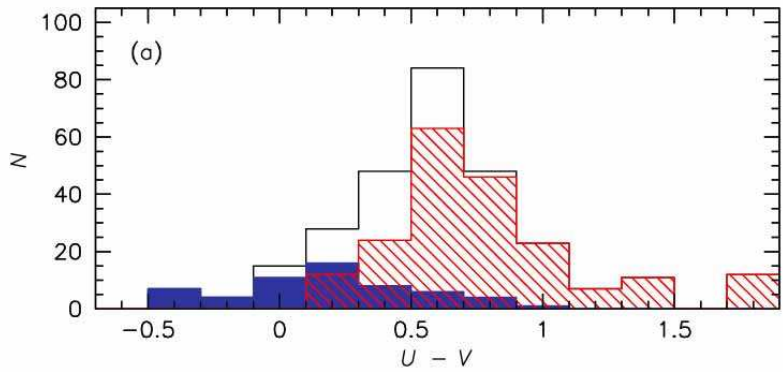,height=2.0in,angle=0}
\caption{The distribution of rest-frame U-V color for a Spitzer-selected
sample of $2<z<$3 galaxies with stellar masses $>10^{11}\,M_{\odot}$
from the analysis of van Dokkum et al (2006). Almost 80\% of this
sample in a quiescent state consistent with having completed the
bulk of their assembly.}
\label{fig:34}      
\end{figure}

\subsection{Lecture Summary}

In this lecture we have demonstrated important new techniques
for estimating the stellar masses of all types of galaxies to impressive
high redshift ($z\simeq$3). The reliability of these techniques
is improved greatly by having spectroscopic, rather than photometric,
redshifts and, for $z>0.7$, by the addition of infrared photometry.
These techniques augment and extend more precise measures
available for restricted classes of galaxies - such as the Fundamental
Plane for pressure-supported spheroidals, and the Tully-Fisher 
relationship for rotationally-supported disks.

These various probes point to a self-consistent, but puzzling,
description of mass assembly since $z\simeq$1.5. During this
era, massive galaxies have hardly grown at all - most of the
star formation and rapid growth is occurring in systems whose
masses are less than $10^{10}\,M_{\odot}$. Remarkably,
the star formation appears to be quenched above a certain
threshold mass whose value, in turn, is declining with time.

Energetic sources such as supernovae or active nuclei may
be responsible for this `downsizing' signature but further
work is needed to verify both the weak environmental
trends seen in the observations and the redshift-dependent
trends in the threshold mass.

Beyond $z\simeq$1.5, the number and distribution of massive
($>10^{11}\,M_{\odot}$ galaxies has led to some surprises.
Although the sheer abundance may not be a problem for
contemporary models, the fact that so many have apparently
completed their star formation is more challenging and
consistent with the slow growth rate observed at later
times. The observational situation is rapidly developing
but consistent with the presence of a surprisingly abundant
and mature population of massive galaxies by $z\simeq$3.

\newpage


\section{Witnessing the End of Cosmic Reionization}

\subsection{Introduction - Some Weighty Questions}

We now turn to the exciting and rapidly developing area of
understanding cosmic reionization. This event, which
marks the end of the so-called `Dark Ages' when the intergalactic
medium became transparent to ultraviolet photons, was
a landmark in cosmic history. In some ways the event might be
considered as important as the epoch of recombination which
isolates the formation of the hydrogen atom, or the asssociated 
surface of last scattering when photons and baryonic matter
decouple. In the case of the era of cosmic reionization, although we
cannot yet be sure, many believe we are isolating that period when 
the first sources had sufficient output to contribute to the energy
balance of the intergalactic medium. Even though some early
luminous forerunners might be present, the epoch of reionization
can be directly connected with cosmic dawn for starlight.
 
It seems an impossible task to give an authoritative observational
account of how to probe this era. So many issues are complete
imponderables! When did reionization occur? Was it a gradual
event made possible by a complex time sequence of sources, 
or was there a spectacular synchronized moment? Can we 
conceive of an initial event, followed by recombination and a
second phase? 

What were the sources responsible? History has shown
the naivety of astronomers in assuming a single population
to be responsible for various phenomena - usually some
complex combination is the answer. And, perhaps most
ambitiously, what is the precise process by which photons
escape the sources and create ionized regions?

Four independent observational methods are
helpful in constraining the redshift range where we might
search for answers to the above questions. In this lecture,
we will explore how these work and the current (and rapidly
changing) constraints they offer. These are:

\begin{itemize}

\item{} Evolution in the optical depth of Lyman series absorption
lines in high redshift quasars (Fan et al 2006a,b). Although QSOs
are only found to redshifts of $z\simeq$6.5, high quality
spectroscopic data gives us a glimpse of a potential change
in the degree of neutrality of the IGM beyond $z\simeq$5.5
such as might be expected if reionization was just ending at
that epoch.

\item{} The ubiquity of metal absorption lines in the IGM
as probed again by various spectra of the highest redshift  
QSOs. Carbon, in particular, is only produced in stellar
nuclei and its presence in all sightlines to $z\simeq$5-6 (Songaila
2004) is highly indicative of a widespread process of earlier
enrichment from the first generation of supernovae.

\item{} The large angular scale power in the temperature
- polarization cross-correlation function seen in the microwave
background (Kogut et al 2003, Spergel et al 2006). This
signal is produced by electron scattering foreground to the
CMB. Depending on models of large scale structure, the
optical depth of scattering gives some indication of the
redshift range where the ionized particles lie. 

\item{} The stellar mass density in assembled sources at
redshifts $z\simeq$5-6, about 1 billion years after the
Big Bang, as probed by the remarkable combination of
HST and Spitzer (Stark \& Ellis 2005, Stark et al 2006a,
Yan et al 2006, Eyles et al 2006). A census of the mass
in stars must be the integral of past activity which can
be compared with that necessary to reionize the Universe.  

\end{itemize}

\subsection{The Gunn-Peterson Test and SDSS QSOs}

In a remarkable paper, long before QSOs were located
at redshifts beyond 2.5, James Gunn and Bruce Peterson
realized (Gunn \& Peterson 1965) that the absence of broad 
troughs of hydrogen
absorption in the spectra of QSOs must indicate intergalactic
hydrogen is ionized. They postulated a future test whereby the spectra
of QSOs of successively higher redshift would be scrutinized
to locate that epoch when the IGM was neutral. 

For an optical spectrum, redshifted to reveal that
portion of the rest-frame UV shortward of the Lyman $\alpha$
emission line of the QSO itself ($\lambda$\,1216 \AA\ ),  the
relative transmission $T$ is defined as

$$T = f(\lambda)\,/\,f_{cont}$$

where $f_{cont}$ represents the continuum radiation from
the QSOs. The transmission is reduced by Lyman $\alpha$ 
absorption in any foreground (lower redshift) clouds of neutral 
hydrogen whose {\it Gunn-Peterson opacity} is then

$$\tau_{GP} = - ln\, T$$

The first `complete troughs' in the absorption line spectra
of distant QSO were presented by Becker et al (2001)
and Djorgovski et al (2001) and a more comprehensive
sampling of 11 SDSS QSOs was presented by Fan et al
(2003). Recently, an analysis of 19 5.74$<z<$6.42 QSOs was
presented by Fan et al (2006b).

Figure 35 illustrates how absorption structures along the 
same line of sight can be independently probed using
Lyman $\alpha$ and higher order lines such as Lyman $\beta$.

\begin{figure}
\psfig{file=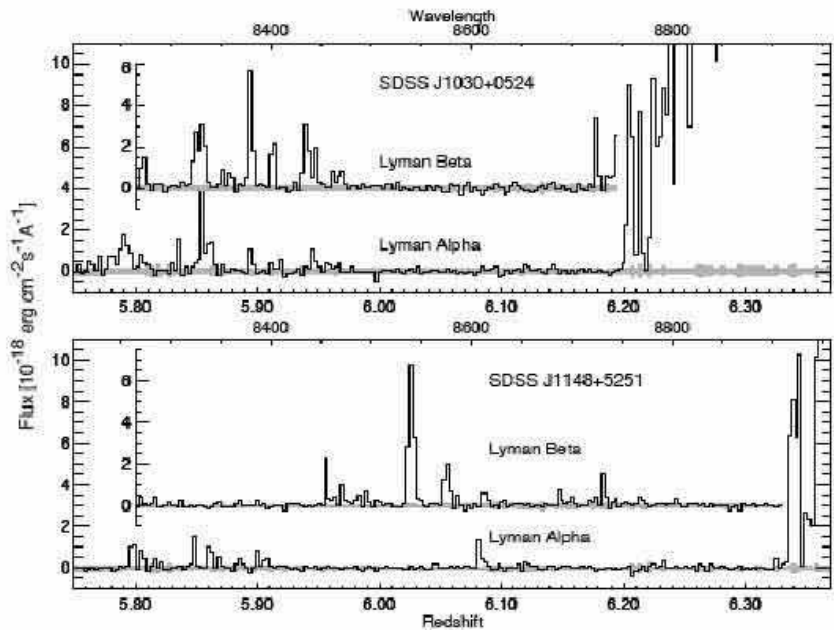,height=3.4in,angle=0}
\caption{The absorption line spectrum of two SDSS QSOs
from the recent study of Fan et al (2006b). Here the structures
in the Lyman series absorption lines $\alpha$ and $\beta$ 
have been aligned in redshift space thereby improving
the signal to noise along a given sightline. The effects
of cosmic variance can clearly be seen by comparing the
structures seen along the two sightlines.}
\label{fig:35}      
\end{figure}

To understand how this is effective, in a uniform medium $\tau_{GP}$ is
related to the abundance of absorbing neutral hydrogen atoms by the
following expression:

$$\tau_{GP} = \frac{\pi  e^2}{m_e c}\,f_{\alpha}\,\lambda_{\alpha}\,H^{-1}(z)\,n_{HI}$$

where $f$ is the oscillator strength of the Lyman $\alpha$ line, $H$ 
is the redshift-dependent Hubble parameter and $n_{HI}$ is the 
neutral number density.

Numerically, this becomes

$$\tau_{GP}= 1.8 \, 10^5 h^{-1}\, \Omega_m^{-\frac{1}{2}}\,\frac{\Omega_b h^2}{0.02}\,(\frac{1\,+\,z}{7})^{3/2}\,\frac{n_{HI}}{n_H}$$

where $x_{HI}=(\frac{1\,+\,z}{7})^{3/2}\,\frac{n_{HI}}{n_H}$ is then the neutral fraction.

Inspection of this equation is quite revealing. Firstly, even a tiny neutral fraction, 
$x_{HI}\sim10^{-4}$, would give a very deep, seemingly complete GP trough;
for reference $x_{HI}\simeq10^{-5}$ today. So clearly the test is not
a very sensitive one in absolute terms.

Secondly, since $\tau_{GP}\,\propto\,f\,\lambda$, for the same $n_H$, the 
optical depth in the higher order Lyman $\beta$ and $\gamma$ lines would 
be $\simeq$6 and 18 times smaller respectively.

In practice, the above relations are greatly complicated by any clumpiness
in the medium. This affects our ability to make direct inferences on $x_{HI}$
as well as to {\it combine} the various Lyman lines into a single test.

Instead, workers have examine the {\it relative} distribution of $\tau_{GP}$
with redshift independently from the various Lyman series absorption statistics.
An increase in $\tau_{GP}$ with redshift could just be a natural thickening
of the Lyman absorption forest and, given its weak connection with $x_{HI}$,
not imply anything profound about cosmic reionization. However, if it can
be shown empirically that the various diagnostics show a {\it discontinuity} in the
$x_{HI}$-redshift trends, conceivably we are approaching the neutral era.

Figure 36 shows that for $z<$5.5, $\tau \propto (\frac{1+z}{5})^{4.3}$
for both the Ly$\alpha$ and Ly$\beta$ forests to reasonable precision.
However, beyond $z\simeq$5.5, both redshift trends are much steeper, 
$\propto(1+z)^{11}$. The dispersion around the trends also increases 
significantly at higher redshift. Taken together, both results 
suggest a qualitative change in nature of the IGM beyond $z\simeq$5.5
(but for an alternative explanation see Becker et al 2006).

\begin{figure}
\centerline{\hbox{
\psfig{file=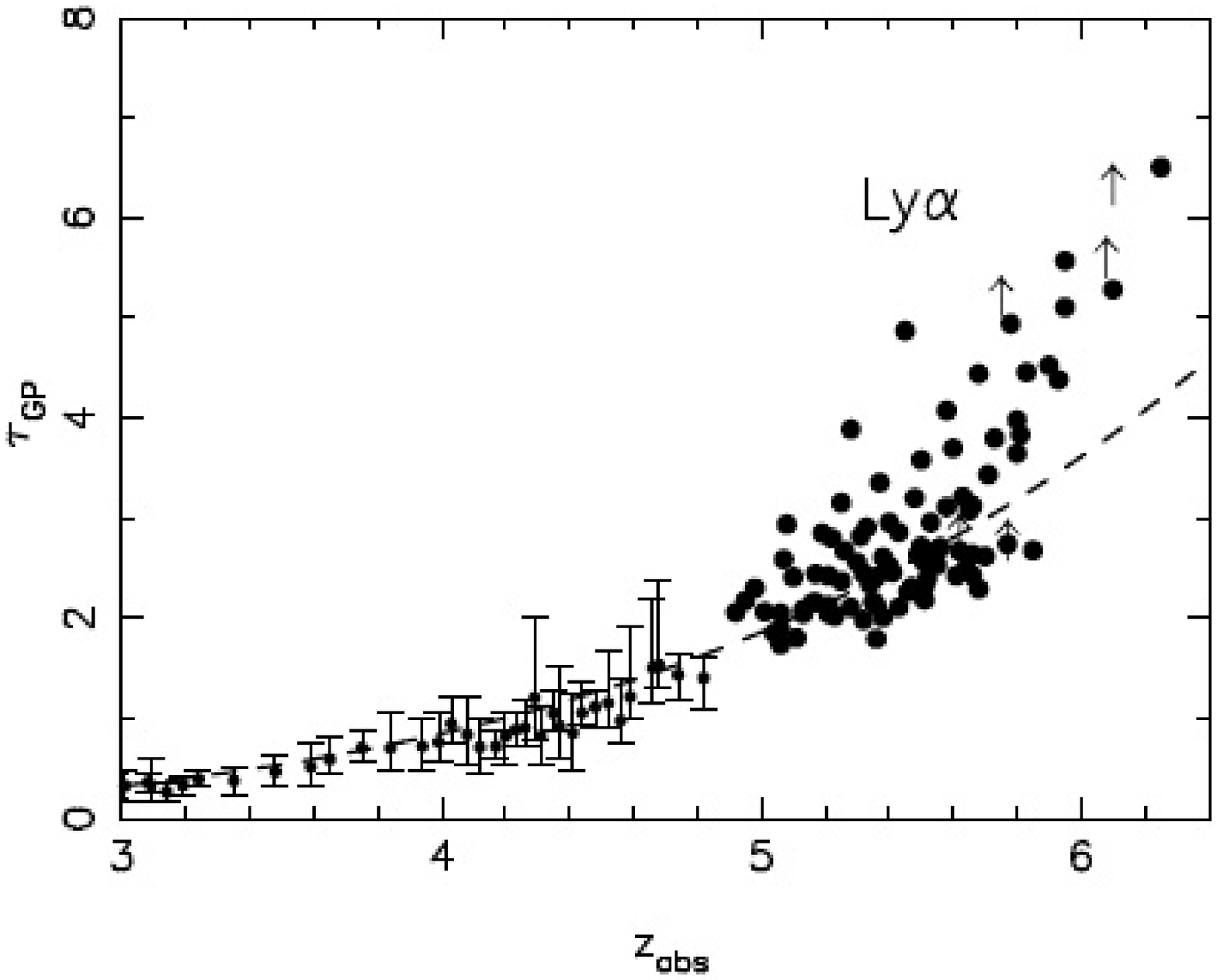,height=1.9in,angle=0}
\psfig{file=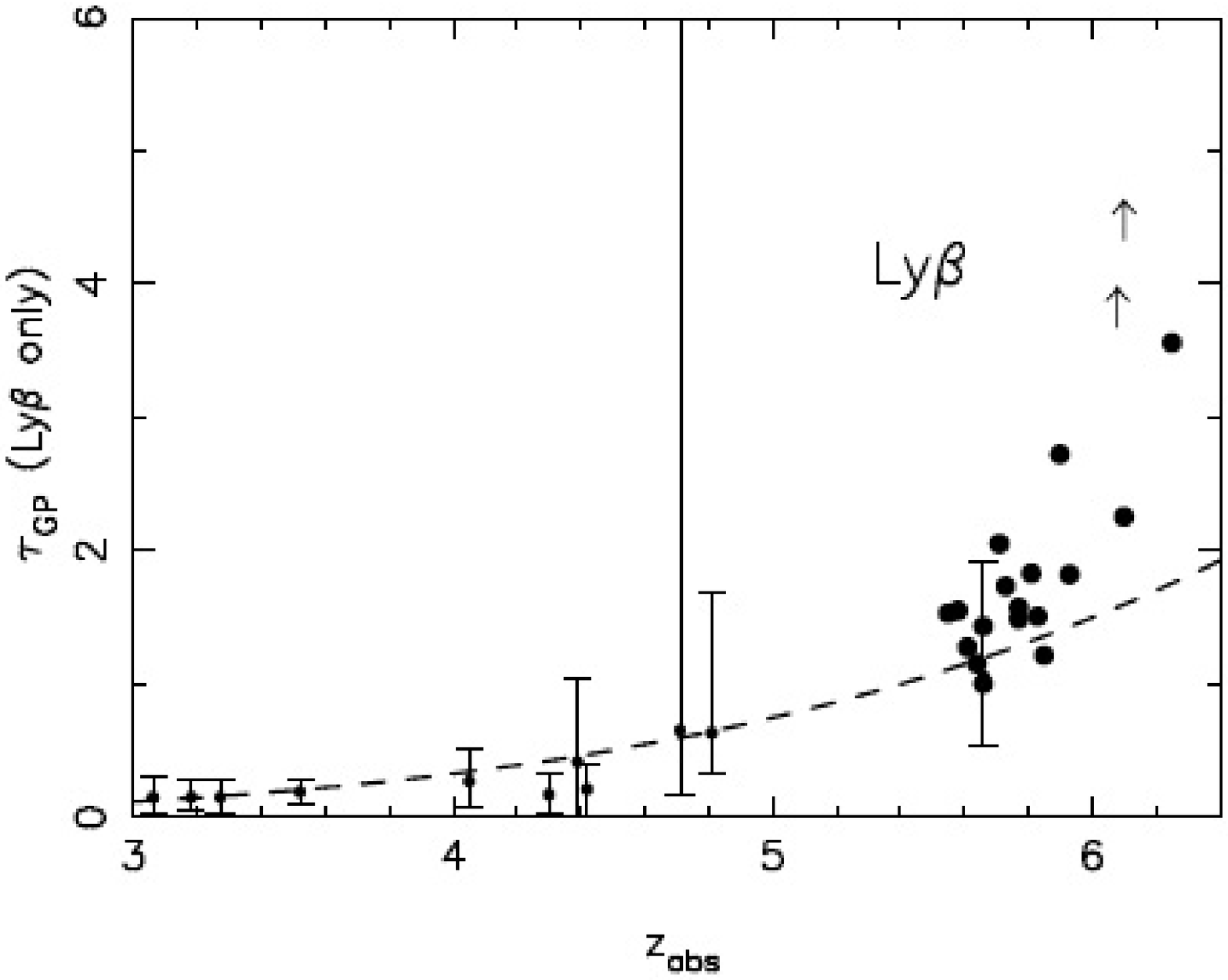,height=1.9in,angle=0}}}
\caption{Evolution in the Gunn-Peterson optical depth, $\tau$, for
both the Lyman $\alpha$ (left) and $\beta$ (right) forests from the
distant SDSS QSO analysis of Fan et al (2006. The dotted lines
represent fits to the data for $z_{abs}<$5.5, beyond which there
is evidence in both species for an upturn in the opacity of the intergalactic
medium.}
\label{fig:36}      
\end{figure}

Fan et al (2006a,b) discuss several further probes
of the nature of the IGM at $z\simeq$6. One relates to the
{\it proximity effect} - the region around each QSO where it
is clear from the spectrum that the IGM is being ionized by
the QSO itself. Although this region is excluded in the analyses
above, the {\it extent} of this region contains valuable information
on the nature of the IGM. It appears that the radius of the region 
affected, $R$ is less at higher redshift according to 
$R\propto[(1\,+\,z)\,x_{HI}]^{-1/3}$ suggesting that the most
distant QSOs in the sample ($z\simeq$6.5) lie in a IGM whose
neutral fraction is $\simeq$14 times higher than those at $z\simeq$5.7.

Another valuable measure is how the regions of complete
absorption, the so-called `dark gaps' in the spectrum where
transmission is effectively zero, are distributed. Fan et al define
a `gap' as a contiguous region in redshift space where $\tau>$3.5.
The distribution of gaps contains some information on
the topology of reionization. We would expect regions of high
transmission to be associated with large HII regions, centered
on luminous star-forming sources. The dark gaps increase
in extent from $\simeq$10 to 80 comoving Mpc over the
redshift range samples suggesting the IGM is still
not neutral at $z\simeq$6.5. Although the Gunn-Peterson and
gap statistics make similar statements about reionization, suggesting
the neutral fraction at $z\simeq$6.2 is 1-4\%, Fan et al 
consider that beyond $z\simeq$6.5, gap statistics will become 
a more powerful probe. This is because the redshift
distribution of those few spectroscopic pixels where the transmission
is non-zero will become the only effective signal.

\subsection{Metallicity of the High Redshift IGM}

A second measure of the redshift range of early star formation is 
contained in the properties of the CIV forest observed in the
spectra of high redshift QSOs (Songaila 2005, 2006).
Carbon is only produced in stellar nuclei (it is not produced in 
the hot Big Bang) and so the ubiquity of CIV along many sightlines 
to $z\simeq$5-6 QSOs is a powerful argument for early
enrichment. 

CIV was seen in the Ly$\alpha$ forest in 1995 (Cowie et al 1995) with N(CIV)/N(HI) $\sim$10$^{-2}$ to 10$^{-3}$. However, it was
subsequently seen in even the weakest Ly$\alpha$ systems
(Ellison et al 2000). This is a particularly powerful point since it 
argues that enrichment is not confined to localized regions of
high column density but is generic to the intergalactic medium
as a whole (Figure 37)

\begin{figure}
\centerline{\psfig{file=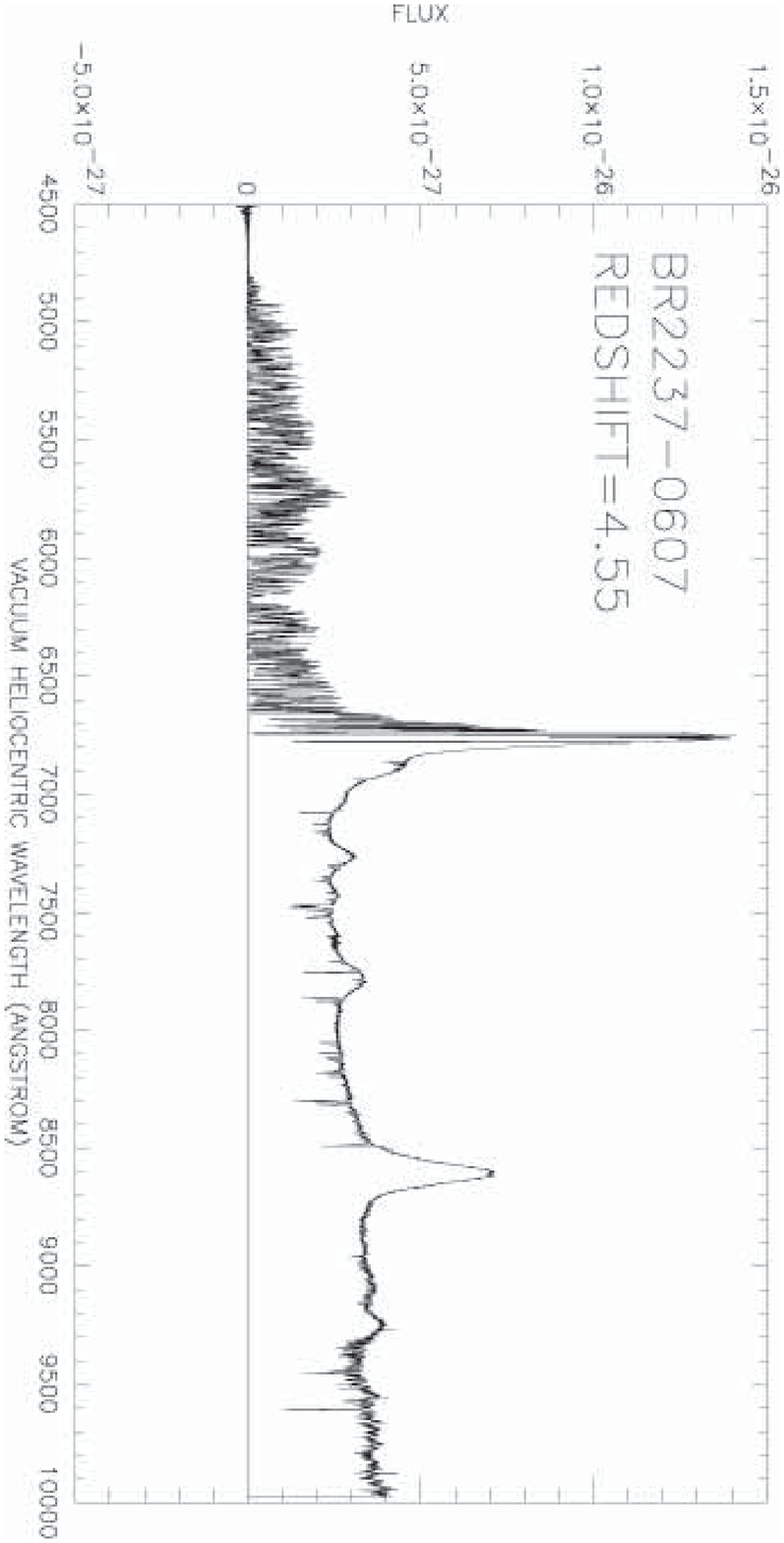,height=4.2in,angle=90}}
\centerline{\psfig{file=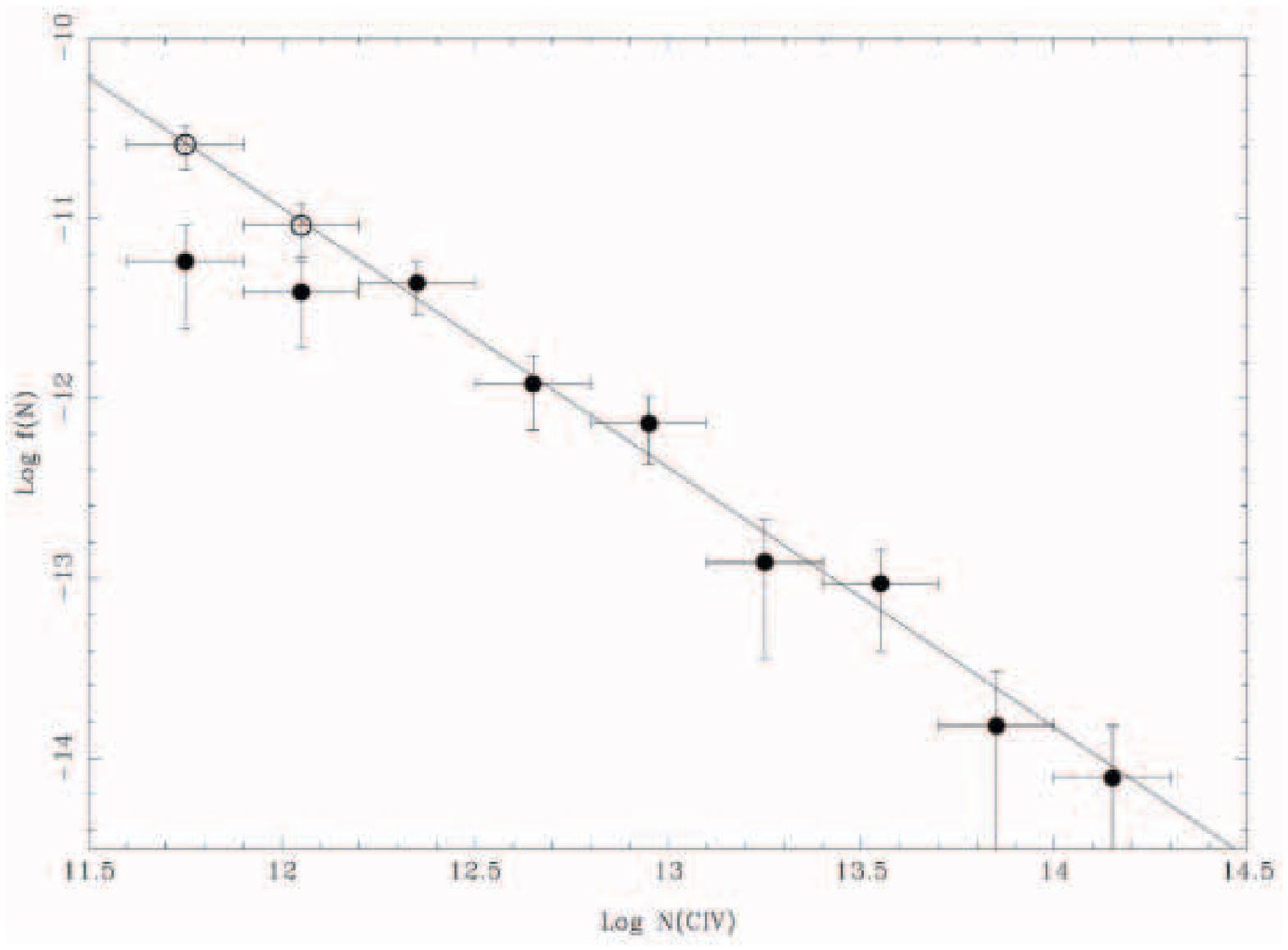,height=2.5in,angle=0}}
\caption{(Left) Keck ESI absorption line spectrum of the z=4.5 QSO
BR2237-0607 from the study of Songaila (2005). A sparsely
populated CIV forest from 7500-8500\AA\ accompanies the
dense Ly$\alpha$ forest seen below 6800 \AA\ . (Right)
Distribution of column densities of CIV absorbers per unit
redshift interval in Q1422+231 from the survey of Ellison et al (2000).}
\label{fig:37}      
\end{figure}

A quantitative interpretation of the CIV abundance, in terms of 
how much early star formation occurred earlier than the
highest redshift probed, relies on locating a `floor' in the 
abundance-redshift relation. Unfortunately, the actual 
observed trend, measured via the contribution of the ion 
to the mass density $\Omega(CIV)$ from $z\simeq$5 to 2, 
does not seem to behave in the manner expected. For
example, there is no strong rise in the CIV abundance
to lower redshift despite the obvious continued star
formation that occurred within these epochs (Songaila 2006,
Figure 38). This is a major puzzle (c.f. Oppenheimer et al 2006).

\begin{figure}
\centerline{\psfig{file=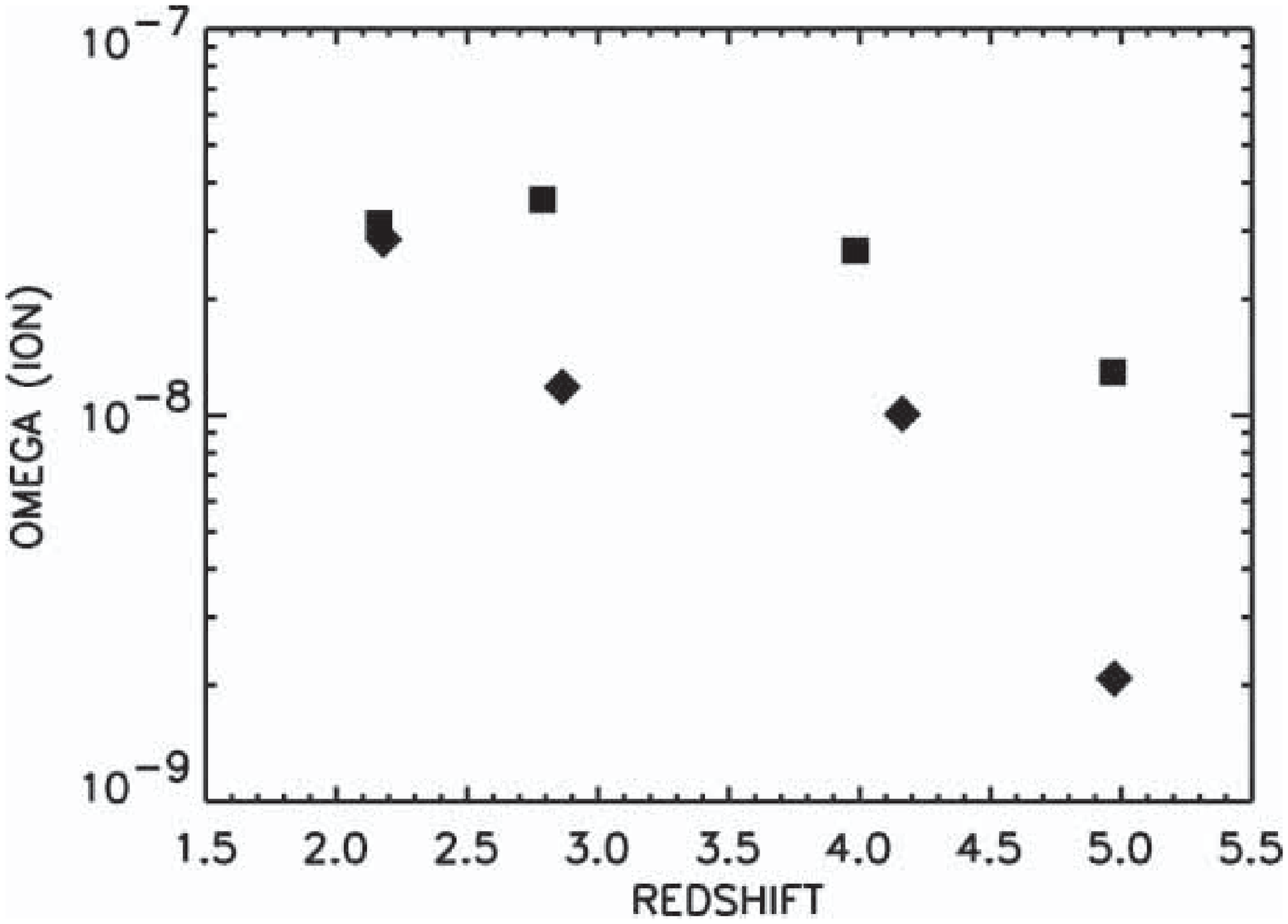,height=2.6in,angle=0}}
\caption{Modest evolution in the contribution of intergalactic
CIV and SIV over 2$<z<$5 as measured in terms of the
ionic contribution to the mass density, $\Omega$ (Songaila
2005).}
\label{fig:38}      
\end{figure}

\subsection{Linear Polarization in the WMAP Data}

In 2003, the WMAP team (Kogut et al 2003) presented the 
temperature-polarization cross angular power spectrum from the
first year's data and located a 4$\sigma$ non-zero signal at very low multipoles ($l<$8) 
which they interpreted in terms of foreground electron scattering of
microwave background photons with an optical depth 
$\tau_e=0.17\pm0.04$ corresponding to ionized structures at $z_{reion}\simeq$20
(Figure 39a). The inferred redshift range depends sensitively on 
the history of the reionization process. Bennett et al (2003) argued 
that if reionization occurred instantaneously it corresponds to a 
redshift $z_{reion}\simeq17\pm5$, whereas adopting a more reasonable  
Press-Schechter formalism and an illustrative cooling and
enrichment model, Fukugita \& Kawasaki (2003) demonstrated
that the same signal can be interpreted with a delayed reionization
occurring at $z_{reion}\simeq$9-10.

Just before the Saas-Fee lectures, the long-awaited third year WMAP data
was published (Spergel et al 2006). A refined analysis significantly
lowered both the normalization of the dark matter power spectrum to
$\sigma_8=0.74\pm0.05$, and the optical depth to electron scattering to 
$\tau_e=0.09\pm0.03$. The same model of instantaneous reionization reduces
the corresponding redshift to $z_{reion}=11\pm3 (2\sigma)$ - a significant
shift from the 1 year data. 

To illustrate the uncertainties, Spergel et al introduce a more realistic 
history of reionization via the ionization fraction $x_e$. Suppose above $z_{reion}$,
$x_e\equiv0$ and below $z$=7, $x_e\equiv1$. Then suppose $z_{reion}$
is defined as that intermediate point when $x_e=x_e^0$ for 7$<z<z_{reion}$.
Figure 39b illustrates the remarkable {\it insensitivity} of $z_{reion}$ to the
adopted value of $x_e^0$ for $x_e^0<$0.5. Despite improved data,
the redshift range implied by the WMAP data spans the full range 10$<z<$20.

\begin{figure}
\centerline{\hbox{
\psfig{file=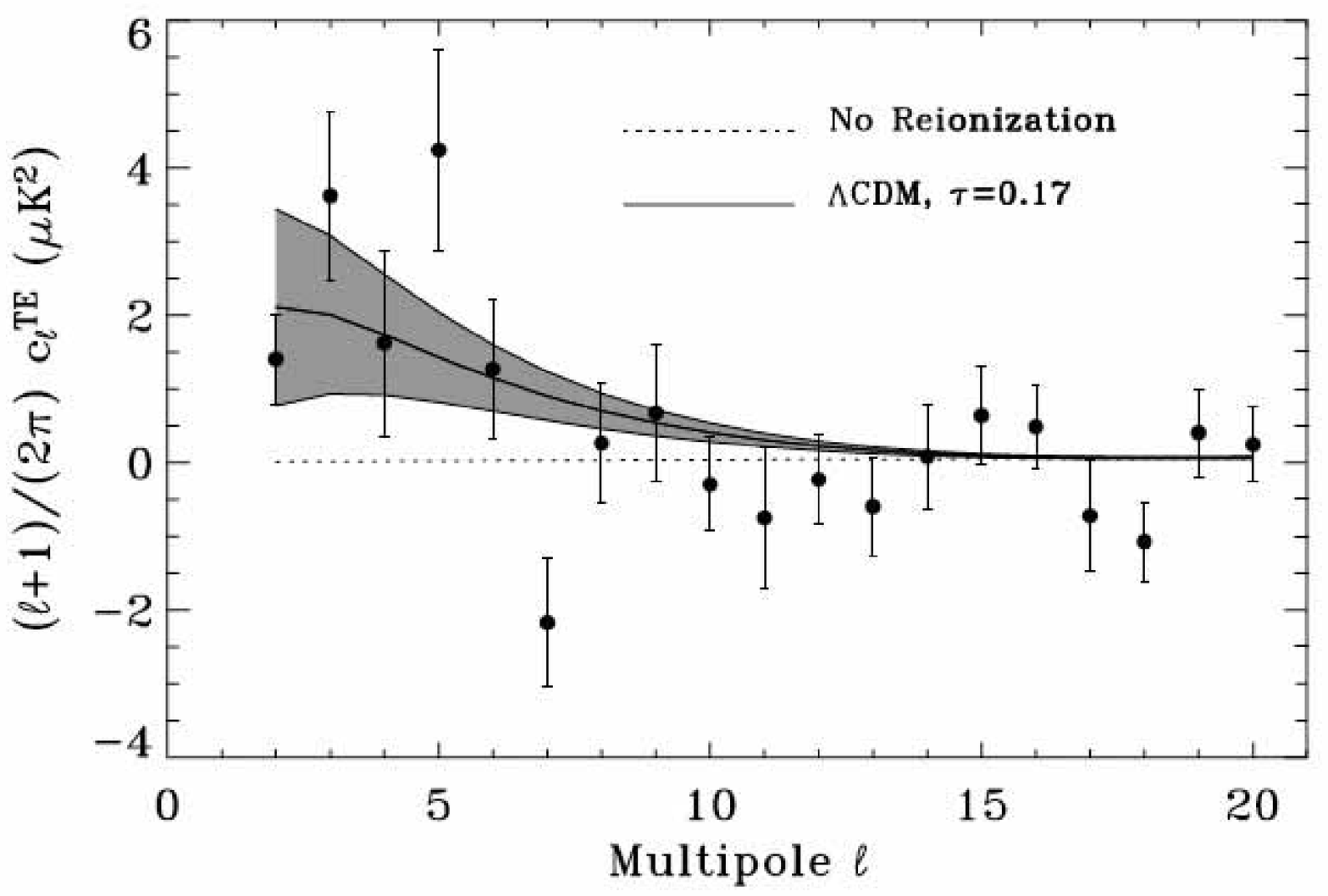,height=2.0in,angle=0}
\psfig{file=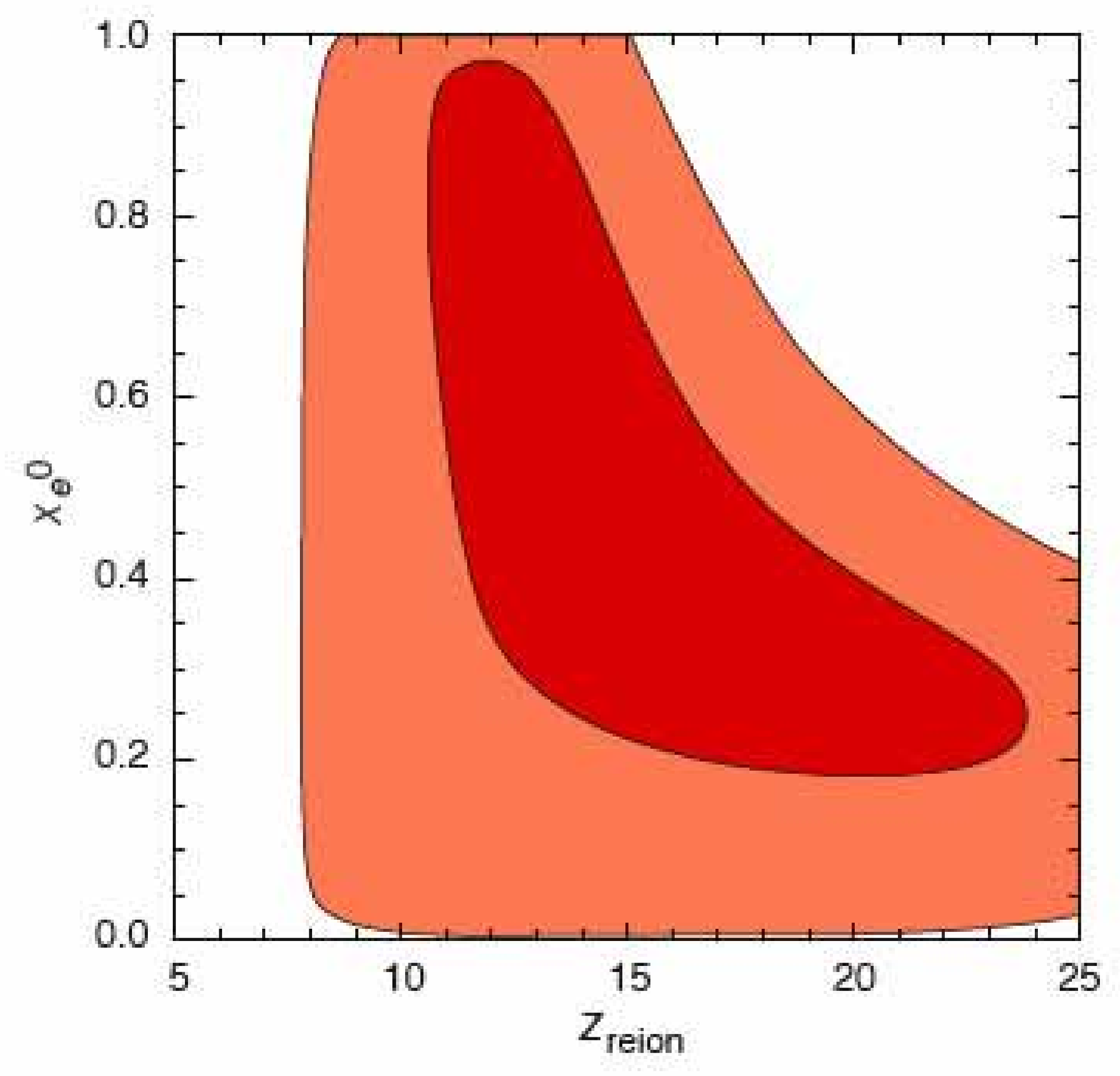,height=2.0in,angle=0}}}
\caption{(Left) The 4$\sigma$ detection of reionization via an excess
signal at large scales in the angular cross correlation power spectrum 
of the temperature and polarization data in the first year WMAP data 
(Kogut et al 2003). (Right) Constraints on the redshift of reionization,
$z_{reion}$, from the third year WMAP data (Spergel et al 2006).
The contours illustrate how the $z_{reion}$ inferred from the lowered
optical depth depends on the history of the ionized fraction $x_e(z)$, see 
text for details.}
\label{fig:39}      
\end{figure}

\subsection{Stellar Mass Density at $z\simeq$5-6}

Neither the Gunn-Peterson test nor the WMAP polarization data 
necessarily demonstrate that reionization was caused by early
star-forming sources; both only provide constraints on when
the intergalactic medium was first reionized. The CIV test
is a valuable complement since it provides a measure of
early enrichment which can only come from star-forming sources.
Unfortunately, powerful though the ubiquitous presence of CIV 
is in this context, as we have seen, quantitative constraints
are hard to derive. We thus seek a further constraint on
the amount of early star formation that might have occurred.

In Lecture 4 we introduced the techniques astronomers are
now using to derive {\it stellar masses} for distant galaxies.
Although the techniques remain approximate, it must follow
that the stellar mass density at a given epoch represents
the integral over time and volume of the past star formation.
Indeed, we already saw a successful application of this in
reconciling past star formation with the local stellar mass density
observed by 2dF (Figure 21).  Specifically, at a particular
redshift, $z$

$$M_{\ast} = \int_\infty^z \rho_{\ast}(z)\,dV(z)$$

Using the techniques described in Lecture 4, stellar mass
estimates have become available for some very high 
redshift galaxies detected by Spitzer (Eyles et al 2005,
Mobasher et al 2005, Yan et al 2005). For the most
luminous Lyman `dropouts', these estimates are quite
substantial, some exceeding $10^{11}\,M_{\odot}$ implying
much earlier activity. Recently, several groups (Stark et al
2006a, Yan et al 2006, Eyles et al 2006)  have been
motivated to provide the first crude estimates on the volume
averaged {\it stellar mass density} at these early epochs.
Part of this motivation is to check whether the massive
galaxies seen at such high redshift can be reconciled
with hierarchical theory, but as Stark \& Ellis (2005) proposed,
the established stellar mass can also be used to probe earlier
star formation and its likely impact upon cosmic reionization.

A very relevant question is whether the observed mass density 
at $z\simeq$5-6 is greater than can be accounted for by the 
observed previous star formation history. We will review
the rather uncertain data on the star formation density 
$\rho_{\ast}(z)$ beyond $z\simeq$6 in the next Lecture. 
However, Stark et al (2006a) find that even taking a reasonably
optimistic measure of $\rho_{\ast}(z)$ from recent compilations
by Bouwens et al (2006) and Bunker et al (2006), it is
hard to account for the stellar mass density at $z\simeq$5
(Figure 40). 

\begin{figure}
\psfig{file=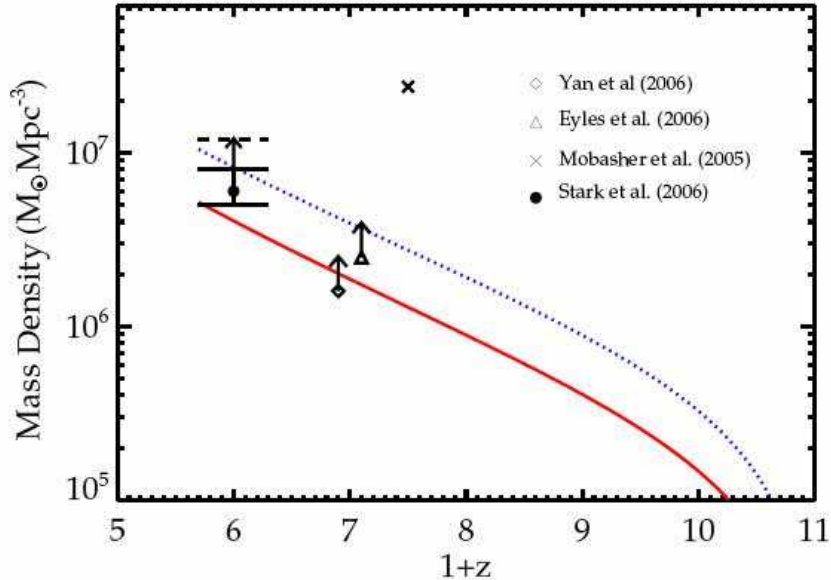,height=3.3in,angle=0}
\caption{A comparison of the assembly history of stellar mass 
inferred from the observed decline in star formation history to 
$z\simeq$10 (solid line) with extant data on the stellar mass
density at $z\simeq$5 and 6 (data points from Stark et al 2006a,
Yan et al 2006, Eyles et al 2006). Different estimates at a given
redshift represent lower limits based on spectroscopically-confirmed
and photometric redshift samples. The red line shows the growth
in stellar mass expected from the presently-observed luminous
star forming galaxies; a shortfall is observed. The blue dotted line 
shows the improvement possible when a dominant component
of high z lower luminosity systems is included. }
\label{fig:40}      
\end{figure}

There are currently two major limitations in this comparison.
First, most of the v- and i-drops are located photometrically; even
a small degree of contamination from lower redshift galaxies
could upward bias the stellar mass density. On the other hand 
only star-forming galaxies are located by the Lyman break technique 
so this bias could easily be offset if there are systems in a quiescent
state as evidenced by the prominent Balmer breaks seen in
many of the Spitzer-detected sources (Eyles et al 2005, 2006). 
This limitation will ultimately be overcome with more careful
selection methods and deeper spectroscopy. Secondly, and more 
profoundly, the precision of the stellar masses may not be up to this 
comparison. Much has to be assumed about the nature of the
stellar populations involved which may, quite reasonably, be 
somewhat different from those studied locally. The discrepancy noted 
by Stark et al is only a factor of $\times$2-3, possibly within the 
range of uncertainty.

Regardless, if this mismatch is reinforced by better data, the
implications are very interesting in the context of reionization.
It could mean early star-forming systems are extincted, lie
beyond $z\simeq$10 where current searches end, or perhaps most
likely that early star formation is dominated by lower luminosity 
systems (Figure 40). By refining this technique and using diagnostics
such as the strength of the tell-tale Balmer break, it may ultimately
be possible to age-date the earlier activity and compare its
efficacy with that required to reionize the Universe.

\subsection{Lecture Summary}

In this lecture we have introduced four very different and independent
probes of cosmic reionization, each of which suggests star formation activity
may extend well into the redshift range 6$\,<z<\,$20. Two of these probes rely on
a contribution from early star formation (the metallicity of the intergalactic
medium and the assembled stellar mass density at $z\simeq$5-6).

The earliest result was the presence of neutral hydrogen troughs
in the spectra of distant QSOs. Although the arguments for reionization
ending at $z\simeq$6 seem compelling at first sight, they ultimately
rely on an empirically-deduced transition in the changes in the opacity
of the Ly$\alpha$ and Ly$\beta$ line below and above $z\simeq$5.5.

The second result - the ubiquity of carbon in even the weakest absorbing
clouds at $z\simeq$5 is firm evidence for early star formation. However, it seems
hard to locate the high redshift `abundance floor' and hence to quantify
whether this early activity is sufficient for reionization. Indeed, a major
puzzle is the lack of growth in the carbon abundance over the redshift
range where galaxies are assembling the bulk of their stars.

The WMAP polarizations results have received the most attention,
mainly because the first year data indicated a surprisingly large
optical depth and a high redshift for reionization. However, there were
some technical limitations in the original analysis and it now seems
clear that the constraints on the redshift range when the foreground
polarization is produced are not very tight.

Finally, an emerging and very promising technique is simply the
census of early star formation activity as probed by the stellar masses
(and ages) of the most luminous dropouts at $z\simeq$5-6. Although
significant uncertaintes remain, the prospects for improving these
constraints are good and, at this moment, it seems there must have
been quite a significant amount of early ($z>$6) star formation activity,
quite possibly in low luminosity precursors.  

\newpage


\section{Into the Dark Ages: Lyman Dropouts}
\label{sec:6}

\subsection{Motivation}

Surveys of galaxies at and beyond a redshift $z\simeq$6 represent
the current observational frontier. We are motivated to search to
conduct a census of the earliest galaxies seen 1 Gyr after the Big Bang 
as well as to evaluate the contribution of early star formation to
cosmic reionization. Although impressive future facilities such as
the next generation of extremely large telescope\footnote{e.g. The 
US-Canadian Thirty Meter Telescope - {\rm http//www.tmt.org}} and the 
James Webb Space Telescope (Gardner et al 2006) are destined to
address these issues in considerable detail, any information we
can glean on the abundance, luminosity and characteristics of
distant sources will assist in planning their effective use.

In this lecture and the next, we will review the current optical
and near-infrared techniques for surveying this largely uncharted
region. They include 

\begin{itemize}

\item{} Lyman dropouts: photometric searches based on 
locating the rest-frame ultraviolet continuum of star-forming
sources introduced in Lecture 3. The key issue here is
reducing contamination from foreground sources since
most sources selected via this technique are too faint for
confirmatory spectroscopy.

\item {}Lyman alpha emitters: spectroscopic or narrow-band
searches for sources with intense Ly$\alpha$ emission. As
the line is resonantly-scattered by neutral hydrogen its profile
and strength gives additional information on the state of the
high redshift intergalactic medium.

\item{} Strong gravitational lensing: by coupling both above
techniques with the magnification afforded by lensing clusters,
it is possible to search for lower luminosity sources at high redshift.
Since the magnified areas are small, the technique is only 
advantageous if the luminosity function has a steep faint end
slope.

\end{itemize}

For the Lyman dropouts discussed here, as introduced in
the last lecture, there is an increasingly important role played
by the Spitzer Space Telescope in estimating stellar masses
and earlier star formation histories.

The key questions we will address in this lecture focus on
the (somewhat controversial) conclusions drawn from the
analyses of Lyman dropouts thus far, namely:

\begin{enumerate}

\item{} How effective are the various high $z$ selection
methods? The characteristic luminosity $L^{\ast}$ at 
$z\simeq$6 corresponds to $i_{AB}\simeq$26 where
spectroscopic samples are inevitably biased to those
with prominent Ly$\alpha$ emission. Accordingly there
is great reliance on photometric redshifts and a real
danger of substantial contamination by foreground red
galaxies and Galactic cool stars.

\item{} Is there a decline in the UV luminosity density, $\rho_{UV}$,
over the range 3$<z<$6? The early results were in some
disagreement. Key issues relate to the degree of foreground
contamination and cosmic variance in the very small deep
fields being examined.

\item{} Is the observed UV density, $\rho_{UV}$, at $z\simeq$6
sufficient to account for reionization? The answer depends
on the contribution from the faint end of the luminosity
function and whether the UV continuum slope is steeper
than predicted for a normal solar-metallicity population.

\item{} Significant stellar masses have been determined
for several $z\simeq$6 galaxies. Are these in conflict
with hierarchical structure formation models?

\end{enumerate}

\subsection{Contamination in $z\simeq$6 Dropout Samples}

The traditional dropout technique exploited very effectively
at $z\simeq$3 (Lecture 3) is poorly suited for $z\simeq$6
samples because the use of a simple $i-z>1.5$ color cut still permits 
significant contamination by passive galaxies at $z\simeq$2 and Galactic stars.
The addition of an optical-infrared color allows some
measure of discrimination (Stanway et al 2005) since a
passive $z\simeq$2 galaxy will be red over a wide range in wavelength,
whereas a star-forming $z\simeq$6 galaxy should be relatively
blue in the optical-infrared color corresponding to its rest-frame
ultraviolet (Figure 41). Application of this two color technique suggests 
contamination by foreground galaxies is $\simeq$10\% 
at the bright end ($z_{AB}<$25.6) but negligible
at the UDF limit ($z_{AB}<$28.5)

\begin{figure}
\psfig{file=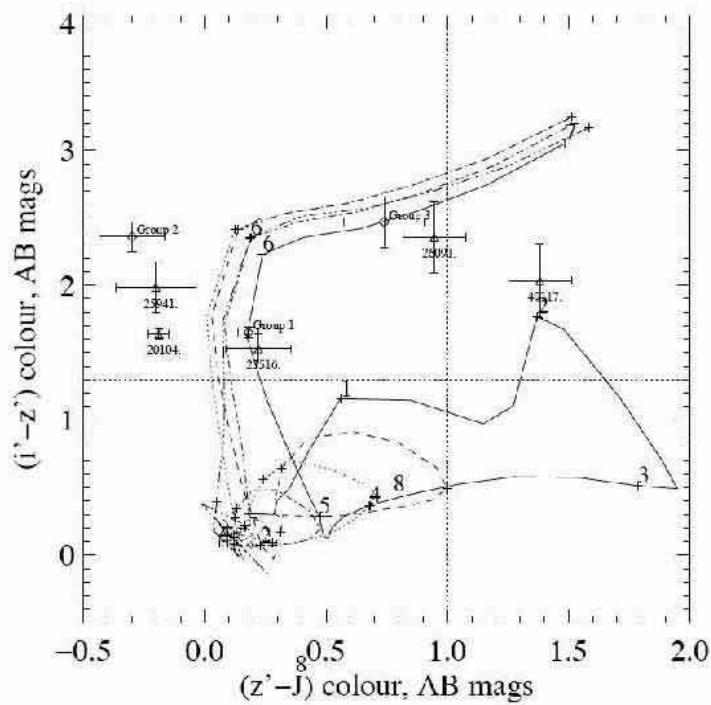,height=3.8in,angle=0}
\caption{The combination of a $i-z$ and $z-J$ color cut
permits the distinction of $z\simeq$5.7-6.5 star forming
and $z\simeq$2 passive galaxies. Both may satisfy the 
$i-z>$1.5 dropout selector, but the former should lie 
blueward of the $z-J$=1.0 divider, whereas $z\simeq$2
are red in both colors. Crosses represent the location of 
candidates in the GOODS field and model tracks illlustrate 
the predicted colors for typical SEDs observed at the 
respective redshifts (Stanway et al 2005).}
\label{fig:41}      
\end{figure}

Unfortunately, the spectral properties of cool Galactic L dwarfs 
are dominated by prominent molecular bands rather than simply 
by their effective temperature. This means that they cannot be 
separated from $z\simeq$6 galaxies in a similar color-color diagram. 
Indeed, annoyingly, these dwarfs occupy precisely the location
of the wanted $z\simeq$6 galaxies (Figure 42)!
The only practical way to discriminate L dwarfs is either via
spectroscopy or their unresolved nature in ACS images.

\begin{figure}
\centerline{\hbox{
\psfig{file=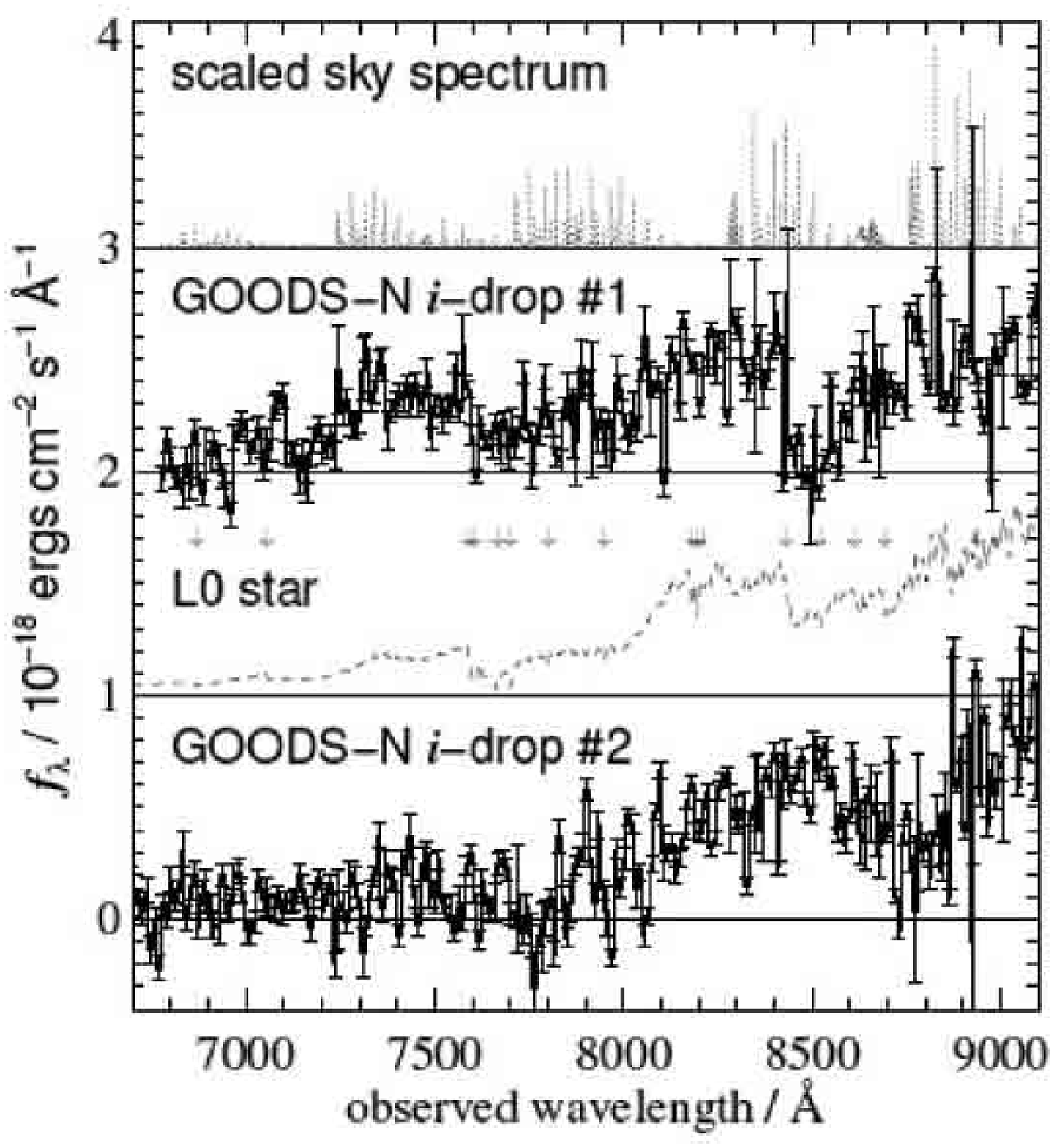,height=2.5in,angle=0}
\psfig{file=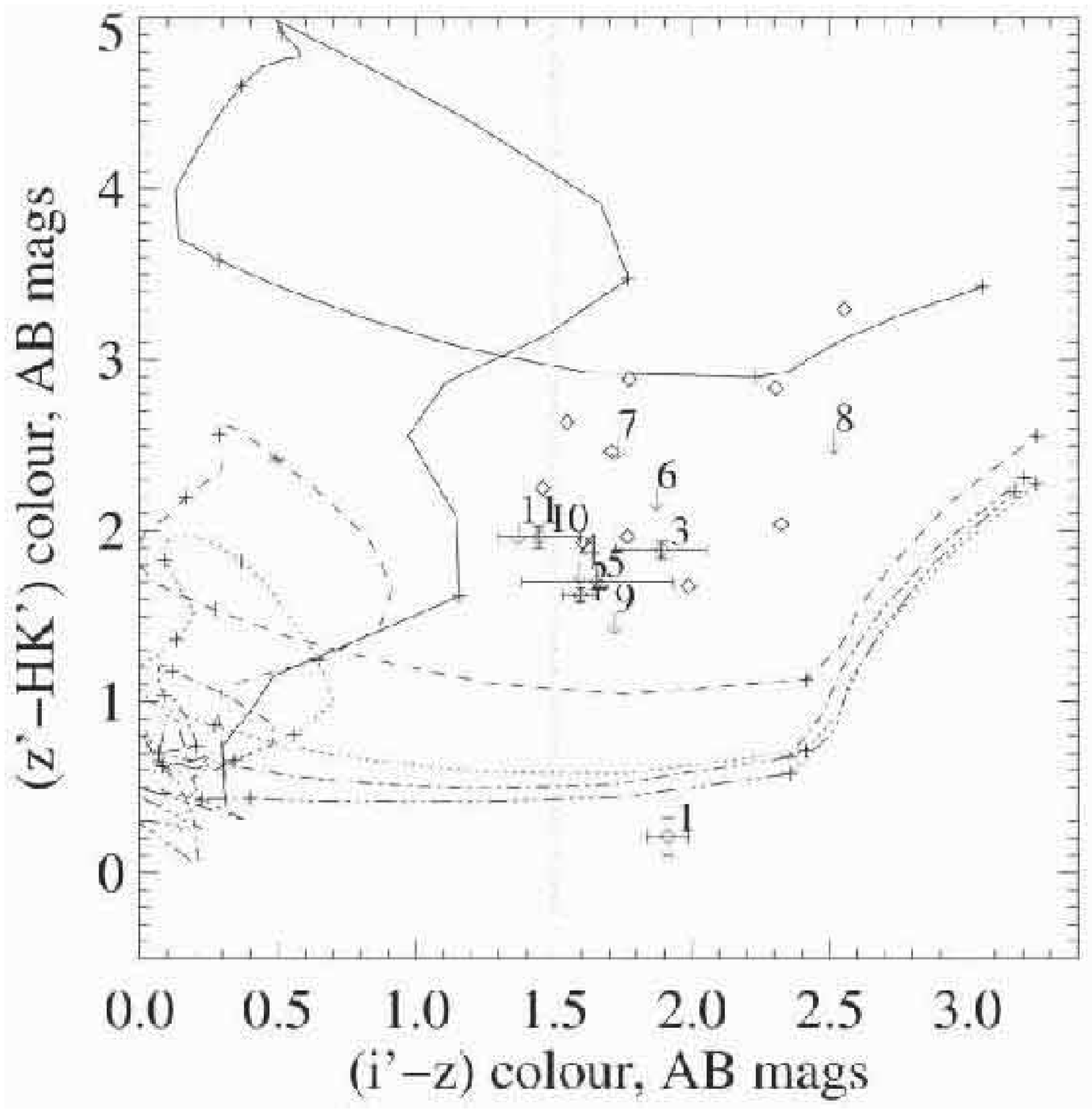,height=2.5in,angle=0}}}
\caption{(Left) Keck spectroscopic verification of two contaminating
L dwarfs lying within the GOODS $i-z$ dropout sample but pinpointed
as likely to be stellar from ACS imaging data. The smoothed
spectra represent high signal to noise brighter examples for comparison
purposes. Strong molecular bands clearly mimic the Lyman dropout
signature. (Right) Optical-infrared color diagram with the dropout
color selector, $i_{AB}-z_{AB}>1.5$, shown as the vertical dotted line. 
Bright L dwarfs (lozenges) frustratingly occupy a similar region of color 
space as the $z\simeq$6 candidates (points with error bars) (Stanway et al
2004).}
\label{fig:42}      
\end{figure}

Stanway et al (2004) conducted the first comprehensive spectroscopic and ACS
imaging survey of a GOODS $i$-drop sample limited at
$z_{AB}<$25.6, finding that stellar contamination at the bright end
of the luminosity function of a traditional ($i_{AB}-z_{AB}>$1.5)
color cut could be as high as 30-40\%. Unfortunately, even with 
substantial 6-8 hour integrations on the Keck telescope, redshift 
verification of the distant population was only possible in those dropout
candidates with Ly$\alpha$ emission. Stanway et al (2005) subsequently 
analyzed the ACS imaging properties of a fainter subset arguing that stellar 
contamination decreases with increasing apparent magnitude.

Further progress has been possible via the use of the ACS grism on
board HST (Malhotra et al 2005). As the OH background is eliminated
in space, despite its low resolution, it is possible even in the fairly low 
signal/noise data achievable with the modest 2.5m aperture of HST to 
separate a Lyman break from a stellar molecular band. It is claimed that 
of 29 $z_{AB}<$27.5  candidates with (i-z)$>$0.9, only 6 are likely to be low redshift interlopers.

Regrettably, as a result of these difficulties, it has become routine to 
rely entirely on photometric and angular size information without 
questioning further the degree of contamination. This is likely one 
reason why there remain significant 
discrepancies between independent assessments of the abundance 
of $z\simeq$6 galaxies (Giavalisco et al 2004, Bouwens et al 2004,
Bunker et al 2004). Although there are indications from the tests
of Stanway et al (2004, 2005) and Malhotra et al (2005) that
contamination is significant only at the bright end, the lack of
a comprehensive understanding of stellar and foreground contamination
remains a major uncertainty.

\subsection{Cosmic Variance}

The deepest data that has been searched for i-band dropouts includes
the two GOODS fields (Dickinson et al 2004) and the Hubble Ultra Deep
Field (UDF, Beckwith et al 2006). As these represent publicly-available fields
they have been analyzed by many groups to various flux limits. The
Bunker/Stanway team probed the GOODS fields to $z_{AB}$=25.6 
(spectroscopically) and 27.0 (photometrically, and the UDF to $z_{AB}$=28.5.
At these limits, it is instructive to consider the comoving cosmic volumes
available in each field within the redshift range selected by the typical
dropout criteria. For both GOODS-N/S fields, the total volume is 
$\simeq$5. 10$^5$ Mpc$^{3}$, whereas for the UDF it is only 2.6 10$^4$ Mpc$^3$.
These contrast with 10$^6$ Mpc$^3$ for a single deep pointing taken
with the SuPrime Camera on the Subaru 8m telescope.

Somerville et al (2004) present a formalism for estimating, for any
population, the fractional uncertainty in the inferred number density from 
a survey of finite volume and angular extent. When the clustering signal 
is measurable, the cosmic variance can readily be calculated analytically. 
However, for frontier studies such as the i-dropouts, this is not the case.
Here Somerville et al propose to estimate cosmic variance 
by appealing to the likely halo abundance for the given observed
density using this to predict the clustering according to CDM models.
In this way, the uncertainties in the inferred abundance of i-dropouts
in the combined GOODS fields could be $\simeq$20-25\% whereas that
in the UDF could be as high as 40-50\%.

It seems these estimates of cosmic variance can only be strict
lower limits to the actual fluctuations since Somerville et al make
the assumption that halos containing star forming sources are visible 
at all times. If,  for example, there is intermittent activity with some 
duty cycle whose "on/off" fraction is $f$, the cosmic variance will be 
underestimated by that factor $f$ (Stark et al, in prep).

\subsection{Evolution in the UV Luminosity Density 3$<z<$10?}

The complementary survey depths means that combined studies of GOODS and 
UDF have been very effective in probing the shape of the UV luminosity 
function (LF) at $z\simeq$6\footnote{In this section we will only refer to 
the {\em observed} (extincted) LFs and luminosity density}. Even so, 
there has been a surprising variation 
in the derived faint end slope $\alpha$. Bunker et al (2004) claim their data 
(54 i-dropouts) is consistent with the modestly-steep $\alpha$=-1.6 found 
in the $z\simeq$3 Lyman break samples (Steidel et 2003), whereas 
Yan \& Windhorst (2004)  extend the UDF counts to $z_{AB}$=30.0 and, 
based on 108 candidates, find $\alpha$=-1.9, a value close to a divergent
function! Issues of sample completeness are central to understanding
whether the LF is this steep. 

In a comprehensive analysis based on all the extant deep data, Bouwens et al 
(2006)  have attempted to summarize the decline in rest-frame UV luminosity 
density over 3$<z<$10 as a function of luminosity (Figure 43). They attribute
the earlier discrepancies noted between Giavalisco et al (2004), Bunker et al (2004)
and Bouwens et al (2004) to a mixture of cosmic variance and differences
in contamination and photometric selection. Interestingly, they claim
a {\it luminosity-dependent} trend in the sense that the bulk of the decline
occurs in the abundance of luminous dropouts, which they attribute to
hierarchical growth. 

A similar trend is seen in ground-based data obtained with Subaru. Although 
HST offers superior photometry and resolution which is effective in
eliminating stellar contamination, the prime focus imager on Subaru
has a much larger field of view so that each deep exposure covers
a field twice as large as both GOODS N+S. As they do not have access
to ACS data over such wide fields, the Japanese astronomers
have approached the question of stellar contamination in an imaginative
way. Shioya et al (2005) used two intermediate band filters at 709nm and 
826nm to estimate stellar contamination in both $z\simeq$5 and $z\simeq$6
broad-band dropout samples. By considering the {\it slope} of the continuum 
inbetween these two intermediate bands, in addition to a standard $i-z$ criterion, 
they claim an ability to separate L and T dwards. In a similar, but independent, study,
Shimasaku et al (2005) split  the $z$ band into two intermediate filters 
thereby measuring the rest-frame UV slope just redward of the Lyman
discontinuity. These studies confirm both the redshift decline and, to a lesser 
extent, the luminosity-dependent trends seen in the HST data. 

\begin{figure}
\psfig{file=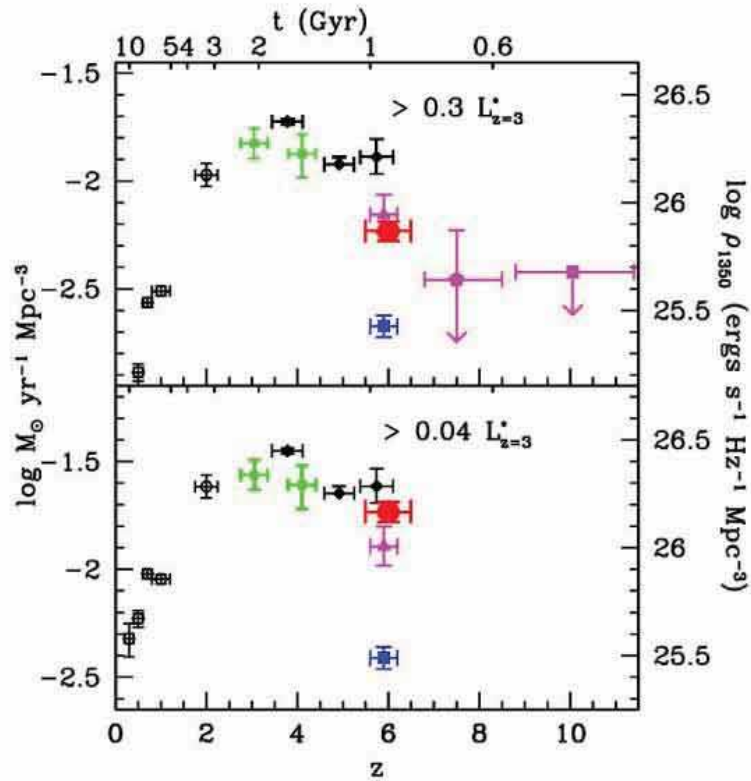,height=4.2in,angle=0}
\caption{Evolution in the rest-frame UV (1350 \AA\ ) luminosity density 
(right ordinate) and inferred star formation rate density ignoring extinction
(left ordinate) for drop-out samples in two luminosity ranges from the
compilation by Bouwens et al (2006). A marked decline is seen over
$3<z<6$ in the contribution of  luminous sources.}
\label{fig:43}      
\end{figure}

Although it seems there is a $5\times$ abundance decline in luminous
UV emitting galaxies from $z\simeq$3 to 6, it's worth noting again that the
relevant counts refer to sources uncorrected for extinction. This is appropriate
in evaluating the contribution of UV sources to the reionization process
but not equivalent, necessarily, to a decline in the star formation rate
density. Moreover, although the luminosity dependence seems similar
in both ground and HST-based samples, it remains controversial
(e.g. Beckwith et al 2006). 

\subsection{The Abundance of Star Forming Sources Necessary for Reionization}

Have enough UV-emitting sources been found at $z\simeq$6-10 to account
for cosmic reionization? Notwithstanding the observational uncertainties
evident in Figure 43, this has not prevented many teams from addressing
this important question. The main difficulty lies in understanding the physical
properties of  the sources in question. The plain fact is that we cannot
predict, sufficiently accurately, the UV luminosity density that is sufficient
for reionization!

Some years ago, Madau, Haardt \& Rees (1999) estimated the star
formation rate density based on simple parameterized assumptions
concerning the stellar IMF and/or metallicity $Z$ essential for converting 
a 1350 \AA\ luminosity into the integrated UV output, the fraction $f_{esc}$
of escaping UV photons, the clumpiness of the surrounding intergalactic hydrogen, 
$C=<\rho_{HI}^2>/<\rho_{HI}>^2$, and the temperature of the
intergalactic medium $T_{IGM}$. In general terms, for reionization
peaking at a redshift $z_{reion}$, the necessary density of sources goes as:

$$\rho \propto  f_{esc}^{-1}\, C\, (1 + z)^3\, (\Omega_B h^2)^2 Mpc^{-3}$$  

For likely ranges in each of these parameters, Stiavelli et al (2004) tabulate
the required source surface density which, generally speaking, lie above
those observed at $z\simeq$6 (e.g. Bunker et al 2004). 

It is certainly possible to reconcile the end of the reionization at $z\simeq$6
with this low density of sources (Figure 43) by appealing to cosmic
variance, a low metallicity and/or top-heavy IMF (Stiavelli et al 2005) or
a steep faint end slope of the luminosity function (Yan \& Windhorst
2004) but none of these arguments is convincing without further proof.
As we will see, the most logical way to proceed is to explore
both the extent of earlier star formation from the mass assembled
at $z\simeq$5-6 (Lecture 5) and to directly measure, if possible,
the abundance of low luminosity sources at higher redshift.

\subsection{The Spitzer Space Telescope Revolution: Stellar Masses at $z\simeq$6}

One of the most remarkable aspects of our search for the most
distant and early landmarks in cosmic history is that a modest cooled
85cm telescope, the Spitzer Space Telescope, can not only assist
but provide crucial diagnostic data! The key instrument is the InfraRed
Array Camera (IRAC) which offers four channels at 3.6, 4.5, 5.8
and 8$\mu$m corresponding to the rest-frame optical 0.5-1$\mu$m
at redshifts $z\simeq$6-7. In the space of only a year, the subject
has progressed from the determination of stellar masses for a
few $z\simeq$5-7 sources to mass densities and direct constraints
on the amount of early activity, as discussed in Lecture 5.

An early demonstration of the promise of IRAC in this area was
provided by Eyles et al (2005) who detected two spectroscopically-confirmed
$z\simeq$5.8 $i$-band dropouts at 3.6$\mu$m, demonstrating
the presence of a strong Balmer break in their spectral energy
distributions (Figure 44). In these sources, the optical
detection of Ly$\alpha$ emission provides an estimate of the
current ongoing star formation rate, whereas the flux longward
of the Balmer break provides a measure of the past averaged
activity. The combination gives a measure of the {\it luminosity
weighted age} of the stellar population. In general terms,
a Balmer break appears in stars whose age cannot, even
in short burst of activity, be younger than 100 Myr. Eyles
et al showed such systems could well be much older (250-650 Myr)
depending on the assumed form of the past activity. As the
Universe is only 1 Gyr old at $z\simeq$6, the IRAC detections
gave the first indirect glimpse of significant earlier star formation -
a glimpse that was elusive with direct searches at the time.

\begin{figure}
\centerline{\hbox{
\psfig{file=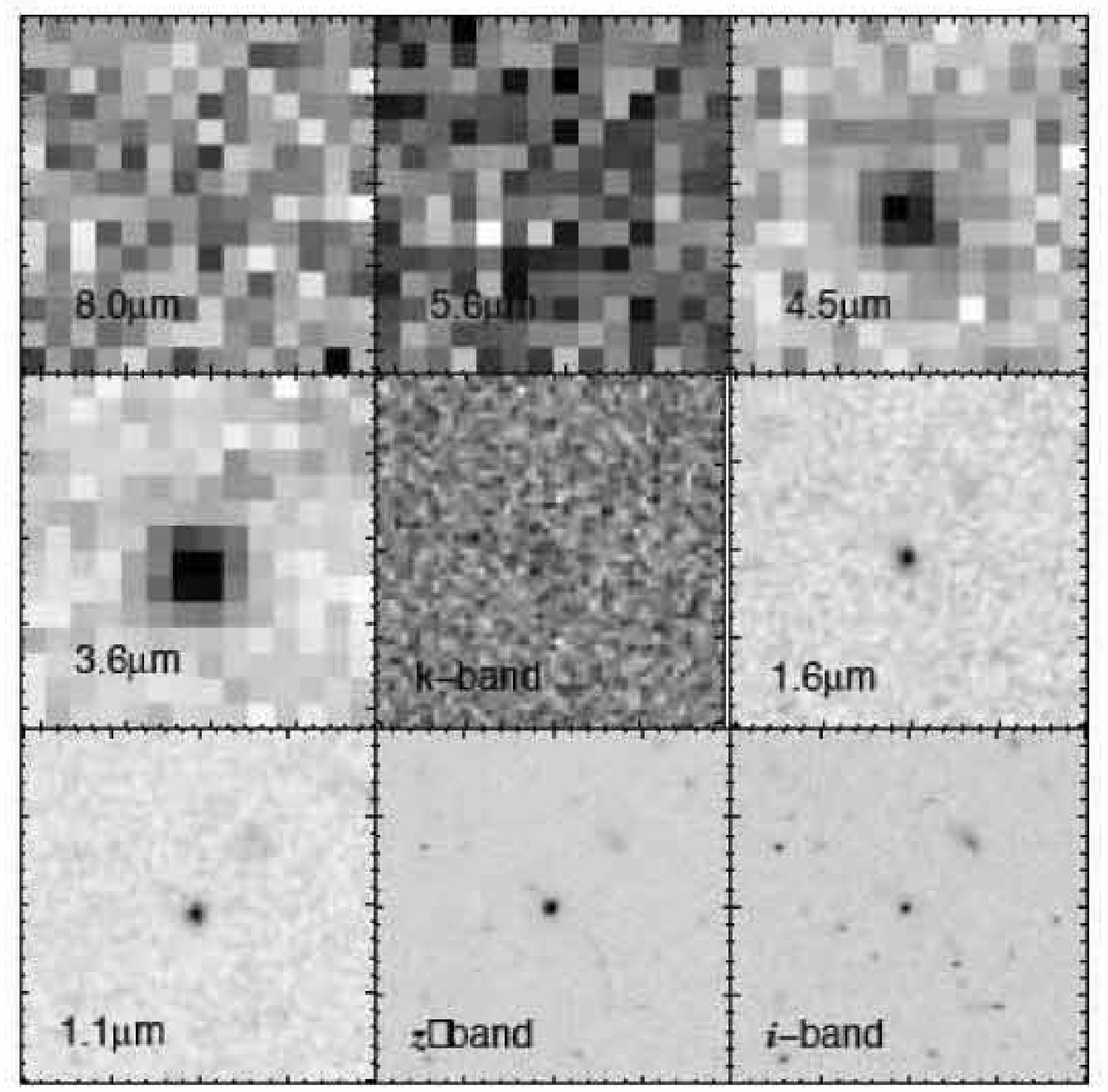,height=2.3in,angle=0}
\psfig{file=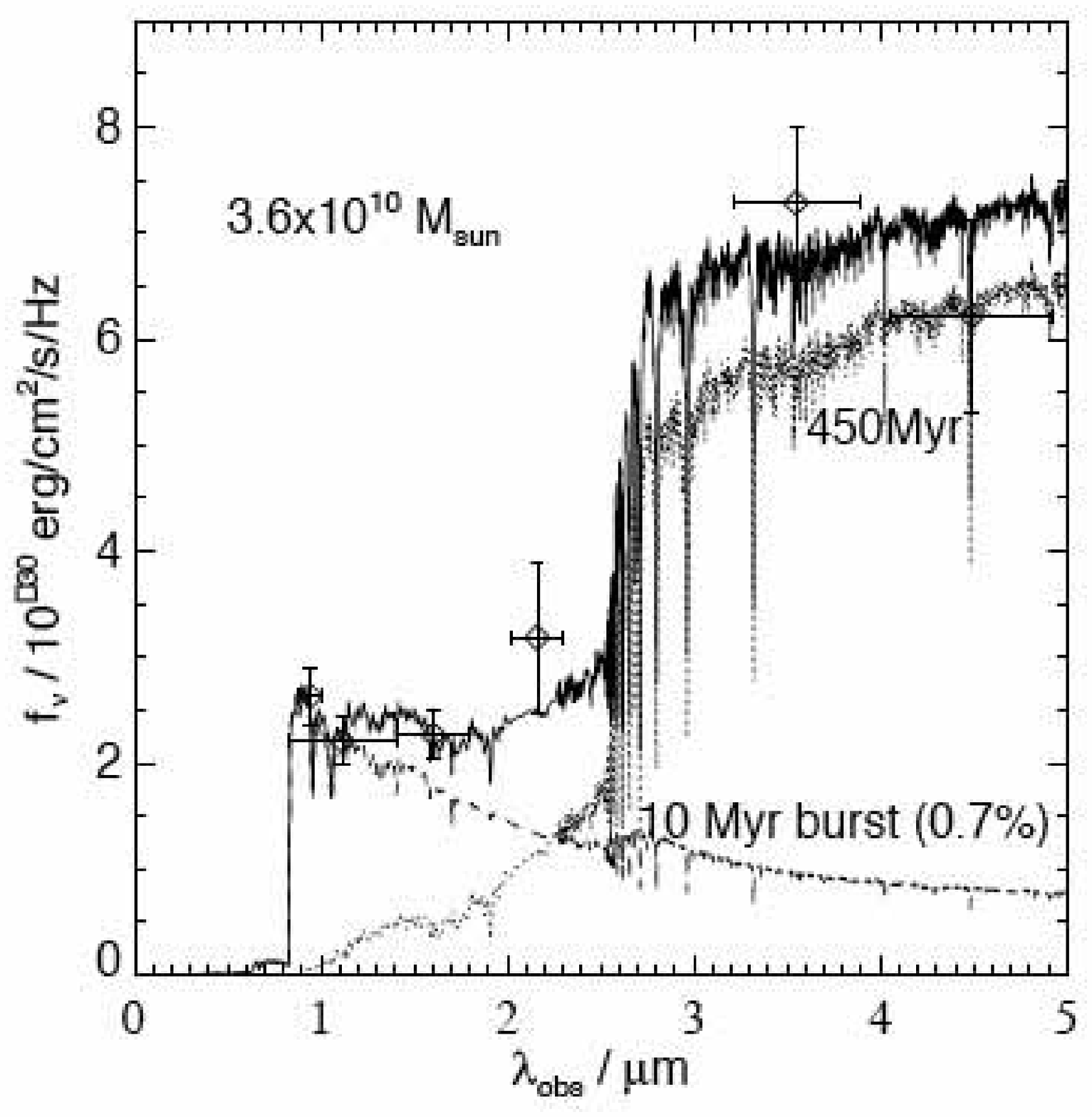,height=2.4in,angle=0}}}
\caption{(Left) Detection of a spectroscopically-confirmed 
$i$-drop at $z$=5.83 from the analysis of Eyles et al (2005). 
(Right) Spectral energy distribution of the same source. Data points 
refer to IRAC at 3.6 and 4.5$\mu$m, VLT (K) and HST 
NICMOS (J,H) overplotted on a synthesised spectrum; note
the prominent Balmer break. Synthesis models indicate
the Balmer break takes 100 Myr to establish. However, the 
luminosity-weighted age could be significantly older depending 
on the assumed past star formation rate. In the example shown,
a dominant 450 Myr component ($z_F\sim$10) is
rejuvenated with a more recent secondary burst whose ongoing
star formation rate is consistent with the Ly$\alpha$ flux observed
in the source.}
\label{fig:44}      
\end{figure}

 Independent confirmation of both the high stellar masses
 and prominent Balmer breaks was provided by the analysis
 of Yan et al (2005) who studied 3 $z\simeq$5.9 sources. Moreover,
 Yan et al also showed several objects had $(z-J)$ colors
 bluer than the predictions of the Bruzual-Charlot models for
 all reasonable model choices - a point first noted by Stanway et al (2004).

Eyles et al and Yan et al proposed the presence
of established stellar populations in $z\simeq$6 $i$-drops and
also to highlight the high stellar masses ($M\simeq1-4\,10^{10} M_{\odot}$)
they derived. At first sight, the presence of $z\simeq$6 sources as massive as the Milky
Way seems a surprising result.  Yan et al discuss the question in 
some detail and conclude the abundance of such massive objects 
is not inconsistent with hierarchical theory. In actuality it is hard to 
be sure because cosmic variance permits a huge range in the 
derived volume density and theory predicts the halo abundance
(e.g. Barkana \& Loeb 2000) rather than the stellar mass density. 
To convert one into the other requires a knowledge of the star
formation efficiency and its associated duty-cycle.

One early UDF source detected by IRAC has been a particular
source of puzzlement. Mobasher et al (2005) found a J-dropout candidate
with a prominent detection in all 4 IRAC bandpasses. Its photometric
redshift was claimed to be $z\simeq$6.5 on the basis of both a Balmer
and a Lyman break. However, despite exhaustive efforts, its redshift 
has not been confirmed spectroscopically. The inferred stellar mass is 
2-7 $10^{11} M_{\odot}$, almost an order of magnitude larger than 
the spectroscopically-confirmed sources studied by Eyles et al and 
Yan et al. If this source is truly at $z\simeq$6.5, finding such a massive galaxy whose 
star formation likely peaked before $z\simeq$9 is very surprising in the 
context of contemporary hierarchical models. Such sources should be
extremely rare so finding one in the tiny area of the UDF is 
all the more puzzling. Dunlop et al (2006) have proposed the
source must be foreground both on account of an ambiguity
in the photometric redshift determination and the absence of
similarly massive sources in a panoramic survey being conducted
at UKIRT (McClure et al 2006).

This year, the first estimates of the {\it stellar mass density}
at $z\simeq$5-6 have been derived (Yan et al 2006, Stark et al
2006a, Eyles et al 2006). Although the independenty-derived results are
consistent, both with one another and with lower redshift estimates (Figure 45)
the uncertainties are considerable as discussed briefly in the previous
lecture. There are four major challenges to undertaking a census
of the star formation at early times.

\begin{figure}
\psfig{file=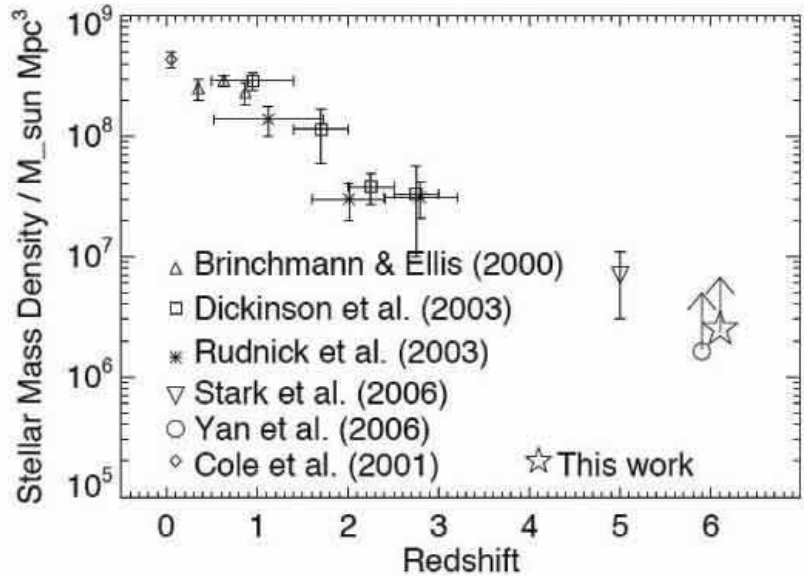,height=3.2in,angle=0}
\caption{Evolution in the comoving stellar mass density
from the compilation derived by Eyles et al (2006). The recent $z\simeq$5-6
estiimates constitute lower limits given the likelihood of
quiescent sources missed by the drop-out selection technique. 
Results at $z\simeq$6 are offset slightly in redshift for clarity.}
\label{fig:45}      
\end{figure}

Foremost, the bulk of the faint sources only have photometric
redshifts. Even a small amount of contamination from foreground
sources would skew the derived stellar mass density upward.
Increasing the spectroscopic coverage would be a big
step forward in improving the estimates.

Secondly, IRAC suffers from image confusion given its lower
angular resolution than HST (4 arcsec c.f. 0.1 arcsec). Accordingly,
the IRAC fluxes cannot be reliably estimates for blended
sources. Stark et al address this by measuring the masses only
for those uncontaminated, isolated sources, scaling up their
total by the fraction omitted. This assumes confused sources
are no more or less likely to be a high redshift.

Thirdly, as only star forming sources are selected using the
$v-$ and $i-$ dropout technique, if star formation is episodic,
it is very likely that quiescent sources are present and thus
the present mass densities represent lower limits. The missing
fraction is anyone's guess. As we saw at $z\simeq$2, the
factor could be as high as $\times$2.

Finally, as with all stellar mass determinations, many assumptions
are made about the nature of the stellar populations involved
and their star formation histories. Until individual $z\simeq$5-6 sources 
can be studied in more detail, perhaps via the location of one or two 
strongly lensed examples, or via future more powerful facilities, this will 
regrettably remain the situation. At present, such density estimates
are unlikely to be accurate to better than a factor of 2. Even so,
they provide good evidence for significant earlier star formation
(Stark et al 2006a, Lecture 5).

\subsection{Lecture Summary}

In this lecture we have discussed the great progress made in using
$v, i, z$ and $J$ band Lyman dropouts to probe the abundance
of star forming galaxies over 3$<z<$10. At redshifts $z\simeq$6
alone, Bouwens et al (2006) discuss the properties of a catalog
of 506 sources to $z_{AB}$=29.5.

In practice, the good statistics are tempered by uncertain contamination
from foreground cool stars and dusty or passively-evolving red 
$z\simeq$2 galaxies and the vagaries of cosmic variance in the
small fields studied. It may be that we will not overcome these
difficulties until we have larger ground-based telescopes.

Nonetheless, from the evidence at hand, it seems that
the comoving UV luminosity density declines from $z\simeq$3 
to 10, and that only by appealing to special circumstances can the
low abundance of star forming galaxies at $z>$6 be reconciled
with that necessary to reionize the Universe. 

One obvious caveat is our poor knowledge of the contribution from lower
luminosity systems. Some authors (Yan \& Windhorst 2004, Bouwens
et al 2006) have suggested a steepening of the luminosity
function at higher redshift. Testing this assumption with lensed
searches is the subject of our next lecture.

Finally, we have seen the successful emergence of the Spitzer
Space Telescope as an important tool in confirming the
need for star formation at $z>$6. Large numbers of $z\simeq$5-6
galaxies have now been detected by IRAC. The prominent
Balmer breaks and high stellar masses argue for much
earlier activity. Reconciling the present of {\it mature} galaxies
at $z\simeq$6 with the absence of significant star formation
beyond, is one of the most interesting challenges at the
present time.

\newpage


\section{Lyman Alpha Emitters and Gravitational Lensing}

\subsection{Strong Gravitational Lensing - A Primer}

Slowly during the twentieth century, gravitational lensing moved from a 
curiosity associated with the verification of General Relativity (Eddington
1919) to a practical tool of cosmologists and those studying distant galaxies.
There are many excellent reviews of both the pedagogical aspects
of lensing (Blandford \& Narayan 1998, Mellier 2000, Refregier 2002)
and a previous Saas-Fee contributor (Schneider 2006).

To explore the distant Universe, we are primarily concerned with
{\em strong} lensing - where the lens has a projected mass density,
above a critical value, $\Sigma_{crit}$, so that multiple images and 
high source magnifications are possible. For a simple thin lens

$$\Sigma_{crit} = \frac{c^2\,D_{OS}}{4\,\pi\,G\,D_{OL}\,D_{LS}}$$

where $D$ represents the angular diameter distance and the subscripts
$O,L,S$ refer to the observer, source and lens, respectively.
Rather conveniently, for a lens at $z\simeq$0.5 and a source at
$z>$2, the critical projected density is about 1 g cm$^{-2}$ - a
value readily exceeded by most massive clusters. The merits of 
exploring the distant Universe by imaging through clusters
was sketched in a remarkably prophetic article by Zwicky (1937).

In lensing theory it is convenient to introduce a {\it source plane}, the
true sky, and an {\it image plane}, the detector at our telescope,
where the multiple images are seen. The relationship between the 
two is then a mapping transformation which
depends on the relative distances (above). Crucially, what the
observer sees depends on the degree of alignment between
the source and lens as illustrated in Figure 46.

An elliptical lens with $\Sigma>\Sigma_{crit}$ for a given source
and lens distance produces a pair of {\it critical lines} in the
image plane where the multiple images lie. These lines map
to {\it caustics} in the source plane. The outer critical
line is equivalent to the Einstein radius $\theta_E$

$$\theta_E = \sqrt{\frac{4\,G\,M\,D_{OS}}{c^2\,D_{OL}\,D_{LS}}}$$

and, for a given source and lens, is governed by the enclosed mass $M$. 
The location of the inner critical line depends on the {\it gradient} of 
the gravitational potential (Sand et al 2005). 

\begin{figure}
\centerline{\psfig{file=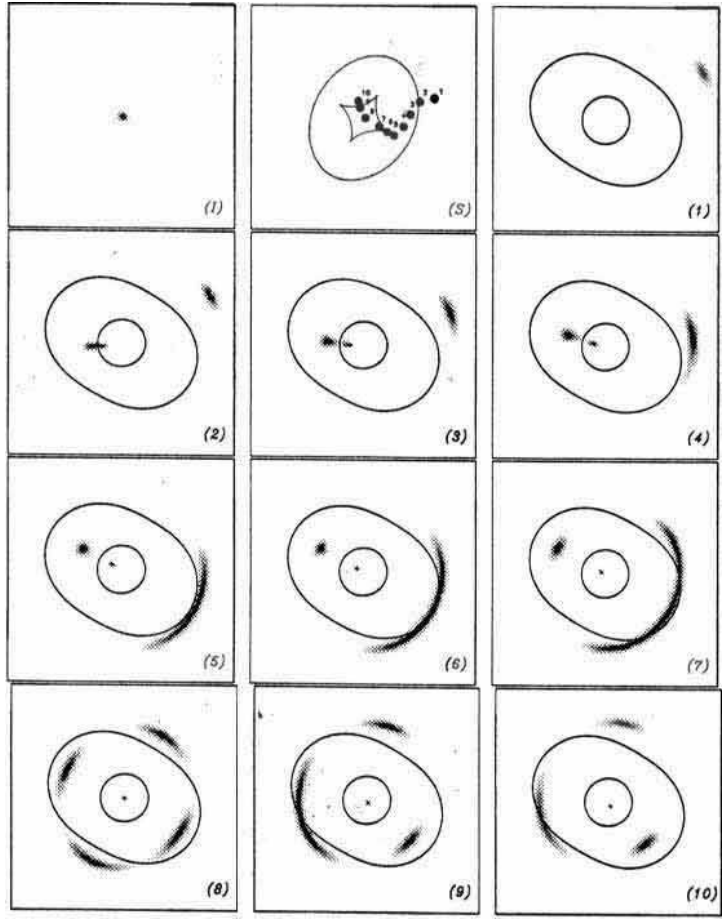,height=4.8in,angle=0}}
\caption{Configurations in the image plane for an elliptical
lens as a function of the degree of alignment between the
source and lens (second panel). Lines in the source
plane refer to `caustics' which map to `critical lines' in
the image plane (see text for details). (Courtesy of 
Jean-Paul Kneib)}
\label{fig:46}      
\end{figure}

The critical lines are important because they represent
areas of sky where very high magnifications can be 
encountered - as high as $\times$30! For $\simeq$20
well-studied clusters the location of these lines can be precisely determined 
for a given source redshift. Accordingly, it is practical to survey 
just those areas to secure a glimpse of otherwise inaccessibly
faint sources boosted into view. The drawback is that, as in an 
optical lens, the sky area is similarly magnified, so the surface 
density of faint sources must be very large to yield any results. Regions 
where the magnification exceed $\times$10 are typically only
0.1-0.3 arcmin$^2$ per cluster in extent in the image plane 
and inconveniently shaped for most instruments (Figure 47). 
The sampled area in the source plane is then ten times smaller so 
to see even one magnified source/cluster requires a surface
density of {\it distant} sources of $\sim$50 arcmin$^{-2}$.

Two other applications are particularly useful in faint
galaxy studies. Firstly, strongly magnified systems at
$z\simeq$2-3 can provide remarkable insight into an
already studied population by providing an apparently
bright galaxy which is brought within reach of superior
instrumentation. cB58, a Lyman break galaxy at 
$z$=2.72 boosted by $\times$30 to $V$=20.6 (Yee et al 1996, 
Seitz et al 1998) was the first distant galaxy to be studied 
with an echellette spectrograph (Pettini et al 2002), yielding
chemical abundances and outflow dynamics of unprecedented
precision.

More generally, a cluster can magnify a larger area of
$\simeq$2-4 arcmin$^2$ by a modest factor, say $\times$3-5.
This has been effective in probing sub-mm source counts
to the faintest possible limits (Smail et al 1997) and the
method shows promise for similar extensions with the IRAC
camera onboard Spitzer.

\begin{figure}
\psfig{file=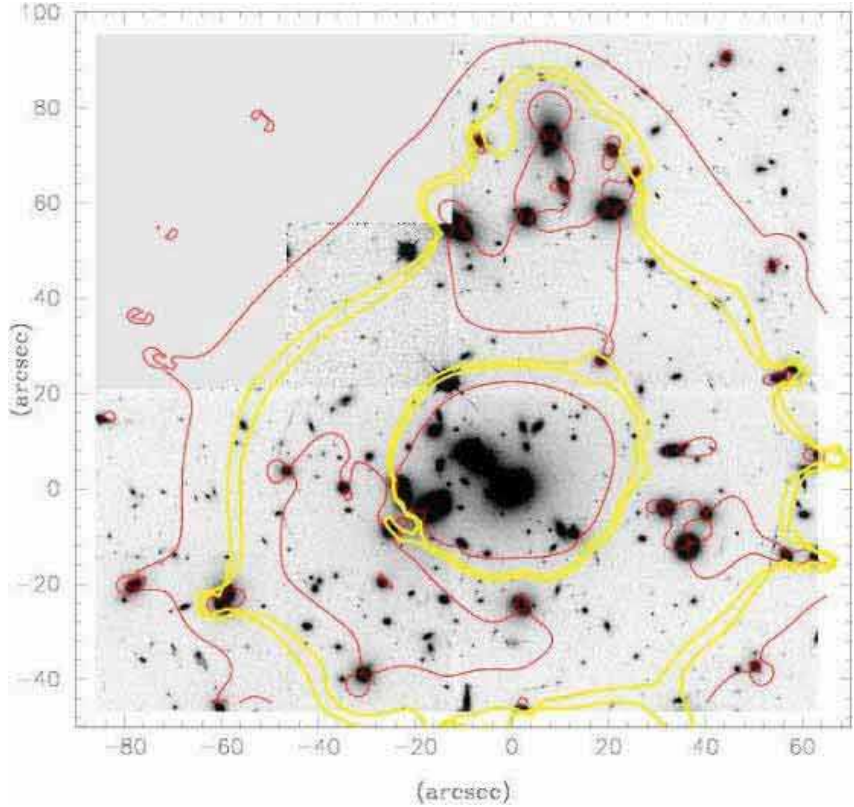,height=4.2in,angle=0}
\caption{Hubble Space Telescope image of the rich cluster
Abell 1689 with the critical lines for a source at $z\simeq$5 overlaid
in yellow. The narrow regions inbetween the pairs of yellow lines
refer to regions where the magnification exceeds $\times$10.}
\label{fig:47}      
\end{figure}

\subsection{Creating a Cluster Mass Model}

In the applications discussed above,  in order to analyze the results, the inferred
magnification clearly has to be determined. This will vary as
a function of position in the cluster image and the relative distances
of source and lens. The magnification follows from the construction
of a {\it mass model} for the cluster.

The precepts for this method are discussed in the detailed analysis of 
the remarkable image of Abell 2218 taken with the WFPC-2 camera 
onboard HST in 1995 (Kneib et al 1996). An earlier image
of AC 114 showed the important role HST would play in the recognition
of multiple images (Smail et al 1995). Prior to HST, multiple images 
could only be located by searching for systems with 
similar colors, using the fact that lensing is an achromatic phenomenon.
HST revealed that morphology is a valuable additional identifier; the improved
resolution also reveals the local shear (see Figure 48). 

\begin{figure}
\centerline{\hbox{
\psfig{file=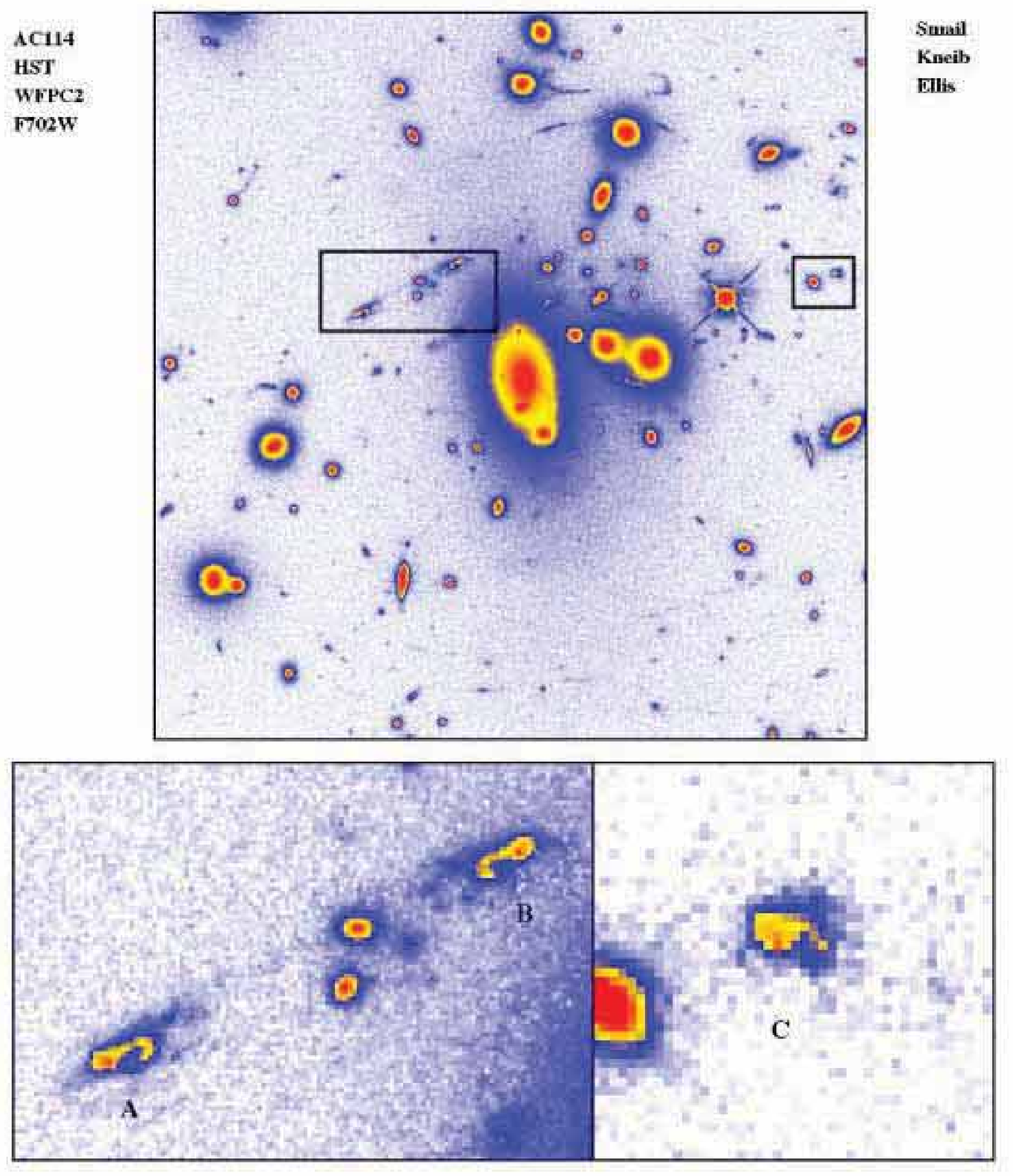,height=2.2in,angle=0}
\psfig{file=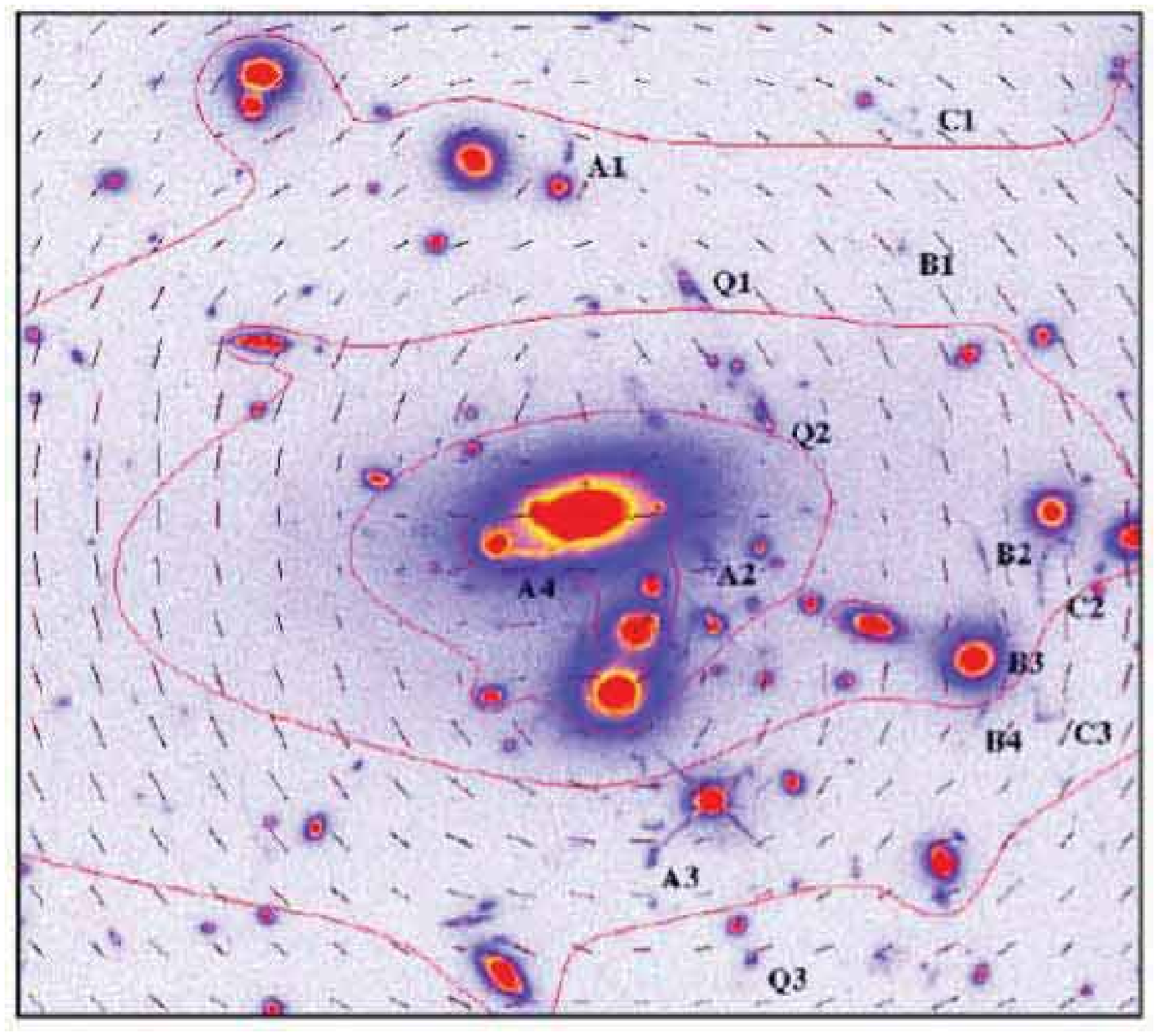,height=2.2in,angle=0}}}
\caption{Hubble Space Telescope study of the rich cluster
AC114 (Smail et al 2005) (Top) Morphological recognition of a 
triply-imaged source. The lower inset panels zoom in on
each of 3 images of the same source.  (Bottom) Construction of
mass contours (red lines) and associated shear (red vectors)
from the geometrical arrangement of further multiple images
labelled A1-3, B1-4, C1-3, Q1-3.}
\label{fig:48}      
\end{figure}

Today, various approaches are possible for constructing precise
mass models for lensing clusters (Kneib et al 1996, Jullo et al 2006,
Broadhurst et al 2005). These are generally based 
on utilizing the geometrical positions of sets of multiply-imaged systems
whose redshift is known or assumed. This then maps the form and diameter
of the critical line for a given $z$. Spectroscopic redshifts are particularly
advantageous, as are pairs that straddle the critical line whose location
can then be very precisely pinpointed. A particular mass model can be 
validated by `inverting' the technique and predicting the redshifts of other 
pairs prior to subsequent spectroscopy (Ebbels et al 1999). 

The main debate among cognescenti in this area lies in the extent to which
one should adopt a parametric approach to fitting the mass distribution,
particularly in relation to the incorporation of mass clumps associated
with individual cluster galaxy halos (Broadhurst et al 2005). Stark et al
(2006b) discuss the likely uncertainties in the mass modeling process
arising from the various techniques.

\subsection{Lensing in Action: Some High $z$ Examples}

Before turning to Lyman $\alpha$ emitters (lensed and unlensed), we
will briefly discuss what has been learned from strongly-lensed dropouts.

Figure 49 shows a lensed pair in the cluster Abell 2218 ($z$=0.18)
as detected by NICMOS onboard HST and the two shortest wavelength
channels of IRAC (Kneib et al 2004, Egami et al 2005). Although no spectroscopic 
redshift is yet available for this source, three images have been located by HST and their
arrangement around the well-constrained $z$=6 critical line suggests
a source beyond $z\simeq$6 (Kneib et al 2004).

\begin{figure}
\psfig{file=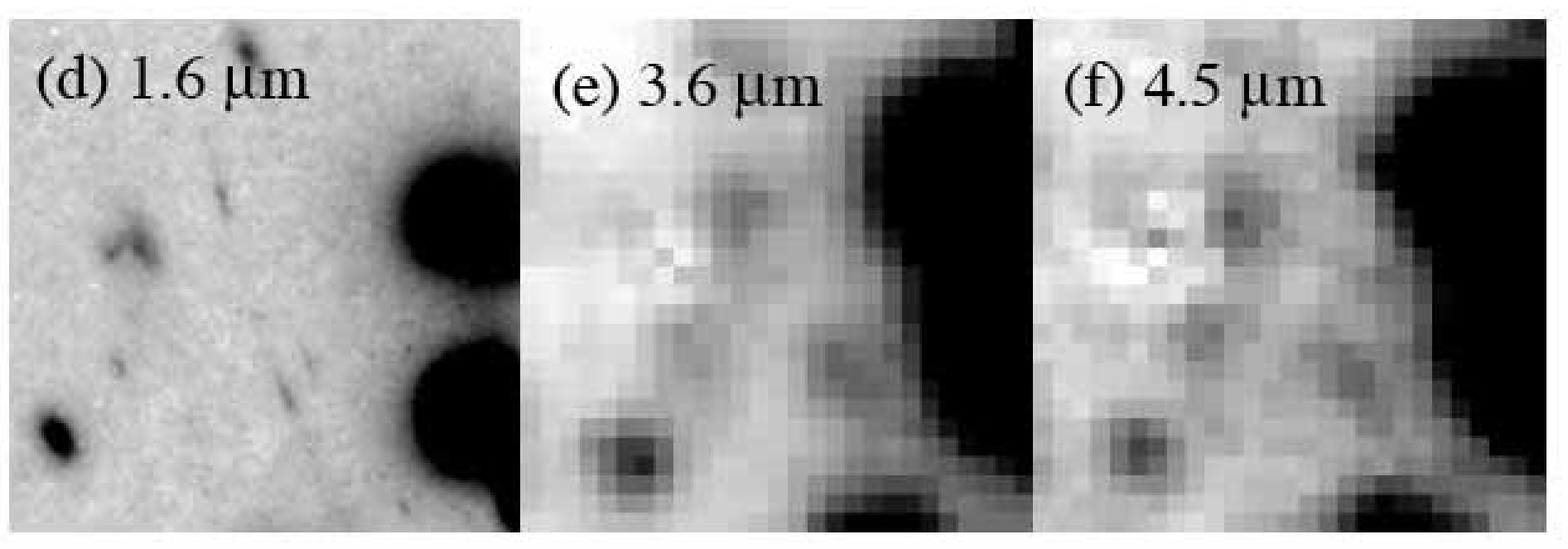,height=1.5in,angle=0}
\psfig{file=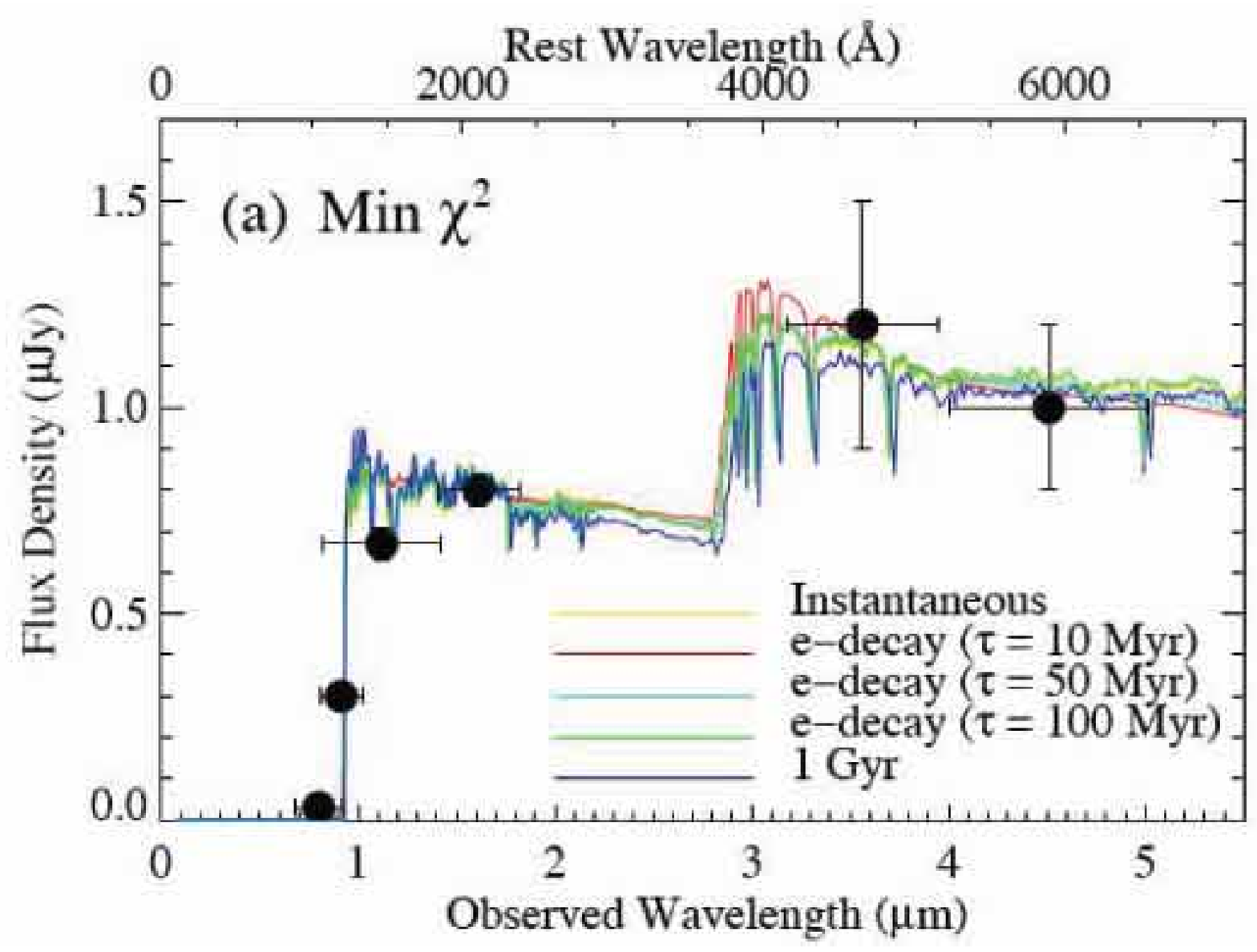,height=3.0in,angle=0}
\caption{(Top) Lensed pair of a $z$=6.8 source as seen by NICMOS and IRAC
in the rich cluster Abell 2218 (Egami et al 2005). The pair straddles the critical line
at $z\simeq$6 and a fainter third image at a location predicted by the lensing model 
has been successfully recovered in the HST data (Kneib et al 2004). (Bottom) 
Spectral energy distribution of the source revealing a significant Balmer break and 
improved estimates of the star formation rate, stellar mass and luminosity
weighted age. }
\label{fig:49}      
\end{figure}

As with the unlensed $i$-band drop out studies by Eyles et al 
(2005) and Yan et al (2005), the prominent IRAC detections 
(Egami et al 2005) permit an improved photometric redshift and important constraints 
on the stellar mass and age. A redshift of $z$=6.8$\pm$0.1 
is derived, independently of the geometric constraints used by
Kneib et al. The  stellar mass is $\simeq$5-10 10$^8 M_{\odot}$ and
the current star formation rate is $\simeq$2.6 M$_{\odot}$ yr$^{-1}$.
The luminosity-weighted age corresponds to anything from 40-450 Myr
for a normal IMF depending on the star formation history. Interestingly,
the derived age for such a prominent Balmer break generally
exceeds the e-folding timescale of the star formation history (Fig.~49)
indicating the source would have been more luminous at redshifts
$7<z<$12 (unless obscured). 

Given the small search area used to locate this object, such low mass 
sources may be very common. Accordingly, several groups are now surveying
more lensing clusters for further examples of $z$ band dropouts and
even $J$-band dropouts (corresponding to $z\simeq$8-10) . Richard et al 
are surveying 6 clusters with NICMOS and IRAC with deep ground-based $K$ band
imaging from Subaru and Keck. In these situations, one has to distinguish
between magnifications of $\times$5 or so expected across the 2-3 arcmin fields
of NICMOS and IRAC, and the much larger magnifications possible
close to the critical lines. Contamination from foreground sources should 
be similar to what is seen in the GOODS surveys discussed in Lecture
6. The discovery of image pairs in the highly-magnified regions would be 
a significant step forward since spectroscopic confirmation of any
sources at the limits being probed ($H_{AB}\simeq$26.5-27.0) will be
exceedingly difficult.

\subsection{Lyman alpha Surveys}

The origin and characteristic properties of the Lyman $\alpha$ emission line
has been discussed by my colleagues in their lectures (see also Miralda-Escude
1998, Haiman 2002, Barkana \& Loeb 2004 and Santos 2004). The $n$=2 to 1 
transition corresponding to an energy difference of 10.2 eV and rest-wavelength of 
$\lambda$1216 \AA\  typically arises from ionizing photons absorbed by nearby 
hydrogen gas. The line has a number of interesting features which make it
particularly well-suited for locating early star forming galaxies as well as
for characterizing the nature of the IGM.

In searching for distant galaxies, emission lines offer far more contrast
against the background sky than the faint stellar continuum of a drop-out.
A line gives a convincing spectroscopic redshift (assuming it is correctly
identified) and models suggest that as much as 7\% of the bolometric
output of young star-forming region might emerge in this line. For a normal
IMF and no dust, a source with a star formation rate (SFR) of 1 $M_{\odot}$ yr$^{-1}$
yields an emission line luminosity of 1.5 10$^{42}$ ergs sec$^{-1}$.

Narrow band imaging techniques (see below) can reach fluxes of $<$
10$^{-17}$ cgs in comoving survey volumes of $\simeq$10$^5$ Mpc$^3$, 
corresponding to a SFR $\simeq$3 $M_{\odot}$ yr$^{-1}$ at $z\simeq$6. 
Spectroscopic techniques can probe fainter due to the improved
contrast. This is particularly so along the critical lines where the additional 
boost of gravitational lensing enables fluxes as faint as 3. 10$^{-19}$ cgs 
to be reached (corresponding to SFR $\simeq$0.1 $M_{\odot}$ yr$^{-1}$).
However, in this case the survey volumes are much smaller ($\simeq$50 Mpc$^3$).
In this sense, the two techniques (discussed below) are usefully complementary.

Having a large dynamic range in surveys for Ly$\alpha$ emission is important
not just to probe the luminosity function of star-forming galaxies but also because
it can be used to characterize the IGM. As a resonant transition, foreground hydrogen 
gas clouds can scatter away Ly$\alpha$ photons in both direction and frequency. 
In a partially ionized IGM, scattering is maximum at $\lambda$1216 \AA\ in the 
rest-frame of the foreground cloud, thus affecting the {\it blue} wing of the 
observed line. However, in a fully neutral IGM, scattering far from resonance
can occur leading to damping over the entire observed line. Figure 50 illustrates
how, in a hypothetical situation where the IGM becomes substantially neutral 
during 6$<z<$7, surveys reaching the narrow-band flux limit would still
find emitters at $z$=7. Their intense emission would only be partially
damped by even a neutral medium. However, lines with fluxes
at the spectroscopic lensing limit would not survive. Accordingly, one possible
signature of reionization would be a significant change in the shape
of the Ly$\alpha$ luminosity function at the faint end (Furlanetto et
al 2005).

\begin{figure}
\psfig{file=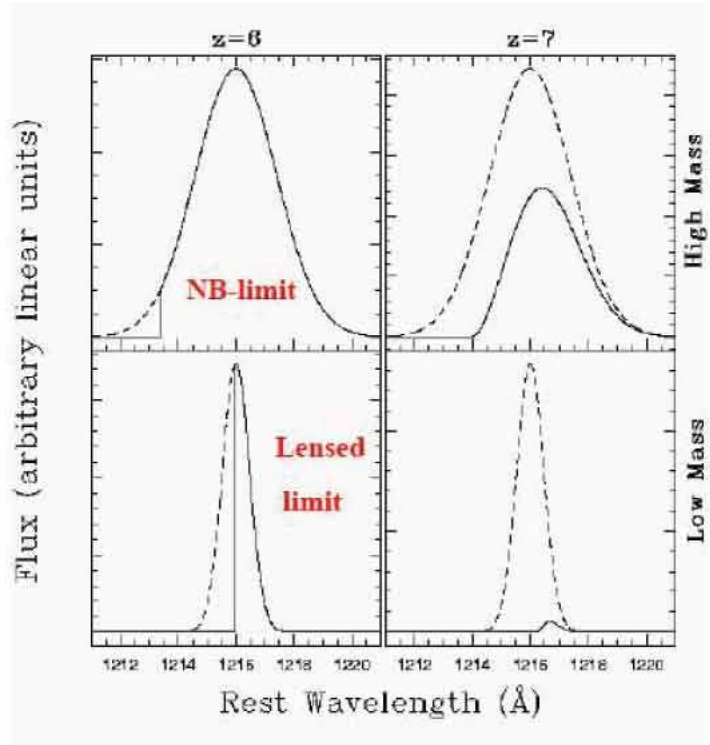,height=3.9in,angle=0}
\caption{The Lyman $\alpha$ damping wing is absorbed
by neutral hydrogen and thus can act as a valuable tracer
of the nature of the IGM. The simulation demonstrates the effect
of HI damping on emission lines in high mass and 
low mass systems (characteristic of sources detected in typical
narrow band and lensed spectroscopic surveys respectively) assuming
reionization ends inbetween $z$=6 and 7. The dramatic change
in visibility of the weaker systems suggests their study with
redshift may offer a sensitive probe of reionization. Courtesy:
Mike Santos. }
\label{fig:50}      
\end{figure}

\subsection{Results from Narrow Band Ly$\alpha$ Surveys}

The most impressive results to date have come from various narrow-band
filters placed within the SuPrime camera at the prime focus
of the Subaru 8m telescope (Kodaira et al 2003, Hu et al 2004, 
Ouchi et al 2005, Taniguchi et al 2005,  Shimasaku et al 2006, Kashikawa 
et al 2006, Iye et al 2006). Important conclusions have also been
deduced from an independent 4m campaign (Malhotra \& Rhoads
2004).

\begin{figure}
\psfig{file=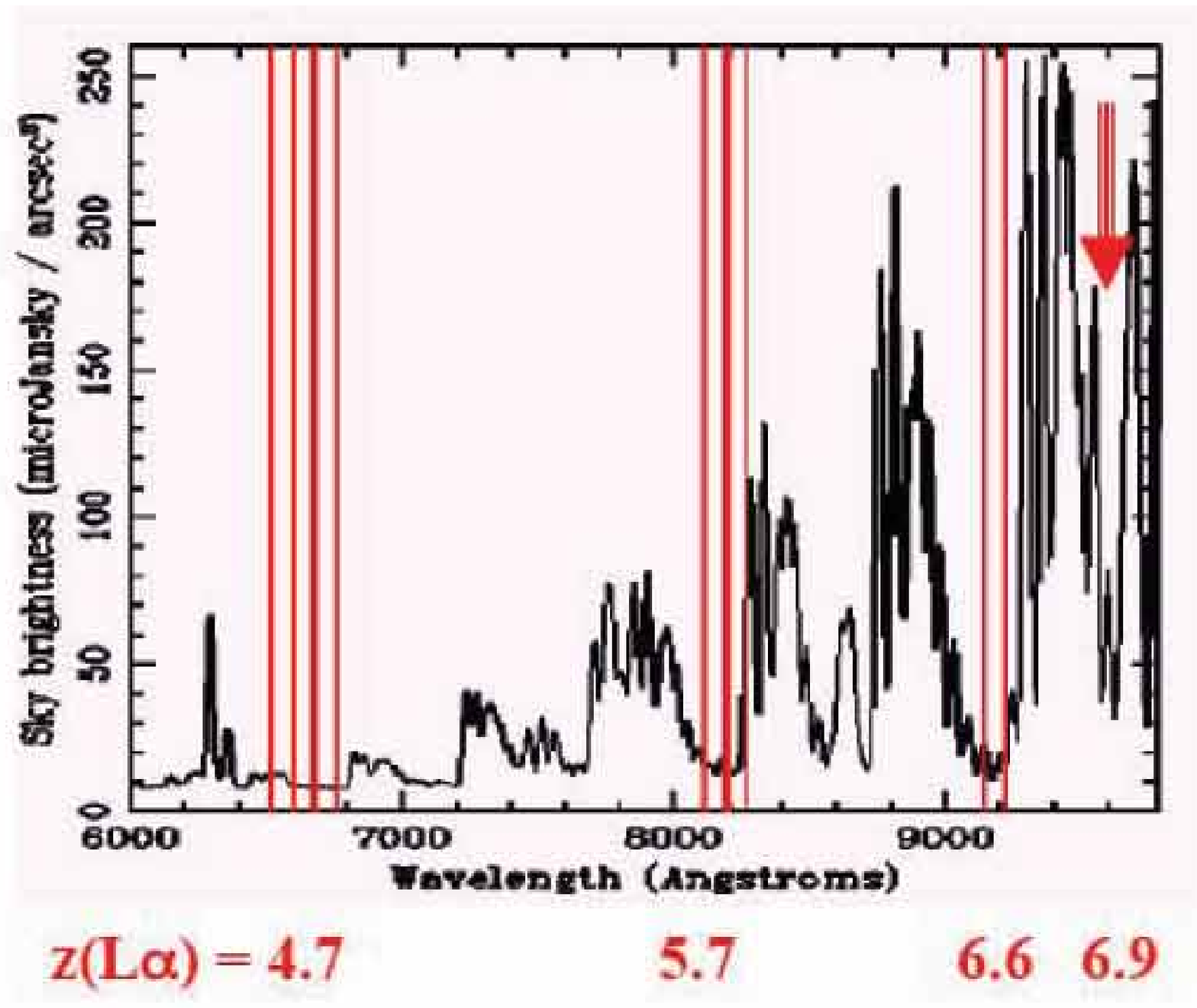,height=3.4in,angle=0}
\caption{Night sky spectrum and the deployment of narrow band
filters in `quiet' regions corresponding to redshifted Lyman $\alpha$
emission as indicated below. The final optical window, corresponding
to $z\simeq$6.9, was successfully exploited by Iye et al (2006) to find
two sources close to $z\simeq$7.
}
\label{fig:51}      
\end{figure}

Narrow band filters are typically manufactured at wavelengths where
the night sky spectrum is quiescent, thus maximizing the contrast. These
locations correspond to redshifts of $z$=4.7, 5.7, 6.6 and 6.9 (Fig. 52). A recent
triumph was the successful recovery of two candidates at $z\simeq$6.96
by Iye et al (2006). Candidates are selected by comparing their
narrow band fluxes with that in a broader band encompassing the
narrow band wavelength range. The contrast can then be used as
an indicator of line emission (Fig. 53). Spectroscopic follow-up is
still desirable as the line could arise in a foregound galaxy with [O II] 3727 \AA\
or [O III] 5007 \AA\ emission. The former line is a doublet and the latter is part
of a pair with a fixed line ratio, separated in the rest-frame by only 60 \AA\
or so, so these contaminants are readily identified. Furthermore, Ly$\alpha$ is often
revealed by its asymmetric profile (c.f. Fig. 51). 

\begin{figure}
\centerline{\hbox{
\psfig{file=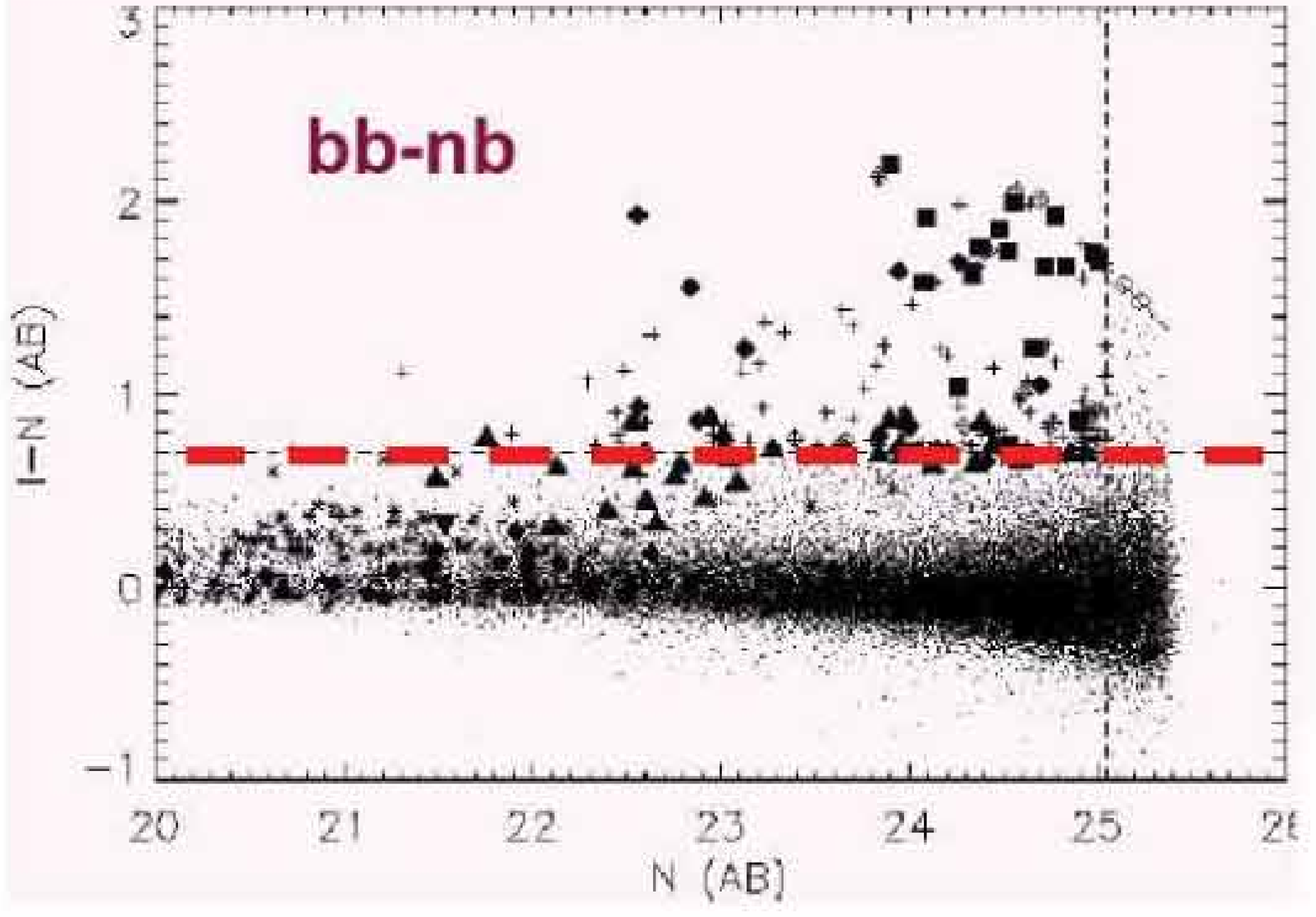,height=1.9in,angle=0}
\psfig{file=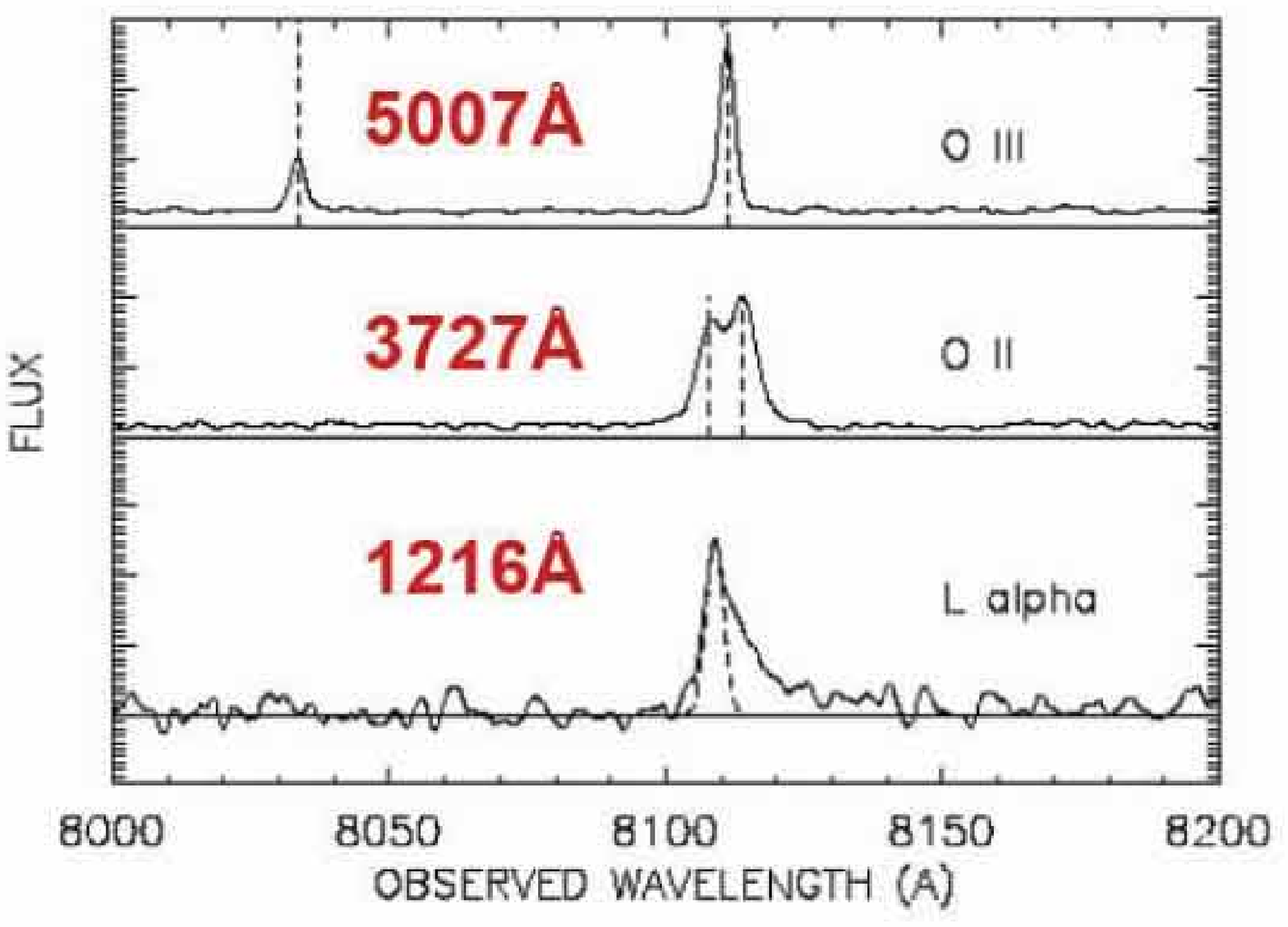,height=1.9in,angle=0}}}
\caption{The two-step process for locating high redshift Lyman
$\alpha$ emitters (Hu et al 2004). (Left) Comparison of broad and 
narrow band magnitudes; sources with an unusual difference in 
the sense of being brighter in the narrow band filter represent 
promising candidates. (Right) Spectroscopic follow-up
reveals typically three possibilities - [O III] or [O II] at
lower redshift, or Ly$\alpha$ often characterized by its
asymmetric line profile. }
\label{fig:52}      
\end{figure}

Spectroscopic follow-up is obviously time-intensive for a large
sample of candidates, hundreds of which can now be found with
panoramic imagers such as SuPrimeCam.  Therefore it is worth
investigating additional ways of eliminating foreground sources. Tanuguchi
et al (2005) combine the narrow band criteria adopted in Fig. 53
with a broad-band $i-z$ drop-out signature. Spectroscopic follow-up
of candidates located via this double color cut revealed a 50-70\% success rate
for locating high $z$ emitters. The drawback is that the sources so
found cannot easily be compared in number with other, more traditional,
methods. Nagao et al (2005) used a narrow  - broad band color
criterion in the opposite sense, locating sources with a narrow
band {\it depression} (rather than excess). Such rare sources 
are confirmed to be sources with extremely intense emission 
elsewhere in the broad-band filter. Such sources, with Lyman
$\alpha$ equivalent widths in excess of several hundred \AA\ 
are interesting because they may challenge what can be produced
from normal stellar populations.

\begin{figure}
\centerline{\psfig{file=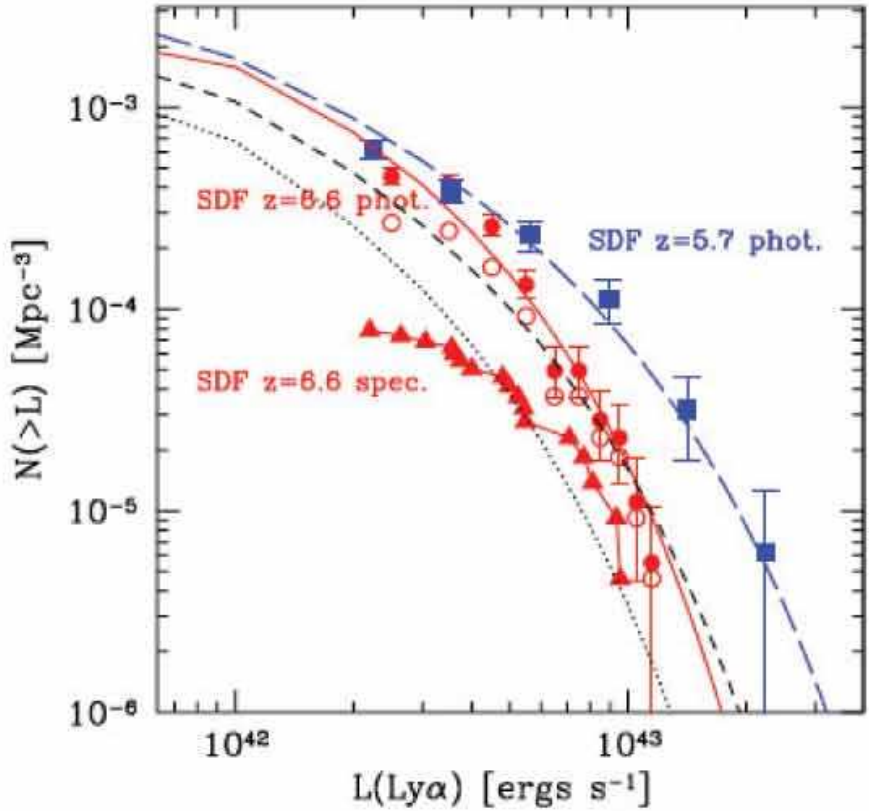,height=4.7in,angle=0}}
\caption{Comparison of the Lyman $\alpha$ luminosity
functions at $z$=5.7 and 6.5 from the surveys of
Kashikawa et al (2006) and Shimasaku et al (2006); both
spectroscopically-confirmed and photometric candidates
are plotted. The decline in luminous emitters
is qualitatively similar to trends seen over 3$<z<$6 in
luminous continuum drop-outs.}
\label{fig:53}      
\end{figure}

Malhotra \& Rhoads (2004) were the first to consider the absence
of evolution in the Ly$\alpha$ LF as a constraint on the neutral
fraction. Although they found no convincing change in the LF 
between $z$=5.7 and 6.5, the statistical uncertainties in both 
LFs were considerable. Specifically, below luminosities of 
$L_{Ly\alpha}\simeq$10$^{42.5}$ ergs sec$^{-1}$ no detections were
then available. Hu et al (2005) have also appealed to the absence
of any significant change in the mean Ly$\alpha$ profile.
Malhotra \& Rhoads deduced the neutral fraction must be $x_{HI}<$0.3 
at $z\simeq$6 supporting early reionization. However, 
Furlanetto et al (2005) reanalyzed this constraint and indicated
that strong emitters could persist even when $x_{HI}\simeq$0.5
(c.f. Fig. 51).

Kashikawa et al (2006) and Shimasaku et al (2006) have
determined statistically greatly improved Lyman $\alpha$ luminosity 
functions and discuss both spectroscopic confirmed and 
photometrically-selected emitters (Fig. 53). No decline
is apparent in the abundance of low luminosity emitters as expected
in an IGM with high $x_{HI}$; indeed the most significant change
is a decline with redshift in the abundance of the most luminous
systems. Although the change seems surprisingly rapid given
the time interval is only 150 Myr, this is consistent with
growth in the halo mass function (Dijsktra et al 2006).

\subsection{Results from Lensed Ly$\alpha$ Surveys}

\begin{figure}
\psfig{file=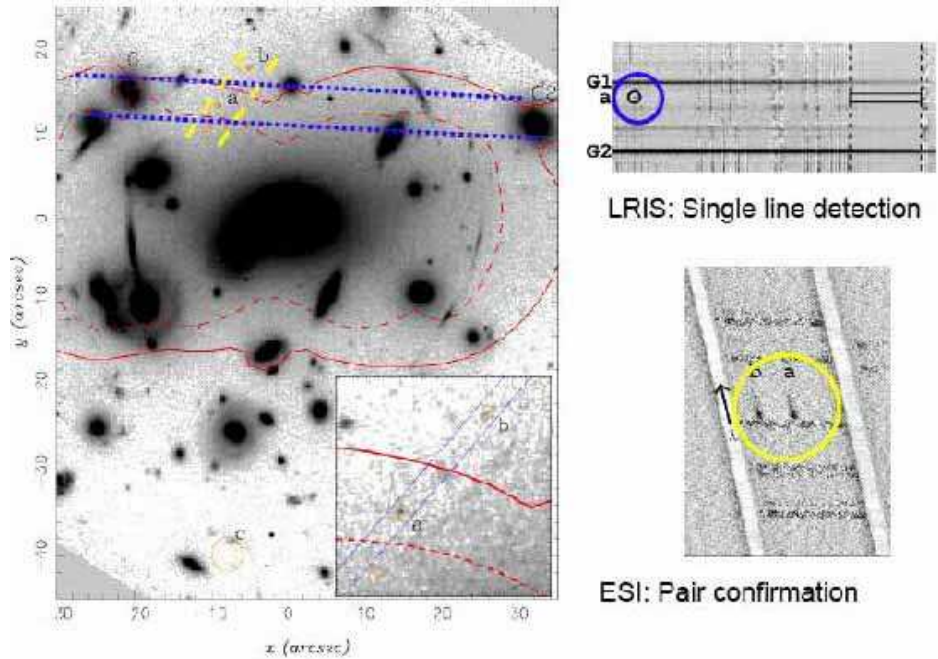,height=3.5in,angle=0}
\caption{Critical line mapping in Abell 2218: how it works (Ellis et al 2001).
The red curves show the location of the lines of very high magnification
for a source at a redshift $z$=1 (dashed) and $z$=6 (solid). Blue
lines show the region scanned at low resolution with a long-slit
spectrograph. The upper right panel shows the detection of an
isolated line astrometrically associated with (a) in the
HST image for which a counter image (b) is predicted and
recovered (see also inset to main panel). A higher dispersion
spectrum aligned between the pair (yellow lines) reveal strong
emission with an asymmetric profile in both (lower right panel).}
\label{fig:54}      
\end{figure}

The principal gain of narrow band imaging over other techniques in locating
high redshift Lyman $\alpha$ emitters lies in the ability to exploit panoramic 
cameras with fields of view as large as 30-60 arcmin. Since cosmic lenses
only magnify fields of a few arcmin or less, lensing searches are only
of practical utility when used in spectroscopic mode. As discussed
above, the gain in sensitivity can be factors of $\times$30 or more, and given
the small volumes explored, they are primarily useful in testing the
faint end of the Ly$\alpha$ luminosity function at various redshifts.
A number of workers (e.g. Barkana \& Loeb 2004) have emphasized
the likelihood that the bulk of the reionizing photons arise from an abundant
population of intrinsically-faint sources, and lensed searches provide
the only practical route to observationally testing this hypothesis.

A practical demonstration of a blind search for lensed
Ly$\alpha$ emitters is summarized in Figure 54. A long slit is
oriented along a straight portion of the critical line (whose location 
depends on the source redshift). The survey comprises several
exposures taken in different positions offset perpendicular to the
critical line. Candidate emission lines are astrometrically located
on a deep HST image and, if a counter-image consistent with the
mass model can be located, a separate exposure is undertaken
to capture both (as was the case in the source located by Ellis
et al 2001). Unfortunately, continuum emission is rarely seen from a faint
emitter and the location of a corresponding second image is
often too uncertain to warrant a separate search. In this case 
contamination from foreground sources has to be inferred from 
the absence of corresponding lines at other wavelengths 
(Santos et al 2004, Stark et al 2006b). 

Using this technique with an optical spectrograph sensitive to
Ly$\alpha$ from 2.2$<z<$6.7, Santos et al (2004) conducted a 
survey of 9 lensing clusters and found 11 emitters probing luminosities 
as faint as  $L_{Ly\alpha}\simeq10^{40}$ cgs, significantly fainter
than even the more recent Subaru narrow band imaging searches
(Shimasaku et al 2006). The resulting luminosity function is
flatter at the faint end than implied for the halo mass function
and is consistent with suppression of star formation in the lowest
mass halos.

\begin{figure}
\psfig{file=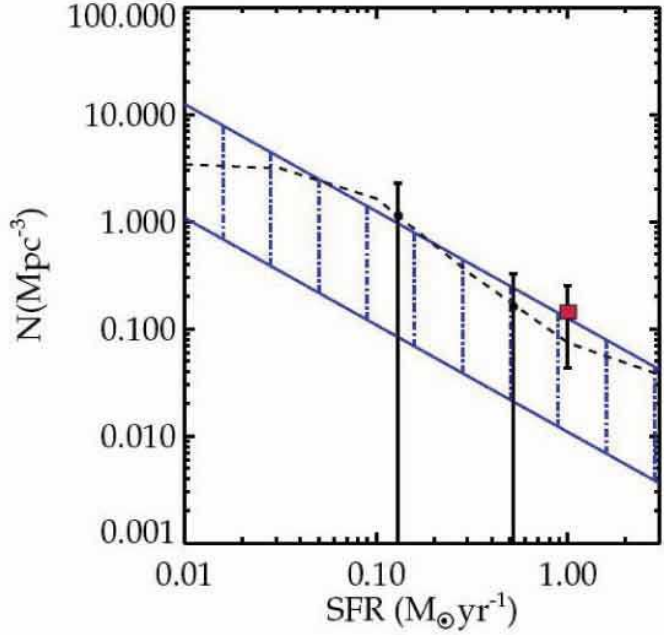,height=3.4in,angle=0}
\caption{The volume density of sources
of various star formation rates at $z\simeq$8-10 
required for cosmic reionization for a range of assumed 
parameters (blue hatched region) compared to the 
inferred density of lensed emitters from the
survey of Stark et al (2006b). The open red symbol
corresponds to the case if all detected emitters
are at $z\simeq$10, the black symbols correspond
to the situation if the two most promising candidates
are at $z\simeq$10, and the dashed line corresponds
to the 5$\sigma$ upper limit if none of the candidates is
at $z\simeq$10.}
\label{fig:55}      
\end{figure}

Stark et al (2006b) have extended this technique to higher
redshift using an infrared spectrograph operating in the $J$
band, where lensed Ly$\alpha$ emitters in the range 8.5$<z<$10.2
would be found. This is a much more demanding experiment
than that conducted in the optical because of the brighter and
more variable sky brightness, the smaller slit length necessitating
very precise positioning to maximize the magnifications and, obviously, 
the fainter sources given the increased redshift. Nonetheless, a 
5$\sigma$ sensitivity limit fainter than 10$^{-17}$ cgs corresponding 
to intrinsic (unlensed) star formation rates of 0.1 $M_{\odot}$ yr$^{-1}$  
is achieved with the 10m Keck II telescope in exposure times
of 1.5 hours per slit position. 

After surveying 10 clusters with several slit positions per
cluster, 6 candidate emission lines have been found and,
via additional spectroscopy, it seems most cannot
be explained as foreground sources. Stark et al estimate
the survey volume taking into account both the spatially-dependent
magnification (from the cluster mass models) and the 
redshift-dependent survey sensitivity (governed by the night sky
spectrum within the spectral band). 

Madau et al (1999) and Stiavelli et al (2004) have
introduced simple prescriptions for estimating the abundance
of star forming sources necessary for cosmic reionizations.
While these prescriptions are certainly simple-minded, given
the coarse datasets at hand, they provide an illustration
of the implications.

Generally, the abundance of sources of a given star formation
rate SFR necessary for cosmic reionization over a time interval
$\Delta\,t$ is

$$ n \propto \frac{B\,n_H}{f_c\,SFR\,\Delta\,t}$$

where $B$ is the number of ionizing photons required to
keep a single hydrogen atom ionized, $n_H$ is the
comoving number density of hydrogen at the redshift
of interest and $f_c$ is the escape fraction of ionizing
photons. Figure 56 shows the upshot of the Stark et al (2006b)
survey for various  assumptions. The detection of even a few convincing
sources with SFR $\simeq$0.1-1 $M_{\odot}$ yr$^{-1}$
in such small cosmic volumes would imply a significant
contribution from feeble emitters at $z\simeq$10. Although
speculative at this stage given both the uncertain nature
of the lensed emitters and the calculation above, it nonetheless
provides a strong incentive for continued searches.

\subsection{Lecture Summary}

In this lecture we have shown how Lyman $\alpha$ emission
offers more than simply a way to locate distant galaxies. The
distribution of line profiles, equivalent widths and its luminosity
function can act as a sensitive gauge of the neutral fraction
because of the effects of scattering by hydrogen clouds. Surveys
have been undertaken using optical cameras and narrow
band filters to redshifts $z\simeq$7.

However, despite great progress in the narrow-band surveys,
as with the earlier $i$-band drop outs, there is some dispute
as to the evolutionary trends being found. Surprisingly strong
evolution is seen in luminous emitters over a very short period
of cosmic time corresponding to 5.7$<z<$6.5. And, to date 
there is no convincing evidence that line profiles are
evolving or that the equivalent width distribution of emitters
is skewed beyond what can be accounted for by normal
young stellar populations. One suspects we will have to
push these techniques to higher redshift which will be hard
given the Ly$\alpha$ line moves into the infrared where no
such panoramic instruments are yet available.

We have also given a brief tutorial on strong gravitational
lensing. In about 20 or so clusters, spectroscopic redshifts
for sets of multiple images has enabled quite precise mass
models to be determined which, in turn, enable accurate
magnification maps to be derived. Remarkably faint sources
can be found by searching along the so-called {\it critical lines}
where the magnification is high. The techniques has revealed
a few intrinsically faint sources and, possibly, the first glimpse of a
high abundance of faint star forming sources at $z\simeq$10
has been secured.

\newpage


\section{Cosmic Infrared Background}

\subsection{Motivation}

In this lecture we examine the role of cosmic backgrounds.  First we make
an important distinction. We are primarily concerned with extragalactic 
backgrounds composed of unresolved faint sources rather than the
fairly isotropic microwave radiation that comes from the recombination of 
hydrogen. Such unresolved source background have played a key
role in astrophysics. 

The pattern of discovery outside the optical and near-infrared spectral windows 
often goes like the following; the background is first discovered by sensitive
detectors which do not have the angular resolution to see if there is any fine
structure from faint sources. There is then some puzzlement as to its origin:
for example, does it arise from an unforeseen population of sources
beyond those already counted? In this manner, the X-ray background was
identified as arising from active galactic nuclei and the sub-mm background
from dusty star-forming galaxies. In each case we are concerned with
separating the counts of resolved sources to some detectability limit
with the measured value of the integrated background. Key issues
relate to the contribution of resolved sources and the removal of
spurious (non-extragalactic) foreground signals.  Excellent
reviews of the subject have been provided by Leinert et al (1998),
Hauser \& Dwek (2001) and Kashlinksy (2005). 

In our case, we are interested in extending source counts
of star forming galaxies beyond $z\simeq$6 and so the question of
a near-infrared extragalactic background signal is of greatest 
relevance. The instruments concerned include cameras on
Hubble Space Telescope and Spitzer SpaceTelescope, but
other space missions such as COBE and the Infrared Telescope
in Space have provided important results.

\subsection{Methodogy}

To understand the distinction between the {\em resolved} and
{\em unresolved} components of the background, it is helpful to derive
some fundamental relations for galaxy counts.

In the magnitude system, the differential count slope $\gamma$ with increasing
magnitude is

$$\gamma = \frac{d\,log\,N(m)}{dm}$$

where $N(m)$ is the differential number of galaxies per unit sky area (e.g. deg$^{-2}$)
in some counting bin $dm$.

Now the contribution to the surface brightness of the extragalactic night sky from sources 
as a given magnitude $m$ is:

$$i_{\nu}(m) = 10^{-0.4(m+const)}\,N(m)$$

where the 0.4 factor arises from the relationship between flux and magnitude.

And the integrated surface brightness (the extragalactic background light, EBL)  is
obtained by extending this integral to infinitely faint limits, viz:

$$I_{\nu} = \int_{-\infty}^{\infty} i_{\nu}(m)\,dm$$

The bolometric equivalent of the EBL is then $\int I_{\nu}\,d\nu$. The EBL is often
alternatively expressed as EBL = $\int \nu\,I_{\nu}\d\,d\nu$.

Surface brightness is usually expressed in nW m$^{-2}$ steradian$^{-1}$ although very
early articles refer to some derivative of magnitudes deg$^{-2}$. As infrared fluxes
are often expressed in Janskies (1 Jansky = 10$^{-26}$ W m$^{-2}$ Hz$^{-1}$, a
useful conversion is 1 nw m$^{-2}$ ster$^{-1}$ = 3000/$\lambda(\mu$\,m) MJy ster$^{-1}$.

Examination of the above relation shows that if $\gamma>$0.4, the surface brightness
contribution from fainter sources outshines that of brighter ones and so the EBL will
diverge. By deduction, therefore, the maximum contribution of resolved sources to the
background will be where $\gamma\simeq$0.4. If the slope of the counts turns 
down below 0.4 at some point, it may seem pointless to speculate that there is
much information from fainter unresolved sources. In the case of searching for
cosmic reionization however, where the first sources may be a distinct population,
continuity in the source counts may not be expected. For this reason, the interesting
quantity is the difference between a measured EBL and the integrated contribution
from the faintest resolved sources.

\begin{figure}
\psfig{file=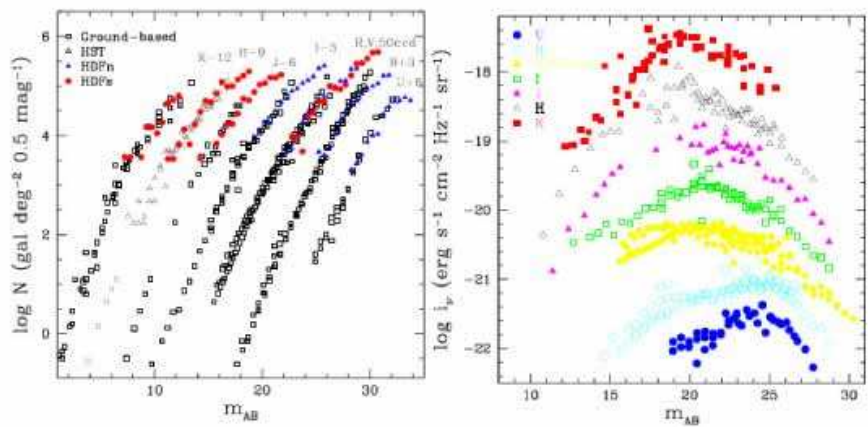,height=2.3in,angle=0}
\caption{(Left) Differential galaxy counts as a function of wavelength from the
compilation of Madau \& Pozzetti (2000). Note the absolute numbers have
been arbitrarily scaled for convenience.  (Right) Magnitude-dependent contribution 
of the counts to the  surface brightness $i_{\nu}$ of the extragalactic night sky. Depending on
the waveband, the peak contribution occurs at $m_{AB}\simeq$20-25. 
}
\label{fig:56}      
\end{figure}

Madau \& Pozzetti (2000) present a careful analysis of the deep optical and near-infrared
counts at various wavelengths, mostly from Hubble Space Telescope data prior
to the Ultra Deep Field (Figure 56). Extrapolation of the counts enables the 
contribution to the integrated light from {\em known populations} to be evaluated 
as a function of wavelength. 
The total EBL from this analysis is 55 nW m$^{-2}$ sr$^{-1}$ of which the dominant
component lies longward of 1$\mu$m. However, it must be remembered that
galaxies are extended objects and so it is possible that significant light is
lost in the outermost regions of each. As surface brightness is relativistically
dimmed by the cosmic expansion, $\propto(1+z)^4$, the contribution from
distant sources could be seriously underestimated (Lanzetta et al 2002).

In a similar fashion to the check we made that the integrated star formation
history produces the present stellar mass density (Lecture 4), so it is possible
to verify that the bolometric output from stellar evolution should be consistent
with the present mass density of stars. 

The bolometric radiation density $\rho_{bol}(t)$ is

$$ \rho_{bol}(t) = \int_{0}^{t} L(\tau) \dot{\rho_s} (t - \tau) d\,\tau$$

where $\dot{ \rho_s}$ is the star formation rate per comoving volume.

The integrated EBL is then

$$I_{EBL} = \frac{c}{4\pi}\,\int_{t_F}^{t_H} \frac{\rho_{bol}(t)}{1+z}\,dt$$

The check is a little bit trickier because a typical star formation history has to be 
assumed and presumably there is a large variety for different kinds of sources 
observed at various redshifts. The issue is discussed in detail by Madau
\& Pozzetti.

\subsection{Recent Background Measurements}

The goal for making extragalactic background light (EBL) measures is thus to
determine the extent to which the measured value (or limit) exceeds
that predicted from extrapolation of the galaxy counts. This might then
provide some evidence for a new distant population such as the sources
responsible for cosmic reionization.

\begin{figure}
\psfig{file=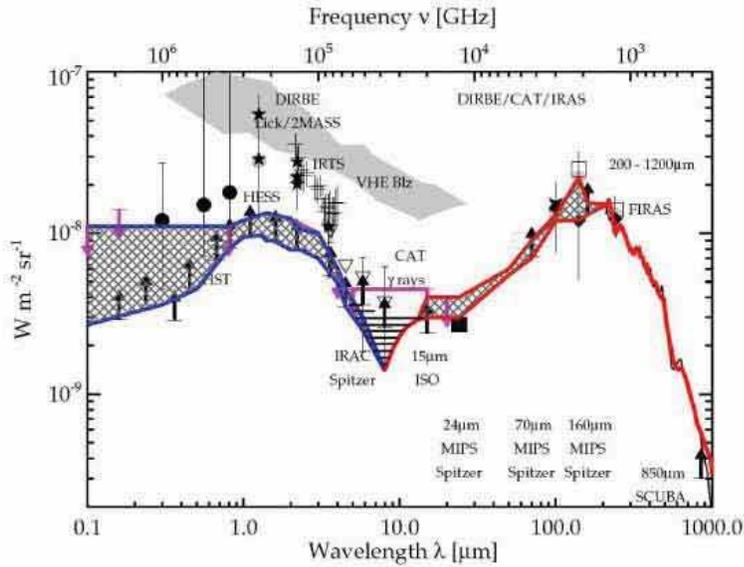,height=3.0in,angle=0}
\caption{Summary of recent measures of the extragalactic background light
from the compilation of Dole et al (2006). The cross-hatched region represents 
the region of claimed excess between the various  background meaasures
(labelled points) and the lower limits to the integrated counts (shown
as upward arrows in the critical optical-near IR region). Interesting excesses
are found in the 0.3 to 5 $\mu$m region.}
\label{fig:57}      
\end{figure}

Various claims of an excess have been made in the 0.3 to 10$\mu$m
wavelength region (Figure 57). Outside of this window the background
appears to be entirely produced by known sources. Such EBL measurements
are extremely difficult to make for several reasons. An accurate absolute 
calibration is essential since the interesting signal is a `DC difference' in
surface brightness. The removal of spurious foregrounds is likewise
troublesome: at some wavelengths, the foreground signal greatly
exceeds the sought-after effect.

In the optical window, careful experiments have been undertaken by 
Mattila (1976), Dube et al (1979) and most recently Bernstein et al
(2002). The most vexing foreground signals at optical wavelengths 
arise from airglow (emission in the upper atmosphere which has a 
time-dependent structure on fine angular scales), zodiacal light 
(scattering of sunlight by interplanetary dust which varies with the 
motion of the Earth along the ecliptic plane) and diffuse Galactic
light or `cirrus' (gas clouds illuminated by starlight).  The wavelength
dependence of these foregrounds (including the microwave background
itself which is considered a `contaminant' in this respect!) is
summarized in Figure 58. For the critical near-infrared region where
redshifted light from early sources might be detected, airglow and
zodiacal light are the dominant contaminants.

\begin{figure}
\psfig{file=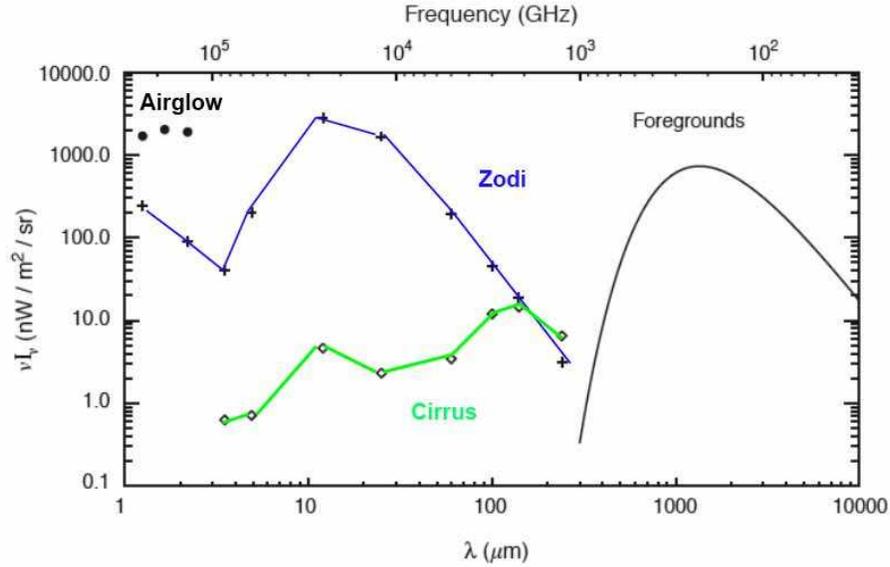,height=3.0in,angle=0}
\caption{Wavelength dependence of the dominant foreground
signals from the compilation by Kashlinsky (2005). Airglow and
zodiacal light dominate at the wavelengths of interest for
stellar radiation from highly redshift sources.}
\label{fig:58}      
\end{figure}

Experiments have differed in the way these foregrounds might
be removed. The use of a space observatory avoids airglow
and observations at different Galactic latitudes 
can be used to monitor or minimize the effect of cirrus. Zodiacal light,
the dominant foreground for all mid-infrared studies and all
space-based near-infrared studies, is the most troublesome.

To first order, the zodiacal light can be predicted depending on
the geometrical configuration of the ecliptical plane, the earth,
sun and target field. Although seasonal variations can be
tracked and the spectrum of the zodiacal light  is solar,
imponderables enter such as the nature and distribution of
the interplanetary dust and its scattering properties. 

The positive EBL excess measures at optical to near-infrared wavelengths 
have remained controversial. This is because all
such experiments search for a `DC signal' - a small absolute difference
between two signals within which foregrounds have to be painstakingly
removed. A brief study of the Bernstein et al (2002) experiment
will make the difficulties clear.

Bernstein et al (2002) claim a significant optical excess at 300, 550
and 800nm from fields observed with Hubble's WFPC-2 and FOS.
Airglow is elminated since HST is above 90km and sources were
removed in each HST image to $V_{AB}$=23.0. The zodiacal
light spectrum was measured simultaneously with a ground-based
optical telescope and iteratively subtracted from the HST data.

At 550nm, the total WFPC-2 background measured after removing
all sources was 105.7 $\pm$ 0.3 units (1 unit = 10$^{-9}$ cgs ster$^{-1}$ \AA\ $^{-1}$).
The measured zodi background was 102.2 $\pm$0.6 units.
Galactic cirrus and faint galaxies beyond $V_{AB}$=23.0 were
estimated as 0.8 and 0.5 units respectively. The claimed excess
signal at this wavelength is 2.7$\pm$1.4 units. Not only is this
only a 2 $\sigma$ detection but it would only require a 2\% 
error in the estimated  Zodiacal light signal to be spurious. 

Studies of the infrared background advanced significantly with the
launch of the COBE satellite which carried DIRBE - a 10 channel
photometer operating in the 1-240 $\mu$m range with 0.7 degree
resolution chopping at 32Hz onto an internal zero flux surface,
and FIRAS - an absolute spectrometer with 7 degree resolution 
operating in the 100 $\mu$m to 5mm range.

DIRBE measured the integrated background at 140 and 240$\mu$m
using an elaborate time-dependent model of the Zodiacal light 
(Kelsall et al 1998, Hauser et al 1998). Schlegel et al (1998)
combined these data with higher resolution IRAS 100$\mu$m maps
to improve removal of Galactic cirrus and to measure the
temperature of the dust emission. Finkbeiner et al (2000)
extended the model of Zodiacal light  to provide the first
detection at 60 and 100$\mu$m and Wright \& Reese (2000) used
2MASS observations to improve Galactic source removal
claiming a detection at 2.2$\mu$m. 

Dole et al (2006) have undertaken a very elegant analysis
of 19000 Spitzer MIPS images. By centering the images on deep
24$\mu$m sources, they can evaluate the statistical contribution of 
otherwise inaccessibly faint 70 and 160 $\mu$m sources.

Matsumoto et al (2005) has extended these detections to shorter 
wavelengths using a spectrometer onboard the Infrared Telescope 
in Space (IRTS). This signal has the tantalising signature of a 
distant star-forming stellar population -  a steeply-rising continuum
down to 1$\mu$m and a discontinuity when extended to include
optical data (Figure 57). Mattila (2006) has argued this signal
represents an artefact arising from incorrect foreground removal.
He has likewise criticised the Bernstein et al (2002) optical
detections (Mattila 2003).

Supposing the DIRBE/2MASS/IRTS detections to be
real, what would this imply? Assuming the J-band background
($>$2.5 nW m$^{-2}$ ster$^{-1}$) arises from $z\simeq$9 Ly$\alpha$
emission, Madau \& Silk (2005) calculate the associated stellar mass
that is produced. They find an embarrassingly high production
rate, corresponding to $\Omega_{\ast}=0.045\,\Omega_B$; in other
words almost all the stars we see today would need to be produced by
$z\simeq$9 to explain the signal. Likewise the ionizing flux
produced by such star formation would be in excess of that
required to explain the WMAP optical depth. 

\subsection{Fluctuation Analyses}

Kashlinksky et al (2002, 2005) have argued that the difficulties
inherent in extracting the EBL signal by the DC method may be alleviated by
considering the angular fluctuation spectrum $\delta\,F(\theta)$
and its 2-D Fourier transform $P_2(q)$.

In more detail, the fluctuation in the measured background is

$$\delta\,F(\theta) = F(\theta) - <F> = (2\pi)^{-2} \int \delta\,F_q exp(=i\,q\,.\,\theta) d^2q$$

and 

$$P_2(q) \equiv <| \delta\,F_q|^2>$$

The success of the method depends on whether the various foreground
contributions to $P_2(q)$ can be readily distinguished from one another.
Although advantageous in using independent information from the
DC measurements, it is unfortunately not easy to intercompare
the experiments or to interpret the fluctuation analyses.

Kashlinksky et al (2005) recently applied this method to deep IRAC
images. Sources were extracted to reasonably faint limits (0.3 $\mu$Jy
or $m_{AB}\simeq$22-25). The pixels associated with each image 
were masked out in a manner that could be adjusted depending on 
the significance of the source over the background and its area.
The data was split into two equal halves (A, B) and the power 
spectrum of the signal ($P_S(q)$, S=A+B) was compared to that
of the noise ($P_N(q)$, N=A-B). Figure 59 illustrates the
residual signal, $P_S(q) - P_N(q)$, and the fluctuations
$(q^2P_2(q)/2\pi)^{\frac{1}{2}}$ in 3 IRAC fields as a function 
of angular scale ($2\pi/q$ arcsec) in the four IRAC channels. 
Although the positive fluctuations on small scales are consistent
with shot noise from the galaxy counts (solid lines in both panels),
the excess is particularly prominent on scales of 1-2 arcmin
and consistently in fields of various Galactic latitudes and
in all four IRAC bandpasses. 

\begin{figure}
\psfig{file=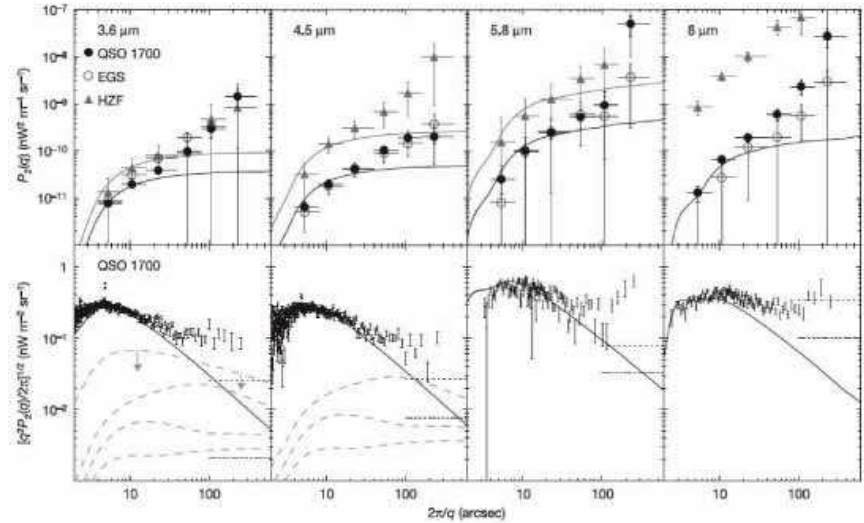,height=2.8in,angle=0}
\caption{Detection of fluctuations in the infrared background from the analysis
by Kashlinky et al (2005). The upper panel shows the angular power
spectrum as a function of scale for 3 IRAC fields plotted for all four
IRAC channels (3.6 through 8 $\mu$m). The solid line shows the effect
of shot noise from the galaxy counts. This fits the fluctuations on
small scales but there is a significant and consistent excess
on large scales. Bottom panels refer to fluctuations for one
IRAC field plotted as $(q^2P_2(q)/2\pi)^{\frac{1}{2}}$ with the
predicted contributions from various populations of unresolved distant 
sources shown as dotted lines.
 }
\label{fig:59}      
\end{figure}

Kashlinksky et al argue that the excess signal is fairly flat
spectrally in $\nu\,I_{\nu}$ ruling out any instrumental or
Galactic contamination. The constancy of the signal over the
wide range in Galactic latitude likewise suggests the signal 
is extragalactic. However, zodiacal light is difficult to eliminate
as a contaminant. Although the signal persists when the Earth
is 6 months further around in its orbit, no analysis of the likely 
zodi fluctuation spectrum is presented. 

If the signal persists, what might it mean for sources fainter
than 0.3 $\mu$Jy? Without a detailed redshift distribution,
it is hard to be sure. The clustering pattern is remarkably
strong given the depth of the imaging data (much deeper
fields have subsequently been imaged but no independent
analyses have yet been published).  In summary, this is
an intriguing result, somewhat unexplained and in need
of confirmation\footnote{Recent developments are discussed
by Sullivan et al 2006 and Kashlinsky et al 2006}

\subsection{EBL Constraints from TeV Gamma Rays}

High energy (TeV) gamma rays are absorbed in the earth's
atmosphere and converted into secondary particles forming
an `air shower'. Cerenkov light is generated - a beam of
faint blue light lasting a few $\times 10^{-9}$ sec is produced
illuminating  an area of 250m in diameter on the ground.

The significance of searching for TeV gamma rays is that they
interact with the sea of 1-10 $\mu$m (infrared background)
photons via the pair-creation process, viz:

$$\gamma\,\gamma \rightarrow e^+\, e^-$$

producing attenuation in distant sources (blazars)
whose spectral energy distributions are assumed to be
power laws.

$$ \frac{d\,N}{d\,E} \propto E^{-\Gamma}$$
 
The strength of the attention, measured as a change in $\Gamma$,
for the most distant accessible blazars is thus a measure of the 
degree of interaction between the gamma rays and the 
ambient infrared background.

The HESS team (Aharonian et al 2006) have recently
analyzed the gamma ray spectrum of a blazar at $z$=0.186
and fitted its energy spectrum. As a result they can predict
an upper limit to the likely infrared background for various
assumptions (Figure 60). The constraint is most useful
in the wavelength range 1-10$\mu$m where the 
associated gamma ray spectral data is of high quality
(top axis of Fig. 60). This new method is intriguing. 
Although in detail the constraint is dependent on an 
assumed form for the TeV energy spectrum of blazars, 
for all reasonable assumptions the acceptable background
is much lower at 1-4 $\mu$m than claimed by the traditional
experiments.

\begin{figure}
\psfig{file=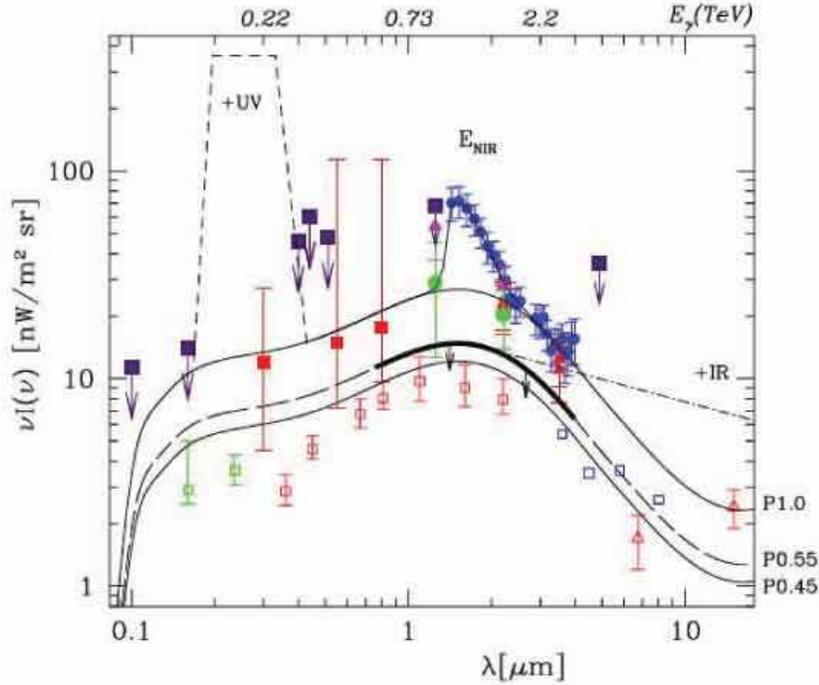,height=3.7in,angle=0}
\caption{Constraints on the cosmic infrared background 
from the degree of attenuation seen in the high energy
gamma ray spectrum of a distant blazar (Aharonian et
al 2006). The black curves indicate likely upper limits
to the infrared background for various assumptions. This
independent method argues the near-IR detections by
Matsumoto et al (2005, blue circles) and others are spurious.}
\label{fig:60}      
\end{figure}

\subsection{Lecture Summary}

In this lecture we have learned that extragalactic background
light measurements in principle offer an important and unique
constraint on undiscovered distant source populations. 

However, the observations remain challenging and 
controversial because of instrumental effects, related
to the precision of absolute calibrations, and dominant
foregrounds. In the 1-5$\mu$m spectral regions
positive detections in excess of extrapolations of
the source counts have been reported. This is interesting
because this is the wavelength range where an abundant and early
population of faint, unresolved $z>$10 sources would produce
some form of infrared background.

At the moment, the claimed detections at 1-5 $\mu$m appear 
unreasonably strong. Too many stars would be produced
by the star formation associated with this background signal
and recent ultrahigh energy spectra of distant blazars
likewise suggest a much weaker infrared background.

IRAC is a particularly promising instrument for the redshift
range 10$<z<$20 and fluctuation analyses offer a valuable
independent probe of the background, although it is hard
to interpret the results and compare them with the more
traditional DC level experiments.

Despite the seemlingly chaotic way in which the subject
of the cosmic extragalactic background light has unfolded
in the past decade, it is worth emphasizing that the
motivation remains a strong one. Any information on
the surface brightness, angular clustering and redshift
range of radiation beyond the detected source counts
will be very valuable information ahead of the discoveries
made possible by future generations of facilities.

\newpage
 

\section{Epilogue: Future Prospects}

\subsection{Introduction}

The students attending this lecture course are living at a special
time! During your lifetime you can reasonably confidently expect to 
witness the location of the sources responsible for cosmic reionization, 
to determine the redshift range over which they were active and perhaps 
even witness directly the `cosmic dawn' when the first stellar systems shone
and terminated the `Dark Ages'!  

We have made such remarkable progress in the past decade, reviewed
here, that such a bold prediction seems reasonable, even for a cautious
individual like myself! We have extended our fundamental knowledge of 
how various populations of galaxies from 0$<z<$3 combine to give us a 
broad picture of galaxy evolution, while extending the frontiers to $z\simeq$7
and possibly beyond. Certainly many issues remain, including the apparent 
early assembly of certain classes of quiescent galaxy, the abundance
patterns in the intergalactic medium and the apparent `downsizing'
signatures seen over a variety of redshifts. However,  the progress has 
been rapid and driven by observations. Accompanying this is a
much greater synergy between theoretical predictions and observations
than ever before.

We have seen that the question of `First Light' - the subject of this course - is
the remaining frontier for observations of galaxy formation. The physical
processes involved are poorly understood and thus observations will
continue to be key to making progress. In this final lecture, I take out
my crystal ball and consider the likely progress we can expect
at optical and near-infrared wavelengths with current and near-term 
facilities ahead of those possible with the more ambitious new observatories 
such as the James Webb Space Telescope (JWST) and a new 
generation of extremely large ground-based telescopes.

\subsection{The Next Five Years}

Panoramic imaging with large optical and near-infrared cameras 
on telescopes such as Subaru, UKIRT and VISTA will enable
continued exploration of the abundance of luminous drop-outs and
Ly$\alpha$ emitters over 5$<z<$7, reducing the currently troublesome
issues of cosmic variance. Deeper data will continue to be
provided from further fields taken with ACS and NICMOS.
With further spectroscopic surveys on large telescopes, we
can expect improved stellar mass density estimates at $z\simeq$5-6
(Eyles et al 2006).  However, strong gravitational lensing may still be the only
route to probing intrinsically fainter sources, particularly beyond
$z\simeq$7.

\begin{figure}
\centerline{\psfig{file=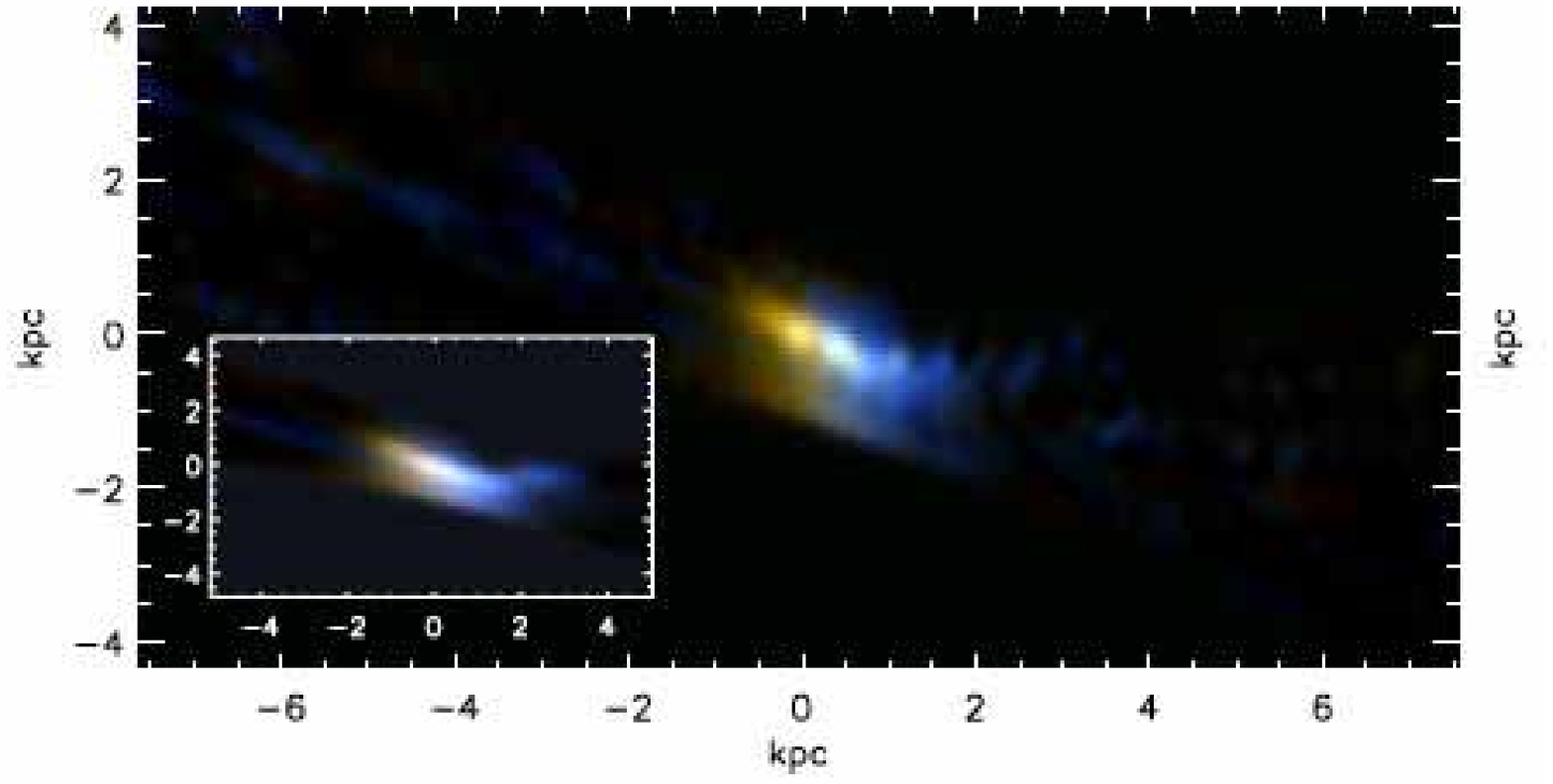,height=2.2in,angle=0}}
\centerline{\psfig{file=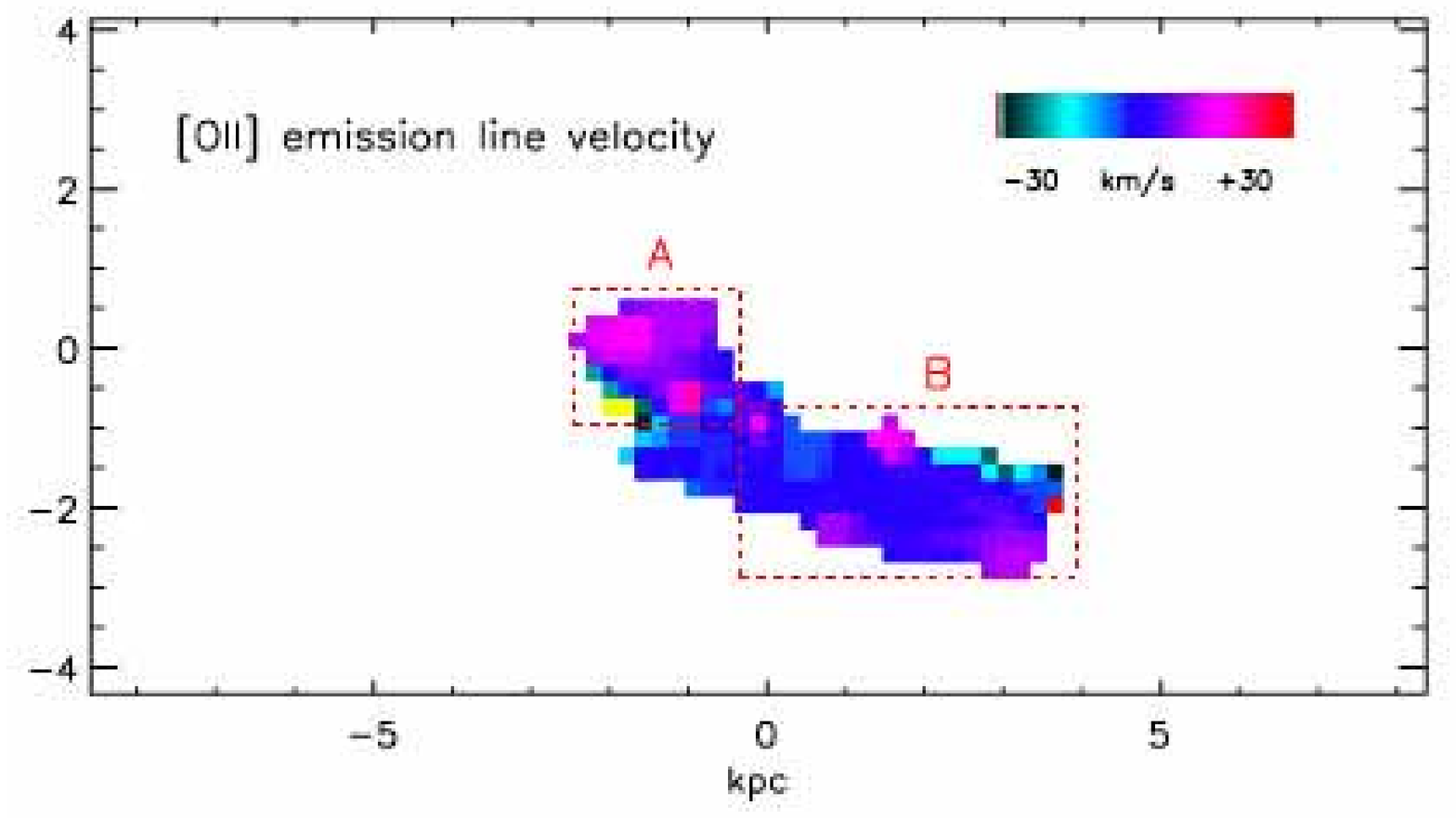,height=2.3in,angle=0}}
\caption{Detailed studies of a lensed $z\simeq$5 galaxy (Swinbank
et al 2006). (Top) Reconstructed color HST VI
image of the $z$=4.88 arc in the cluster RCS0224-002. The inset 
shows the effect of 0.8 arcsec seeing on the reconstruction, thereby
demonstrating the advantage of lensing. In the source plane, the galaxy is
2.0$\times$0.8 kpc. (Bottom) [O II] velocity field obtained 
during a 12 hr VLT SINFONI exposure without AO.
Spatial comparison with the Ly$\alpha$ field gives clear
evidence of significant bi-polar outflows.}
\label{fig:61}      
\end{figure}

More detailed characterization of the properties of the
most massive galaxies at $z\simeq$5-6 will also be
worthwhile, in addition to continuing to deduce statistical properties
such as abundances and luminosity functions.. Laser-guide
star adaptive optics is now in widespread use on our 8-10 meter
class telescopes and, with integral field spectrographs, can be 
used to probe the resolved dynamics of distant galaxies
(Genzel et al 2006). Application of these techniques to the
most intense star-forming lensed systems at $z>$4 (Swinbank
et al 2006, Figure 61) is already providing unique insight into the physical
state of the earliest galaxies. Extending the same techniques with adaptive
optics will advance progress and address process on remarkably 
small scales ($\simeq$200 pc).

A number of special purpose ground-based instruments are also being 
developed, both with and without adaptive optics, to probe
for $z>$7 dropouts and Lyman $\alpha$ emitters. 
These include:

\begin{itemize}

\item{} The Dark Age Z Lyman Explorer (DAZLE, Horton et al 
2004)\footnote{\rm http://www.ast.cam.ac.uk/~optics/dazle/},  
a moderate field (7 arcmin) infrared narrow-band  imager
with airglow discrimination whose goal is to reach a sensitivity
of $\simeq10^{-18}$ cgs in a night of VLT time. This corresponds 
to a limiting star formation rate of $\simeq1 M_{\odot}$ yr$^{-1}$ in 
quiet regions of the airglow spectrum at  $z\simeq$7.7
and 9.9. Given the modest field and the need for dedicated
spectroscopic follow-up, a long term campaign is envisaged
to conduct a reliable census.

\item{} The Gemini Genesis Survey (GGS
\footnote{\rm http://odysseus.astro.utoronto.ca/ggs-blog/?page\_id=2})
which uses F2T2, a clone of the tunable filter destined for JWST (Scott 
et al 2006). This instrument is planned to work behind the Gemini
Observatory's multi-conjugate adaptive optics facility and
image, in steps of wavelength, regions lensed by foreground
clusters. The success of such a strategy will depend on both
the angular size and line widths of $z>$7 Lyman $\alpha$ emitters and
the abundance of intrinsically faint examples. Ellis et al
(2001) found the lensed emitter at $z\simeq$5.7 was less
than 30 milli-arcsec across with a line width of only 100-200 
kms sec$^{-1}$.  If such tiny emitters are the norm at $z>$7, the 
GGS may go much deeper than extant
spectroscopic lensed searches (Stark et al 2006b).

\end{itemize}

The exciting redshift range 7$<z<12$ will also be the
province of improved drop-out searches using the
instrument WFC3 slated to be installed on Hubble
Space Telescope in 2008-9\footnote{\rm http://www.stsci.edu/hst/wfc3}.
The infrared channel of this instrument spans 850 - 1170 nm
with a field of $\simeq$ 2 arcmin at an angular resolution of 
0.13 arcsec pixel$^{-1}$. This resolution is coarser than 
that of adaptive optics-assisted instruments such as F2T2. 
The principal gain over ground-based instruments will be 
in deep broad-band imaging free from airglow. The survey 
efficiency is about an order of magnitude better than that of 
NICMOS. WFC3 also has two infrared grisms which will be 
helpful in source discrimination.

A major stumbling block at the moment, even at $z\simeq$5-6,
is efficient spectroscopic follow-up of dropout candidates.
As we discussed in \S6, photometric redshifts have unfortunately become
{\it de rigeur} in statistical analyses of luminosity densities
and luminosity functions (Bouwens et al 2006), yet their precision
remains controversial. For z- and J-band dropous beyond $z\simeq$7, photometric
redshifts will be even less reliable. Spitzer detections will
be harder and fewer, and the typical source may have only
2-3 detected bands. Candidates may be found in abundance
but how will they be confirmed?

The various 8-10m telescopes are now building a new
generation of cryogenic near-infrared multi-slit spectrographs. It
can be hoped that long exposures with these new instruments
will be sufficient to break this impasse.

\subsection{Beyond Five Years}

A number of facilities are being planned, motivated by the
progress discussed in these lectures. These include:

\begin{figure}
\centerline{\psfig{file=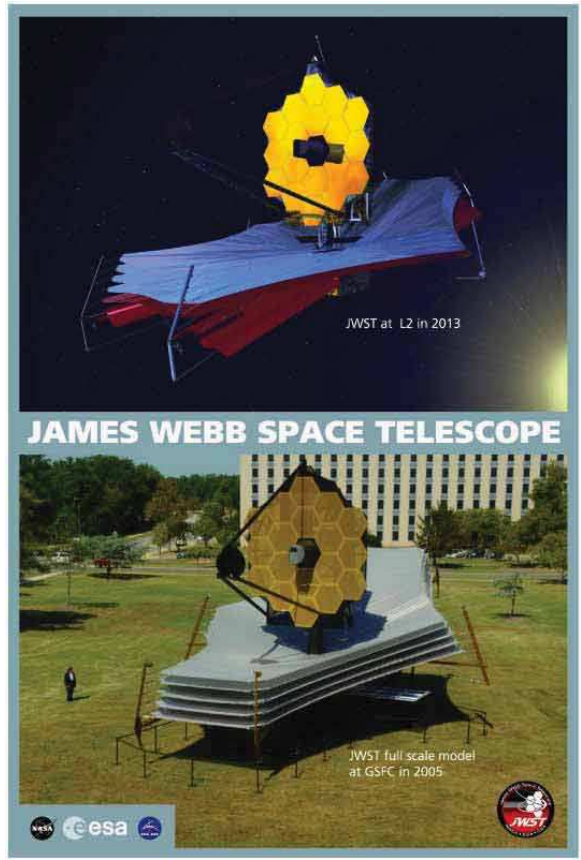,height=4.5in,angle=0}}
\caption{The 6.5m James Webb Space Telescope: Then (2013 in
orbit) and  and Now (2005, full scale model).}
\label{fig:62}      
\end{figure}

\begin{figure}
\centerline{\psfig{file=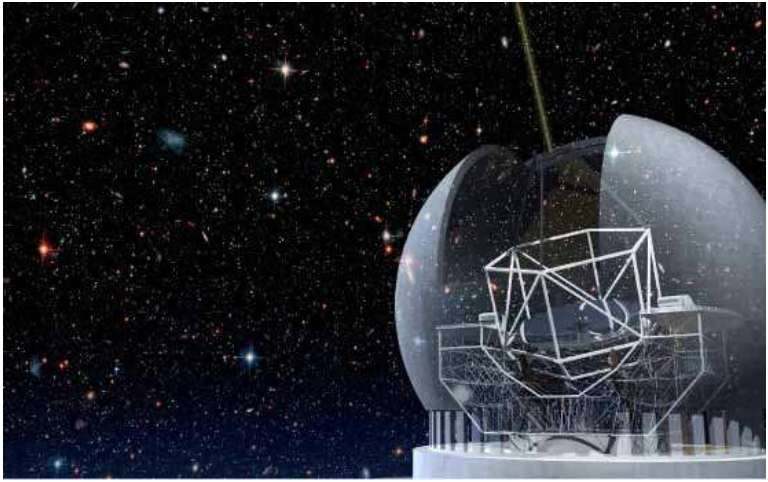,height=2.5in,angle=0}}
\caption{The proposed US-Canadian Thirty Meter Telescope ({\rm www.tmt.org}) 
now in the detailed design phase.}
\label{fig:63}      
\end{figure}

\begin{itemize}

\item{} James Webb Space Telescope - a 6.5 meter
optical-infrared space observatory (Gardner et al 2006) for which a primary
mission is the detailed study of the earliest star-forming
galaxies (Figure 62). The facility is currently planned to
have a near-infrared imager NIRCam, a spectrograph
NIRSpec with both integral field and limited multi-object
modes, a tunable filter (as F2T2) and a mid-IR imager, MIRI. 
Stiavelli (2002) has presented a cogent summary of the likely 
strategies of using this facility, due to be launched in 2013, for 
studies of the earliest systems.

\item{} Extremely Large Telescopes on the ground including
the US-Canadian Thirty Meter Telescope (TMT, see {\rm www.tmt.org}, 
Figure 63) which will have a diffraction-limited near-infrared 
imager/spectrograph (IRIS) and a adaptive-optics assisted
spectrograph (IRMOS) with multiple deployable {\it integral field units} - 
area mapping units which can be arranged not only to scan regions 
surrounding luminous sources but also to undertake multi-object studies
in crowded regions (Figure 64). At the time of writing, TMT is expected
to be operational from 2016 onwards.

\end{itemize}

These will complement redshifted 21cm line tomography
with radio facilities and address the key questions of
the escape fraction of photons from star-forming sources
and how they create ionized bubbles which merge to
cause reionization. 

It is quite likely that, by 2013, the redshift range containing the
earliest galactic sources, estimated at present to be
10$<z<$20 perhaps, will have been refined sufficiently 
by special-purpose instruments on our existing 8-10 meter class
telescopes. Thus one can surmise that that both
JWST and future ELTs will be used for much more
challenging work related to the physical process
of reionization, as well as the chemical maturity
of the most luminous sources found at high redshift.

\begin{figure}
\centerline{\psfig{file=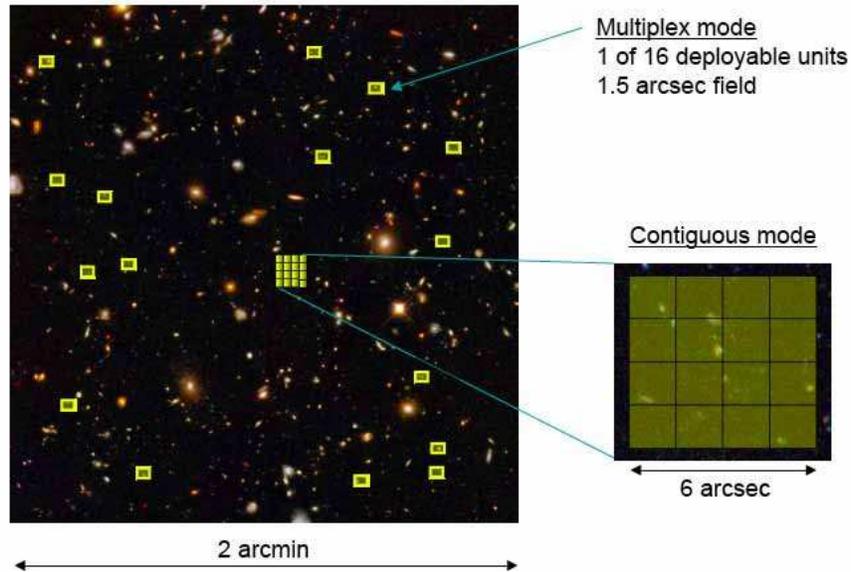,height=3.3in,angle=0}}
\caption{Flexibility in the deployment strategy for the multiple
integral field units (yellow squares) for the proposed TMT 
infrared multi-object adaptive optics assisted instrument 
IRMOS. The IFUs can be distributed around the full
field of 2 arcmin in classic multi-object mode. Alternatively,
the IFUs can be configured together to map a small
6 $\times$ 6 arcsec field at high angular resolution in `blind' 
mode. Such flexibility will be important in mapping Ly$\alpha$ emission
at high redshift in various situations.}
\label{fig:64}      
\end{figure}

An obvious partnership between JWST and TMT,
for example, which would complement the 21cm
studies, would be to (i) search for the extent and
topology of faint Ly$\alpha$ emission in ionized
bubbles around JWST-selected luminous star
forming galaxies and, (ii) pinpoint early sources
for spectroscopic scrutiny so as to identify
signatures of Population III stars.  

At the present time there are so many imponderables in our 
knowledge of the earliest sources, that even the design 
parameters for the ELT instruments is a considerable challenge. 
How big are the faintest Ly$\alpha$ emitters? What are
the typical line widths in km sec$^{-1}$? How big are
the ionized bubbles at a given redshift? And, crucially,
what is the surface density of various types of star-forming
galaxies. Flexibility in the design of survey strategies
will be crucial with instruments such as TMT's IRMOS 
(Fig.~64).

{\it Any} information we can glean on the properties
of $z\simeq$10 sources in the next 5 years will
be valuable in optimizing how to move forward when
these magnificent new generation telescopes are
made available to us. Indeed, it is foolhardy to wait!
Time and again, we can retrospectively look back
at what we thought we would accomplish with our
planned facilities and we always find that we achieved
more than we expected! 

\section{Acknowledgements}

I thank Daniel Schaerer, Denis Puy and Angela Hempel for inviting 
me to give these lectures in such a magnificent location with
an enthusiastic group of students. I also thank my fellow lecturers, 
Avi Loeb and Andrea Ferrara and all of the foregoing for their 
patience in waiting for the completion of my written lectures. 
I thank my close colleagues Kevin Bundy, Sean Moran, Mike Santos
and Dan Stark for their help and permission to show results in progress as
well as Ivan Baldry, Jarle Brinchmann, Andrew Hopkins, John Huchra 
and Jean-Paul Kneib  for valuable input. Finally, I thank 
Ray Carlberg  and his colleagues for their hospitality of the Astronomy 
Department at the University of Toronto where the bulk of these 
lectures notes were completed. 

\newpage

\section{References}



\noindent  Aaronson, M 1978 Ap J 221, L103

\noindent Aharonian, F. et al 2006 Nature 440, 1018

\noindent Adelberger, K et al 1998 Ap J 505, 18

\noindent Astier, P. et al 2006 Astron. Astrophys., 447, 31

\noindent Bahcall, N. et al 1999 Science 284, 1481

\noindent Baldry, I. \& Glazebrook, K. 2003 Ap J 593, 258

\noindent Barkana, R. \& Loeb, A 2000 Phys Rep. 349, 125 

\noindent Baugh, C. et al 1996 MNRAS 283, 1361

\noindent Baugh, C. et al 1998 Ap J 498, 504

\noindent Baugh, C. et al 2005a MNRAS 305, L21

\noindent Baugh, C. et al 2005b MNRAS 356, 1191

\noindent Becker, G et al 2006 Ap J in press, (astro-ph/0607633)

\noindent Becker, R.H. et al 2001 AJ 122, 2850

\noindent Beckwith, S. et al 2006 AJ 132, 1179

\noindent Bell, E. et al 2004 Ap J 608, 752

\noindent Bell, E. et al 2006 Ap J 640, 241

\noindent Bender, R. et al 1992 Ap J 399, 462

\noindent Bennett, C. et al 2003 Ap J Suppl., 148, 97

\noindent Benson, A. et al 2002 MNRAS 336, 564

\noindent Benson, A. et al 2003 Ap J 599, 38

\noindent de Bernadis, P. et al 2000 Nature 404, 955

\noindent Bernstein, R. et al 2002 Ap J 571, 56

\noindent Blain, A. et al 2002 Phys. Rep. 369, 111

\noindent Blain, A. et al 2004 Ap J 611, 725

\noindent Blandford, R. \& Narayan, R. 1992 ARAA 30, 311

\noindent Blanton, M.R. et al 2001 A.J. 121, 2358

\noindent Blumenthal, G. et al 1984 Nature 311, 517

\noindent Bolton, A. et al 2006 Ap J 638, 703

\noindent Bouwens, R. et al 2004 Ap J 606, L25

\noindent Bouwens, R. et al 2006 Ap J in press (astro-ph/0509641)

\noindent Bower, R. et al 1992 MNRAS 254, 601

\noindent Bower, R. et al 2006 MNRAS 370, 645

\noindent Brinchmann, J. \& Ellis, R.S. 2000 Ap J 536, L77

\noindent Broadhurst, T. et al 1992 Nature 355, 55

\noindent Broadhurst, T. et al 2005 Ap J 621, 53

\noindent Bruzual, G. 1980 Ph.D. thesis, UC Berkeley

\noindent Bundy, K. et al 2004 Ap J 601, L123

\noindent Bundy, K. et al 2005 Ap J 625, 621

\noindent Bundy, K et al 2006 Ap J 651, 120

\noindent Bunker, A. et al 2004 MNRAS 355, 374

\noindent Bunker, A. et al 2006 New AR 50, 94

\noindent Calzetti, D. et al 2000 Ap J 533, 682

\noindent Caroll, S. et al 1992 ARAA 30, 499

\noindent Chabrier, G. 2003 PASP 115, 763

\noindent Chapman, S. et al 2003 Nature, 422, 695

\noindent Chapman, S. et al 2005 Ap J 622, 772 

\noindent Conselice, C. et al 2005 Ap J 628, 160

\noindent Cowie, L. et al 1995 AJ 109, 1522

\noindent Cowie, L. et al 1996 AJ 112, 839

\noindent Cole, S. et al 2000 MNRAS 319, 168

\noindent Cole, S. et al 2001 MNRAS 326, 255

\noindent Cole, S. et al 2005 MNRAS 362, 505

\noindent Colless, M. et al 2001 MNRAS 328, 1039

\noindent Couch, W. et al 1998 Ap J 497, 188

\noindent Croton, D. et al 2006 MNRAS 365, 11

\noindent Daddi, E. et al 2004 Ap J 617, 746

\noindent Davies, R. L. et al 2001 Ap J 584, L33

\noindent de Lucia, G. et al 2006 MNRAS 366, 499

\noindent Dickinson, M. et al 2003 Ap J 587, 25

\noindent Dickinson, M. et al 2004 Ap J 600, L99

\noindent Djorgovski, S. \& Davis, M. 1987 Ap J 313, 59

\noindent Djorgovski, S. et al 2001 Ap J 560, L5

\noindent Dole, H. et al 2006 Astron. Astrophys. 451, 417

\noindent Dressler, A. 1980 Ap J 236, 351

\noindent Dressler, A. et al 1987 Ap J 313, 42

\noindent Dressler, A.. et al 1997 Ap J 490, 577

\noindent Driver, S. et al 1995 Ap J 453, 48

\noindent Drory, N. et al 2005 Ap J 619, L131

\noindent Dube, R.R. et al 1979 Ap J 232, 333

\noindent Dunlop, J. et al 2006 MNRAS in press (astro-ph/0606192)

\noindent Ebbels, T. et al 1998 MNRAS 295, 75

\noindent Eddington, A.S. 1919 Obs. 42, 119

\noindent Egami, E. et al 2005 Ap J 618, L5

\noindent Erb, D. et al 2006 Ap J 644, 813

\noindent Ellis, R.S. 1997 ARAA 35, 389

\noindent Ellis, R.S. et al 1996 MNRAS 280, 235

\noindent Ellis, R.S. et al 2001 Ap J 560, L119

\noindent Ellison, S. et al 2000 AJ 120, 1175
 
\noindent Eyles, L. et al 2005 MNRAS 364, 443

\noindent Eyles, L. et al 2006 MNRAS, in press (astro-ph/0607306)

\noindent Fall, S. \& Efstathiou, G. 1980 MNRAS 193, 189

\noindent Fall, S. et al 1996 Ap J 464, L43

\noindent Fan, X. et al 2003 AJ 125, 1649

\noindent Fan, X. et al 2006a ARAA 44, 415

\noindent Fan, X. et al 2006b AJ 132, 117

\noindent Felton, J. E. 1977, AJ 82, 861

\noindent Finkbeiner, D et al 2000 Ap J 544, 81

\noindent Fontana, A. et al 2004 Astron. Astrophys., 424, 23. 

\noindent Franx, M. et al 2003 Ap J 587, L79

\noindent Frayer, D. et al 2000 AJ 120, 1668

\noindent Freedman, W. et al 2001 Ap J 553, 47

\noindent Fukugita, M. \& Kawasaki, M. 2003 MNRAS 343, L25 

\noindent Fukugita, M. \& Peebles, P.J.E. 2004 Ap J 616, 643

\noindent Fukugita, M. et al 1998 Ap J 503, 518

\noindent Furlanetto, S. et al 2005 Phys. Reports 433, 181

\noindent Gardner, J. et al 2006 Space Sci. Rev. in press (astro-ph/0606175)
 
\noindent Genzel, R. et al 2006 Nature 442, 786
 
\noindent Gerhard, O. et al 2001 AJ 121, 1936

\noindent Giavalisco, M. et al 2004 Ap J 600, 103

\noindent Glazebrook, K. et al 1995 MNRAS 275, L19

\noindent Glazebrook, K. et al 2004 Nature 430, 181

\noindent Gradshteyn, I.S. \& Ryzhik, I.M. 2000, 
{\it Tables of Integrals, Series and Products}, Academic Press 

\noindent Granato, G.L. et al 2000 Ap J 542, 710

\noindent Gunn, J.E. \& Peterson, B.A. 1965 Ap J 142, 1633

\noindent Haiman, Z. 2002 Ap J 576, 1

\noindent Hanany, et al 2000 Ap J 545, 5

\noindent Hauser, M \& Dwek, E 2001 ARAA 39, 249

\noindent Hauser, M. et al 1998 Ap J 508, 25

\noindent Hoekstra, H. et al 2005 Ap J 647, 116

\noindent Hopkins, A.M.  2004 Ap J 615, 209

\noindent Hopkins, A.M. \& Beacom, J. 2006 Ap J 651, 142 

\noindent Horgan, J. 1998 The End of Science, Addison-Wesley

\noindent Horton, A. et al 2004 in {\it Ground-based Instrumentation
for Astronomy}, S.P.I.E. 5492, 1022

\noindent Hu, E. et al 2004 AJ 127, 563

\noindent Hubble, E. 1936 The Realm of the Nebulae, Yale Univ. Press

\noindent Hughes, D. et al 1998 Nature 394, 241

\noindent Iye, M. et al 2006 Nature 443, 186

\noindent Jorgensen, I. et al 1996 MNRAS 280, 167

\noindent Jullo, E et al 2006, Astron. Astrophys, submitted

\noindent Juneau,  S. et al 2005 Ap J 619, L135

\noindent Kaiser, N. et al 1987 MNRAS 277, 1

\noindent Kauffmann, G. \& Charlot, S. 1998 MNRAS 297, L23

\noindent Kauffmann, G. et al 1993 MNRAS 264, 201

\noindent Kauffmann, G. et al 1996 MNRAS 283, L117

\noindent Kauffmann, G. et al 1999 MNRAS 303, 188

\noindent Kauffmann, G. et al 2003 MNRAS 341, 33

\noindent Kashikawa, N. et al 2006 Ap J 648, 7

\noindent Kashlinsky 2005 Phys. Reports 409, 361

\noindent Kashlinsky et al 2002 Ap J 579, 53

\noindent Kashlinsky et al 2005 Nature 438, 45

\noindent Kashlinsky et al 2006 Ap J Lett in press (astro-ph/0612445)


\noindent Kelsall, T. et al 1998  Ap J 508, 44

\noindent Kennicutt, R. 1983 Ap J 272, 54

\noindent Kennicutt, R. 1998 ARAA 36, 189

\noindent Kneib, J-P. et al 1996 Ap J 471, 643

\noindent Kneib, J-P. et al 2004 Ap J 607, 697

\noindent Kodaira, I. et al 2003 PASJ 55, L17

\noindent Kogut, A. et al 2003 Ap J Suppl., 148, 161.

\noindent Kong et al 2006 Ap J 638, 72

\noindent Koo, D. \& Kron, R. 1992 ARAA 30, 613. 

\noindent Koopmans, L. \& Treu, T. 2003 Ap J 583, 606

\noindent Kriek, M. et al 2006 Ap J 649, L71

\noindent Kroupa, P. et al 1993 MNRAS 262, 545

\noindent Kroupa, P. 2001 MNRAS 332, 231

\noindent Lanzetta, K. et al 2002 Ap J 570, L492

\noindent Leinert, C. et al 1995 Astron. Astrophys. Suppl. 112, 99

\noindent Le Fevre, O. et al 2000 MNRAS 311, 565

\noindent Lilly, S.J. et al 1996 Ap J 460, L1

\noindent McCarthy, P. 2004 ARAA 42, 477

\noindent McCarthy, P.  et al 2004 Ap J 614, L9

\noindent McClure, R. et al 2006 MNRAS 372, 357

\noindent Madau, P. \& Pozzetti, L. 2000 MNRAS 312, L9

\noindent Madau, P. \& Silk, J. 2005 MNRAS 359, L37

\noindent Madau, P. et al 1996 MNRAS 283, 1388

\noindent Madau, P. et al 1998 Ap J 498, 106

\noindent Madau, P. et al 1999 Ap J 514, 648

\noindent Maddox, S. et al 1990 MNRAS 242, 43

\noindent Malhotra, S. \& Rhoads, J. 2004

\noindent Malhotra, S. et al 2005 Ap J 627, 666

\noindent Malin, D. \& Carter, D 1980 Nature 285, 643

\noindent Mather, J. et al 1990 Ap J 354, 37

\noindent Mattilia, K. 1976 Astron. Astrophys. 47, 77

\noindent Mattila, K.  2003 Ap J 591, 119

\noindent Mattila, K.  2006 MNRAS 372, 1253

\noindent Matsumoto, T. et al 2005  Ap J 626, 31

\noindent Meier, D.L. 1976 Ap J 207, 343

\noindent Mellier, Y. 2000 ARAA 37, 127

\noindent Miller, G.E. \& Scalo, J. 1979 Ap J Suppl. 41, 513

\noindent Miralda-Escude, J. et al 1998 Ap J 501, 15

\noindent Mobasher, B. et al 2005 Ap J 635, 832

\noindent Nagamine, et al 2004 Ap J 610, 45

\noindent Nagao, T. et al 2005 Ap J 634, 142

\noindent Newman J. \& Davis, M 2000 Ap J 534, L11

\noindent Norberg, P et al 2002 MNRAS 336, 907

\noindent Oppenheimer, B.D. et al 2006 in {\it Chemodynamics: First
Stars to Local Galaxies}, in press (astro-ph/0610808)

\noindent Ostriker, J. \& Peebles, P. 1973 Ap J 186, 467

\noindent Ostriker, J \& Steinhardt, P. 1996 Nature 377, 600 

\noindent Ouchi, M. et al 2005 Ap J 620, L1

\noindent Papovich, C. et al 2006 Ap J 640, 92

\noindent Partridge, B. \& Peebles, P.J.E. 1967 Ap J 147, 868

\noindent Peacock et al 2001 Nature 410, 169

\noindent Peebles, P. 1980 {\it Large Scale Structure of the
Universe}, Univ. Chicago.

\noindent Penzias, A. \& Wilson, 1965 Ap J 142, 419

\noindent Perlmutter, S. et al 1999 Ap J 517, 565

\noindent Pettini, M. et al 2002 Astrophys.  Sp. Sci. 281, 461

\noindent Postman, M. et al 2005 Ap J 623, 721

\noindent Pritchet, C. 1994 PASP 106, 1052

\noindent Reddy, N. et al 2005 Ap J 633, 748

\noindent Refregier, A. 2003 ARAA 41, 645

\noindent Richard, J. et al 2006 Astron. Astrophys. 456, 861

\noindent Riess, A. et al 1995 AJ 116, 1009

\noindent Rowan-Robinson, M. 1985 The Cosmic Distance Scale (Freeman)

\noindent Rubin, V. 2000 PASP 112, 747

\noindent Rubin, V. et al. 1976 AJ 81, 687

\noindent Rudnick, G. et al 2003 Ap J 599, 847

\noindent Salpeter, E. 1955 Ap J 121, 161

\noindent Sand, D.J. et al 2005 Ap J 627, 32

\noindent Sandage, A. 2005 ARAA 43, 581

\noindent Santos, M. 2004 MNRAS 349, 1137

\noindent Santos, M. et al 2004 Ap J 606, 683

\noindent Sargent, M.T. et al Ap J in press (astro-ph/0609042) 

\noindent Seitz, et al 1998  MNRAS 298, 945

\noindent Scalo, J. 1986 Fund. Cosmic Phys. 11, 1

\noindent Schechter, P. 1976 Ap J 203, 297

\noindent Schlegel, D. et al 1998 Ap J 500, 525

\noindent Schneider, P. 2006 in {\it Gravitational Lensing: Strong, Weak \& Micro}, Saas-Fee
Advanced Course 33

\noindent Scott, A. et al 2006 in {\it Ground-based and Airborne Instrumentation for
Astronomy}, S.P.I.E. 6269, 176 

\noindent Shapley, A. et al 2001 Ap J 562, 95

\noindent Shapley, A. et al 2003 Ap J 588, 65

\noindent Shapley, A. et al 2005 Ap J 626, 698

\noindent Shimasaku, K. et al 2005 PASJ  57, 447

\noindent Shioya Y. et al 2005 PASJ 57, 287 

\noindent Smail, I. et al 1995 Ap J 440, 501

\noindent Smail, I. et al 1997 Ap J 490, L5

\noindent Smith, G. et al 2005 Ap J 620, 78

\noindent Smoot, G. et al 1992 Ap J 396, 1

\noindent Somerville, R. \& Primack, J. 1999 MNRAS 310, 1087

\noindent Somerville, R. et al 2001 MNRAS 320, 504

\noindent Somerville, R. et al 2004 Ap J 600, L171

\noindent Songaila, A. 2004 AJ 127, 2598

\noindent Songaila, A. 2005 AJ 130, 1996

\noindent Songaila, A. 2006 AJ 131, 24

\noindent Spergel, D. et al 2003 Ap J Suppl. 148, 175

\noindent Spergel, D. et al 2006 Ap J in press (astro-ph/0603449)

\noindent Springel, V. et al; 2005 Nature 435, 629

\noindent Stanway, E. et al 2004 Ap J 607, 704.

\noindent Stanway, E. et al 2005 MNRAS 359, 1184

\noindent Stark, D. \& Ellis, R.S. 2005 New AR 50, 46

\noindent Stark, D. et at 2006a Ap J in press (astro-ph/0604250)

\noindent Stark, D. et al 2006b Ap J submitted

\noindent Stiavelli, M. 2002 in {\it Future Research Directions \&
Visions for Astronomy}, S.P.I.E. 4835, 122

\noindent Stiavelli, M. et al 2004 Ap J 600, 508 

\noindent Steidel, C. et al 1996 Ap J 492, 428

\noindent Steidel, C. et al 1999a Ap J 519, 1

\noindent Steidel, C. et al 1999b Phil Trans R. Soc. 357, 153

\noindent Steidel, C. et al 2003 Ap J 592, 728

\noindent Struck-Marcell, \& Tinsley, B. 1978 Ap J 221, 562

\noindent Sullivan, I. et al 2006 Ap J in press (astro-ph/0609451)

\noindent Sullivan, M. et al 2000 MNRAS 312, 442

\noindent Sullivan, M. et al 2001 Ap J 558, 72

\noindent Swinbank, M. et al 2006 Ap J submitted

\noindent Taniguchi, Y. et al 2005 PASJ 57, 165

\noindent Tinsley, B. 1980 Fund. Cosmic Phys. 5, 287

\noindent Toomre, A. \& Toomre, J 1972 Ap J 178, 623

\noindent Tran, K.-V. et al 2005 Ap J 627, L25

\noindent Treu, T. et al 2005 Ap J 622, L5

\noindent Treu, T et al 2006 Ap J 640, 662 

\noindent Tully, B. \& Fisher, 1977 Astron. Astrophys. 54, 661

\noindent van der Wel, A. et al 2005 Ap J 631, 145

\noindent van der Wel, A. et al 2006 Ap J 636, L21

\noindent van Dokkum, P.  2005 AJ 130, 2647

\noindent van Dokkum, P. \& Ellis, R.S. 2003 Ap J 592, L53

\noindent van Dokkum, P. et al 2003 Ap J 587, L83

\noindent van Dokkum, P. et al 2006 Ap J 638, L59

\noindent Vogt, N. et al 1996 Ap J 465, L15

\noindent Vogt, N. et al 1997 Ap J 479, L121

\noindent Weiner, B. et al 2005 Ap J 620, 595

\noindent Wright, E. \& Reese, E.D. 2000 Ap J 545, 43 

\noindent Yan, H. \& Windhorst, R. 2004 Ap J 612, 93

\noindent Yan, H. et al 2005 Ap J 634, 109

\noindent Yan, H. et al 2006 Ap J 651, 24

\noindent Yee, H. et al 1996 AJ 111, 1783 

\noindent York, D. et al 2001 AJ 120, 1579

\noindent Zwaan, M.A. et al 2003 AJ 125, 2842

\noindent Zwicky, F. 1933 Helv. Physica Acta, 6, 110

\noindent Zwicky, F. 1937 Phys Rev. 51, 290 




\end{document}